\documentclass[12pt, a4paper]{extbook} 
\usepackage[a4paper,headheight=15pt,left=25mm,top=25mm,bottom=25mm,right=25mm]{geometry}

\usepackage[T1]{fontenc}
\usepackage[utf8]{inputenc}
\usepackage[cotutelle]{psl-cover}

\usepackage{float}
\usepackage{amsmath}
\usepackage{amsfonts}
\usepackage{emptypage} 
\usepackage{lipsum}
\usepackage{listings}
\usepackage{graphicx}
\usepackage{adjustbox}
\usepackage{array} 
\usepackage{booktabs}
\usepackage[resetlabels]{multibib}
\usepackage{longtable}
\usepackage{mfirstuc}
\usepackage{bm}
\usepackage[toc]{appendix}
\usepackage[acronym]{glossaries}
\usepackage{xcolor}
\usepackage{afterpage}
\usepackage{comment}
\usepackage{fancyhdr}
\DeclareMathOperator{\sinc}{sinc}

\newcommand{\mywidth}{0.7\textwidth}
\usepackage[caption=false,font=footnotesize]{subfig}
\usepackage{import}
\usepackage{soul}
\usepackage{dirtree}
\usepackage{algorithm}
\usepackage{algorithmic}
\usepackage{caption}
\usepackage{nameref}
\usepackage{pgfgantt}
\usepackage{CJKutf8}

\title{Nano-émetteurs photoniques couplés à des guides d'ondes pour l'optique non-linéaire et l'optique quantique}
\institute{l'Ecole Normale Supérieure de Paris}
\doctoralschool{Physique en Île-de-France}{564}
\specialty{Optique}

\author{Chengjie Ding}
\date{26 janvier 2021}
\jurymember{1}{Isabelle Philip}{Directeure de recherche à l'Université de Montpellier }{Rapporteure}
\jurymember{2}{Sylvie Lebrun}{Maître de conférences à l'Institut d'Optique}{Rapporteure}
\jurymember{3}{Sylvain Ravets}{Chargé de recherche à Centre de Nanosciences et de Nanotechnologies - Université Paris-Saclay}{Examinateur}
\jurymember{4}{Wenxue Li}{Professeure à East China Normal University}{Examinatrice}
\jurymember{5}{Kun Huang}{Professeure à East China Normal University}{Examinateur}
\jurymember{6}{Sile Nic Chormaic}{Professeure à Okinawa Institute of Science and Technology}{Examinatrice}
\jurymember{7}{Alberto Bramati}{Professeur à Sorbonne Université}{Co-directeur}
\jurymember{8}{Quentin Glorieux}{Maître de conférences à Sorbonne Université}{Co-directeur}

\frabstract{
  Cette thèse couvre différents sujets de l'optique non linéaire et quantique, étudiés dans des systèmes avec des échelles plus petites ou comparables à la longueur d'onde d'intérêt.
   Le manuscrit est divisé en deux parties.
   La première partie du manuscrit traite de plusieurs applications des nanofibres optiques en tant que plateforme d'interaction lumière-matière. En particulier nous explorerons : le couplage entre d'un dipôle posé à la surface de la nanofibre ; le couplage entre deux nanofibres parallèles ; ainsi que les propriétés mécaniques d'une nanofibre optique.
   Dans la deuxième partie du manuscrit, nous nous concentrerons sur la résolution de deux principaux défis lorsque l'on travail avec des émetteurs : l'un est le rendement quantique de photoluminescence, l'autre est l'atténuation lors du transport des photons générés dans réseaux télécom fibrés.
}
\enabstract{
This thesis covers different subjects of nonlinear and quantum optics, studied in systems with scales smaller or comparable to the wavelength of interest.
    The manuscript is divided into two parts.
    The first part of the manuscript deals with several applications of optical nanofibers as a platform for light-matter interaction. In particular we will explore: the coupling between a dipole placed on the surface of the nanofiber; the coupling between two parallel nanofibers; as well as the mechanical properties of an optical nanofiber.
    In the second part of the manuscript, we will focus on solving two main challenges when working with emitters: one is the quantum efficiency of photoluminescence, the other is the attenuation during transport of the photons generated in fiber telecom networks.

}
\frkeywords{Guide d'ondes optique, Interaction lumière-matière, Nanophotonique et Effet plasmonique}
\enkeywords{Optical waveguide, Light-matter interaction, Nanophotonics and Plasmonic effect}

\entitle{Waveguide-coupled nano-emitters for\\ non-linear and quantum optics}
\otherinstitute{East China Normal University}
\logootherinstitute{East_China_Normal_University_logo}

\begin{document}
\maketitle{}
    

\chapter*{Acknowledgments}
The present thesis is accomplished within the cooperation program between l’École Normale Supérieure de Paris (ENS – Paris) in France and East China Normal University (ECNU) in China.
I want to sincerely thank all the professors and colleagues who worked on this project or helped me during my Ph.D.

First of all, my work is inseparable from the patient guidance of my supervisor Prof. Alberto Bramati (Laboratoire Kastler Brossel, UPMC, Paris France), Prof. Quentin Glorieux (Laboratoire Kastler Brossel, UPMC, Paris France) and Prof. E Wu (State Key Laboratory of Precision Spectroscopy, ECNU, Shanghai, China).
With the guidance and support of them, I got the chance to work on interesting research projects during these years 2016-2020 and accomplish my PhD thesis.
I feel lucky to have the opportunity to meet many talented and nice person thanks to this experience.

In 2013, I joined the group of Prof. Wu as a master student in East China Normal University (ECNU), which opens a gate for me towards the world of optics.
During these three years, I gradually expanded my knowledge about laser physics, fluorescence and manipulating optics. 
I learned the techniques about confocal microscope, atomic force microscope, single photon generation and detection.
I would like to thank the colleagues worked with me in Prof. Wu's group, Kun Huang, Yan Liu, Min Song, Gengxu Chen, Yingxian Xue, Youying Rong and the other students of Prof. Wu during these years, Ruikai Tang, Xueting Ci, Qian Zhou, Wenjie Wu, Jianhui Ma, Huiqin Hu, Yu Chen, Chengda Pan, Guangjian Xu, Qiang Ma, Zhiping Ju, Linxiao Chen, Shikang Liu, Weijie Cai, Mengyao Qin and the some new numbers of our group.

In 2016, I came to the Laboratoire Kastler Brossel in Paris and started to work in the group of Prof. Bramati thanks to the cooperation program between l’École Normale Supérieure de Paris (ENS – Paris) in France and East China Normal University (ECNU) in China.
There are quite a lot of wonderful memories during my PhD and I'm glad to be one of the group members.
During my PhD, the professors and the other colleagues in our group helped me both in science and daily life.
I would like to thank Prof. Bramati.
As my supervisor in France, he offered me the chance to come to France for studying as a PhD student and offered me interesting research projects.
Besides, whenever I have problem and ask him for help, he always responds in the first minutes and try his best to solve the problem, which I'm really appreciate.
I would like to thank Prof. Glorieux, who gave me a lot of suggestion and guidance in the experiment, paper writing and my thesis.
I also want to thank Prof. Wu, who is my supervior in ECNU.
Although I spent most of my PhD in France, she is always following my experiment progress and my life and offering suggestions and help no matter where I am.

I would like to thank the doctor school for organizing a lot of courses and Prof. Glorieux for offering me the chance to attend the summer school in Les Houches organized by him.
I also want to thank all the professors who gave interesting lectures and explained the physics clearly during the lectures.

I must also appreciate my research partners Doctor Maxime Joos, who I worked with on several projects and helped me a lot especially at the beginning of my PhD and explained me a lot of details about the experiment in terms of theoretical and technical aspects, Post-doc Simon Pigeon, Tom Bienaime and Rajiv Boddeda, with whom I discussed the theoretical problem I met during the data analysis, and Post-doc Vivien Loo and Thomas Boulier, from whom I learned experiment techniques and ideas.

I also want to record the colleagues in our group, Stefano Pierini, Quentin Fontaine, Murad Abuzarli, Anne Maitre, Ferdinand Claude, Marianna Damato and Wei Liu.
All the nice persons around me are the reasons that I feel warm and accompanied in a country ten thousands kilometer away from my hometown.

I would like to thank the professors in LKB that offered help or answered my questions patiently, Pierre Claude, Tristan Briant, François Nez, and Michel Brune.

I would like to thank the people working in the electronic workshop and the mechanical workshop, . 
I'm also appreciate the people working on administration, Laetitia Morel, Romain Rybka,
Thierry Tardieu and Nora Aissous. 
They offered a lot of help during my stay in LKB.
Also the ones who work in ENS, Auguste Filippi, Stéphane Emery, in Sorbonne, Delphine Dieudonne and in doctor school EDPIF, Jean-Francois Allemand, Laura Baron-Ledez. 
Besides, I want to thank Xiaoling Liu and Xiaoyan Liu in ECNU, who is working on the administrative stuff of the cooperation program between ENS and ECNU.

I need to thank the Chinese Scholarship Council who offered four years of scholarship during my PhD and cares about our study and life in France.

I need to thank Professor Nicolas Joly and Doctor Jonas Hammer who showed me the experiment and explain me the theory and technique during my visit in Max Planck Institute for the Science of Light in Germany.

I must thank all the jury of my thesis, Isabelle Philip, Sylvie Lebrun, Sylvain Ravets, Sile Nic Chormaic, Wenxue Li and Kung Huang.

In the end, I would like to thank my family members, the ones who loves me and I love. 
They are always supporting me, giving me courage and willing to offer help.

\vfill
\begin{center}
	Chengjie Ding \\
	Paris, France \\
	30.11.2020
\end{center}
\chapter*{Abstract}
This thesis covers different subjects of nonlinear and quantum optics, studied in systems with scales smaller or comparable to the wavelength of interest.
The manuscript is divided into two parts.
The first part of the manuscript deals with several applications of optical nanofibers as a platform for light-matter interaction. In particular we will explore: the coupling between a dipole placed on the surface of the nanofiber; the coupling between two parallel nanofibers; as well as the mechanical properties of an optical nanofiber.

The research on the coupling of linear dipoles on the surface of nanofibers aims to study the polarization characteristics of the dipoles coupled to the guiding mode of the fiber.
Due to the chirality of the guiding mode of the nanofiber, we can detect the polarization properties of the dipole by monitoring the polarization at the fiber output. Full polarization including linear polarization, circular polarization and elliptical polarization are available through the coupling of dipole and optical nanofiber.
Complete polarization control can be achieved by controlling the azimuth angle of the dipole and the orientation of the gold nanorod.

The research on the coupling between two parallel optical nanofibers aims to build models to describe the photonic structures of nanofibers, including two different rings.
This provides a reference for the construction of integrated optical devices (Sagnac interferometer and Fabry-Perot resonator) based on nanofibers.

Finally, we combined the optical and mechanical properties of the nanofiber to achieve a displacement detection system with an accuracy of 1.2~nm/$\sqrt{\text{Hz}}$.

In the second part of the paper, we will focus on solving the problem of the limited photoluminescence quantum efficiency faced by single-photon emitters in the attenuation of photon transmission in optical fiber telecommunication networks.
We use frequency conversion to convert the visible photons generated by the single-photon emitters to the low-loss telecommunication band photons transmitted in the optical fiber.
Based on single-photon sources such as diamond color centers and CdSe-CdS quantum dots, we propose numerical simulations and possible experimental solutions for quantum frequency conversion. 
And we used laser to simulate the zero phonon line of the diamond nitrogen vacancy color center and realized efficient frequency down conversion.

\chapter*{R\'esum\'e}

Cette thèse couvre différents sujets de l'optique non linéaire et quantique, étudiés dans des systèmes avec des échelles plus petites ou comparables à la longueur d'onde d'intérêt.
Le manuscrit est divisé en deux parties.
La première partie du manuscrit traite de plusieurs applications des nanofibres optiques comme plate-forme d'interaction lumière-matière. En particulier, nous explorerons: le couplage entre un dipôle placé à la surface de la nanofibre; le couplage entre deux nanofibres parallèles; ainsi que les propriétés mécaniques d'une nanofibre optique.

La recherche sur le couplage de dipôles linéaires à la surface de nanofibres vise à étudier les caractéristiques de polarisation des dipôles couplés au mode de guidage de la fibre.
En raison de la chiralité du mode de guidage de la nanofibre, nous pouvons détecter les propriétés de polarisation du dipôle en surveillant la polarisation à la sortie de la fibre. La polarisation complète, y compris la polarisation linéaire, la polarisation circulaire et la polarisation elliptique sont disponibles grâce au couplage du dipôle et de la nanofibre optique.
Un contrôle de polarisation complet peut être obtenu en contrôlant l'angle azimutal du dipôle et l'orientation de la nanorode d'or.

La recherche sur le couplage entre deux nanofibres optiques parallèles vise à construire des modèles pour décrire les structures photoniques des nanofibres, comprenant deux anneaux différents.
Cela constitue une référence pour la construction de dispositifs optiques intégrés (interféromètre Sagnac et résonateur Fabry-Pérot) à base de nanofibres.

Enfin, nous avons combiné les propriétés optiques et mécaniques de la nanofibre pour obtenir un système de détection de déplacement avec une précision de 1.2~nm/$\sqrt{\text{Hz}}$.

Dans la deuxième partie de l'article, nous nous concentrerons sur la résolution du problème de l'efficacité quantique de photoluminescence limitée rencontrée par les émetteurs à photon unique dans l'atténuation de la transmission de photons dans les réseaux de télécommunication à fibre optique.
Nous utilisons la conversion de fréquence pour convertir les photons visibles générés par les émetteurs à photon unique en photons de bande de télécommunication à faible perte transmis dans la fibre optique.
Sur la base de sources à photon unique telles que les centres de couleur du diamant et les points quantiques CdSe-CdS, nous proposons des simulations numériques et des solutions expérimentales possibles pour la conversion de fréquence quantique.
Et nous avons utilisé le laser pour simuler la ligne zéro phonon du centre de couleur de vide d'azote diamant et réalisé une conversion efficace vers le bas de fréquence.

\begin{CJK*}{UTF8}{gbsn}   
   \chapter*{摘要}
本论文在标度小于或等于目标波长的波导系统中对非线性光学和量子光学进行了研究。
当光纤的直径减小到亚波长区域时，被引导的光的边界条件发生改变。
一部分光在光纤表面的消逝场中传播。
在消逝场中传播的光被高度限制在光纤表面附近的横向平面中。
纳米光纤周围的消逝场产生更大的光学深度。
因此，利用纳米光纤，可以实现将光与原子，纳米颗粒或者共振腔在光纤表面耦合，从而为研究光与物质相互作用提供了平台，有非常广泛的理论研究和应用的空间。

论文分为两部分。
论文的第一部分介绍了纳米光纤作为光和物质相互作用平台的几种应用。 我们研究了置于纳米光纤表面的偶极子之间的耦合； 两根平行的纳米光纤之间的耦合； 以及纳米光纤的机械性能。

其中纳米光纤表面的偶极子之间的耦合的研究旨在研究偶极子发光耦合至光纤的传导模式后的偏振特性。
由于纳米光纤传导模式中的手性特性，我们通过激发与纳米光纤表面耦合的金纳米棒可以在光纤出射端探测到偶极子的偏振性能，并能通过偶极子与光纤耦合获得包括线偏振，圆偏振以及椭圆偏振。
通过对偶极子方位角和相对光纤轴的夹角的控制可以实现完全的偏振控制。

对于两根平行的纳米光纤之间的耦合的研究旨在构建模型以描述纳米光纤构成的光子结构，包括两种不同的扭结。
这为基于纳米光纤构建集成化的光学器件(Sagnac干涉仪以及Fabry-Perot共振腔)提供了参考。

最后，我们将纳米光纤的光学特性与机械特性结合，实现了精度为 1.2 nm$/\sqrt{\text{Hz}}$的位移探测系统。
该光学系统拥有足够的灵敏度，可将纳米光纤作为探针用来测量辐射力。

在论文的第二部分中，我们将着重解决单光子发射器面临的有限的光致发光的量子效率在光纤电信网络中产生的光子传输过程中的衰减的问题。

我们通过频率转换，将单光子发射器产生的可见光子转换至在光纤中传输具有低损耗的通信波段光子从而实现减少传输损耗的目的。
我们基于金刚石色心以及CdSe-CdS量子点等单光子源，提出了用于量子频率转换的数值模拟和可能的实验解决方案，并利用激光器模拟金刚石氮空穴色心的零声子线实现了高效的频率下转换。

其中，在利用PPLN波导之外，我们介绍了一种利用悬浮纤芯光纤中进行四波混频的单光子频率转换实验方案。
相位匹配可以通过控制悬浮芯光纤的孔中注入的稀有气体的压力来调节。这个方法可以实现可调谐的单光子频率转换。
\end{CJK*}
\chapter*{Publications}
\label{chap:publications}

The work presented in this thesis has given rise to the following publications:

\begin{description}
	\item \text{[1]} Joos, M.$^*$, Ding, C.$^*$, Loo, V., Blanquer, G., Giacobino, E., Bramati, A., Krachmalnicoff, V. and Glorieux, Q. (2018). Polarization control of linear dipole radiation using an optical nanofiber. Physical Review Applied, 9(6), 064035.
	\item \text{[2]} Ding, C., Loo, V., Pigeon, S., Gautier, R., Joos, M., Wu, E., Giacobino, E., Bramati, A. and Glorieux, Q. (2019). Fabrication and characterization of optical nanofiber interferometer and resonator for the visible range. New Journal of Physics, 21(7), 073060. 
	
	\item \text{[3]} Ding, C., Joos, M., Bach, C., Bienaimé, T., Giacobino, E., Wu, E., Bramati, A. and Glorieux, Q. (2020). Nanofiber based displacement sensor. Applied Physics B, 126(6).
	
	\item \text{[4]} Xue, Y., Ding, C., Rong, Y., Ma, Q., Pan, C., Wu, E., Wu, B. and Zeng, H. (2017). Tuning plasmonic enhancement of single nanocrystal upconversion luminescence by varying gold nanorod diameter. Small, 13(36), 1701155.
	
	\item \text{[5]} Chen, X., Ding, C., Pan, H., Huang, K., Laurat, J., Wu, G., and Wu, E. (2017). Temporal and spatial multiplexed infrared single-photon counter based on high-speed avalanche photodiode. Scientific Reports, 7, 44600.
	
	\item \text{[6]} Ding, C. J., Rong, Y. Y., Chen, Y., Chen, X. L., and Wu, E. (2019). Direct Measurement of Non-Classical Photon Statistics with a Multi-Pixel Photon Counter. Journal of Electronic Science and Technology, 17(3), 204-212.
	
	\item \text{[7]} Chen, Y., Balasubramanian, P., Cai, Y., Rong, Y., Liu, Y., Jelezko, F., Ding, C., Chen, X. and Wu, E. (2020). Quantum Calibration of Multi-pixel Photon Counter and Its Application in High-sensitivity Magnetometry with NV Center Ensemble. IEEE Journal of Selected Topics in Quantum Electronics, 26(3), 1-7.
	
\end{description}


\tableofcontents




\chapter*{Introduction}
\markboth{Introduction}{}
\addcontentsline{toc}{chapter}{Introduction} 
This thesis focuses on the research and application of optical waveguides and nano emitters in nonlinear and quantum optics.

The study of optical waveguides has been rising quickly in recent years. 
Their potential as a platform for quantum optics and nonlinear media is clear.
Among the different types of optical waveguides, we study mostly the optical nanofiber.

The high transmission of the optical fibers are realized by strict control of the surface roughness.
The high transmission glass fibers is observed by Kao and Hockham in 1966 \cite{kao1966dielectric}, which is always a competitive point for the fused silica optical fiber. 
This advantage has been inherited for optical nanofiber.

\section*{Fabrication technique of optical nanofibers}

In 2003, Limin Tong published his work on fabrication of silica wires with diameter only 50 nm by two-step drawing while keeping low loss \cite{tong2003subwavelength}.
The first step is to reduce the fiber diameter to few micrometers with a flame.
In the second step, the silica fiber is coiled around a sapphire tip and heated to melting temperature.
The fiber is then pulled in a direction perpendicular to the axis of the sapphire tip with diameter controlled by the pulling speed.

Later, another way of fabricating optical nanofiber using heat-and-pull method introduced by Warken \cite{warken2007ultradunne} in 2007 and optimized by Hoffman in 2014 \cite{hoffman2014ultrahigh}, which stretches a softened fused silica optical fiber to subwavelength region with only one step.
The technique is now quite mature.
With a profile offering adiabatic transition, an optical nanofiber can have transmission above 95\%.
The fiber waist has only a bare silica core with a diameter typically smaller than guiding wavelength and the surrounding air acts as the fiber cladding.
While the two sides of the optical nanofiber keep the structure of original optical fiber.
With this technique, the fiber waist length can be from several millimeter to one centimeter and a waist radius down to 100 nm.
Since the side of the optical fiber remains unmodified, it can be directly connected to other fiber system.

Electrospinning is normally used to produce polymer nanofibers, but it can also be used on silica fiber fabrication \cite{praeger2012fabrication}.
By controlling the movement of collector stage, we can get aligned nanofibers.

Laser spinning is a method that can generate glass nanofibers with lengths up to several centimeters \cite{quintero2009laser}. It is realized by melting glass with a focused high power laser. The melted material is blew by a gas jet with high velocity.

Another new technique based on laser melting can produce continuous glass fibers of 1 m long with diameter down to 300 nm \cite{quintero2020continuous}.

Among all the techniques for glass nanofiber fabrication, one step heat-and-pull method is the most attractive due to its simplicity and high transmission.
Especially for optical applications, it can be adapted to other fiber system.
This is also the technique we are using in our lab to fabricate optical nanofibers.

\section*{Application of optical nanofibers}

When the diameter of optical fibers is reduced to sub-wavelength region, the boundary condition of the guided light changes. 
Part of the light propagates in the evanescent field at the fiber surface. 
The light propagating in the evanescent field is highly confined in the transverse plane near the fiber surface.
The evanescent field around the optical nanofiber produces larger optical depth.
Therefore, with optical nanofiber platform, it is possible to couple light with atoms \cite{nayak2007optical,vetsch2010optical, nayak2008single,corzo2019waveguide, nieddu2016optical}, other small particles \cite{schroder2012nanodiamond, liebermeister2014tapered, yalla2012efficient,fujiwara2011highly} to study the interaction between light and matters.

\begin{figure}
    \centering
    \includegraphics[width=0.8\linewidth]{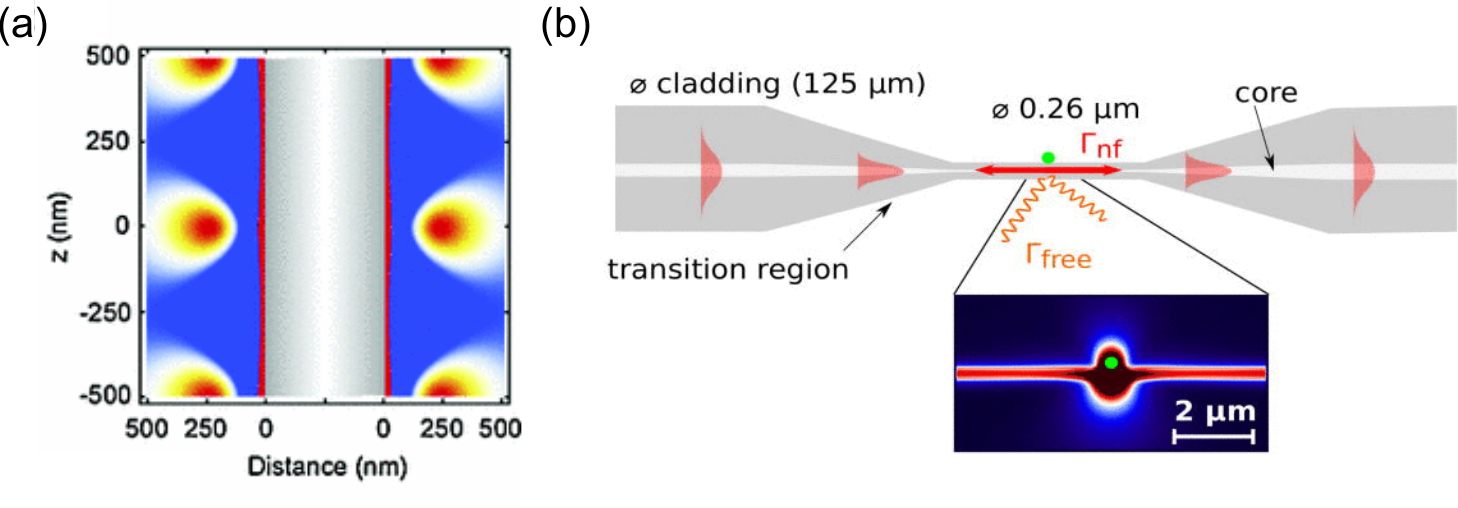}
    \caption{(a) Cesium atoms are localized about 200 nm away from the nanofiber surface trapped by the evanescent field adapted from \cite{vetsch2010optical}. (b) Tapered optical nanofiber couplied with single nitrogen vacancy center adapted from \cite{liebermeister2014tapered}. }
    \label{fig:atom and NV on TOF}
\end{figure}

Once we have atoms or nanoparticles close enough to the surface of optical nanofiber, the emission or the scattered photons can be coupled into the evanescent field of the guided mode.
We can use following methods to detect the light-matter interaction.
First, we can send laser through the optical nanofiber, and monitor the transmitted signal to get the information of the photons absorbed by atom or nanoparticles.
Second, we can send laser through the optical nanofiber and detect the emission or scattered photons from the atom or nanoparticles in the free space around the optical nanofiber.
Third, we can send a laser beam on the atoms or nanoparticles from the free space and detect the emission or scattered photons coupled into the guided mode of optical nanofiber.

Since the optical nanofiber can be made in the lab with a simple heat-and-pull system and can be connected to other quantum information devices with fiber connection to generate hybrid quantum systems that can be used for quantum processing or quantum memory \cite{corzo2019waveguide}.

Optical nanofibers can also be used to couple light into resonators \cite{pollinger2009ultrahigh} by placing the fiber surface next to a cavity.

Optical nanofibers are also used in sensing and detection systems \cite{russell2011sub, zhang2011micro,han2014side}.

In consequence a strong evanescent field in the air is present along the fiber surface.
The possibility of coupling to nano emitters on the fiber surface with coupling efficiency up to 20\% opens gates to many applications in quantum optics.

\section*{Plan of the thesis:}

Chapter 1 is the introduction about the fundamental background knowledge of optical nanofibers.

The first part of the manuscript, including Chapter 2-4, is about several applications of the optical nanofiber considering the coupling between the dipole on the nanofiber surface, the coupling between two parallel nanofibers and the mechanical property of optical nanofiber.

In these experiments, we used gold nanorods or gold nanospheres as nano emitters which exhibit dipolar emission pattern.
When point-like metallic nanoparticles are deposited on the optical nanofiber, the spin and orbital angular momentum got coupled. 
The guided light shows polarization corresponding to the dipole orientation and azimuthal position.

In Chapter 2, we observed the emission of various elliptical polarizations, including circularly polarized light by coupling a linear dipole with a tapered nanofiber without the need for birefringent components.
We experimentally analyzed the light directly collected in the guided mode of the nanofiber in regard to the azimuthal position and orientation of the dipole (gold nanorod), observed by means of scanning electron microscopy. 
We demonstrated a mapping between purely geometrical degrees of freedom of a light source and all the polarization states. 
This could open the way to alternative methods for polarization control of light sources at the nanoscale.
The result of this experiment has been the subject of the publication \cite{joos2018polarization}.

In Chapter 3, we fabricated and characterized two types of photonic structures based on tapered nanofiber, which are a ring and a knot.
In the framework of the linear coupling theory, we studied the coupling between two parallel nanofibers.
We proposed a general approach to predict the properties of these structures.
We found that although both nanofiber systems contain ring structures, they appear to operate in different ways due to the difference in geometry, one refers to a Sagnac interferometer and another refers to a Fabry-Perot resonator.
In addition, we described a new source of birefringence due to the ovalization of a nanofiber under strong bending, known in mechanical engineering as the Brazier effect.
The Fabry-Perot Cavity made of optical nanofiber can couple to atoms or luminescence nanoparticles with evanescent wave at the surface of the fiber. 
The system is both integrated and has a relatively small mode volume with a finesse of 8 and a quality factor of 1300.
The result of this experiment has been the subject of the publication \cite{ding2019fabrication}.

In Chapter 4, we take advantage of the optical nanofiber's small mass to implement a highly sensitive sensor of vibrations and weak forces.
We combine the optical and mechanical properties of a nanofiber to create a sensing device for displacement measurements and optomechanical applications.
We coupled a gold nanosphere with optical nanofiber and placed it within a optical standing wave to form a position sensing system with a resolution down to 1.2~nm/$\sqrt{\text{Hz}}$.
To calibrate the sensitivity of this displacement system towards radiation pressure measurement, we proposed a mechanical model to estimate the response of our sensor to optically-induced force.
We found that the sensitivity corresponding to the sensibility of an externally applied force at the nanofiber waist is 1~pN.
The result of this experiment has been the subject of the publication \cite{ding2020nanofiber}.

The second part of the manuscript includes Chapter 5-7, in which we describe quantum and nonlinear optics with single photon nano emitters.
In this part, we focus on solving two main challenges of emitters, one is the photoluminescence quantum yield, another is the attenuation during propagation of the generated photons.
The nano emitters we have studied include quantum emitters like defects in nanodiamond which are promising single photon sources at room temperature, lanthanide-doped NaYF$_4$ nanoparticles, metallic nanoparticles, and quantum dots (CdSe-CdS dot-in-rods).

In Chapter 5, we introduced the enhancement of luminescence from single lanthanide-doped upconversion nanoparticle with surface plasmon resonance.
With the help of atomic force microscope manipulation, we assemble a gold nanorod with a Yb$^{3+}$/Er$^{3+}$/Mn$^{2+}$ co-doped NaYF$_4$ nanocrystal into a hybrid dimer.
By comparing the upconversion transmission from nanocrystals with and without the gold nanorod, we revealed the plasmonic enhancement up to 110 times with gold nanorod.
The result of this experiment has been the subject of the publication \cite{xue2017tuning}.

For the typical fused silica glass fibers we use in this work, it is especially advantageous for long-distance communications with infrared light which has loss generally below 1 dB/km and a minimum loss at 1550 nm (0.2 dB/km). It is a much lower attenuation compared to electrical cables. 
This allows to span long distances with few repeaters.
Here comes the problem that the communication between quantum nodes is suffering from the mismatch between the frequency of the emitted photons and the low-loss telecommunication-band region of silica, for which the quantum frequency conversion can be a solution.
The wavelength of the generated photons in the visible range can be converted to communication-friendly telecommunication band while preserving the other optical and quantum properties.

In Chapter 6 and 7, we introduced the numerical calculation and possible experimental solutions based on different quantum emitters for quantum frequency conversion.

In Chapter 6, we numerically studied the difference frequency generation with defects in diamond, specifically Nitrogen vacancy and Silicon vacancy color centers in nanodiamond.
We calculated the phase matching condition of these two types of defects in diamond while using two different nonlinear medias, the periodically poled lithium niobate (PPLN) and periodically poled potassium titanyl phosphate (PPKTP).
Since the light refraction in the two crystals is different (lithium niobate is uniaxial crystal and potassium titanyl phosphate is biaxial crystal), we want to understand how different nonlinear crystals affect the phase matching condition and whether we find a numerical solution to increase the conversion efficiency and bandwidth compression.

Finally in Chapter 7, we introduced a solution for single photon frequency conversion with four wave mixing in suspended core fiber.
The suspended core fiber has a cross section with a sub-micro core surrounded by three holes.
The interesting point of using the suspended core fiber as the non-linear waveguide is that the surrounding holes can be filled with gas to induce non-linearity.
The type and the pressure of gas modify the nonlinear refractive index of the fiber, therefore providing on demand phase matching.

In general, this manuscript covers different subjects of nonlinear and quantum optics, studied in systems with scales smaller or comparable to the wavelength of interest.
Perspectives and experiments proposals conclude this manuscript.

\chapter{Subwavelength optical nanofiber}
\label{chap:introduction}
Waveguide is a physical structure that guides electromagnetic waves.
The first study of dielectric waveguide appeared in 1910 by Hondros and Debye who described the mode propagation within a dielectric rod.
In the late 1930's, people found that the evanescent field of the wave propagates outside the rod and provides highly directional radiation at the bend, which leads to the development in microwave antennas.
Since late 1950's, the light guiding properties of dielectric waveguides for optical wavelength was gradually being studied.
In 2009, a Nobel Prize was awarded for the groundbreaking achievements by Charles K. Kao concerning the transmission of light in glass fiber waveguides for optical communication.
The development of optical waveguides brings the possibility for "integrated optics" and long distance communication with high transmission.

The optical waveguide has a dielectric core with higher refractive index than the outside cladding.
In the higher index region of a waveguide, light propagates with total internal reflection.
In the lower index region, the fields decays exponentially, which is defined as the evanescent wave.
Thanks to the properties of optical waveguides especially the high transmission and evanescent field, optical waveguides shows high potential as a platform to investigate light-matter interactions and quantum optics.
Quantum optics deals with quantum states of light, as opposed to Poissonian distributed states (lasers) or super-Poissonian distributed states (thermal light). 
Single photons Fock states, NOON states or squeezed states, for instance, are quantum states of light.

This manuscript is divided into two parts, the first part will be about the application of the optical nanofibers, the second part will be about the nano emitters.

In this Chapter, we will introduce the optical nanofiber.

\begin{figure}
    \centering
    \includegraphics[width=0.6\linewidth]{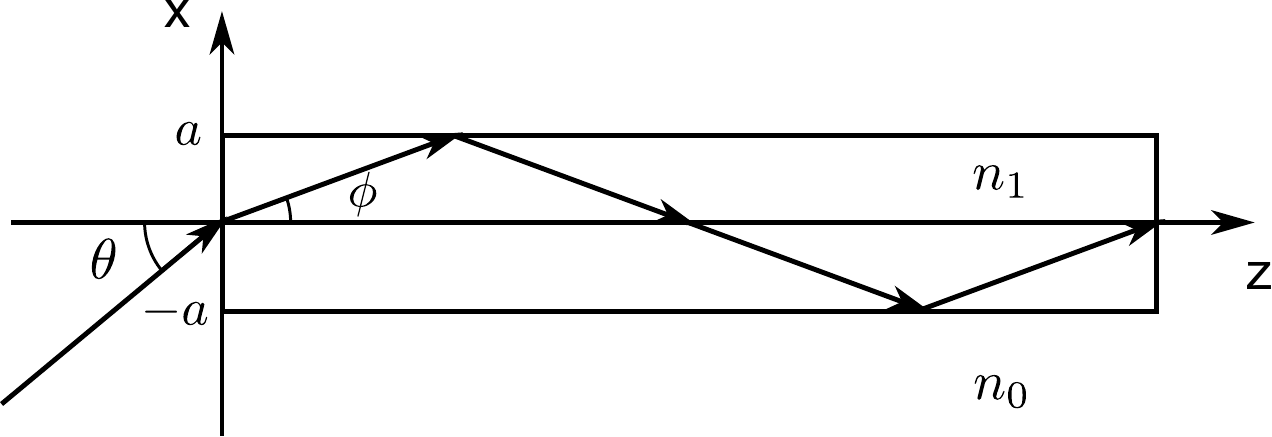}
    \caption{Basic structure of an optical waveguide. $a$ is the fiber radius. $\theta$ and $\phi$ represent the incident angle and the angle of the light propagating in the waveguide respectively. The refractive index of the core and cladding is marked with  $n_1$ and $n_0$.}
    \label{fig:basic structure of optical waveguide}
\end{figure}

The optical waveguides have a structure that consist of a dielectric core and a substrate around it.
The light coupled to the end surface of the waveguide is confined by total internal reflection since the refractive index of the cladding $n_0$ is lower than the refractive index of the core $n_1$.
The refractive index mismatch between core and cladding is defined as:
\begin{equation}\Delta=\frac{n_{1}^{2}-n_{0}^{2}}{2 n_{1}^{2}}.\end{equation}
This important figure of merit tells apart the weakly and strongly guiding regimes.
Weakly guiding fibers are the optical fibers with small refractive index mismatch (substantially less than 1\%).
Otherwise, it will be regarded as a strongly guiding fiber.

At the incident plane, we have the relationship between the incident angle $\theta$ and the light propagating angle in the waveguide $\phi$, shown in Fig. \ref{fig:basic structure of optical waveguide}, described as $\sin \theta=n_{1} \sin \phi$.
The maximum angle for the propagating light within the core is given by $\phi_{\max } = \sqrt{2 \Delta} $, and the maximum incident angle or numerical aperture is given by $\theta_{\max } = \sin ^{-1} \sqrt{n_{1}^{2}-n_{0}^{2}} $.

The wavelength and the wavenumber of light in the core are given by $\lambda / n_{1}$ and $k n_{1}$ with $k=2 \pi / \lambda$.
The propagation constant $\beta$ along $z$ is expressed by:
\begin{equation}\beta=k n_{1} \cos \phi .\end{equation}

In this manuscript, our experiments are related to three different types of waveguides, the optical nanofiber, the suspended core fiber, and periodically-poled lithium niobate (PPLN) waveguide.
In this chapter, we are going to introduce an interesting optical waveguide, the subwavelength optical nanofiber.
The optical nanofiber with subwavelength diameter can be used as a platform for coupling nanoparticles or another optical nanofiber with the strong evanescent field at the air-cladding fiber surface.
Several subsections will describe how these waveguides work and how we design and make them. Another subsection will be dedicated to the coupling between these waveguides and different nanoparticles, as this will be important for chapters 2 and 4.

The suspended core fiber has a core equivalent to an optical nanofiber with diameter down to 800 nm and a length up to kilometer.
It has both potential in particle coupling and the application in third-order nonlinear effect of silica.

The PPLN waveguides is interesting because the confinement of light within the waveguide offers higher conversion efficiency compare to the traditional PPLN crystal.

Optical nanofiber is an air-cladding fused silica waveguide with circular cross section.
A homemade optical nanofiber can have a diameter of about 200 nm with high transmission, typically above 95\%.
A bare optical nanofiber with low surface roughness can have very high damage threshold.
In the work of Hoffman, a transmission power of more than 400 mW in high vacuum conditions was recorded \cite{hoffman2014ultrahigh}.
When the optical nanofiber has a subwavelength diameter, the guided modes can extend outside the fiber, in the air; the field is then both propagating and evanescent.
With further decreasing of the fiber diameter, the enhanced evanescent field leads to the connectivity to other system for research on light-matter interaction and precise sensing systems.
By monitoring the transmitted signal at the fiber output, one can realize several systems applied for quantum technology by coupling to atom cloud, molecules and particles. 
For example, the emission fluorescence detection, absorption detection and laser-induced fluorescence detection introduced in \cite{morrissey2013spectroscopy}.

\section{Light propagating in an optical nanofiber}
\label{Light propagating in an optical nanofiber}
\begin{figure}
    \centering
    \includegraphics[width=\mywidth]{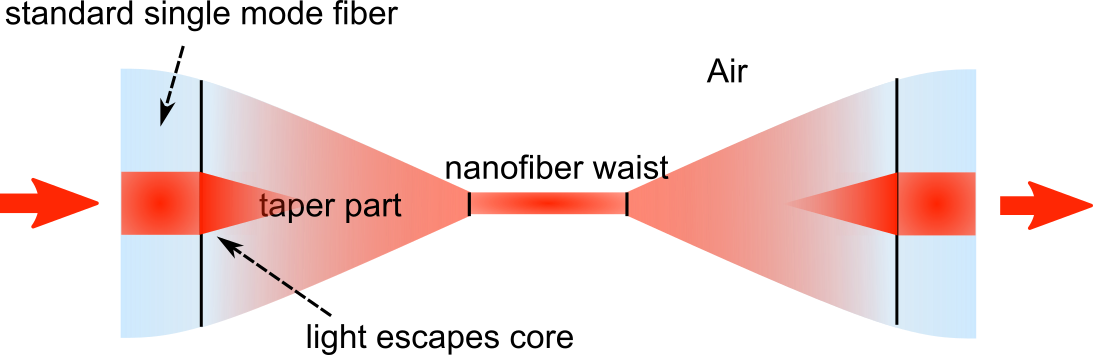}
    \caption{Propagation of light through a tapered nanofiber.}
    \label{fig:light_propagation}
\end{figure}

The propagation of light within a tapered subwavelength optical nanofiber consists of three distinct propagation regions (see in Fig.\ref{fig:light_propagation}): 
\begin{itemize}
    \item the fiber input and output parts with standard optical fiber in weakly-guiding region, where $\Delta$ is small, for example only 0.006 for a Thorlabs HP630 fiber;
    \item the nanofiber waist with air cladding in strongly-guiding region, where in the case for an air-cladding optical nanofiber $\Delta$ is about 0.26;
    \item the tapered region connecting both.
\end{itemize}

Initially the light was launched into optical fiber core from the left. The taper part adiabatically converts the light from unmodified fiber core to the nanofiber waist. 
The shape of the taper part is defined with angle $\Omega$ varying along the fiber longitudinal axis.
For the light propagating in the unmodified fiber and waist region, the two-layer system has only two refractive indexes, $n_\text{core}$ and $n_\text{cladding}$ or $n_\text{cladding}$ and $n_\text{air}$.
Within the tapered region, the effect of finite cladding radius becomes strong and the calculations needs to consider the third medium, normally $n_\text{core}$, $n_\text{cladding}$ and $n_\text{air}$ \cite{hoffman2015rayleigh}.

The transmission rate of the whole system is mostly defined by the taper design, as we will see in subsubsection \ref{Adiabatic profile}.

HE and EH refer to the hybrid modes of optical waveguides with circular symmetry.
In EH modes the axial magnetic field $H_z$ is
relatively strong, whereas in HE modes the axial electric filed $E_z$ is relatively strong.
To characterize which modes are guided by an nanofiber, we introduce the normalized frequency $V$. It depends on the radius $a$, the wavelength of interest lambda and the refractive indexes of the core and the cladding, given by:
\begin{equation}
    V=\frac{2\pi}{\lambda}a\sqrt{n_\text{core}^2-n_\text{cladding}^2}.
\end{equation}

\begin{figure}
    \centering
    \includegraphics[width=0.7\linewidth]{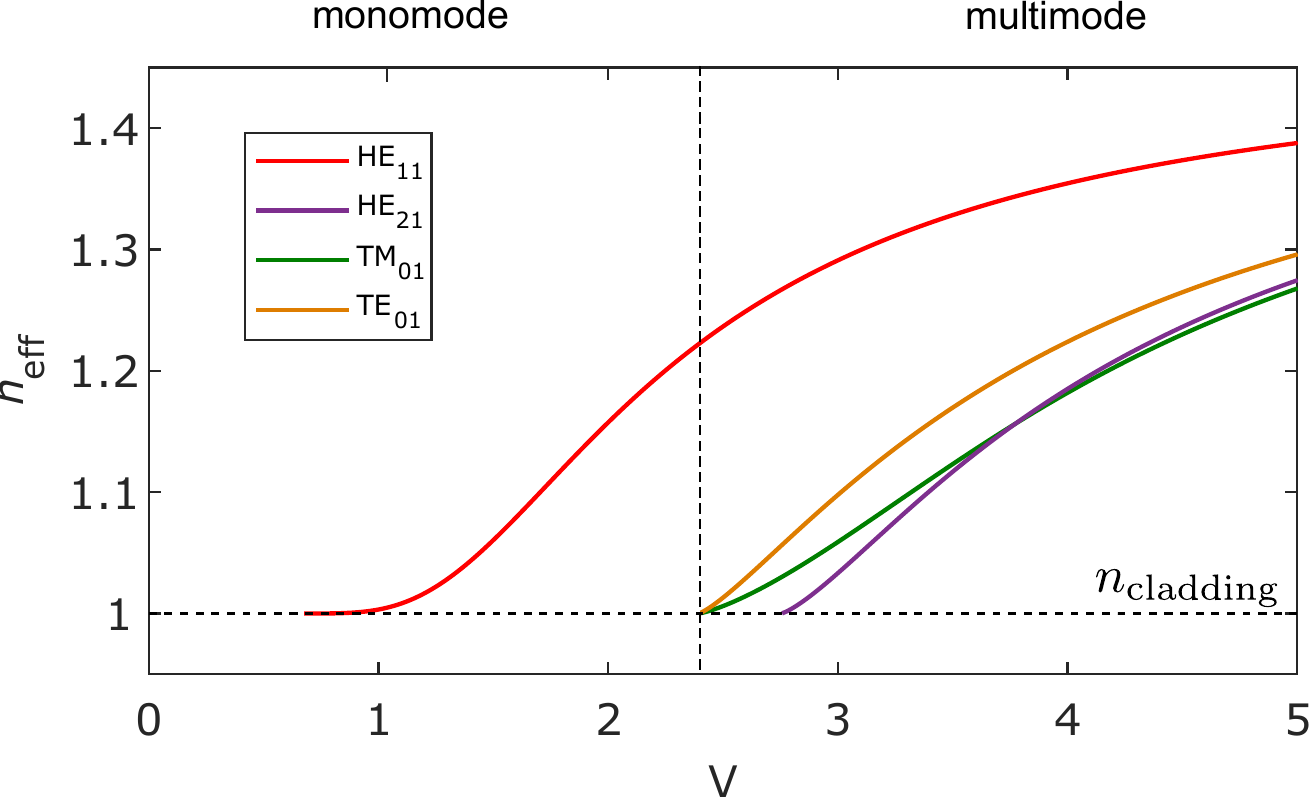}
    \caption{The effective refractive index $n_\text{eff}=\beta / k_0$, where $k_0=2 \pi / \lambda$ as a function of $V$. The boundary of single mode and multi mode is marked with a vertical dashed line. With $\lambda=637$ nm, we have the value of $V$ at the boundary is 2.4. For $V < V_\text{boundary}$, the optical nanofiber propagates only HE$_\text{11}$ mode.}
    \label{fig:neff}
\end{figure}

Fig.\ref{fig:neff} shows the effective refractive index $n_\text{eff}=\frac{\beta}{k}$ as a function of the normalized frequency $V$ for various modes.
The value of $n_\text{eff}$ is between $n_\text{core} =1.45$ (fused silica) and $n_\text{cladding} =1$ (air).
The boundary of single mode and multi mode is marked with a vertical dashed line in Fig.\ref{fig:neff}.
With $\lambda=637$ nm, the value of $V$ at the boundary is 2.4, defined as $V_\text{boundary}$.
Here we show only the lower order modes, HE$_\text{11}$, HE$_\text{21}$, TM$_\text{01}$, TE$_\text{01}$.
The lowest-order mode in an optical waveguide is HE$_\text{11}$, which is a mode without cutoff and is defined as the fundamental mode of a optical fiber.
When $V$ of the nanofiber system is sufficiently small ($V < V_\text{boundary}$), the fundamental mode HE$_\text{11}$ will be the only mode guided in the nanofiber.
The mode structure of HE$_\text{11}$ will be detailed in the next section.

Air-cladding nanofibers show different guiding properties compared to a weakly guiding fiber.
Due to the large refractive index mismatch, the HE$_\text{11}$ can not be approximated to linearly polarized (LP) mode. 
This leads to the interesting properties of the subwavelength air-cladding optical fibers.

\begin{figure}
    \centering
    \includegraphics[width=0.7\linewidth]{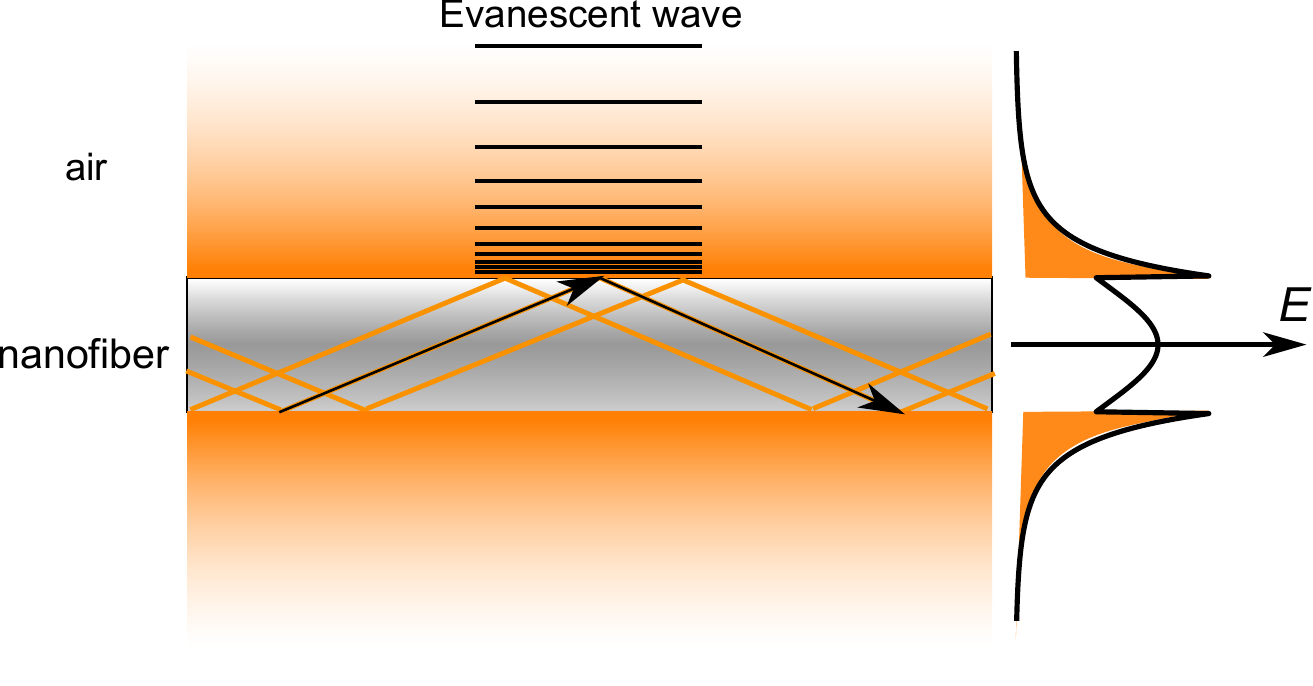}
    \caption{Schematic of the evanescent wave around a subwavelength optical nanofiber.}
    \label{fig:evanescent field}
\end{figure}

Nanofibers show lossless propagating evanescent fields\footnote{Gold films can also be used as waveguides by using surface plasmons, which also exhibit evanescent fields, but the losses in the metal forbid any long distance propagation.}
When the light is propagating inside a higher refractive index medium and is reflected at the surface, the reflected wave interferes with the incident wave to form a standing wave near the interface, and a very small part of the energy will penetrate into the lower refractive index medium. 
The electromagnetic field will penetrate a certain distance and propagate along the interface, as shown in Fig.\ref{fig:evanescent field}. 
This is the evanescent wave.
The  evanescent  field  decays  exponentially from the  surface.
In air-cladding optical nanofibers, the decay parameter of the evanescent field in the cladding becomes smaller due to the larger refractive index mismatch.
The evanescent field becomes significant.
When the fiber diameter is small enough corresponding to the guiding wavelength, the proportion of the evanescent field can be higher than the wave guided inside the nanofiber.

The transverse components of the electric field outside the fiber becomes highly azimuthal dependent.
The $z$ component of the electric field both inside and outside the fiber are substantial.
The reason will be further explained in the next section.

\subsection{Optical nanofiber electronmagnetic modes} 
\label{Optical nanofiber electronmagnetic modes}
The solution of the Maxwell equations in cylindrical coordinates leads to the following expressions for the field components along the horizontal ($x$), vertical ($y$), and longitudinal ($z$) directions inside and outside of the fiber core.
Different modes of a cylindrical core step index waveguide are described by Bessel functions in the core, and modified Bessel functions outside the core.
Assuming the core radius of the nanofiber is $a$, refractive index of the core $n_1$ (fused silica), and the refractive index of the cladding $n_2$ (air).
The fundamental mode of an optical fiber (HE$_{11}$ mode) is given by the three components of electric and magnetic fields $E_{x},E_{y},E_{z}$ and $H_{x},H_{y},H_{z}$ in a cylindrical coordinates ($r$,$\theta$,$z$):

For $r<a$ (inside the fiber),
\begin{equation}
\begin{aligned}
E_{x}=&-j A \beta \frac{a}{u}\left[\frac{(1-s)}{2} J_{0}\left(\frac{u}{a} r\right) \cos \psi-\frac{(1+s)}{2} J_{2}\left(\frac{u}{a} r\right) \cos (2 \theta+\psi)\right] ,\\
E_{y}=&j A \beta \frac{a}{u}\left[\frac{(1-s)}{2} J_{0}\left(\frac{u}{a} r\right) \sin \psi+\frac{(1+s)}{2} J_{2}\left(\frac{u}{a} r\right) \sin (2 \theta+\psi)\right] ,\\
E_{z}=&A J_{1}\left(\frac{u}{a} r\right) \cos (\theta+\psi) ,\\
H_{x}=&-j A \omega \varepsilon_{0} n_{1}^{2} \frac{a}{u}\left[\frac{\left(1-s_{1}\right)}{2} J_{0}\left(\frac{u}{a} r\right) \sin \psi\right. 
\left.+\frac{\left(1+s_{1}\right)}{2} J_{2}\left(\frac{u}{a} r\right) \sin (2 \theta+\psi)\right] ,\\
H_{y}=&-j A \omega \varepsilon_{0} n_{1}^{2} \frac{a}{u}\left[\frac{\left(1-s_{1}\right)}{2} J_{0}\left(\frac{u}{a} r\right) \cos \psi\right. 
\left.-\frac{\left(1+s_{1}\right)}{2} J_{2}\left(\frac{u}{a} r\right) \cos (2 \theta+\psi)\right] ,\\
H_{z}=&-A \frac{\beta}{\omega \mu_{0}} s J_{1}\left(\frac{u}{a} r\right) \sin (\theta+\psi).
\end{aligned}
\end{equation}

For $r>a$ (outside the fiber),
\begin{equation}
\begin{aligned}
E_{x}=&-j A \beta \frac{a J_{1}(u)}{w K_{1}(w)}\left[\frac{(1-s)}{2} K_{0}\left(\frac{w}{a} r\right) \cos \psi+\frac{(1+s)}{2} K_{2}\left(\frac{w}{a} r\right) \cos (2 \theta+\psi)\right] ,\\
E_{y}=& j A \beta \frac{a J_{1}(u)}{w K_{1}(w)}\left[\frac{(1-s)}{2} K_{0}\left(\frac{w}{a} r\right) \sin \psi-\frac{(1+s)}{2} K_{2}\left(\frac{w}{a} r\right) \sin (2 \theta+\psi)\right] ,\\
E_{z}=& A \frac{J_{1}(u)}{K_{1}(w)} K_{1}\left(\frac{w}{a} r\right) \cos (\theta+\psi) ,\\
H_{x}=&-j A \omega \varepsilon_{0} n_{0}^{2} \frac{a J_{1}(u)}{w K_{1}(w)}\left[\frac{\left(1-s_{0}\right)}{2} K_{0}\left(\frac{w}{a} r\right) \sin \psi-\frac{\left(1+s_{0}\right)}{2} K_{2}\left(\frac{w}{a} r\right) \sin (2 \theta+\psi)\right],\\
H_{y}=&-j A \omega \varepsilon_{0} n_{0}^{2} \frac{a J_{1}(u)}{w K_{1}(w)}\left[\frac{\left(1-s_{0}\right)}{2} K_{0}\left(\frac{w}{a} r\right) \cos \psi+\frac{\left(1+s_{0}\right)}{2} K_{2}\left(\frac{w}{a} r\right) \cos (2 \theta+\psi)\right] ,\\
H_{z}=&-A \frac{\beta}{\omega \mu_{0}} s \frac{J_{1}(u)}{K_{1}(w)} K_{1}\left(\frac{w}{a} r\right) \sin (\theta+\psi),
\end{aligned}\end{equation}\label{eq:electromagnetic field}
where $\psi$ is the polarization angle, $s=\left[(w)^{-2}+(u)^{-2}\right] /\left[J_{1}^{\prime}(u) / u J_{1}(u)+K_{1}^{\prime}(w) / w K_{1}(w)\right]$, $s_1=\beta^2 s/(k^2 n_1^2)$, $s_0=\beta^2 s/(k^2 n_2^2)$, $\beta$ is the propagation constant, $u=a \sqrt{n_{1}^{2} k^{2}-\beta^{2}}$ and $w=a \sqrt{\beta^{2}-n_{2}^{2} k^{2}}$ ($u^2 + w^2 = V^2$, $V$ is the normalized frequency). $J_n$ and $K_n$ are the Bessel functions of the first kind and the modified Bessel functions of the second kind, respectively, and the prime stands for the derivative.

For weakly guiding fiber, the value of $s$ can be approximated to -1.
For air-cladding fused silica fiber, refractive index mismatch $\Delta \gg 1\%$, the light is guided in strongly guiding region.
We need to use the full expression of $s$.
The decay parameter of the evanescent field $w$ is based on the refractive index of air, which is small.
The terms $K_{1}\left(\frac{w}{a} r\right)$ and $K_{2}\left(\frac{w}{a} r\right)$ are significant.
The terms $K_{2}\left(\frac{w}{a} r\right)$ applied on $\cos (2 \theta+\psi)$ and $\sin (2 \theta+\psi)$ make $E_{x}$ and $E_{y}$ components highly azimuthal dependent.
The terms $K_{1}\left(\frac{w}{a} r\right)$ gives substantial $E_{z}$ component outside the fiber.
The $\frac{a}{u}$ is parameter describing the field within fiber will be comparable to propagation constant $\beta$.
Therefore, we have non-negligible $z$ component inside the nanofiber compared to the transverse components.

\begin{figure}
    \centering
    \includegraphics[width=\linewidth]{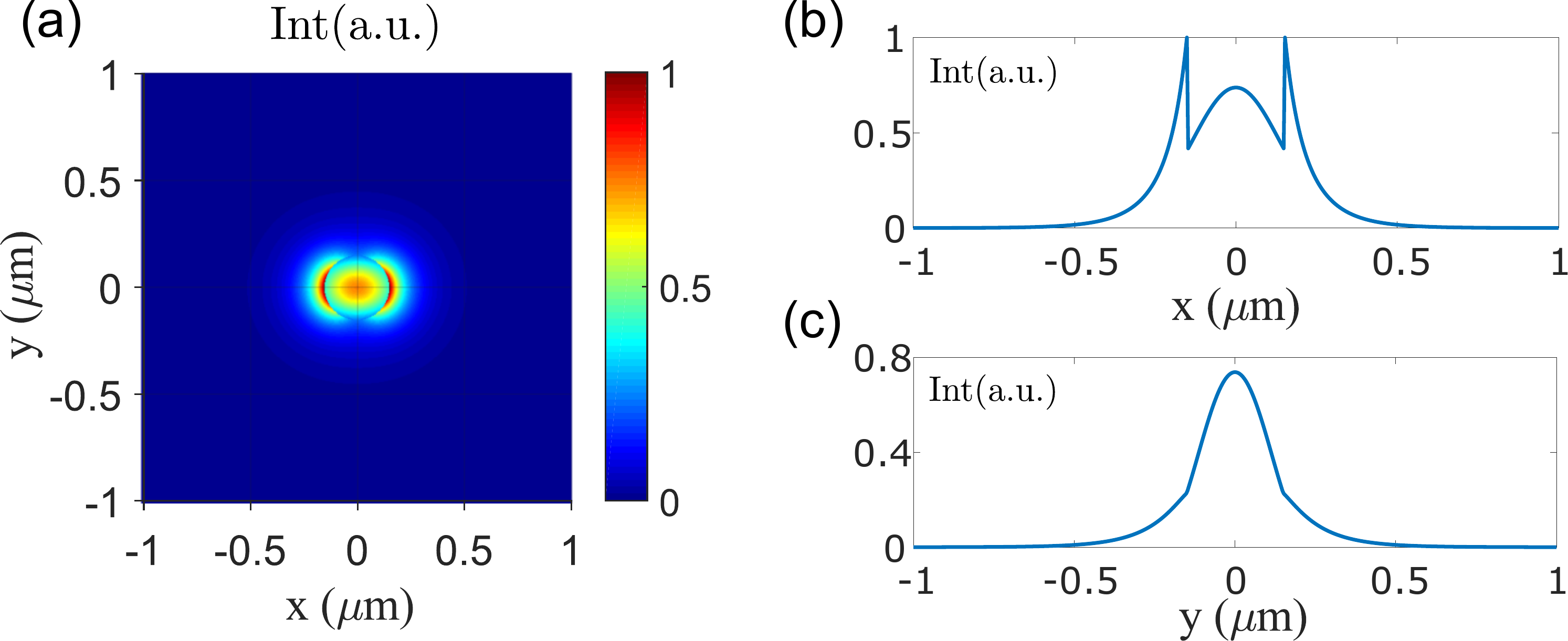}
    \caption{Fundamental mode (HE$_{11}$) at 637 nm in a 300 nm diameter nanofiber with intensity normalized to its maximum value. a) the total intensity normalized to its maximum intensity. b) the intensity distribution along the axis $y=0$. c) the intensity distribution along the axis $x=0$.}
    \label{fig:HE11}
\end{figure}

\begin{figure}
    \centering
    \includegraphics[width=\linewidth]{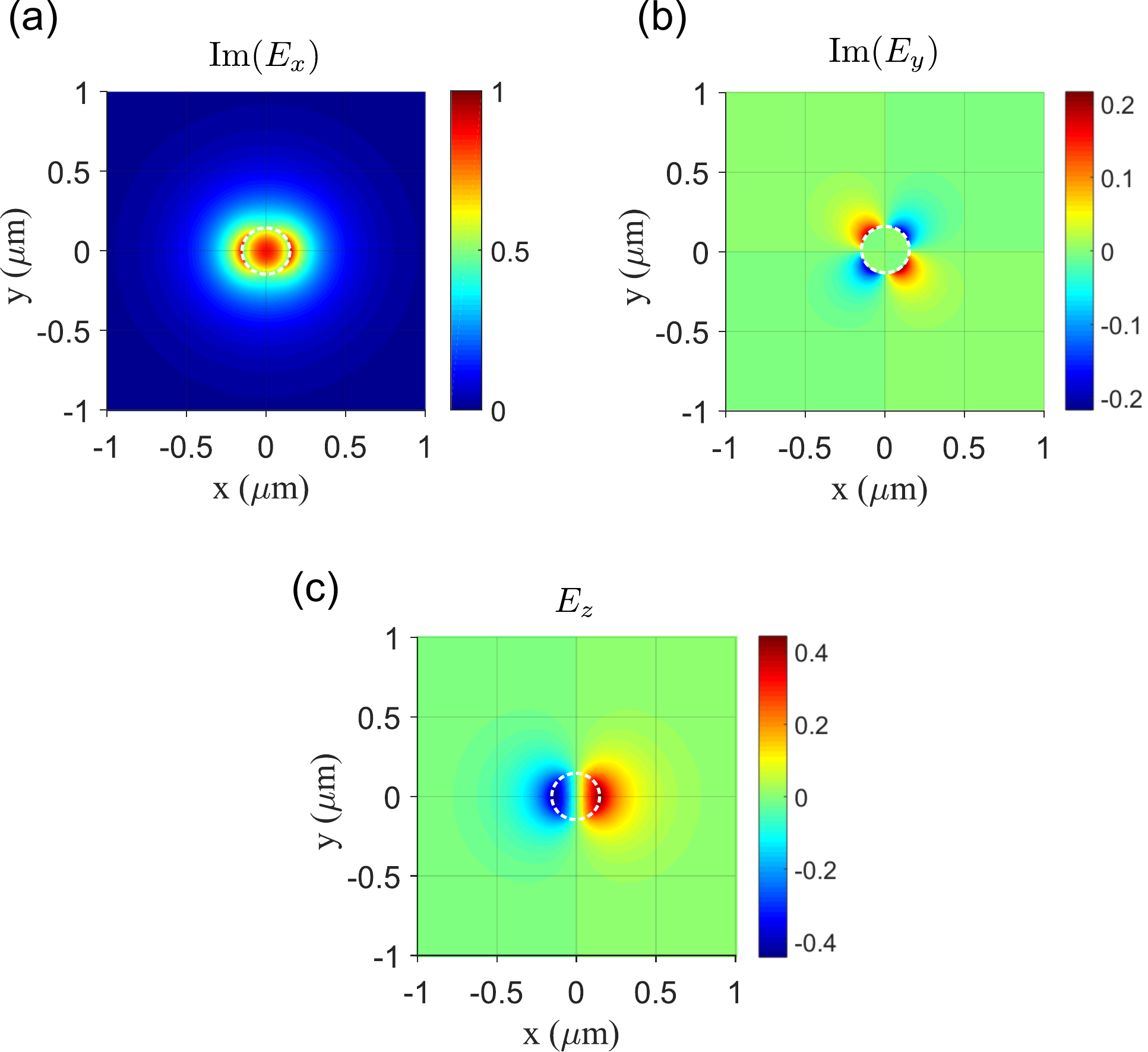}
    \caption{Fundamental (HE$_{11}$) mode structure of a 300 nm diameter optical nanofiber.  The value of the three components Im($E_x$), Im($E_y$) and $E_z$ are normalized to the maximum intensity. The wavelength of the guided light is 637 nm.}
    \label{fig:plotmodefield}
\end{figure}

The Fig. \ref{fig:plotmodefield} shows the three normalized components Im($E_x$), Im($E_y$) and $E_z$ of the electric field of the 637 nm guided light with linear-polarization along x-axis in a 300 nm diameter optical nanofiber.
The component along z-axis is strong compare to linearly polarized light in free space.

With the increasing fiber radius $a$, higher-order modes are supported in optical nanofiber waveguides.
The higher-order modes will have complex evanescent field around the nanofiber, which enables the exploration of interaction with the surrounding medium \cite{solano2017optical}.

\subsection{Adiabatic profile of the transitions in tapered fiber}
\label{Adiabatic profile}
The transmission rate of the whole system is defined by the taper region, where guidance switch from core-cladding regime to cladding-air regime.
Here we are going to explain in detail how the taper profile is calculated.

For plane waves in homogeneous transparent media, the refractive index $n$ can be used to quantify the increase in the wavenumber (phase change per unit length) caused by the medium.
The wavenumber is $n$ times higher than it would be in vacuum.
The effective refractive index $n_{eff}$ has the analogous meaning for light propagation in a waveguide with restricted transverse extension: the $\beta$ value (phase constant) of the waveguide for a given wavelength is the effective refractive index times the vacuum wavenumber.
\begin{equation}
    \beta = n_{eff} \frac{2 \pi}{\lambda}
\end{equation}

The geometry at the tapered region brings optical losses when the fundamental mode is coupled to higher order modes which can not be carried by the nanofiber.
In order to avoid the losses, the profile of the tapered fiber must meet the adiabatic criterium \cite{love1986quantifying}.
The coupling coefficient between two different modes is given by \cite{snyder2012optical}:
\begin{equation}
\label{coupling coefficient between modes}
C_{j l}=\frac{k}{4}\left(\frac{\varepsilon_{0}}{\mu_{0}}\right)^{1 / 2} \frac{1}{\beta_{j}-\beta_{l}} \int_{A} \hat{\mathbf{e}}_{j}^{*} \cdot \hat{\mathbf{e}}_{l} \frac{\partial n^{2}}{\partial z} \mathrm{d} A, \quad j \neq l\end{equation}
where 
the $\beta_{j}$ and $\beta_{l}$ are the propagation constant of the two modes and the $\hat{\mathbf{e}}_{l}$ and $\hat{\mathbf{e}}_{j}$ are the electric fields of the two modes.
The overlap between these two modes are integrated in transverse plane $A$.
If we assume that the cross-section of fiber taper is axisymmetric, LP$_{01}$ mode can couple only to modes with the same azimuthal symmetry, which means to the higher order LP$_{0m}$ modes.
For the calculation and optimization of the fiber profile to minimize the losses when coupling from the fundamental mode, it is intuitive to focus on the higher order mode with the closest propagation constant to the fundamental mode dominating the coupling. 
For the tapered nanofiber, since the propagation constants are close, the coupling between the fundamental mode HE$_{11}$ and HE$_{12}$ is the strongest \cite{love1986quantifying}.
In equation \ref{coupling coefficient between modes}, $j=1$ represents the fundamental mode HE$_{11}$ and $l=2$ represents the higher order mode HE$_{12}$.

\begin{figure}
    \centering
    \includegraphics[width=0.6\linewidth]{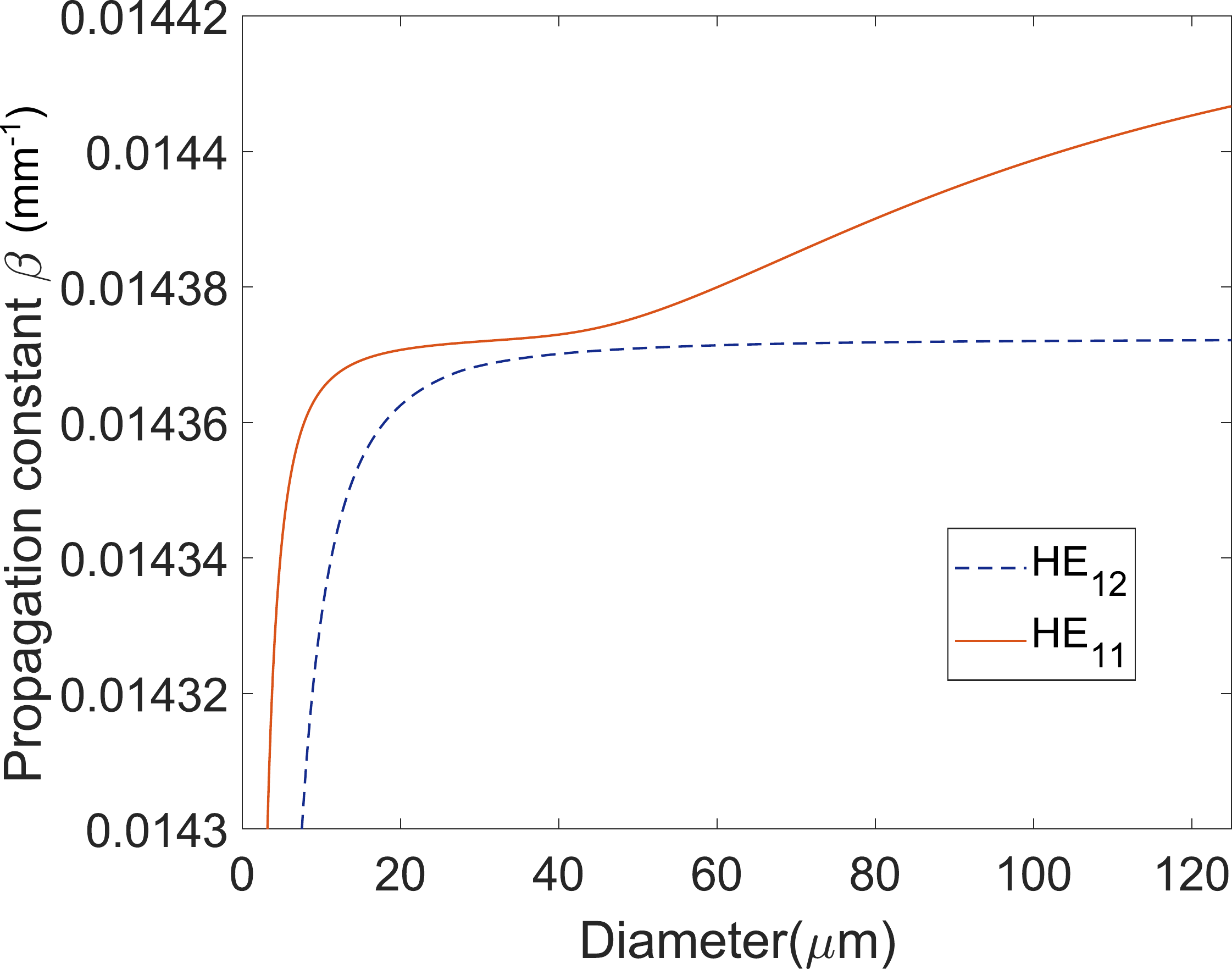}
    \caption{The propagation constants of HE$_{11}$ mode and HE$_{12}$ mode as a function of fiber diameter.}
    \label{fig:propagation constant for HE11 and HE12}
\end{figure}

It is worthwhile to examine their propagation constants more closely.
In Fig.\ref{fig:propagation constant for HE11 and HE12}, we plot the propagation constants of HE$_{11}$ mode and HE$_{12}$ mode as a function of fiber diameter.
Strictly speaking, there is a three-layer system. 
To simplify matters, however, depending on the fiber thickness, a system of core and jacket or of jacket and air is assumed. 
In the area around r = 20 $\mu$m, the HE$_{11}$ mode changes from the core to the cladding, and the difference to the propagation constant of the HE$_{12}$ mode is smallest, which is why the two modes couple there the strongest.

Two coupled modes periodically exchange energy on the length of their coupling region:
\begin{equation}\Delta l=\frac{2 \pi}{\beta_{1}-\beta_{2}}\end{equation}

\begin{figure}
    \centering
    \includegraphics[width=0.6\linewidth]{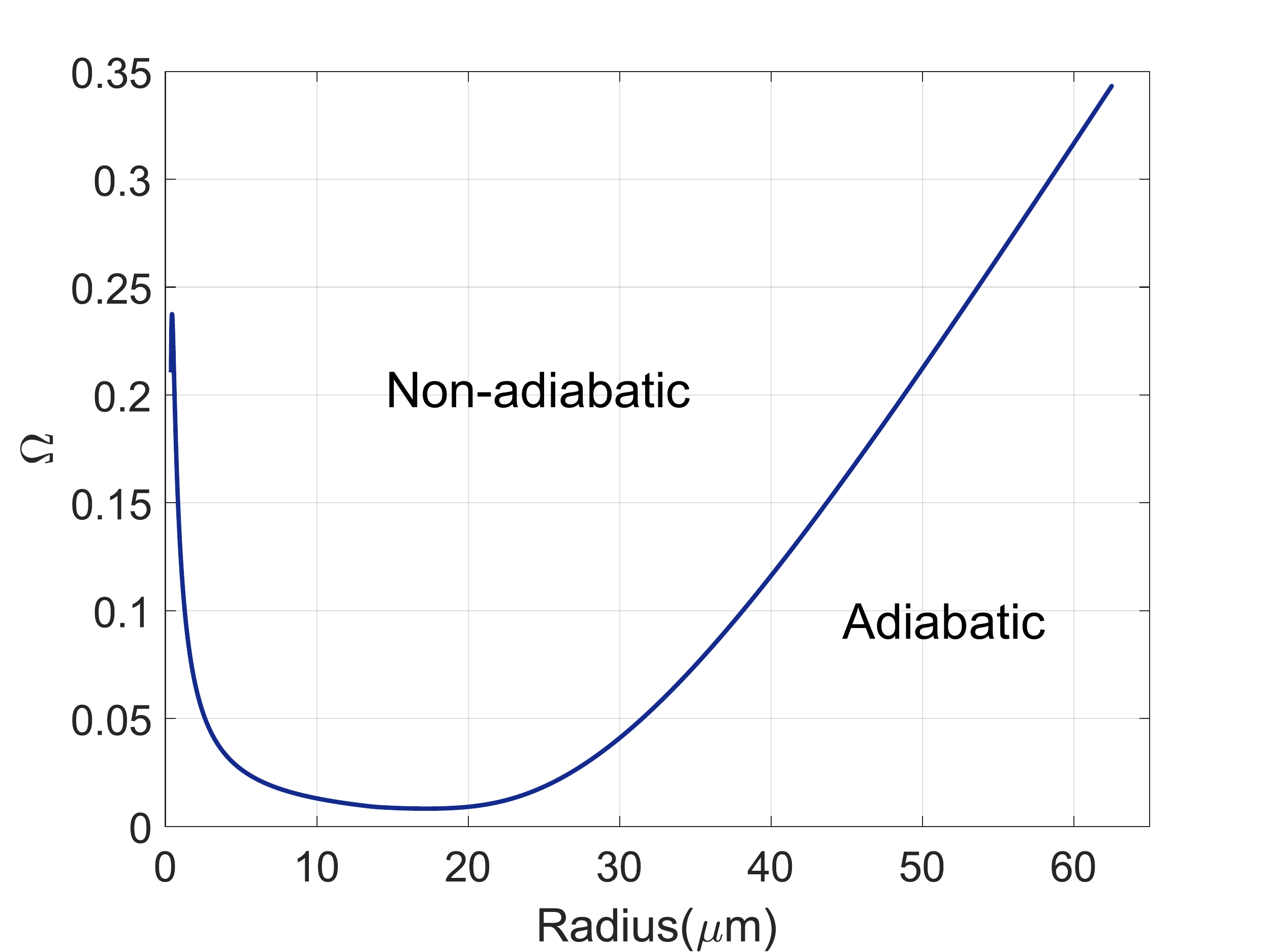}
    \caption{The optimized slope angle of the transition region as a function of the fiber radius for minimizing the transmission loss.}
    \label{fig:omega}
\end{figure}

To have adiabatic transition, the slope angle of the transition must be small at each location $z$ compared to the ratio of the radius $r$ to the length of the coupling region:
\begin{equation}\Omega(z)<\frac{r(z)}{\Delta l}=\frac{r(z)}{2 \pi} \left(\beta_{1}-\beta_{2}\right)\end{equation}
where $r(z)$ is the local core radius and $\beta_{1}(z)$ and $\beta_{2}(z)$ are the local propagation constants of the HE$_{11}$ mode and HE$_{12}$ mode respectively.
In Fig.\ref{fig:omega}, we show the slope angle $\Omega$ of the boundary between adiabatic and non-adiabatic transition when the guiding light is at 637 nm wavelength. 
In order to minimize the coupling towards the higher order mode, we need to choose the slope angle at different fiber radius in the adiabatic region (below the boundary).

The propagation constant of fundamental mode HE$_{11}$ and higher order mode HE$_{11}$ are the closest for a fiber radius of between 10 and 30 $\mu$m approximately. 
Thus, the coupling coefficient $C_{12}$ tends to be bigger in this zone, which can be compensated by smoothing the slope ($\Omega$) of the profile.
For the region where the propagation constants differ more strongly, for example at the beginning and end of a transition, the angle can be larger than in the middle of the transition. 
Such a profile can be approximated experimentally by means of three areas with two linear slopes with a mean slope and one exponential slope, as shown in Fig.\ref{fig:fiber radius along z}.

In the next section, we are going to describe how to fabricate tapered nanofibers in detail.

\section{Fabrication of the optical nanofiber}
\label{Fabrication of the optical nanofiber}
The fabrication of the optical nanofiber is realized by a homemade pulling system.
It consists of an oxyhydrogen flame that can bring the fused silica to its softening point (1585 $^{\circ}$C) and two translation stages for holding and pulling fiber ends, as shown in Fig.\ref{fig:Fiber_tapering_system}. This fabrication method was originally designed by J. E. Hoffman \cite{hoffman2014ultrahigh}.
During the pulling process, the oxyhydrogen flame is fixed. 
The movement of translation stages are controlled by a computer software based on computed adiabatic criterium introduced in the previous section. 
The translation stages movement can be decomposed in two components. 
First, they move away from each other, to stretch the fiber and make it thinner. 
Second, they move in the same direction, displacing the heated portion of the fiber of the fixed heat source. 
This portion is roughly the width of the flame. 
The speed of the translation stages defines the fiber narrowing, indeed, if a given portion of the fiber stays longer over the flame, it will become softer and elongates more easily.
The control of the speed and the moving range at each pass defines the final envelope of the nanofiber.
Here, we use a Matlab script that produces the control parameters for motors to fabricate an optical nanofiber with a user defined taper geometry is written by J. E. Hoffman (https://drum.lib.umd.edu/handle/1903/15069) based on the algorithm of Florian Warken \cite{warken2007ultradunne}.

\begin{figure}
    \centering
    \includegraphics[width=0.6\linewidth]{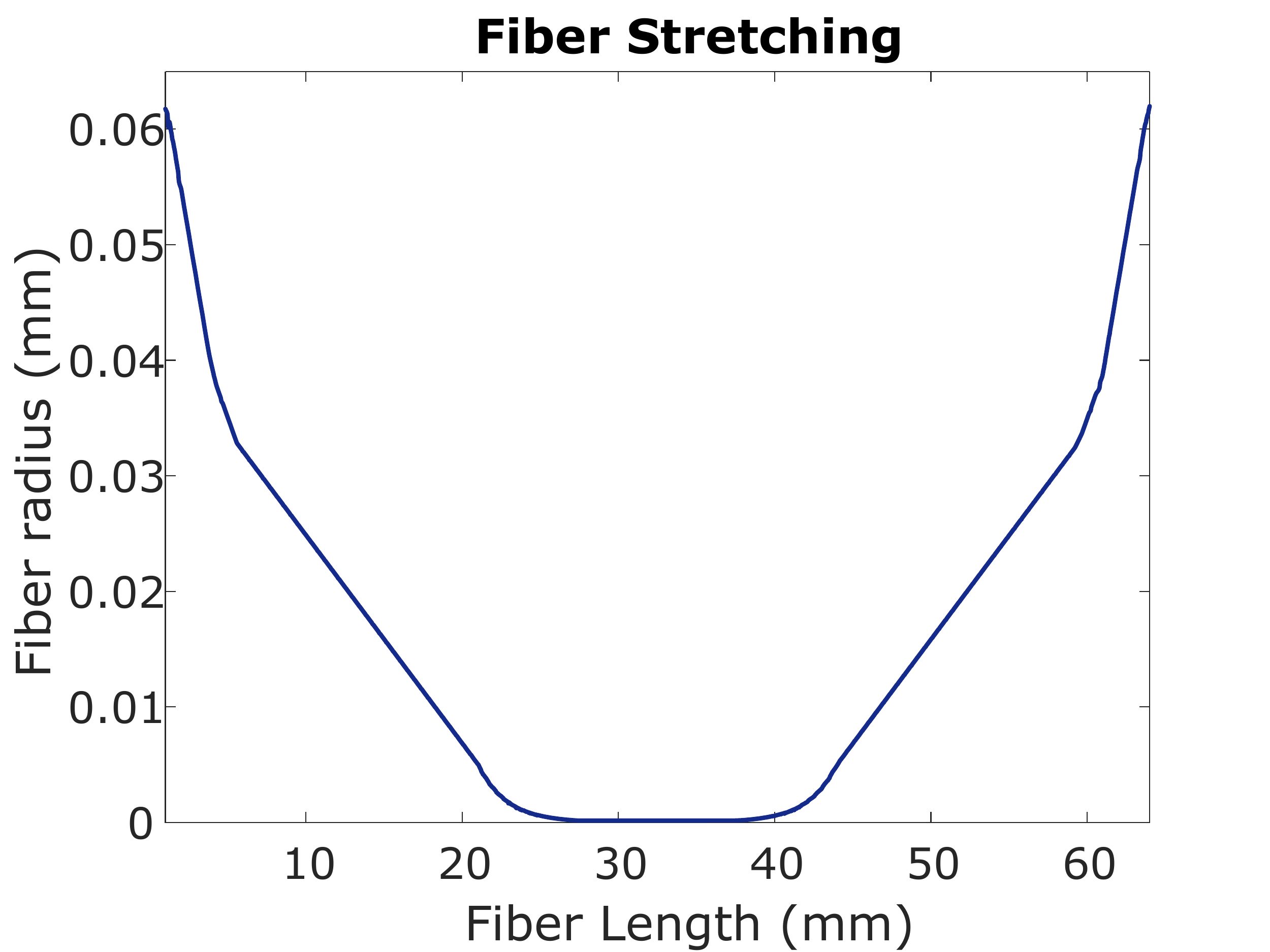}
    \caption{Simulated shape of nanofiber taper described with nanofiber radius along the longitudinal axis of the fiber. The guiding wavelength is 637 nm.}
    \label{fig:fiber radius along z}
\end{figure}

\begin{figure}
    \centering
    \includegraphics[width=0.6\linewidth]{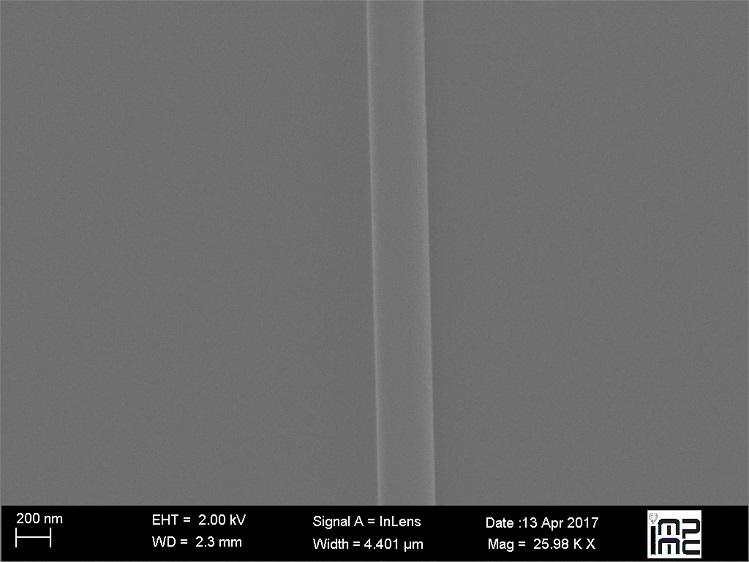}
    \caption{Nanofiber image taken with scanning electronic microscope.}
    \label{fig:SEM of nanofiber}
\end{figure}

\begin{figure}
    \centering
    \includegraphics[width=0.8\linewidth]{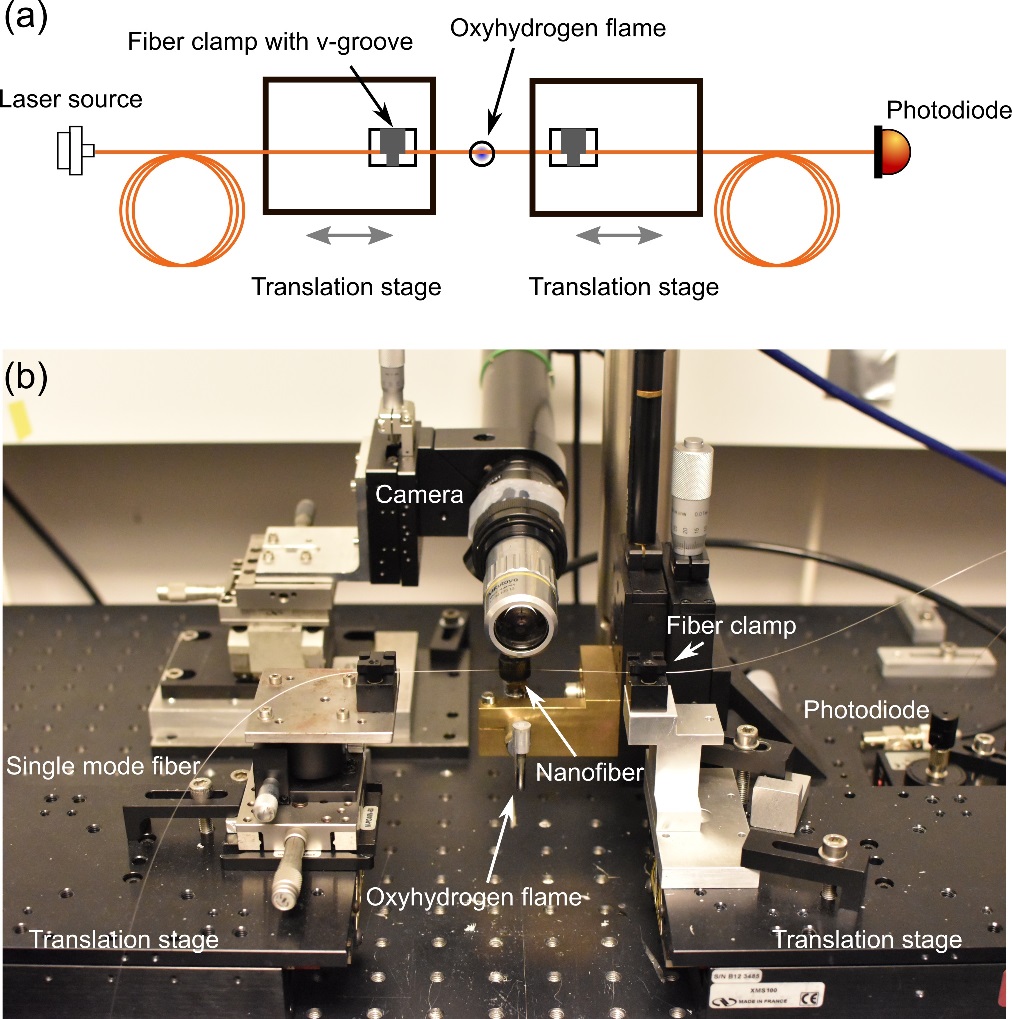}
    \caption{Schematic(a) and photo(b) of the optical nanofiber fabrication system. The two sides of the optical fibers are fixed on the translation stage with v-groove clamps. The oxyhydrogen flame is obtained by burning a mixture of hydrogen and oxygen mixed in a ratio of 2:1. The hydrogen is generated with a hydrogen generator. The oxygen is from a oxygen cylinder. We control the flow rate of the two gases to obtain the best mixing ratio. The transmission of the optical nanofiber is detected with a photodiode. A real time detection is available during the pulling process.}
    \label{fig:Fiber_tapering_system}
\end{figure}

With the size and refractive index of the core and cladding of the original fiber (SM600-Thorlabs) we used to guide the light with wavelength at 637 nm, the simulated nanofiber radius along the longitudinal axis is shown in Fig. \ref{fig:fiber radius along z}. 
In practice, this simulated radius at the nanofiber waist matches well ($\pm 5~\text{nm}$) with the measurement of fiber radius by scanning electronic microscope (SEM), as shown in Fig.\ref{fig:SEM of nanofiber}.
The SEM image is observed by focusing an electron beam on the sample plane and detecting the elastic scattering, emission of secondary electrons by inelastic scattering and the emission of electromagnetic radiation to show the topography and composition at the surface with nanometer resolution.

Producing optical nanofiber requires clean environment and careful fiber cleaning before pulling.
Before the tapering of optical fiber starts, the plastic jacket needs to be removed over the distance separating the fiber clamps (see Fig.\ref{fig:Fiber_tapering_system}).
Any dusts from the environment fell on to the fiber, the remaining detritus of fiber jacket or grease on fingers left on the fiber will burn when it's brought to the flame and create some defects on the fiber surface during pulling.
Therefore, the pulling system is installed in a box with air-flow.
Clean room gloves \footnote{As opposed to standard chemistry protection gloves.} are necessary to avoid contamination by the operator.
The optical fiber is cleaned with a few wipes of isopropanol on lens tissue to remove most of the particles on the surface, followed with a wipe of acetone to dissolve small remaining pieces, and finalized with a wipe of isopropanol to remove the remaining acetone. 
The clean environment can also avoid the dusts attaching to the nanofiber after tapering, which will cause the scattering of light and lower the entire transmission rate.

Then, we mount the cleaned fiber into the v-grooves on the clamps.

A camera continuously monitors the nanofiber from the side.
The focus plane of the camera is adjusted to be on the fiber.
The center of the flame needs to be aligned with the fiber based on the focus plane of the camera.

During the entire fabrication process, we monitor the transmission of the fiber by sending a few $\mu$W of light at the working wavelength of the nanofiber.
The transmission is detected with a photodiode, as shown in Fig. \ref{fig:Fiber_tapering_system}, and recorded.
The output signal is normalized to the initial laser power before pulling.

During the pulling, the hot air from the flame will slowly lift the nanofiber since the weight of the fiber region on top of the flame drops.
It is critical because the distance between flame and the fiber becomes larger than the setting.
The temperature of the fiber might not reach the softening point of silica.
The distance between the two transverse stage increases during the pulling process, and this will break the fiber rather than narrowing it. 
To avoid this problem, we move the flame up five times following the rising of fiber with 100 $\mu m$ at each step.

\begin{figure}
    \centering
    \includegraphics[width=0.6\linewidth]{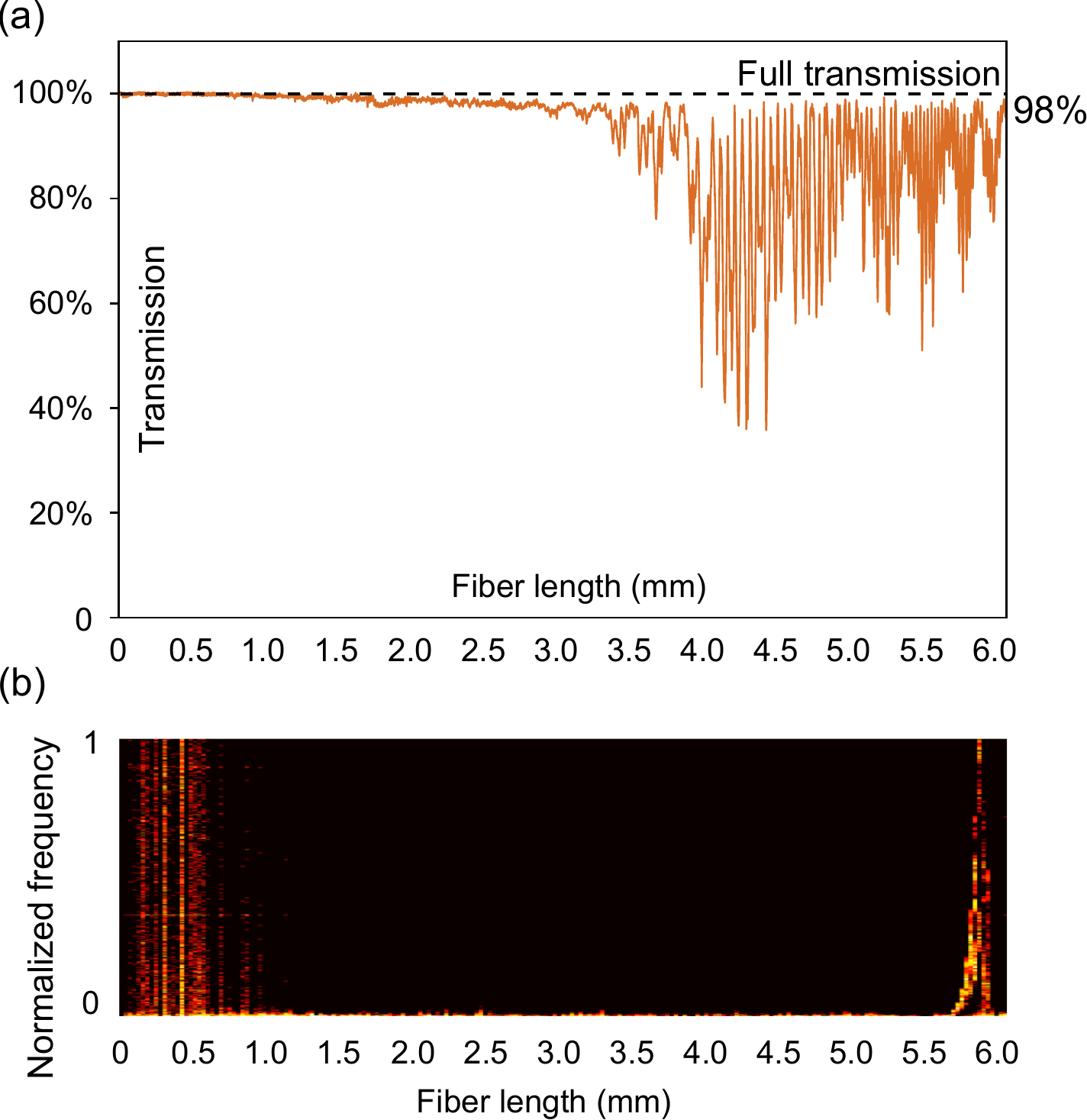}
    \caption{The transmission curve detected with the photodiode (a) and corresponding spectrum (b) through fast Fourier transform during the tapering process.}
    \label{fig:transmission_spectrum}
\end{figure}

Fig.\ref{fig:transmission_spectrum} shows the typical transmission curve as a function of time and corresponding spectrum during the tapering process with the profile shown in Fig.\ref{fig:fiber radius along z} to reach a final waist radius of 150 nm, with a fiber waist length of 10 mm.
The sinusoidal oscillation measured in the transmission represents the energy transfer between the two modes HE$_{11}$ and HE$_{12}$. 
The frequency of the oscillation depends on the difference in the propagation constants.
Therefore, the oscillation frequency increases while tapering since propagation constant decreases when fiber radius decreases, as shown in the fast Fourier transform at the end of the pulling.
In the end, this optical nanofiber achieves a transmission of 98\%.

With this fabrication technique, the transmission of a fabricated optical nanofiber is typically above 95 \% transmission.

\section[Evanescent field coupling]{The evanescent field coupling of optical nanofibers}
Optical nanofiber can be used as a platform for guiding luminescence from the particle on the nanofiber surface into the optical fiber thanks to its evanescent field.
In section \ref{Light propagating in an optical nanofiber}, we introduced the electronic field of the light guided by an optical nanofiber.
In this section, we are going to talk about how we couple the nanoparticles, for example the gold nanospheres and gold nanorods, with the evanescent part of the electromagnetic field outside the fiber.

First, we will numerically explain the dependence of evanescent field at the fiber surface on the wavelength and fiber radius.
Because we deposit the gold nanoparticles directly on the nanofiber surface, the coupling will be decided by the electromagnetic field at the fiber surface where the evanescent field is the strongest.
As introduce in section \ref{Light propagating in an optical nanofiber}, the evanescent field decays exponentially from the surface of the fiber to a few hundreds of nm in the air.
Then, we will introduce the optical response of metallic nanospheres and nanorods.
A gold nanoparticle acts similarly as a dipole when it scatters light.
When single dipole is coupled to optical nanofiber, the guided light shows chiral properties in polarization.
This phenomenon brings the possibility of polarization control of single dipole. 

\begin{figure}
    \centering
    \includegraphics[width=0.6\linewidth]{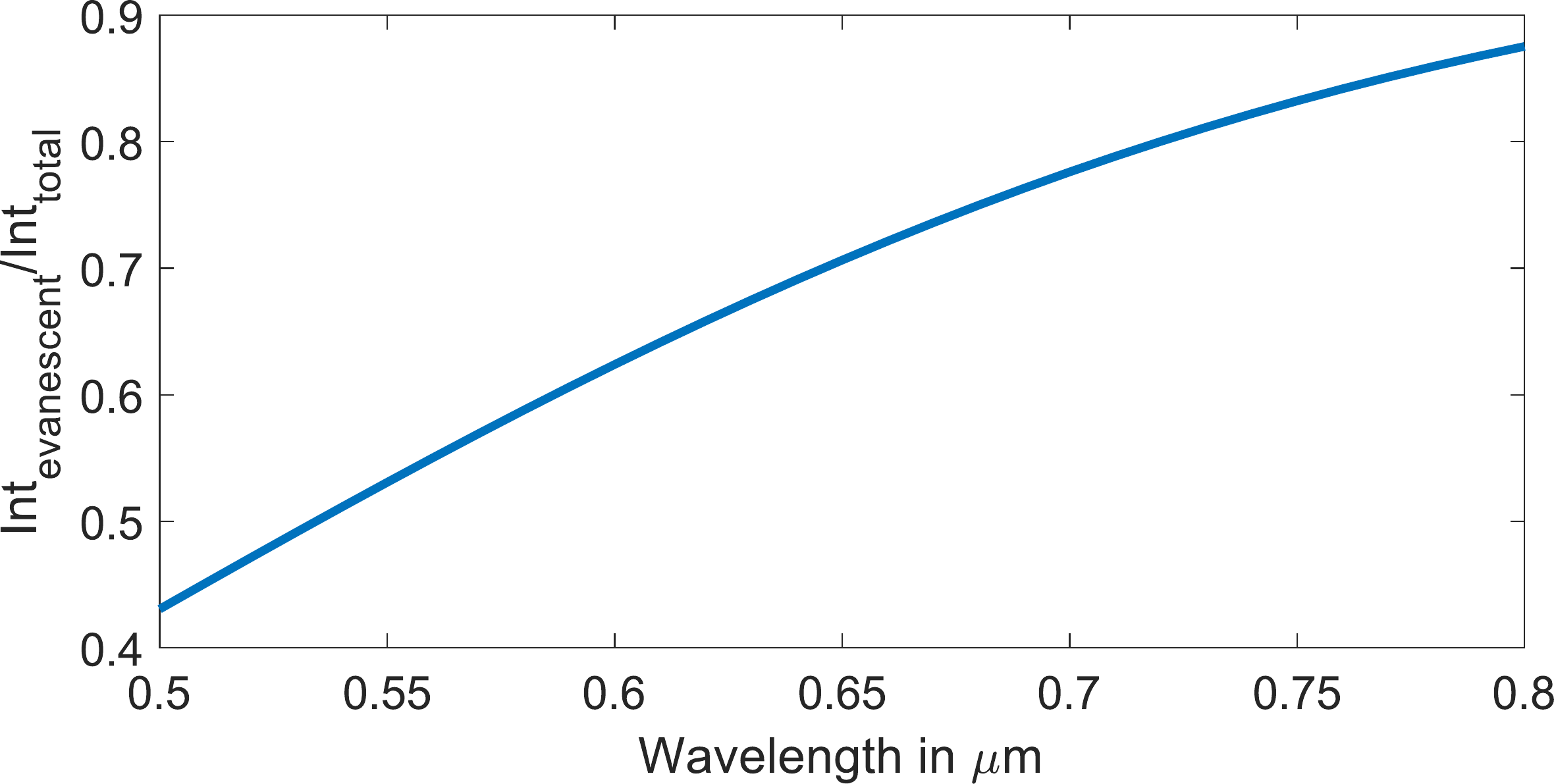}
    \caption{Ratio of the evanescent field among the total intensity of guided light as a function of guiding wavelength $\lambda$. The fiber radius is set as 150 nm.}
    \label{fig:eva_lambda}
\end{figure}

\begin{figure}
    \centering
    \includegraphics[width=0.6\linewidth]{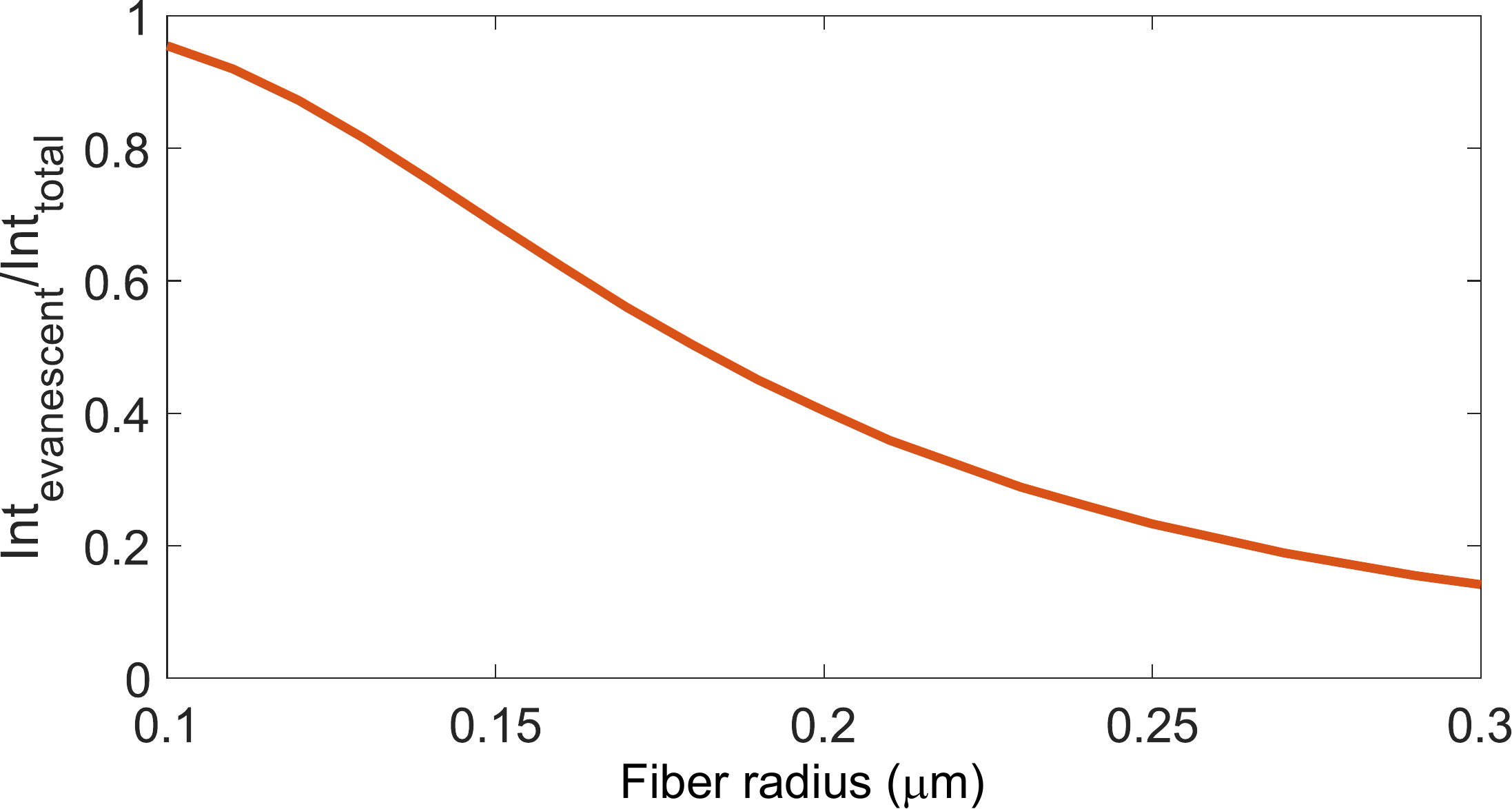}
    \caption{Ratio of the evanescent field among the total intensity of guided light as a function of fiber radius $a$. The guiding wavelength is set as 637 nm.}
    \label{fig:eva_a}
\end{figure}

Depending on the wavelength of the guided light and the fiber radius, the proportion of the light propagating in the evanescent field varies.
In Fig.\ref{fig:eva_lambda}, we show the ratio of the evanescent field $I_\text{evanescent}$ by the total intensity of guided light $I_\text{total}$ increases with longer guiding wavelength $\lambda$.
In Fig. \ref{fig:eva_a}, we show the ratio of the evanescent field by the total intensity of guided light $I_\text{evanescent}/I_\text{total}$ drops with larger fiber radius $a$.
As shown in Fig.\ref{fig:HE11}, the evanescent field of the nanofiber is mostly around the fiber surface.
The ratio of the evanescent field among the total intensity of guided light $I_\text{evanescent}/I_\text{total}$ describes the ratio of the light propagating outside the nanofiber.

\begin{figure}
    \centering
    \includegraphics[width=\linewidth]{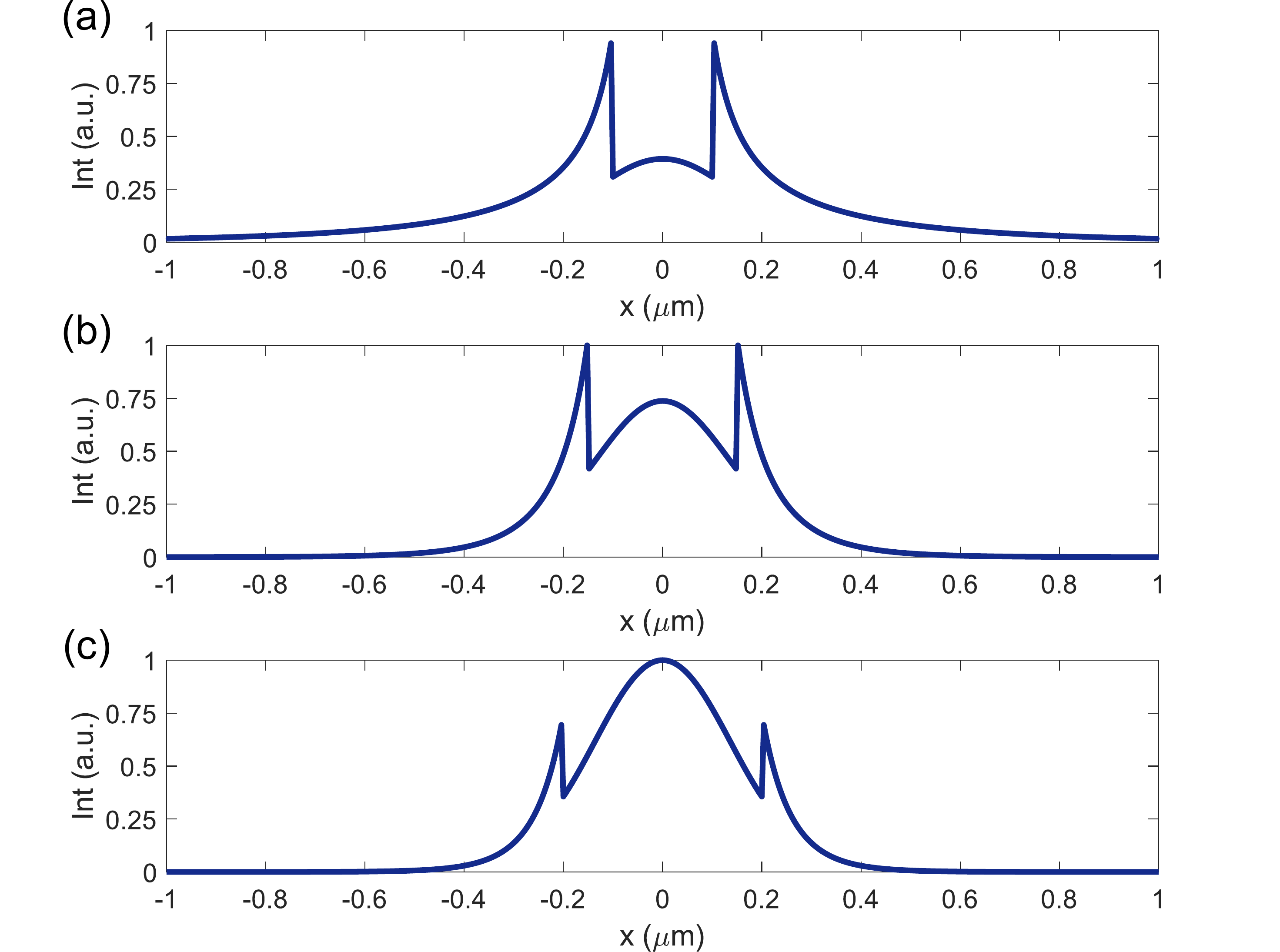}
    \caption{The intensity profile of the mode along x axis with x linearly polarized light at 637 nm with different fiber radius (a) 100 nm, (b) 150 nm and (c) 200 nm.}
    \label{fig:Intensity at the fiber surface}
\end{figure}

However, it's not enough to describe the possible coupling efficiency between nanofiber and emitters deposited on the fiber surface.
Most of the evanescent field is distributed along the axis of the polarization, as shown in Fig.\ref{fig:HE11}.
Therefore, we plot the axis along the linear polarization only in Fig. \ref{fig:Intensity at the fiber surface} to show the evanescent field at the fiber surface while varying the fiber radius. 
For guiding wavelength at 637 nm, the coupling efficiency reaches it maximum with fiber radius around 150 nm.
When the fiber radius is large, as shown in Fig. \ref{fig:Intensity at the fiber surface}-a with fiber radius of 100 nm, most of the light will be guided within the fiber and evanescent part of the mode profile will be small, which is predicted in the ratio of the evanescent field among the total intensity of guided light as a function of fiber radius $a$, as shown in Fig. \ref{fig:eva_a}.
Most of the time, a small fiber radius will offer higher surface coupling efficiency.
But it's not always the case.
When the fiber radius is too small, the evanescent field will not be confined at the fiber surface, as shown in Fig. \ref{fig:Intensity at the fiber surface}-a with fiber radius of 100 nm.
The coupling efficiency between fiber and emitters on the fiber surface will also drop.
Therefore, the best fiber radius needs to be chosen to achieve highest efficiency for surface coupling.

When we find the best radius for the wavelength of interest, we can have the mode confined within the evanescent field in the transverse plane.
Sending a few milliwats in fiber results in high intensity at the fiber surface.
Besides, the nanofiber region of a tapered fiber can easily reach 10 mm with high transmission.
Therefore, the nanofiber offers a long interaction length.
This is very useful in an atomic system because it can provide higher optical depth.

In this work, we study the coupling between the optical nanofiber and single dipoles (gold nanoparticle).
Two types of gold nanoparticles are used in our experiment, the gold nanosphere and the gold nanorods. 
In this section, we will first introduce the coupling between dipole to the evanescent field theoretically, and then introduce the deposition procedure of gold nanoparticles on an optical nanofiber.

\subsection{Coupling between dipole to the evanescent field}
The surrounding evanescent field makes the optical nanofiber a platform to couple with atoms or single particles at the fiber surface.
In our experiment, gold nanospheres and gold nanorods are used to couple with the evanescent field.
In free space, the scattered light from a point like metallic nanoparticle exhibits dipolar emission pattern.
When single metallic nanoparticle is coupled to optical nanofiber, because of the strong transverse confinement of the guided photons around the fiber surface, their internal spin and their orbital angular momentum get coupled \cite{petersen2014chiral}.
In this section, we are going to introduce how the gold nanosphere and gold nanorods are coupled to the evanescent field of optical nanofiber as a dipole theoretically.

\subsubsection{Gold nanoparticles}
The gold nanoparticles we used as scatterer have spatial dimensions of about 50 nm, which is small compare to the wavelength.
The solution of Maxwell’s equations for spherical particles (or infinitely long cylinders) is named after the physicist Gustav Mie and the expansion for elliptical particles became known as Gans or Mie-Gans theory.

First, we use the Mie theory for the analytical solution of spherical particles.
Scalar wave equation in spherical coordinates $\psi(r, \theta, \varphi)$ is given by:
\begin{equation}\left[\frac{1}{r^{2}} \frac{\partial}{\partial r}\left(r^{2} \frac{\partial}{\partial r}\right)+\frac{1}{r^{2} \sin \theta} \frac{\partial}{\partial \theta}\left(\sin \theta \frac{\partial}{\partial \theta}\right)+\frac{1}{r^{2} \sin ^{2} \theta} \frac{\partial^{2}}{\partial \varphi^{2}}+k^{2}\right] \psi=0\end{equation}

\begin{equation}\begin{array}{l}
u=\mathrm{e}^{-\mathrm{i} \omega t} \cos (\varphi) \sum_{l=1}^{\infty}(\mathrm{i})^{l} a_{l} \frac{2 l+1}{l(l+1)} P_{l}^{1}(\cos \theta) h_{l}^{(2)}\left(k_{\mathrm{out}} r\right) \\
v=\mathrm{e}^{-\mathrm{i} \omega t} \sin (\varphi) \sum_{l=1}^{\infty}(\mathrm{i})^{l} b_{l} \frac{2 l+1}{l(l+1)} P_{l}^{1}(\cos \theta) h_{l}^{(2)}\left(k_{\mathrm{out}} r\right)
\end{array}\end{equation}
where $a_{l}$ and $b_{l}$ are Mie scattering coefficients for mode $l$, given by:
\begin{equation}\begin{array}{l}
a_{l}=\frac{n_{r} \psi_{l}\left(n_{r} x\right) \psi_{l}^{\prime}(x)-\mu_{r} \psi_{l}(x) \psi_{l}^{\prime}\left(n_{r} x\right)}{n_{r} \psi_{l}\left(n_{r} x\right) \xi_{l}^{\prime}(x)-\mu_{r} \xi_{l}(x) \psi_{l}^{\prime}\left(n_{r} x\right)} ;
b_{l}=\frac{\mu_{r} \psi_{l}\left(n_{r} x\right) \psi_{l}^{\prime}(x)-n_{r} \psi_{l}(x) \psi_{l}^{\prime}\left(n_{r} x\right)}{\mu_{r} \psi_{l}\left(n_{r} x\right) \xi_{l}^{\prime}(x)-n_{r} \xi_{l}(x) \psi_{l}^{\prime}\left(n_{r} x\right)}.
\end{array}\end{equation}

Once we have the Mie coefficients, we can calculate the extinction, absorption and scattering cross sections or the electromagnetic fields inside and outside the spherical particle.
The cross sections is the net rate of the electromagnetic energy crossing a sufficiently large sphere surface surrounding the particle.

The scattering, extinction and absorption cross sections of a spherical particle is given by:
\begin{equation}C_{\mathrm{sca}}=\frac{2 \pi}{k_{\mathrm{out}}^{2}} \sum_{l=1}^{\infty}(2 l+1)\left(\left|a_{l}^{2}\right|+\left|b_{l}^{2}\right|\right),\end{equation}
\begin{equation}C_{\mathrm{ext}}=\frac{2 \pi}{k_{\mathrm{out}}^{2}} \sum_{l=1}^{\infty}(2 l+1) \Re \mathrm{e}\left(a_{l}+b_{l}\right),\end{equation}
and
\begin{equation}C_{\mathrm{abs}}=C_{\mathrm{ext}}-C_{\mathrm{sca}}.\end{equation}

For metallic nanorods, which can be regarded as elliptical spheroidal particles, the analytical result is solved with Mie-Gans solution.
In this case, the polarizations is given by three values $\alpha_1$, $\alpha_2$, $\alpha_3$ of the polarizability tensor:
\begin{equation}\alpha_{i}=9 V\left(L_{i}+\frac{1}{\varepsilon_{r}-1}\right)^{-1},\end{equation}
where $L_i$ are geometrical factors decided by the shape of the particle and need to meet the rule $L_{1}+L_{2}+L_{3}=1$.
It's also suitable for the spherical particles, in which case, $L_{1}=L_{2}=L_{3}=1/3$.

The shape of gold nanorods is close to elliptical spheroidal particles with lengths of the three axis satisfy $a>b=c$. 
The geometrical factor along the long axis are given by:
\begin{equation}L_{1}=\frac{1-e^{2}}{e^{2}}\left(-1+\frac{1}{2 e} \ln \frac{1+e}{1-e}\right),\end{equation}
where
\begin{equation}e=1-\left(\frac{b^{2}}{a^{2}}\right).\end{equation}
With $L_{1}+L_{2}+L_{3}=1$ and $L_{2}=L_{3}$, we can get the value of $L_{2}$ and $L_{3}$.

Then, we can get the polarizability tensor along three axis, $\alpha_1$, $\alpha_2$, $\alpha_3$.

\subsubsection{Dipole coupled to nanofiber}
\label{Dipole coupled to nanofiber}

Gold nanorods is close to a linear dipole polarized along the longitudinal axis of the nanorod.
In this subsection, we explain the coupling between a linear dipole laying on the fiber surface. 
To make it clear, we define a section of the optical nanofiber with cartesian coordinates ($x$, $y$, $z$), where in the perpendicular plane, $x$ defines the horizontal direction, $y$ defines the vertical direction and $z$ defines the longitudinal direction of the nanofiber. 
The dipole (gold nanorod) is included in a plane tangent to the surface of the fiber which limits its degrees of freedom to two: the angle $\alpha$ defines its azimuthal position around the fiber and the angle $\theta$ defines its orientation relative to the fiber axis.

We introduced a new cartesian coordinates ($x^{\prime}$, $y^{\prime}$, $z$) by rotating the cartesian coordinates ($x$, $y$, $z$) by an angle $\alpha$.
It is convenient to describe the guided electric field $\mathbf{E}$ on the basis of quasi-linear modes $\left\{\mathbf{H E}_{11}^{x^{\prime}}, \mathbf{H E}_{11}^{y^{\prime}}\right\}$. 

\begin{equation}
    \mathbf{E}=C\mathbf{H E}_{11}^{x^{\prime}}+D\mathbf{H E}_{11}^{y^{\prime}}
    \label{eq:guided_field}
\end{equation}

The electric field of air-cladding nanofiber is described in the cylindrical coordinates ($r$, $\varphi$, $z$), as introduced before in \ref{Optical nanofiber electronmagnetic modes}. 
$\varphi_{0}$ defines the polarization orientation of the guided light: $\varphi_{0}=0$ for $x$-polarized light and $\varphi_{0}=\pi/2$ for $y$-polarized light. 

The dipole is located on the surface of nanofiber. In the cartesian coordinates ($x^{\prime}$, $y^{\prime}$, $z$), the $r=a$ and $\varphi=\pi/2$. Thus, the amplitude $C$ and $D$ are given by:
\begin{equation}
\begin{array}{l}
{C=\mathbf{d} \cdot \mathbf{H E}_{11}^{x^{\prime}}(a, \pi / 2)} ,\\
{D=\mathbf{d} \cdot \mathbf{H E}_{11}^{y^{\prime}}(a, \pi / 2)} .
\end{array}
\end{equation}
The dipole moment $\mathbf{d}$ in the cartesian coordinates ($x^{\prime}y^{\prime}z$) is given by:
\begin{equation}
\label{eq:dipole moment}
    \begin{aligned}
    \mathbf{d}_{x}=& d\sin \theta e^{-i \omega t},\\
    \mathbf{d}_{y}=& 0,\\
    \mathbf{d}_{z}=& d\cos \theta e^{-i \omega t},
\end{aligned}
\end{equation}
where $\theta$ is the angle between the dipole and the nanofiber axis $z$, and $d$ is the amplitude of the dipole moment. The component according to $y^\prime$ of the dipole is zero because the dipole is supposed to be tangent to the fiber.

The electric field produced by the dipole $\mathbf{H E}_{11}^{x^{\prime}}(a, \pi / 2)$ and $\mathbf{H E}_{11}^{y^{\prime}}(a, \pi / 2)$ is given by using Equations \ref{eq:electromagnetic field}, with $\varphi_{0}=0$ for $\mathbf{H E}_{11}^{x^{\prime}}(a, \pi / 2)$ and $\varphi_{0}=\pi/2$ for $\mathbf{H E}_{11}^{y^{\prime}}(a, \pi / 2)$.

Therefore, we have the generated electric field by the dipole described as:
\begin{equation}
\label{eq:generated electric field}
    \begin{aligned}
    \mathbf{H E}_{11}^{x^{\prime}}(a, \pi / 2)=(\epsilon_{1}, 0, 0),\\
    \mathbf{H E}_{11}^{y^{\prime}}(a, \pi / 2)=(0, \epsilon_{2}, \epsilon_{3}).
\end{aligned}
\end{equation}

and
\begin{equation}
\begin{aligned}
\epsilon_{1}=& -\mathrm{i}A \frac{\beta}{2 q} \frac{J_{1}(h a)}{K_{1}(q a)}\left[(1-s) K_{0}(q a)-(1+s) K_{2}(q a)\right], \\
\epsilon_{2}=& -\mathrm{i}A \frac{\beta}{2 q} \frac{J_{1}(h a)}{K_{1}(q a)}\left[(1-s) K_{0}(q a)+(1+s) K_{2}(q a)\right], \\
\epsilon_{3}=& A J_{1}(h a).
\end{aligned}
\end{equation}

Based on dipole moment equations (\ref{eq:dipole moment}) and equations (\ref{eq:generated electric field}), we can get the amplitude of the electric field in cartesian coordinates ($x^{\prime}y^{\prime}z$).
In equation (\ref{eq:guided_field}), the amplitude $C=d\epsilon_{1}\sin \theta$, which is an imaginary number, and the amplitude $D=d\epsilon_{3}\cos \theta$, which is a real number. 
The guided electric field $\mathbf{E}$ can be described as:
\begin{equation}
    \mathbf{E}=d\epsilon_{1}\sin \theta\mathbf{H E}_{11}^{x^{\prime}}+d\epsilon_{3}\cos \theta\mathbf{H E}_{11}^{y^{\prime}}
\label{eq:guidedE}
\end{equation}

This equation gives the expression of guided polarization as a function of $\theta$. 
As we can conclude from the equation (\ref{eq:guidedE}), only the $x$ component of x-polarized light and $z$ component of y-polarized light emitted from the dipole on the nanofiber surface will be guided.

The normalised Stokes parameters of the guided light are given by:
\begin{equation}
\begin{split}
S_1 &= (|C|^2 - |D|^2)/S_0, \\
S_2 &= 2\ \text{Re} (C^*D)/S_0, \\
S_3 &= 2\ \text{Im} (C^*D)/S_0, 
\end{split}
\label{eq:8}
\end{equation}
where $S_0 = |C|^2 + |D|^2$, $C=d\epsilon_{1}\sin \theta$, $D=d\epsilon_{3}\cos \theta$. 
The degree of polarization is described with $\sqrt{S_1^2 + S_2^2 + S_3^2}$, it is equal to 1 for purely polarized light.

\begin{figure}
    \centering
    \includegraphics[width=0.7\linewidth]{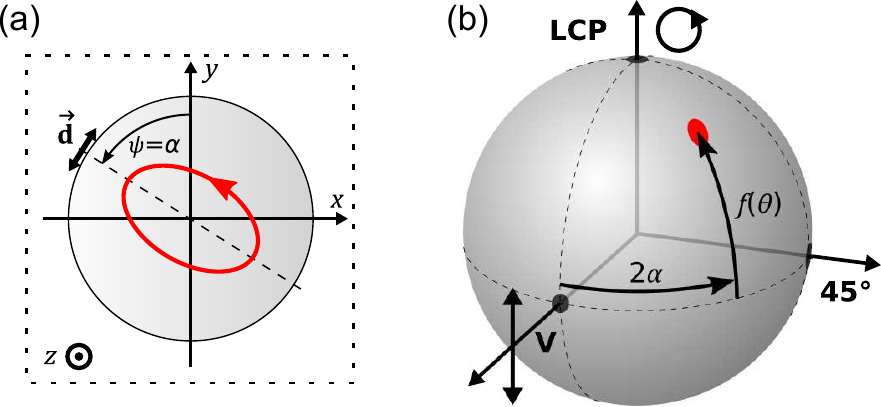}
    \caption{
    Mapping between dipole geometry and emitted polarization. (a) Cross-section of the nanofiber. The azimuth of the dipole $\alpha$ defines the orientation of the polarization ellipse $\psi=\alpha$. (b) Poincar\'e sphere. A dipole with geometrical parameters ($\alpha$,$\theta$) gives rise to a guided polarization represented by a point with coordinate (2$\alpha$,$f(\theta)$) on the Poincar\'e sphere. V (or $S_1$), $45^\circ$ (or $S_2$), and LCP (or $S_3$) axis represent vertically linearly, linearly at $\psi=45^\circ$, and left-circularly polarized light, respectively.
    }
    \label{fig:sphere}
\end{figure}

In Fig.\ref{fig:sphere}, we map the emitted polarization on a Poincar\'e sphere.
The Poincaré sphere is the parametrisation of the last three Stokes' parameters in spherical coordinates.
The latitude of the Poincar\'e sphere shows the ellipticity of guided polarization as the function of $\theta$. 
The longitude of the Poincar\'e sphere shows the information of fast axis of polarization with the angle $\alpha$.

\subsection{Particle deposition procedure}
\label{particle deposition}
In this section, we introduce in detail how we deposit single gold nanoparticle on the optical nanofiber experimentally.
Since the optical nanofiber is extremely fragile and the transmission is highly influenced by the unexpected dusts on the surface.
The deposition of single nanoparticles is realized with special technique using nanoparticle suspension.

\begin{figure}
    \centering
    \includegraphics[width=0.8\linewidth]{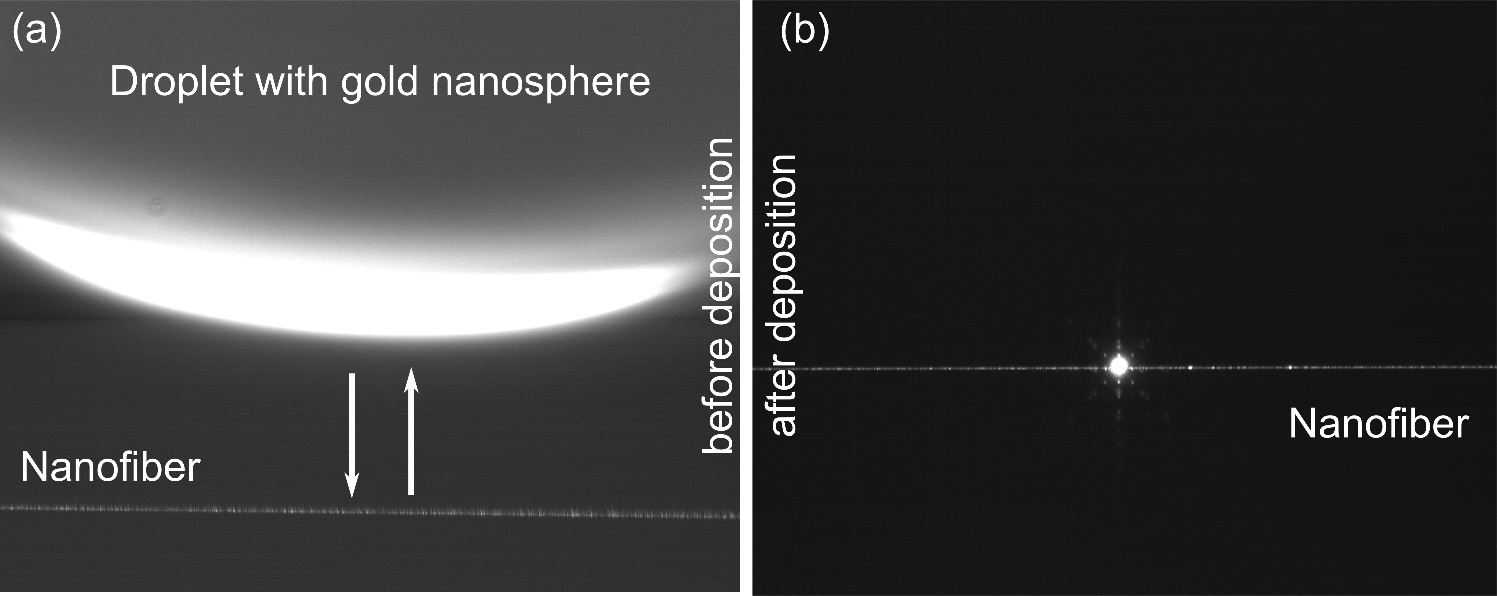}
    \caption{Nanoparticle deposition. a, before deposition. b, A gold nanosphere deposited on the optical nanofiber.}
    \label{fig:deposition}
\end{figure}

The experimental procedure is as following:

First, we prepare a highly diluted gold nanoparticle suspension in water.
The concentration of the nanoparticle suspension needs to be low enough to ensure that the number of possible nanoparticle deposited on the optical fiber for each contact is less than 1.
In practice, this technique effectively avoids depositing clusters.

Then, we take 5 $\mu$L of gold nanoparticle suspension in water from the evenly distributed suspension with a pipette tip (10 $\mu$L volume).
Leaving the sample in ultrasonic bath for ten minutes before deposition will be helpful to distribute the nanoparticles homogeneously in the suspension.
We carefully squeeze the nanoparticle suspension out of the pipette tip and create a droplet of about 3 mm in diameter.
The droplet needs to be big enough so that we can slightly tilt the pipette to move it to the side of the pipette tip.
And at the same time, it can not be too big so that due to the surface tension of the water, the small droplet can maintain its shape without dripping.

To precisely control the approaching of the droplet, we mount the pipette on a three-dimensional displacement stage.
We adjust the focus plane of the camera filming the nanofiber from the side to be on the fiber.
In this way, we can move the droplet to the same focus plane and see clearly the distance between curved droplet surface and the nanofiber.
As shown in Fig. \ref{fig:deposition}-a, the arc on the top of the image is the reflection from the edge of the droplet, and the thin line at the bottom is the scattering from the defects on the nanofiber.

By controlling the space between the droplet and the nanofiber, we move the droplet vertically towards the nanofiber.
The instant when the droplet touches the nanofiber is easily detectable, as light is strongly scattered at the contact.
Then, we quickly lift the droplet to disengage it from the nanofiber.

After several “attaching and disengaging" cycles, a single nanoparticle is deposited on the nanofiber.
On the camera, it appears to be a spot with strong scattering, as shown in Fig.\ref{fig:deposition}-b.
At the same time, the transmission drops about 20 \%.

\section*{Conclusion}
\addcontentsline{toc}{section}{Conclusion} 
Optical nanofibers with subwavelength diameter can be used as a platform for light matter interaction thanks to the transverse confinement of the evanescent field.

In this section, we first introduced the light guiding by optical nanofibers and the fabrication of optical nanofiber with adiabatic profile.

Then, we introduced the coupling between gold nanoparticles and optical nanofiber with surface evanescent field both in detail.
The scattering from gold nanoparticles exhibits dipolar like emission.
When gold nanoparticles coupled to an optical nanofiber as dipole, the guided light shows chiral property.

\chapter{Polarization control of linear dipole radiation using an optical nanofiber}
\markboth{POLARIZATION CONTROL OF LINEAR DIPOLE}{}
\label{chap:Polarisation}

Publication: Joos, M., Ding, C., Loo, V., Blanquer, G., Giacobino, E., Bramati, A., Krachmalnicoff, V. and Glorieux, Q. (2018). Polarization control of linear dipole radiation using an optical nanofiber. Physical Review Applied, 9(6), 064035.

In free space, a linear dipole will emit linearly polarized light. 
However, when we combine a linearly polarized dipole with a subwavelength waveguide, this can change dramatically. 
A subwavelength waveguide has a strong longitudinal component along the light propagation direction, therefore, the guiding of light shows totally property than in the free space.
This phenomena has been reported in our previous work \cite{joos2018polarization} by coupling a gold nanorod to a nanofiber waveguide, and also by Martin Neugebauer and all \cite{neugebauer2019emission} by coupling to  crossing waveguides.
In this chapter, we explain this mechanism and show that a linear dipole is not restricted to emit linearly polarized light, and our experiment provides the evidence that the design of nanophotonic environment will strongly modify the emission diagram and polarization.

\section{Linear dipole}
A nanoparticle has, by definition, much smaller dimensions than the wavelength of visible light with which it interacts which could naively suggest that its shape does not play an important role in phenomena such as absorption or diffusion. 
It is in fact otherwise; the optical response of metallic nanoparticles largely depends on their shapes and is significantly different from the macroscopic properties of the metal that constitutes them.

A rigorous treatment of the diffraction of a monochromatic plane wave by a metallic nanoparticle involves solving Maxwell's equations as well as the boundary conditions for the system studied. 
This tedious work presents analytical solutions only for a few simple geometries such as the cylinder, the sphere and the ellipsoid. 
In addition, this technique does not allow us to develop intuition for cases where there are no analytical solutions but which however interest us. 
Thus, we introduce a classical microscopic model, the Lorentz model \cite{saleh2019fundamentals, born1999principles}, to describe qualitatively the interaction of light with a metallic nanoparticle. In addition to fueling intuition, this model qualitatively predicts the optical properties of gold nanoparticles as we will present in this chapter.

\section{Birefringence effect in a single mode fiber}
Although in principle, a single mode fiber is cylindrically symmetric, it is visually birefringent. 
This birefringence comes from slight anisotropies of the system, which can be specific to the fiber and its manufacture - we then speak of intrinsic birefringence - or caused by the manipulation of the fiber - we then speak of extrinsic birefringence. 
Among the causes of intrinsic birefringence, we can cite the residual ellipticity of the fiber core or internal mechanical tensions. 
The extrinsic causes of birefringence are diverse: pressures, twists, curvatures of the fiber and the variation of temperature, etc.

Extrinsic birefringence is not so much a disadvantage for the experiments, but rather an advantage because it is thus possible, in an integrated manner, to control the polarization at the fiber output by applying the adequate constraints on the fiber. 
Many commercial solutions exist for the integrated polarization control in single-mode fibers. Depending on the nature of the anisotropy introduced, the birefringence may be linear, that is to say that the two transverse components E$_x$ and E$_y$ of the electric field are out of phase with one another, or circular, that is to say that the main axis of polarization rotates - we then speak of rotary power. 
It generally follows that the polarization of the light propagating in a standard single-mode fiber is not maintained. 

\section{Compensate birefringence in a single mode fiber}
Controlling the polarization at the nanofiber waist is a necessary step for controlling the coupling with the evanescent field.
If we want to control the polarization at the nanofiber waist, where the phase of light is already shifted due to the birefringence and rotary power of the single mode fiber, we need to find a way to compensate the phase shift introduced during light propagation from fiber input to nanofiber waist. 
The polarization of the propagating light is represented by Stokes vectors. 

Stokes vector is the combine of Stokes parameters ($S_0$, $S_1$, $S_2$ and $S_3$) that describe the polarization state of electromagnetic radiation.
\begin{equation}\vec{S}=\left(\begin{array}{l}
S_{0} \\
S_{1} \\
S_{2} \\
S_{3}
\end{array}\right)=\left(\begin{array}{l}
I \\
Q \\
U \\
V
\end{array}\right)
=\left(\begin{array}{c}
\text { Intensity } \\
1\left(0^{\circ}\right)-I\left(90^{\circ}\right) \\
I\left(45^{\circ}\right)-I\left(135^{\circ}\right) \\
I(\text{RCP})-I\text{(LCP})
\end{array}\right)
\end{equation}

\begin{figure}
    \centering
    \includegraphics[width=0.8\linewidth]{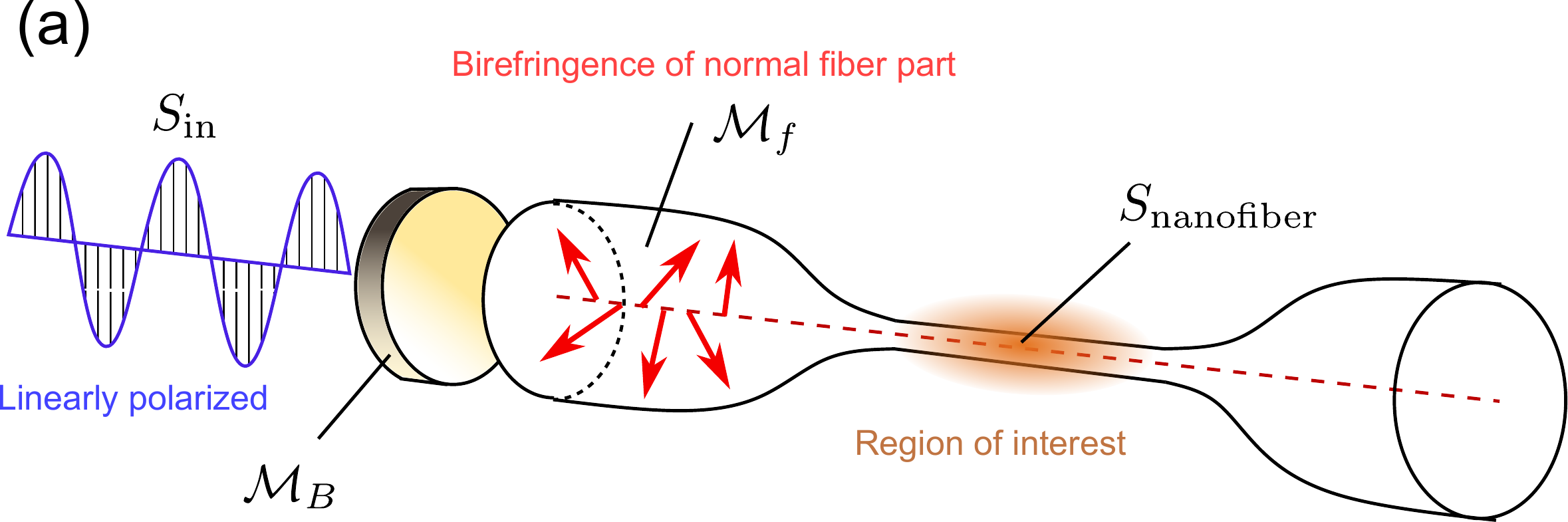}
    \caption{Schematic of the polarization transformation in the system composed of a Berek compensator and a tapered optical fiber. $\mathcal{M}_{B}$ and $\mathcal{M}_{f}$ describes the birefringence and optical rotation induced by a Berek compensator and an arbitrary bent optical fiber respectively. The polarization at the input and at the nanofiber part are written as $\bm{S}_\text{in}$ and $\bm{S}_\text{nanofiber}$.}
    \label{fig:fiber_birefringence}
\end{figure}

Let's assume that the Stokes vector of the light at fiber input is $\bm{S}_\text{in}$, the Stokes vector of the light at nanofiber is $\bm{S}_\text{nanofiber}$. 
We describe the transformation of polarization in the system composed of the fiber and the Berek compensator with Stokes-Müller formalism \cite{olivard1999measurement}:
\begin{equation}
    \bm{S}_\text{nanofiber}=\mathcal{M}_{f}\mathcal{M}_{B}\bm{S}_\text{in}
\end{equation}
The matrix $\mathcal{M}_{f}$ describes the birefringence and optical rotation induced by an arbitrary bent optical fiber. The birefringence will change the ellipticity of the polarization and the optical rotation will rotate the polarization axes.
\begin{equation}
\mathcal{M}_{f}=\mathcal{S}(\theta) \mathcal{G}(\delta, \phi)
\end{equation}
where $\mathcal{S}(\theta)$ is the rotation matrix:
\begin{equation}
\mathcal{S}(\theta)=\left(\begin{array}{cccc}
{1} & {0} & {0} & {0} \\
{0} & {\cos (2 \theta)} & {-\sin (2 \theta)} & {0} \\
{0} & {\sin (2 \theta)} & {\cos (2 \theta)} & {0} \\
{0} & {0} & {0} & {1}
\end{array}\right)
\end{equation}
and $\mathcal{G}(\delta, \phi)$ describes a wave retarder with phase shift $\delta$ and polarization axes at angle $\phi$:
\begin{equation}
\mathcal{G}(\delta, \phi)=\mathcal{S}(\phi)\left(\begin{array}{cccc}
{1} & {0} & {0} & {0} \\
{0} & {1} & {0} & {0} \\
{0} & {0} & {\cos (\delta)} & {-\sin (\delta)} \\
{0} & {0} & {\sin (\delta)} & {\cos (\delta)}
\end{array}\right) \mathcal{S}(-\phi)
\end{equation}

Based on the equations above, the birefringence and optical rotation induced by an arbitrary bent optical fiber can be described with three parameters: the rotation angle $\theta$ of the rotator, the phase shift $\delta$ and polarization axes angle $\phi$ of the retarder.

The goal of the compensation procedure is to induce a birefringence opposite to the equivalent retarder represented by $\mathcal{G}(\delta, \phi)$. 
Here, a Berek compensator was used. 
The Berek compensator is an optical device that is capable of quantitatively determining the wavelength retardation of fiber.
By adjusting the two degrees of freedom of the Berek compensator, we will be able to generate a birefringence $\mathcal{M}_{B}=\mathcal{G}(\delta, \phi)^{-1}$.

In this way, the whole system can be simplified into a rotater with the rotation angle $\theta$:
\begin{equation}
\mathcal{M}_{f} \mathcal{M}_{B}=\mathcal{M}_{f} \mathcal{G}(\delta, \phi)^{-1}=\mathcal{S}(\theta)
\end{equation}

To find the setting of the Berek compensator which compensates the birefringence of the fiber system, it is necessary to be able to compare the polarization at the fiber input with that in the nanofiber.

It is possible to obtain partial information on the polarization in the nanofiber region by studying the diffusion by the nanofiber of the guided light. 
Observation under an optical microscope suggests that the scattering originates from defects (localized impurities or inhomogeneities of the refractive index) whose extent is much less than the observation wavelength. 
This observation is consistent with the assumption generally accepted in the literature \cite{vetsch2012nanofiber} according to which the diffusion is of Rayleigh type, and thus preserves the polarization of the incident beam. 
We can therefore consider each point diffuser as a local dipole probe of the fundamental mode of the nanofiber.

\begin{figure}
    \centering
    \includegraphics[width=0.5\linewidth]{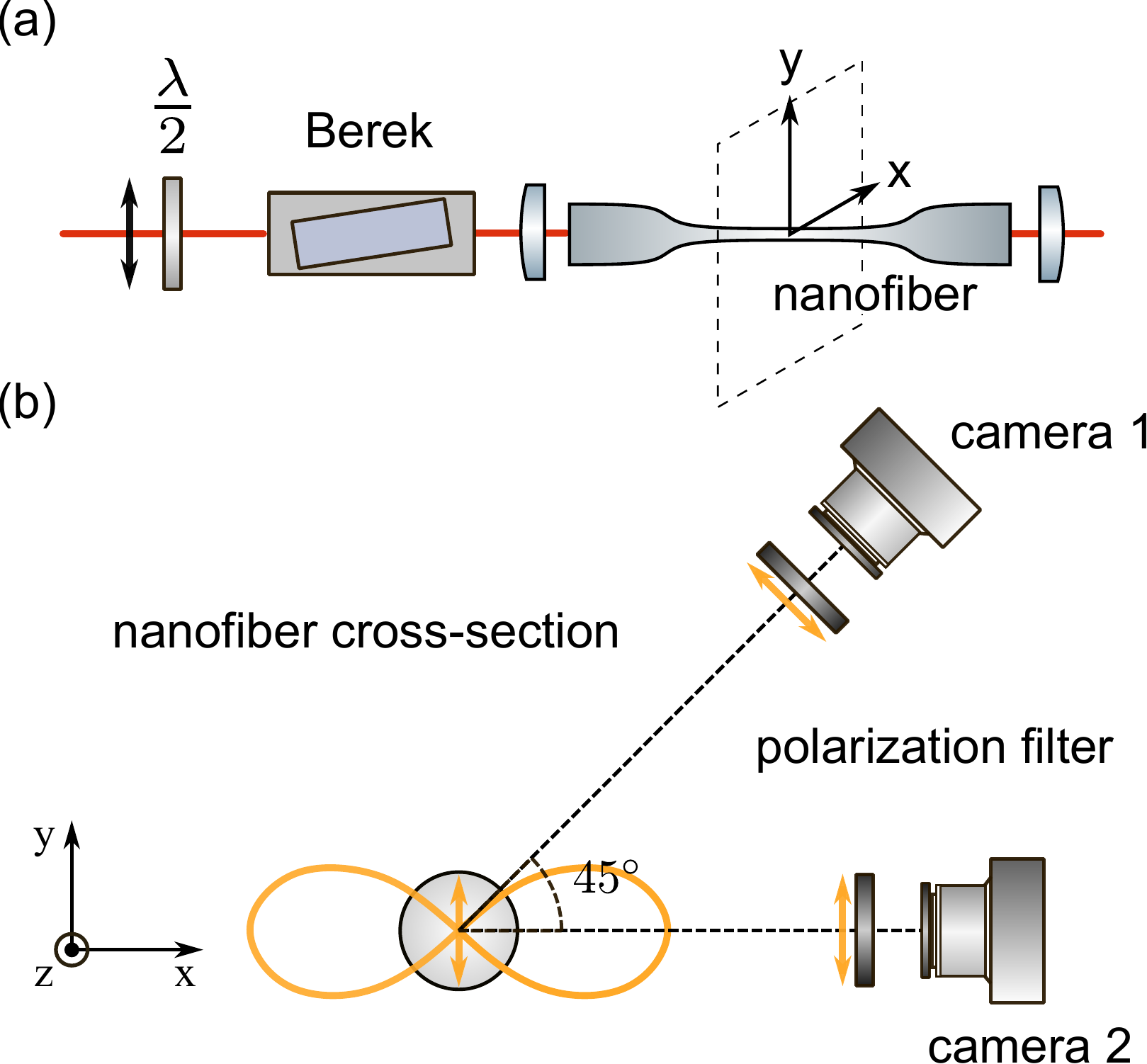}
    \caption{Birefringence compensator. (a) Experiment setup for the Berek compensator. (b) Nanofiber cross-section. Assuming a vertical polarization in the nanofiber, the dipole radiation (Rayleigh) of the silica defects is detected with two cameras from two different orientations with polarization filters.}
    \label{fig:birefringence_compensator}
\end{figure}

The diffusion of the nanofiber is detected with two cameras from two different orientations, as shown in Fig.\ref{fig:birefringence_compensator}. 
One is installed horizontally, another is observing the image of the same region with 45$^{\circ}$ angle.
Both of the cameras are equipped with polarizing filters allowing only the sagittal component of the diffusion to be detected.

To compensate the birefringence, we send a linearly polarized light at the fiber entrance. The orientation of the polarization can be rotated with a half-waveplate, as shown in Fig.\ref{fig:birefringence_compensator}(a). The dipole radiation of the silica defects is detected with two cameras installed horizontally ($I_1$ for camera 1) and with a 45$^{\circ}$ angle ($I_2$ for camera 2), as shown in Fig.\ref{fig:birefringence_compensator}(b).
The detected dipole radiation $I_{1,2}$ oscillates following the polarization orientation at the fiber entrance with a period of $\pi$ respectively.
The phase difference in between is defined as $\phi$ corresponding to the angle between the orientations of the two cameras.
We define the visibility of polarization at camera 1 and camera 2 as:
\begin{equation}
V_{1,2}=\frac{I_{1,2}^{\max }-I_{1,2}^{\min }}{I_{1,2}^{\max }+I_{1,2}^{\min }}
\end{equation}

To induce a birefringence opposite to the one induced by optical fiber, the Berek compensator needs to be finely adjusted with its two degrees of freedom. 
The appropriate setting is confirmed by checking the curve of $I_{1,2}$ plotted as a function of polarization orientation at the region of interest at the nanofiber part and also the polarization at the fiber entrance to meet following requirements:
\begin{itemize}
    \item The visibility of the two cameras ($V_{1,2}$) both reach their maximum (ideal visibility is 1).
    \item The polarization in the nanofiber rotates in the same direction as at the input of the system.\\
    In Fig.\ref{fig:birefringence_compensator}-b, if the polarization at the fiber entrance rotates clockwise, the phase shift $\phi$ should be about 45$^{\circ}$.
\end{itemize}

\section[Dipole coupling]{Relation between a linear dipole on nanofiber and the guided polarization}

\begin{figure}
    \centering
    \includegraphics[width=0.8\linewidth]{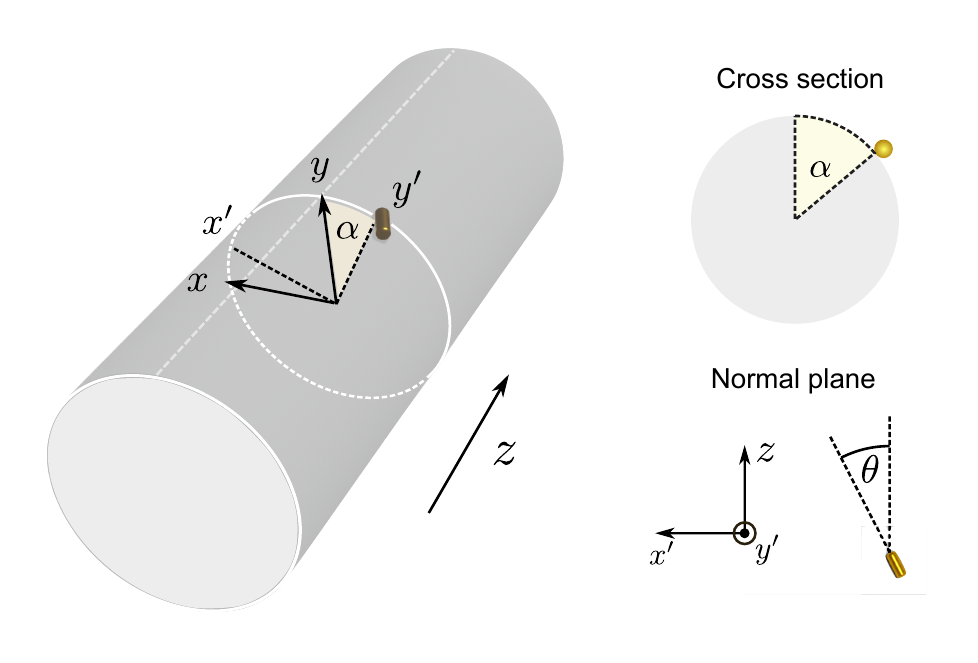}
    \caption{Coordinates definition of a gold nanorod (linear dipole) on the nanofiber. The azimuthal position of the dipole is defined with angle $\alpha$ between vertical axis $y$ and normal vector axis $y^\prime$. The orientation of dipole relative to the fiber axis $z$ on the normal plane is defined with angle $\theta$.}
    \label{fig:fiber}
\end{figure}

In Fig.\ref{fig:fiber}, we define a section of the optical nanofiber with cartesian coordinates ($x$, $y$, $z$), where in the perpendicular plane, $x$ defines the horizontal direction, $y$ defines the vertical direction and $z$ defines the longitudinal direction of the nanofiber. The dipole (gold nanorod) is included in a plane tangent to the surface of the fiber which limits its degrees of freedom to two: the angle $\alpha$ defines its azimuthal position around the fiber and the angle $\theta$ defines its orientation relative to the fiber axis.

We introduced a new cartesian coordinates ($x^{\prime}$, $y^{\prime}$, $z$) by rotating the cartesian coordinates ($x$, $y$, $z$) by an angle $\alpha$.
The guided electric field $\mathbf{E}$ is described on the basis of quasi-linear modes $\left\{\mathbf{H E}_{11}^{x^{\prime}}, \mathbf{H E}_{11}^{y^{\prime}}\right\}$. 

\begin{equation}
    \mathbf{E}=C\mathbf{H E}_{11}^{x^{\prime}}+D\mathbf{H E}_{11}^{y^{\prime}}
    \label{eq:guided_field_2}
\end{equation}


The $\mathbf{E}_{x}$, $\mathbf{E}_{y}$, $\mathbf{E}_{z}$ electric field outside of an optical nanofiber with radius $a$ are given by \cite{le2004field}:
\begin{equation}
    \begin{aligned}
    \mathbf{E}_{x}=&-\mathrm{i} A \frac{\beta}{2 q} \frac{J_{1}(h a)}{K_{1}(q a)}\left[(1-s) K_{0}(q r) \cos \varphi_{0}+(1+s) K_{2}(q r) \cos \left(2 \varphi-\varphi_{0}\right)\right] \mathrm{e}^{\mathrm{i}(\omega t-\beta z)} ,\\
    \mathbf{E}_{y}=&-\mathrm{i} A \frac{\beta}{2 q} \frac{J_{1}(h a)}{K_{1}(q a)}\left[(1-s) K_{0}(q r) \sin \varphi_{0}+(1+s) K_{2}(q r) \sin \left(2 \varphi-\varphi_{0}\right)\right] \mathrm{e}^{\mathrm{i}(\omega t-\beta z)} ,\\
    \mathbf{E}_{z}=& A \frac{J_{1}(h a)}{K_{1}(q a)} K_{1}(q r) \cos \left(\varphi-\varphi_{0}\right) \mathrm{e}^{\mathrm{i}(\omega t-\beta z)} .
\end{aligned}
\label{eq:electric_field_2}
\end{equation}

Here $s=\left[(q a)^{-2}+(h a)^{-2}\right] /\left[J_{1}^{\prime}(h a) / h a J_{1}(ha)+K_{1}^{\prime}(q a) / q a K_{1}(q a)\right]$, $h=\sqrt{n_{1}^{2} k^{2}-\beta^{2}}$ and $q=\sqrt{\beta^{2}-n_{2}^{2} k^{2}}$. 
$J_n$ and $K_n$ are the Bessel functions of the first kind and the modified Bessel functions of the second kind, respectively, and the prime stands for the derivative.

The electric field of air-cladding nanofiber is described in the cylindrical coordinates ($r$, $\varphi$, $z$). $\varphi_{0}$ defines the polarization orientation of the guided light: $\varphi_{0}=0$ for $x$-polarized light and $\varphi_{0}=\pi/2$ for $y$-polarized light. 

As we introduced in the "Coupling between dipole to the evanescent field",
the guided electric field $\mathbf{E}$ can be described as:
\begin{equation}
    \mathbf{E}=d\epsilon_{1}\sin \theta\mathbf{H E}_{11}^{x^{\prime}}+d\epsilon_{3}\cos \theta\mathbf{H E}_{11}^{y^{\prime}},
\label{eq:guidedE_2}
\end{equation}
which gives the expression of guided polarization as a function of $\theta$.

\subsection{Role of the dipole orientation $\theta$}
\begin{figure}
    \centering
    \includegraphics[width=0.8\linewidth]{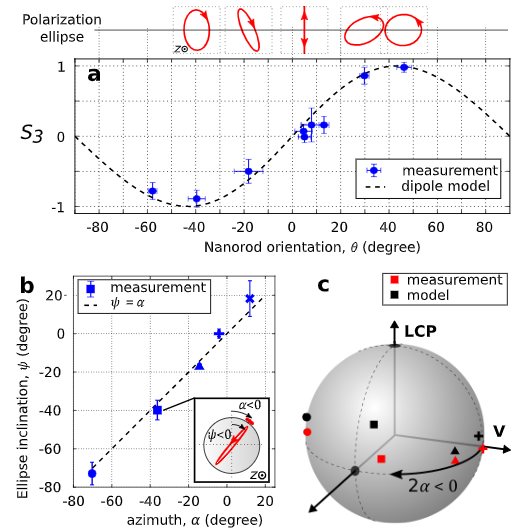}
    \caption{
    Measured state of polarization as a function of the orientation $\theta$ and the azimuth $\alpha$ of the nanorod. (a) Ellipticity as a function of $\theta$. (Graph) Measured Stokes parameter $S_3$ for nine different nanorods compared to the expected dipole model. We plot the measured polarization ellipse for characteristic nanorods. 
    (b) Measured ellipse orientation $\psi$ for five nanorods with different azimuths. (Inset) Cross-section of the nanofiber showing the azimuth of a specific nanorod and the associated polarization ellipse. 
    (c) Poincar\'e sphere representation of the measured polarization for four nanorods from (b).
    }
    \label{fig:polarization}
\end{figure}

The situation presented above evokes another optical device: a quarter-wave plate acting on a linearly polarized beam. 
In the latter case, the amplitude of the components in phase quadrature is determined by the angle between the incident polarization and the optical axis of the plate. 
In the case of our system, the amplitude of the components in phase quadrature is given by the angle between the dipole and the nanofiber.
Thus, in the same way as a quarter-wave plate, it is possible to control the ellipticity of a beam. 
The orientation of the dipole defines the ellipticity of the transverse polarization of the guided light. 
If the azimuthal position of the dipole is known, there is a bijection between the angle $\theta$ and the state of the polarization of the guided light. 
It is possible to represent this relation in different ways: 
In Fig.\ref{fig:polarization}a, we calculated the Stokes parameter $S_3$ characterizing the ellipticity as a function of the orientation of the dipole. 
We see that for an orientation of the dipole $\theta \in [-\theta_{circ}, \theta_{circ}]$, the ellipticity is unique. 
Equivalently, on the Poincar\'e sphere the orientation of the dipole defines the latitude that we have noted $f(\theta)$ and which is approximately equal to:
\begin{equation}
f(\theta) = \arcsin S_3 \approx \frac{90^\circ}{\theta_{circ}}\theta \approx 2\theta.
\end{equation}

According to the orientation angle, we can divide the coupling of the dipole and the nanofiber into several representative cases:
\begin{itemize}
    \item ($\theta=0^\circ$, $f(\theta)=0^\circ$)~~~~~~ The dipole is aligned with the nanofiber (quasi-linearly polarized along $x'$ axis);
    \item ($\theta=\pm 45^\circ$, $f(\theta)=\pm 90^\circ$)~~~~~~ The dipole is vertical to the nanofiber (left or right circularly polarized);
    \item ($\theta=\pm 90^\circ$, $f(\theta)=\pm 180^\circ$)~~~~~~ The dipole is aligned with the nanofiber (quasi-linearly polarized along $y'$ axis);
    \item The orientation of dipole lays in between (elliptically polarized).
\end{itemize}



\subsection{Role of the dipole azimutal position $\alpha$}

Let us now consider the second degree of freedom of the dipole defined by the azimuthal position around the fiber.

For reasons of symmetry, the position of the dipole has no influence on the ellipticity of the polarization. 
The only effect of the position of the dipole is the rotation of the coordinate system which implies an identical rotation of the polarization. 
The azimuthal position $\alpha$ of the dipole therefore defines the inclination $\psi$ of the polarization ellipse as illustrated in Fig.\ref{fig:polarization}b. 
Equivalently, on the Poincaré sphere, the azimuthal position defines the longitude as shown in Fig.\ref{fig:polarization}c.
Thus, the two degrees of freedom of the dipole $\theta$ and $\alpha$ act independently on the ellipticity and the inclination of the polarization and allow in theory to generate in a deterministic way all the possible polarizations.

We measure the nanorod orientation $\theta$ with a scanning electronic microscope (SEM) image.
This measurement was done by collecting the nanofiber part on a silicon wafer after the polarization measurement was finished.

\subsection{Experimental approach}

\begin{figure}
    \centering
    \includegraphics[width=0.6\linewidth]{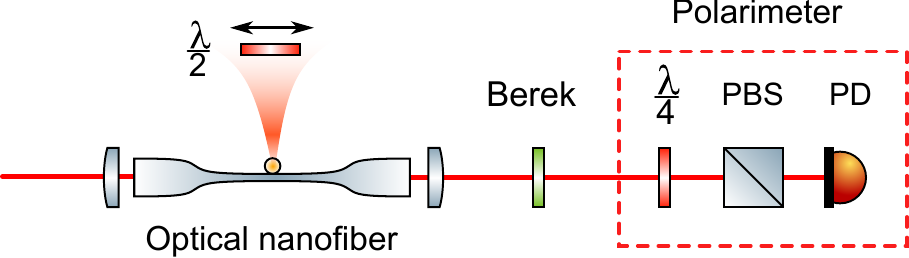}
    \caption{Experimental setup. A single gold nanorod is deposited on the surface of an optical nanofiber. A focused laser beam illuminates the particle from the top of the nanofiber and scatters light in the fundamental mode of the nanofiber. The guided light is then analyzed with a polarimeter allowing to measure Stokes parameters.}
    \label{fig:Experiment_dipole}
\end{figure}

The experimental setup is shown in Fig.\ref{fig:Experiment_dipole}.
The demonstration of the dependence between the nanorod geometry and the polarization of the guided light requires an ensemble of nanorods with various different positions and orientations.
The guided polarization state of each nanorod is analyzed with the following procedure:

First, we deposit a gold nanorod on the surface of the air-cladding optical nanofiber.

Second, as we introduced in section 4.3, we use the Berek compensator to compensate the birefringence induced by the optical fiber between the particle and the polarimeter system.

Third, we illuminate the particle with a focused laser beam from the top of the nanofiber and align the excitation polarization along the nanorod.

Forth, we measure Stokes parameters of the guided light with a polarimeter system.

The second particle is deposited next to the first one.
Since there are already particles on the same nanofiber, the guided polarization state of the nanorod under investigation can be influenced by the other particles on the way to the detection system.
To prevent this issue, each new particle is deposited on the nanofiber between the previously deposited one and the detection system. 
Besides, sufficient space between the different particles is needed for the excitation beam to address nanorods individually.
This ensures that the light emitted by the last deposited nanorod encounters no obstacles when guided towards the detection system.
We also record the exact relative positions of the particles with respect to each other. 
This is meant to facilitate the finding of the nanorods when we ultimately observe the nanofiber with a SEM.

On the SEM, we identify the various depositions, disregard the clusters or the rods with odd shape to record only the azimuth $\alpha$ and the orientation $\theta$ of \textit{proper} single nanorods. The orientation is determined from analyzing SEM images.

The Müller matrix was measured with an optical polarimeter.
The output polarization decoding system is composed of a quarter-wave plate, followed by a polarized beamsplitter and a photon detector.


\begin{figure}
    \centering
    \includegraphics[width=0.8\linewidth]{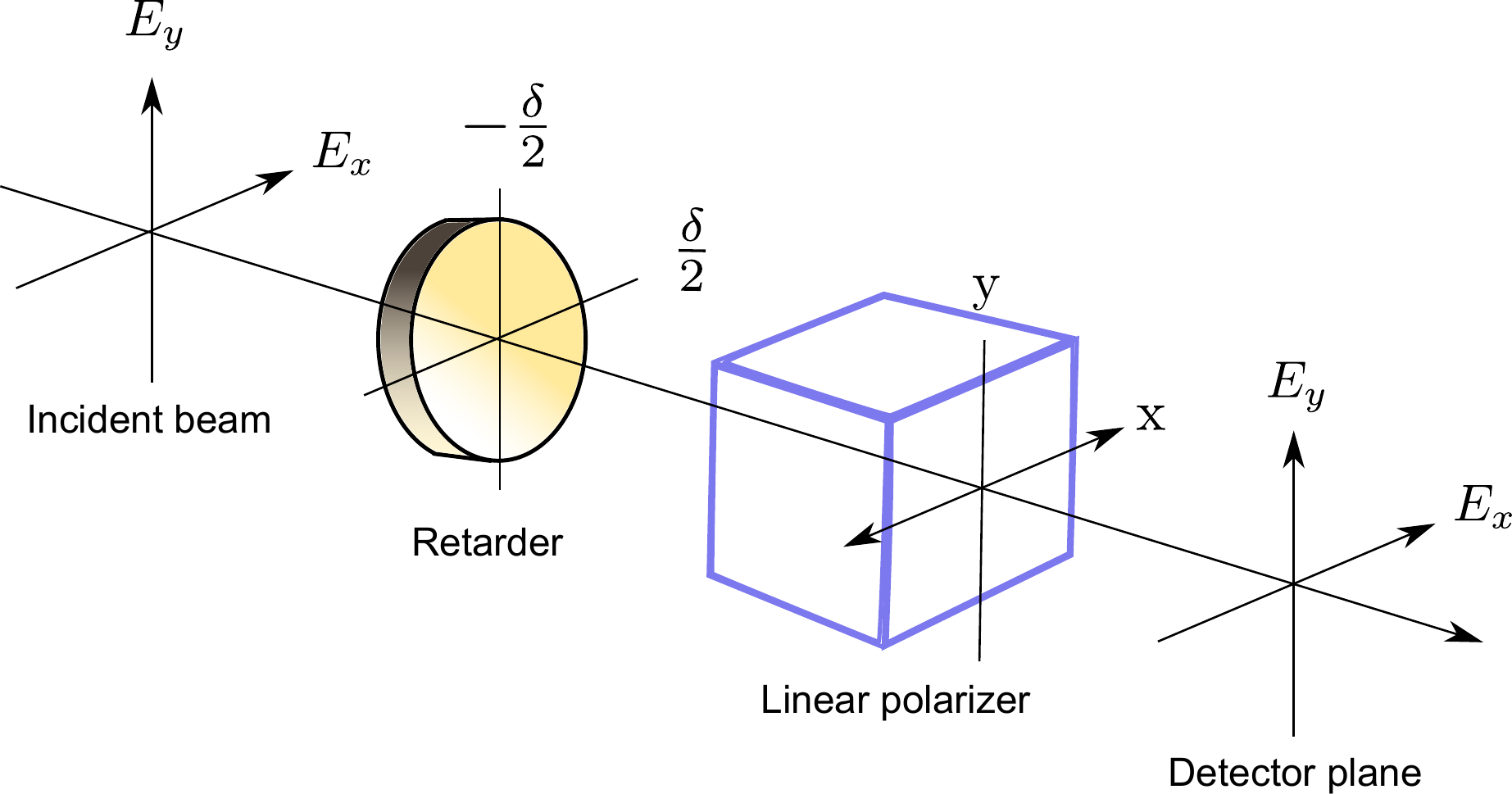}
    \caption{Stokes parameters measurement with polarimeter.}
    \label{fig:polarimeter}
\end{figure}

Assuming the Stokes vector of this beam is represented by:
\begin{equation}\vec{S}=\left(\begin{array}{l}
S_{0} \\
S_{1} \\
S_{2} \\
S_{3}
\end{array}\right)
\label{eq:S_incident}
\end{equation}

The Müller matrix of a retarder with its fast axis at 0$^\circ$ is:

\begin{equation}\mathbf{M}=\left(\begin{array}{cccc}
1 & 0 & 0 & 0 \\
0 & 1 & 0 & 0 \\
0 & 0 & \cos \delta & \sin \delta \\
0 & 0 & -\sin \delta & \cos \delta
\end{array}\right)
\label{eq:M_retarder}
\end{equation}

The Stokes vector of the beam after the retarder is obtained by multiplication of Equations (\ref{eq:S_incident}) and (\ref{eq:M_retarder}), which becomes:
\begin{equation}\mathbf{S}^{\prime}=\left(\begin{array}{c}
S_{0} \\
S_{1} \\
S_{2} \cos \delta+S_{3} \sin \delta \\
-S_{2} \sin \delta+S_{3} \cos \delta
\end{array}\right)\end{equation}

The beam then passes through a linear horizontal polarizer.
The measured intensity $I(\delta)$ at the detector as a function of the rotation angle $\delta$ of the retarder is calculated using Mueller calculus:
\begin{equation}I(\delta)=\frac{1}{2} \cdot[A_1-A_2 \cdot \sin (2 \delta)+A_3 \cdot \cos (4 \delta)+A_4 \cdot \sin (4 \delta)] , \end{equation}
where we obtain the Stokes parameters: 
$$\begin{array}{l}
S_{0}=A_1-A_3 ,\\
S_{1}=2 A_3 ,\\
S_{2}=2 A_4 ,\\
S_{3}=A_2.
\end{array}$$

$S_0$ is equivalent to the total intensity of a light beam. 
$S_1$ is the amount of linearly horizontal or vertical polarization.
$S_1$ is the amount of linearly polarization at $\pm 45^\circ$.
$S_3$ describes the ellipticity of the guided polarization state. 

We deposited and tested several nanorods with various positions and orientations on the nanofiber surface.
In Fig.\ref{fig:polarization}(a), we plot the value of $S_3$ as a function of $\theta$ and its corresponding polarization ellipse.
In Fig.\ref{fig:polarization}(b), we plot the measured ellipse orientation $\psi$ for five nanorods with different azimuths $\alpha$.
In Fig.\ref{fig:polarization}(c), we plot the depiction of the guided polarization states for different nanorods on Poincar\'e sphere.

\section*{Conclusion}
\addcontentsline{toc}{section}{Conclusion} 
\begin{itemize}
    \item We have experimentally demonstrated the capacity for a linear dipole located in an evanescent field to emit an elliptical polarization. 
    For example, circularly polarized light can be emitted by a linear dipole.
    \item We can compensate the effect of birefringence and preserve the polarization from a coupled dipole.
    \item We have thus shown that there is a deterministic relationship between the geometry of the dipole / nanorod and the polarization emitted:\\
    - The orientation of the dipole / nanorod defines the ellipticity of the polarization of the guided light.\\
    - The azimuthal position controls the main axis of polarization.
\end{itemize}

In the future, if we can control the rod orientation and azimuthal position, we can control the dipole polarization, since we realized the mapping between the dipole geometry and the guided polarization.

\chapter{Nanofiber interferometer and resonator}
\markboth{POLARIZATION CONTROL OF LINEAR DIPOLE}{}
\label{chap:Nanofiber_interferometer_and_resonator}
Publication: Ding, C., Loo, V., Pigeon, S., Gautier, R., Joos, M., Wu, E., Giacobino, E., Bramati, A. and Glorieux, Q. (2019). Fabrication and characterization of optical nanofiber interferometer and resonator for the visible range. New Journal of Physics, 21(7), 073060. 

\section{Introduction}

Due to their low losses, optical fibers are undoubtedly a medium of choice to transport optical information, making them critical to current telecommunication networks and to the future quantum internet \cite{kimble2008quantum}.
However collecting a specific state of light, for example a single photon in a fiber with a good coupling efficiency is not an easy task. 
A typical way to couple light into a fiber is to place  the emitter directly at one end of the fiber with or without additional optical elements \cite{albrecht2013coupling}.
An alternative approach recently raised significant interest, by collecting light from the side of the fiber \cite{nayak2007optical,le2004atom}. 
Indeed, stretching down the fiber diameter to the wavelength scale allows for a coupling between the fiber guided mode and an emitter in its vicinity \cite{ding2010ultralow}. 
In such a nanofiber, the fundamental propagating mode has a significant evanescent component at the glass/air interface, which allows for interacting with emitters on the surface \cite{yalla2012efficient,schroder2012nanodiamond,vetsch2010optical,nayak2007optical,le2004atom,goban2012demonstration,joos2018polarization}, as described in chapter 1.

Collection efficiency is limited so far to 22.0~$\pm$~4.8\% for a bare nanofiber \cite{yalla2012efficient}. 
Maximizing this coupling is a challenging task as it requires simultaneously a fine-tuning of the fiber size and the largest possible cross-section for the emitter. 
To render this "injection by the side" technique more attractive, the collection efficiency has to be increased.
One approach to do so is to enhance the effective light-matter interaction. 
It is commonly done by reducing the mode volume using confined modes of the electromagnetic field rather than propagating modes. 
It leads, via the Purcell effect,  to an increase of the spontaneous emission within the nanofiber confined mode and therefore to an increase of the emitter-fiber coupling \cite{solano2017optical}.  
A detailed model predicts more than 90\% collection efficiency if one adds an optical cavity of moderate finesse to the nanofiber \cite{solano2017optical}. 
Diverse strategies have been investigated to do so.
 One is to fabricate two mirrors directly in the fiber to add a Fabry-Perot cavity within the nanofiber itself \cite{nayak2014optical}.
This strategy requires advanced nanofabrication methods such as femtosecond laser ablation to modify the fiber index. 
Using a Talbot interferometer, it has been possible to fabricate two fiber Bragg gratings and form an optical cavity with a transmission of 87\% for a finesse of 39 \cite{nayak2014optical}. 
A similar strategy, called nanofiber Bragg cavity, where a focused ion beam mills the nanofiber to create mirrors has shown a Purcell factor and coupling efficiency of 19.1 and 82\% respectively \cite{takashima2016detailed,schell2015highly}. 
Another solution relies on coupling the nanofiber with a whispering gallery mode resonator with very high quality factor up to $10^9$ \cite{cai2000observation,aoki2006observation}. 
With this strategy, at the difference of previous ones, the cavity is exterior to the nanofiber. 

\section{\label{sec:level1}Effective coupling theory approach}

The manufacturing of nanofibers is a well-controlled process, and it is possible to fabricate fibers with a diameter down to 200 nm \cite{ward2014contributed,ding2010ultralow}, see chapter 1. 
At this size, only the core of the fiber remains, and the surrounding air acts as a cladding. Consequently, there is a strong evanescent field extending around the surface of the nanofiber.
The fundamental mode does not correspond anymore to the standard linearly polarized mode LP$_{01}$. Nevertheless, using Maxwell's equations the correct propagating mode profile can be precisely characterized \cite{le2004field}.
We will consider single mode air-cladding nanofibers only, that is, nanofibers in which the fundamental mode HE$_{11}$ is the only propagating solution \cite{le2004field}. This is the case if the normalized frequency $V$ with $V\equiv ka\sqrt{n^2-1}$ is lower than the cutoff normalized frequency $V_{c}=2.405$, where $k$ is the wavevector, $a$ is the fiber radius, and $n$ is the fiber index \cite{le2004field}.

\begin{figure}[htbp]
\centering\includegraphics[width=0.8\linewidth]{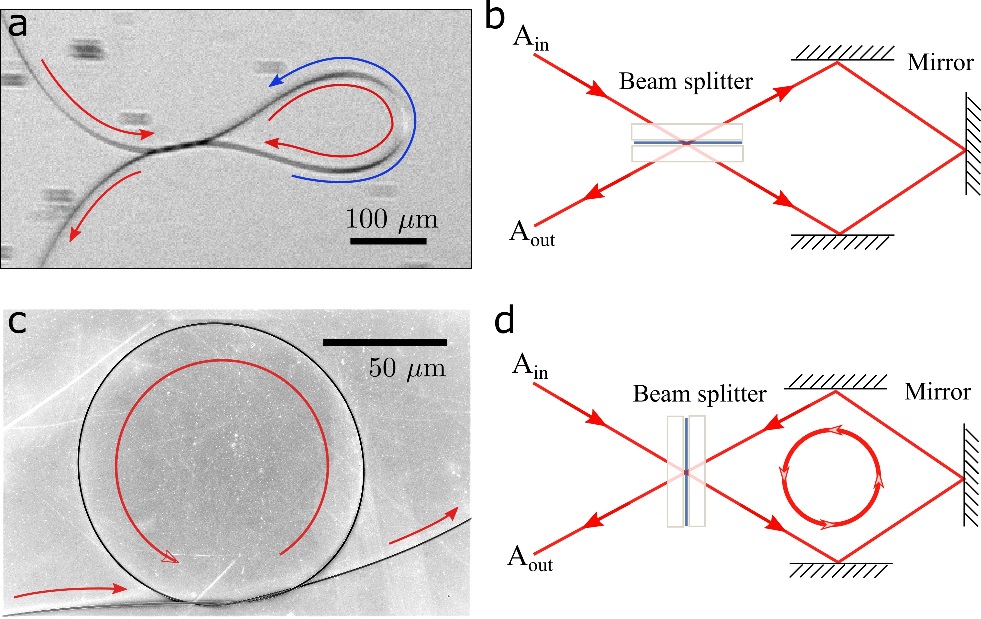}
\caption{Optical nanofiber structures.
a- Nanofiber twisted loop: optical microscope image. 
b, Sagnac interferometer equivalent optical setup: the light emerging from the port $A_{\text{out}}$ consists of two reflections or two transmissions of the light from incident port $A_{\text{in}}$ through the beamsplitter. 
c- Nanofiber knotted loop: scanning electron microscope image. 
d- Fabry-Perot ring resonator equivalent optical setup: light coming from $A_{\text{in}}$ that is not directly reflected to $A_{\text{out}}$ by the beam splitter, is trapped in the cavity formed by the beam splitter and the mirrors.}\label{figure3-1}
\end{figure}

We have bent and twisted manually such nanofibers with great care to realize two miniaturized optical devices: a fiber loop and a fiber knot (Fig. \ref{figure3-1}).
The common feature of these two structures is that they both present a section where the two parts of the nanofiber touch each other as shown on Fig. \ref{figure3-1}. 
However, fiber knots and fiber loops are topologically distinct, as it will be detailed later.

\begin{figure}[htbp]
\centering\includegraphics[width=0.4\linewidth]{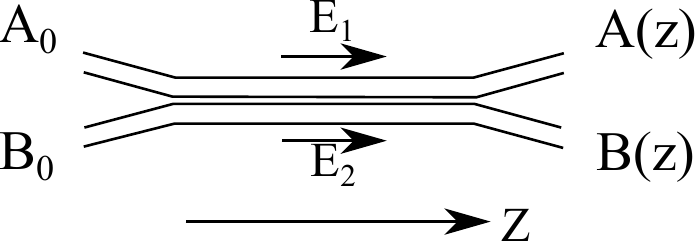}
\caption{Schematic of codirectional couplers. $A_0$ and $B_0$ are the amplitude of electric fields at the input of the two fibers. The outputs of each fiber are labelled $A(z)$ and $B(z)$.}
\label{figure0}
\end{figure}

Given the strong evanescent field of the propagating mode, the contact between different nanofiber regions leads to a coupling between the HE$_{11}$ propagating modes. 
To model this coupling, let us consider two parallel nanofibers nearby to each other, as represented in Fig.~\ref{figure0}.
 
Assuming the eigen modes of the waveguides before coupling are ${\mathbf{E}}_{p}$ and $\mathbf{H}_{p}$ where $p=1~\text{or}~2$ corresponding to the specific waveguide. According to Maxwell's equations, we have:
 \begin{equation}
 \left\{\begin{array}{l}
\nabla \times \tilde{\mathbf{E}}_{p}=-j \omega \mu_{0} \tilde{\mathbf{H}}_{p} \\
\nabla \times \tilde{\mathbf{H}}_{p}=j \omega \varepsilon_{0} N_{p}^{2} \tilde{\mathbf{E}}_{p}
\end{array}\right.
\end{equation}
where $N_{p}^{2}$ represents the refractive-index distribution of each waveguide.

 The electromagnetic fields of the coupled waveguide 1 and waveguide 2 can be expressed as the sum of the eigen modes in each waveguide:
\begin{equation}
\left\{\begin{array}{l}
\tilde{\mathbf{E}}=A(z) \tilde{\mathbf{E}}_{1}+B(z) \tilde{\mathbf{E}}_{2} \\
\tilde{\mathbf{H}}=A(z) \tilde{\mathbf{H}}_{1}+B(z) \tilde{\mathbf{H}}_{2}
\end{array}\right.
\label{eq:5.2}
\end{equation}
 
 The electromagnetic fields in the coupled waveguide also satisfy Maxwell's equation:
\begin{equation}
\left\{\begin{array}{l}
\nabla \times \tilde{\mathbf{E}}=-j \omega \mu_{0} \tilde{\mathbf{H}} \\
\nabla \times \tilde{\mathbf{H}}=j \omega \varepsilon_{0} N^{2} \tilde{\mathbf{E}}
\end{array}\right.
\label{eq:5.3}
\end{equation}
where $\omega=kc$ is the angular frequency, $k$ is the wavenumber and $c$ is the speed of light. 

We do the curl at both side of \ref{eq:5.2}, and got the following relation:
\begin{equation}
\nabla \times(A(z) \tilde{\mathbf{E}_{1}})+\nabla \times(B(z) \tilde{\mathbf{E}_{2}})=\nabla \times(A(z) \tilde{\mathbf{E}}).
\end{equation}

Together with the vector formula:
\begin{equation}\nabla \times(A(z) \tilde{\mathbf{E}})=A(z) \nabla \times \tilde{\mathbf{E}}+\nabla A(z) \times \tilde{\mathbf{E}}=A(z) \nabla \times \tilde{\mathbf{E}}+\frac{d A}{d z} \mathbf{u}_{z} \times \tilde{\mathbf{E}},\end{equation}
and Maxwell's equation, we get:
\begin{equation}
    \frac{d A}{d z} (\mathbf{u}_{z} \times \tilde{\mathbf{E}_{1}}) + A(z) (-j \omega \mu_{0} \tilde{\mathbf{H}_{1}}) +\frac{d B}{d z} (\mathbf{u}_{z} \times \tilde{\mathbf{E}_{2}}) + A(z) (-j \omega \mu_{0} \tilde{\mathbf{H}_{2}})
    =-j \omega \mu_{0} \tilde{\mathbf{H}}.
\end{equation}

Since $\tilde{\mathbf{H}}=A(z) \tilde{\mathbf{H}}_{1}+B(z) \tilde{\mathbf{H}}_{2}$ , we obtained the equation:
\begin{equation}
\left(\mathbf{u}_{z} \times \tilde{\mathbf{E}}_{1}\right) \frac{d A}{d z}+\left(\mathbf{u}_{z} \times \tilde{\mathbf{E}}_{2}\right) \frac{d B}{d z}=0.
\label{eq:5.5}
\end{equation}

\begin{figure}
    \centering
    \includegraphics[width=0.8\linewidth]{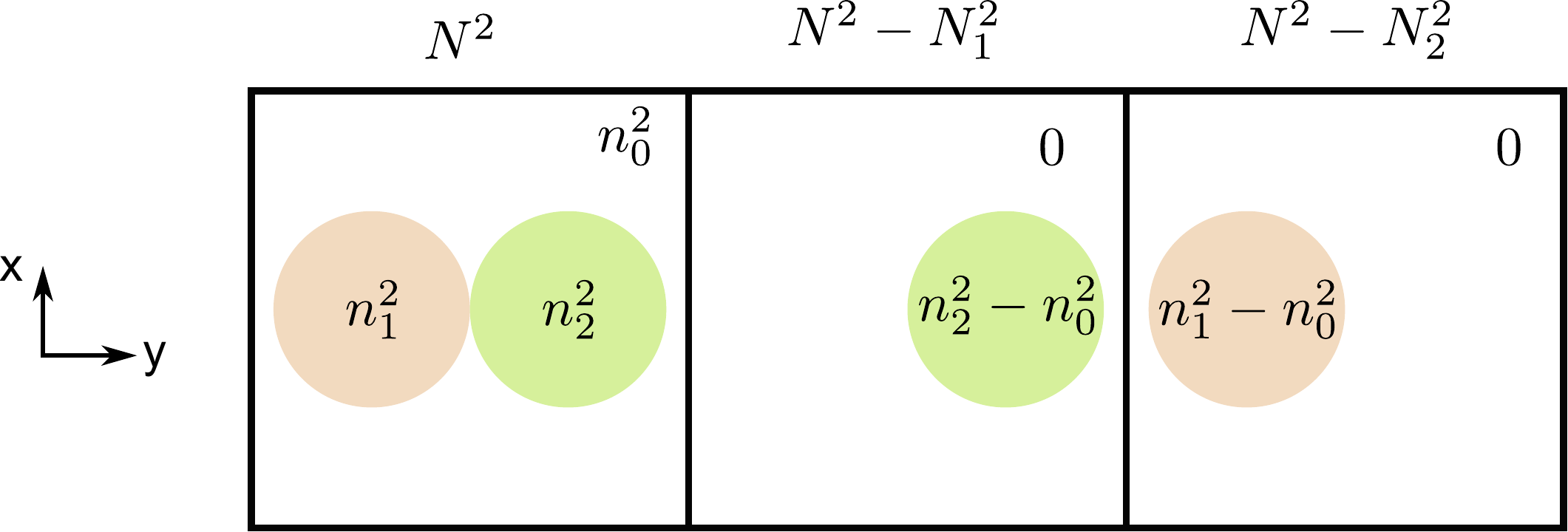}
    \caption{Refractive-index distribution $N^2(x,y)$, and difference of refractive index distribution $N^2-N_1^2$, $N^2-N_2^2$ of the coupled waveguides.}
    \label{fig:refractive-index distribution}
\end{figure}

With similar process with magnetic field, we obtain the equation:
\begin{equation}\begin{array}{l}
\left(\mathbf{u}_{z} \times \tilde{\mathbf{H}}_{1}\right) \frac{d A}{d z}
-j \omega \varepsilon_{0}\left(N^{2}-N_{1}^{2}\right) A(z) \tilde{\mathbf{E}}_{1} 
+\left(\mathbf{u}_{z} \times \tilde{\mathbf{H}}_{2}\right) \frac{d B}{d z}
-j \omega \varepsilon_{0}\left(N^{2}-N_{2}^{2}\right) B(z) \tilde{\mathbf{E}}_{2}=0.
\end{array}
\label{eq:5.6}
\end{equation}
$N^2$, $N_{1,2}^2$ are the refractive index distribution for the coupled waveguides, waveguide 1 and waveguide 2 respectively, as shown in Fig. \ref{fig:refractive-index distribution}.
$N^2-N_1^2$ and $N^2-N_2^2$ are the difference of refractive index distribution. For $N^2-N^2_{p}~(p=1;q=2~or~p=2;q=1)$, the refractive distribution is $n^2_{q}-n^2_{0}$ within waveguide q and is 0 at the other region.

Depending on the propagation direction towards $z$ or $-z$, we have
\begin{equation}
    \left\{\begin{array}{l}\tilde{\mathbf{E}}_{p}=\mathbf{E}_{p} \exp \left(-j \beta_{p} z\right) \\ \tilde{\mathbf{H}}_{p}=\mathbf{H}_{p} \exp \left(-j \beta_{p} z\right)\end{array},\right.
    {or}
    \left\{\begin{array}{l}\tilde{\mathbf{E}}_{p}=\mathbf{E}_{p} \exp \left(j \beta_{p} z\right) \\ \tilde{\mathbf{H}}_{p}=\mathbf{H}_{p} \exp \left(j \beta_{p} z\right)\end{array}.\right.
\end{equation}
where $\beta$ is the phase constant of the waveguide for certain wavelength, which is given by the multiplication of the effective refractive index and the wavenumber.
\begin{equation}
    \beta = n_{eff} \frac{2 \pi}{\lambda}
\end{equation}

Then, we can get the relation between output amplitudes:
\begin{equation}\begin{array}{l}
\frac{d A}{d z}+c_{12} \frac{d B}{d z} \exp \left[-j\left(\beta_{2}-\beta_{1}\right) z\right]+j \chi_{1} A+j \kappa_{12} B \exp \left[-j\left(\beta_{2}-\beta_{1}\right) z\right]=0 ,\\
\frac{d B}{d z}+c_{21} \frac{d A}{d z} \exp \left[+j\left(\beta_{2}-\beta_{1}\right) z\right]+j \chi_{2} B+j \kappa_{21} A \exp \left[+j\left(\beta_{2}-\beta_{1}\right) z\right]=0,
\end{array}
\label{eq:coupling equation with three parameters}
\end{equation}
where:
\begin{equation}\kappa_{p q}=\frac{\omega \varepsilon_{0} \int_{-\infty}^{\infty} \int_{-\infty}^{\infty}\left(N^{2}-N_{q}^{2}\right) \mathbf{E}_{p}^{*} \cdot \mathbf{E}_{q} d x d y}{\int_{-\infty}^{\infty} \int_{-\infty}^{\infty} \mathbf{u}_{z} \cdot\left(\mathbf{E}_{p}^{*} \times \mathbf{H}_{p}+\mathbf{E}_{p} \times \mathbf{H}_{p}^{*}\right) d x d y},\end{equation}

\begin{equation}c_{p q}=\frac{\int_{-\infty}^{\infty} \int_{-\infty}^{\infty} \mathbf{u}_{z} \cdot\left(\mathbf{E}_{p}^{*} \times \mathbf{H}_{q}+\mathbf{E}_{q} \times \mathbf{H}_{p}^{*}\right) d x d y}{\int_{-\infty}^{\infty} \int_{-\infty}^{\infty} \mathbf{u}_{z} \cdot\left(\mathbf{E}_{p}^{*} \times \mathbf{H}_{p}+\mathbf{E}_{p} \times \mathbf{H}_{p}^{*}\right) d x d y},\end{equation}

\begin{equation}\chi_{p}=\frac{\omega \varepsilon_{0} \int_{-\infty}^{\infty} \int_{-\infty}^{\infty}\left(N^{2}-N_{p}^{2}\right) \mathbf{E}_{p}^{*} \cdot \mathbf{E}_{p} d x d y}{\int_{-\infty}^{\infty} \int_{-\infty}^{\infty} \mathbf{u}_{z} \cdot\left(\mathbf{E}_{p}^{*} \times \mathbf{H}_{p}+\mathbf{E}_{p} \times \mathbf{H}_{p}^{*}\right) d x d y}.\end{equation}

There are two cases: $p=1;q=2$ and $p=2;q=1$, which corresponding to the coefficiencies of coupling from waveguide 1 to waveguide 2 and coupling from waveguide 2 to waveguide 1 respectively.

$\kappa$ is the mode coupling coefficient of the directional coupler.
$c$ represents the butt coupling coefficient between the two waveguides. For two adjacent waveguides, the electromagnetic field in the cladding of one waveguide p excites the eigen mode of the second waveguide q. This excitation efficiency is $c_{p q}$ \cite{okamoto2006fundamentals}.
$\chi$ is the self-coupling coefficient, which describes the change of the electromagnetic energy of the modes of original waveguide due to the presence of an adjacent waveguide.

For the HE$_{11}$ mode, which is the fundamental mode of an optical fiber, the electromagnetic field within and outside the fiber core is introduced in section \ref{Light propagating in an optical nanofiber}.

\begin{figure}
    \centering
    \includegraphics[width=0.8\linewidth]{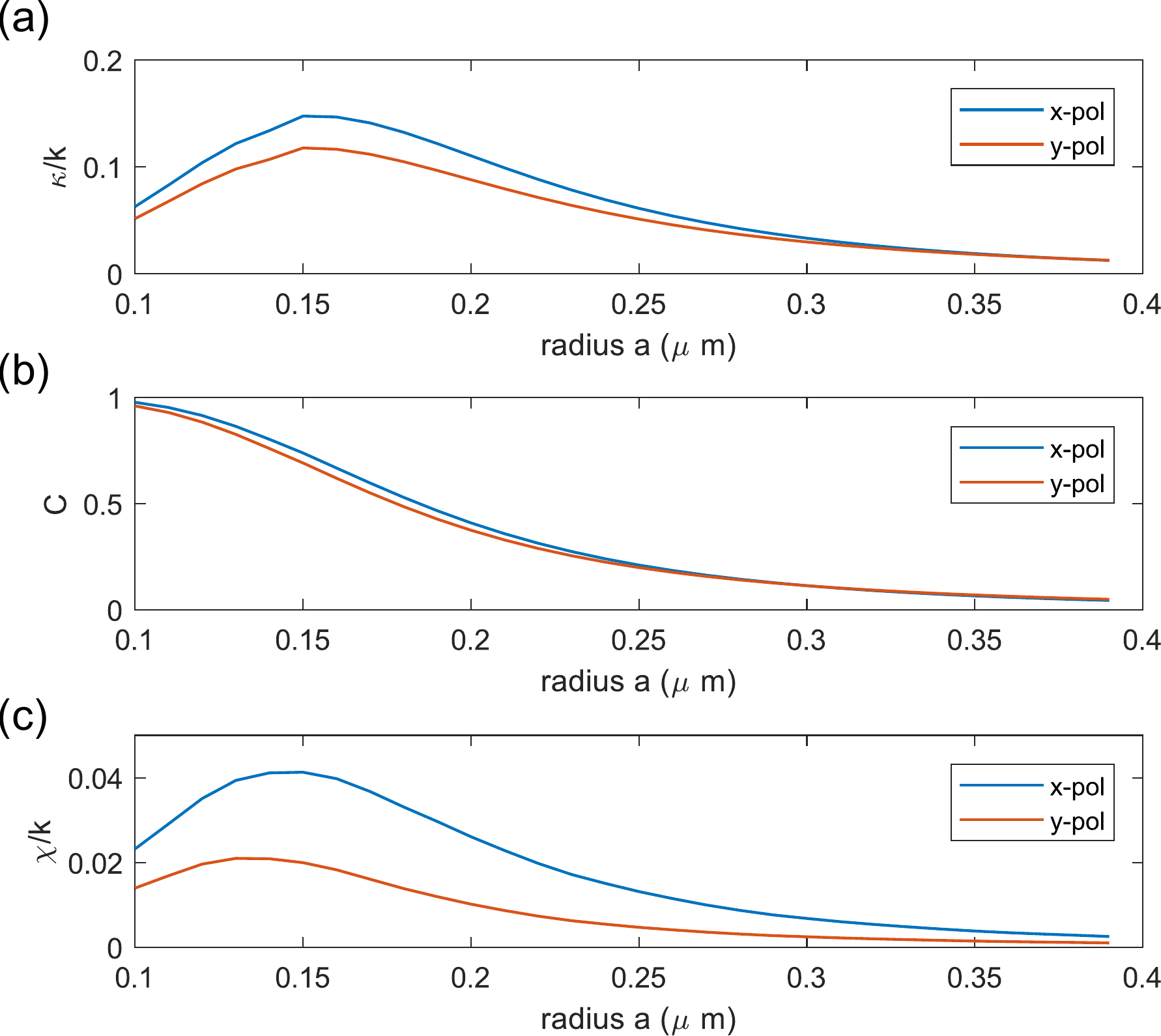}
    \caption{(a), (b) and (c) are Coupling coefficient $\kappa$, butt coupling coefficient $c$, and self-coupling coefficient $\chi$ as functions of the fiber radius $a$ for x-linearly polarized or y-linearly polarized light in both fiber respectively. The coefficients $\kappa$ and $\chi$ are normalized to the wave number $k = 2 \pi/ \lambda$ of light. The wavelength of light is $\lambda$ = 800 nm. The separation distance between the fibers is $d$ = 0.}
    \label{fig:coefficient}
\end{figure}

We plot coupling coefficient, butt coupling coefficient and self-coupling coefficient as a function of fiber radius in Fig.\ref{fig:coefficient}.
We assume the polarization of mode in two fibers are both x-polarized (blue curve) or y-polarized (red curve).
We normalize the coefficients $\kappa$ and $\chi$ to the wave number $k = 2 \pi/ \lambda$ of light.
The wavelength of light is $\lambda$ = 800 nm chosen based on the center wavelength of the white laser.
The distance between two fibers are 0.

According to the calculation, the coupling coefficient increases with fiber radius $a$ when $a$ is small, reaches its maximum at around 150 nm and decreases when fiber radius is higher.
A peak also appears in the self-coupling coefficient as the function of fiber radius.
For comparison, the butt coupling coefficient decreases fast at small radius and gradually approaching 0 when fiber radius increases.
In this curve, we show the effect of this three coefficient at different fiber radius.
When the fiber radius is big, the coupling coefficient is the dominant factor for the mode coupling.
When the fiber radius is small, the butt coupling coefficient and self-coupling coefficient show their effect.

In most of the case, the value of $c_{p q}$ and $\chi_{p}$ are small compared to $\kappa$ and are neglected.
For full calculation, especially for the cases when waveguides have small radius, these two parameters have significant effects on the mode coupling.
 
For the case when butt coupling coefficient $c$ and self-coupling coefficient $\chi$ are neglected, to simplify the equation, we consider only the coupling coefficient $\kappa$. In this scheme, the two fibers exchange energy in the contact region of length $z$ with coupling coefficient $\kappa$.  
Therefore, from equation \ref{eq:coupling equation with three parameters}, given the input amplitudes $A_0$ and $B_0$, the output amplitudes are:
\begin{align}
 A(z) &= A_0\cos(\kappa z)  - i B_0\sin(\kappa z)  \label{Az}\\
 B(z) &= B_0 \cos(\kappa z)  - i A_0 \sin(\kappa z)\; . \label{Bz}
\end{align}

The coupling coefficient $\kappa$ depends on the overlap of the coupled modes \cite{okamoto2006fundamentals}:
\begin{equation}
 \kappa=\frac{\omega \epsilon_0 \iint (N^2-N_1^2) \mathbf{E}_1^* \cdot \mathbf{E}_2 \ \text{d}x \text{d}y}{\iint \mathbf{u}_z \cdot (\mathbf{E}_1^* \times \mathbf{H}_1 + \mathbf{E}_1 \times \mathbf{H}_1^*) \  \text{d}x \text{d}y} \label{k}
\end{equation}
where $N_1$ is the refractive index distribution for the fiber 1, $\mathbf{E}_i$ and $\mathbf{H}_i$ are respectively the electric and magnetic components of the modes propagating in the nanofiber labeled $i$, $\mathbf{u}_z$ is the unitary vector directed toward the propagation direction and $\omega$ the field frequency.

The full calculation of coupling between two parallel fibers where the butt coupling coefficient and self-coupling coefficient is also taken into consideration has been introduced in \cite{kien2020coupling} by Fam Le Kien et all.
This calculation is interesting because when the radius of nanofiber is sufficiently small, the butt coupling coefficient and the self-coupling coefficient play important roles in the mode coupling process, which can not be described with the simplified equations introduced here.

According to Eqs. (\ref{Az}) and (\ref{Bz}), we can regard the light going through the fiber 1 as transmitted, and the light going from fiber 1 to fiber 2 as reflected. 
The system acts as a beam-splitter of transmission coefficient $t=\cos(\kappa z)$, and reflection coefficient $r=-i\sin(\kappa z)$. 
For example, when the system has an input $A_0=1$ and $B_0=0$, the output intensities appear to be $|A(z)|^2=\cos^2(\kappa z)$ and $|B(z)|^2=\sin^2(\kappa z)$, and complete power transfer occurs when $\kappa z=(2p+1) \pi / 2$, $p$ being an integer.
Consequently the quantity $ \pi / 2\kappa$ is equivalent to a coupling length.

Interestingly the orientation of the effective beam splitter depends on the topology of  the structure. 
In the case of the twisted loop represented in Fig.~\ref{figure3-1}-b, the beam splitter is equivalent to a Sagnac interferometer allowing for only one lap in the structure. 
Whereas in the case of the knot (Fig.~\ref{figure3-1}-d), the effective beam splitter allows for multiple laps inside the setup and therefore is equivalent to a ring resonator.

\begin{figure}[t]
\centering\includegraphics[width=0.8\linewidth]{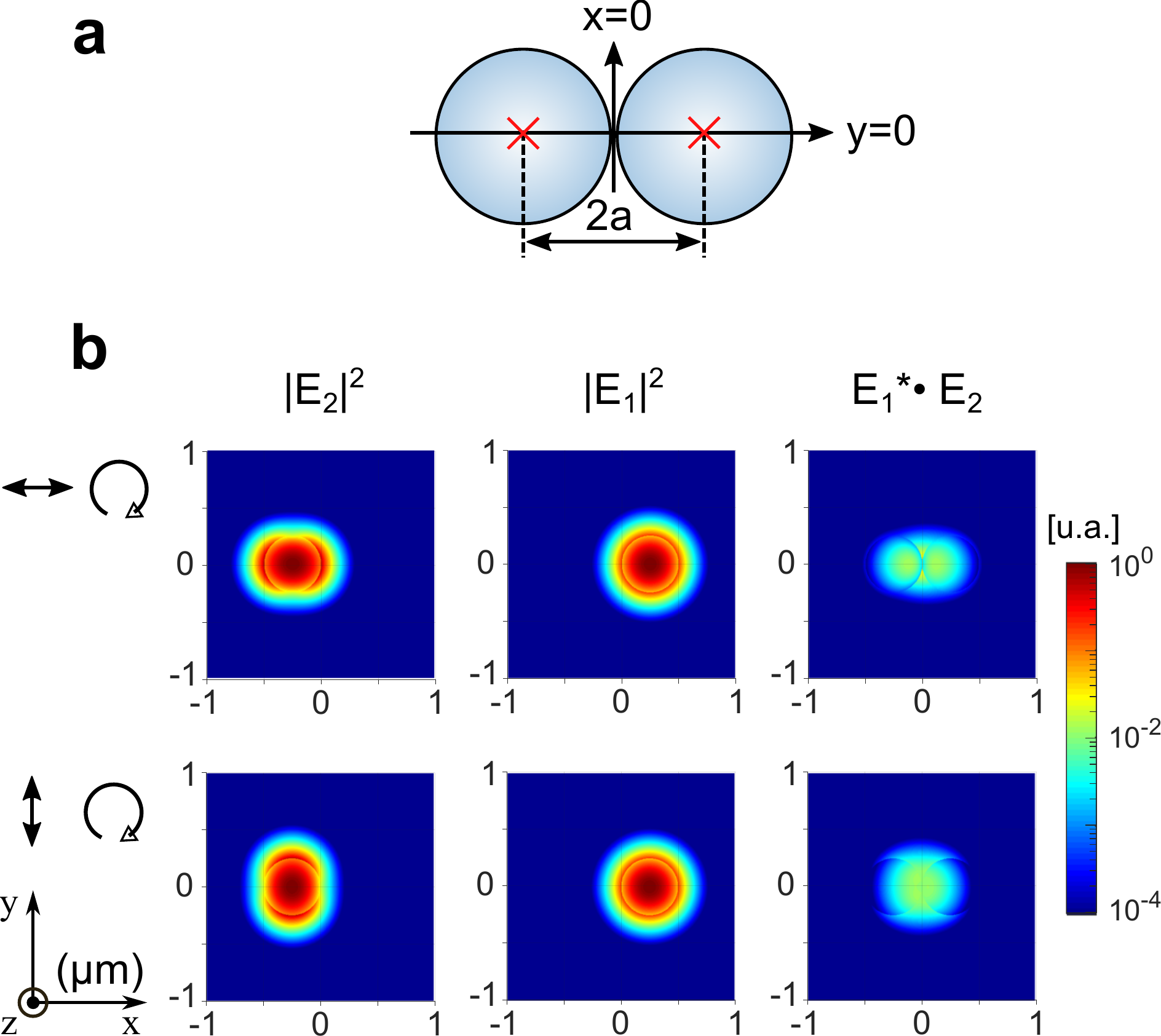}
\caption{Field distribution and overlap for two adjacent nanofibers (fiber radius $a$ = 250 nm). a, The centers of two nanofibers are located on axis y=0. 
Fiber 1 is located at position ($a$,0), and fiber 2 is located at ($-a$,0). 
b, The light in the fiber 2 corresponds to an HE$_{11}$ and  linearly polarized as visible in the left panel (upper left panel for the horizontally polarized and lower left panel for vertically polarized). 
The field is circularly polarized in fiber 1 as visible in the central panels. 
Right panels represent the overlap between the two fields directly related to the coupling strength $\kappa$ given in Eq.~(\ref{k}).
}\label{figure3-2}
\end{figure}

\begin{figure*}[tb] 
\centering
 \makebox[\textwidth]{\includegraphics[width=\linewidth]{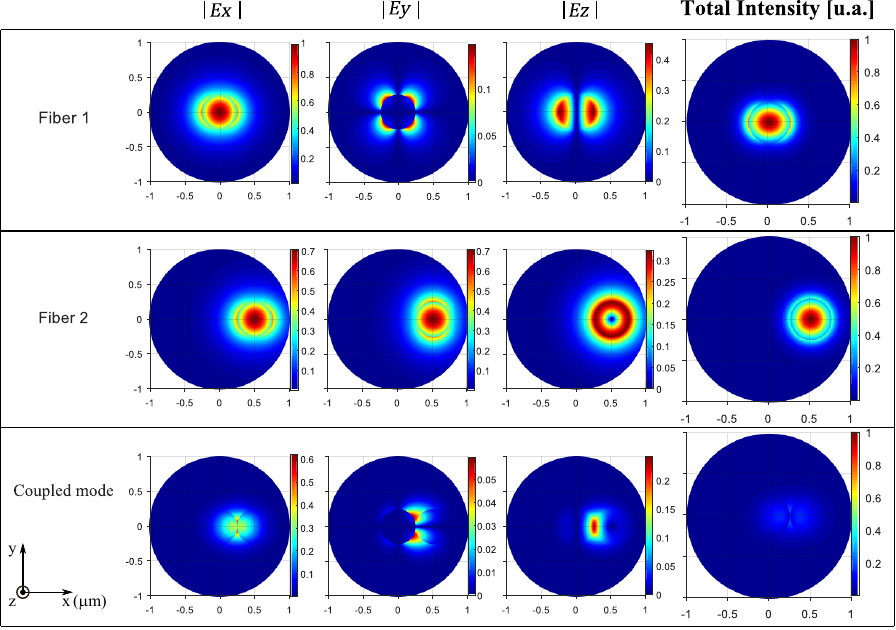}}
\caption{Three components of field distribution and overlap for two adjacent nanofibers (fiber radius $a$ = 250 nm) guiding x-linearly and circularly polarized light respectively.
}\label{coupledmode}
\end{figure*}

As mentioned above, we place ourselves in conditions under which only one propagating mode exists: HE$_{11}$.
To estimate the coupling coefficient $\kappa$, we computed Eq. (\ref{k}) using the exact profile of modes HE$_{11}$ \cite{le2004field}. 
In order to study the coupling coefficient dependence on the polarization, we assume that nanofibers are identical and in contact at (0,0) on $x-y$ plane, as shown in Fig.~\ref{figure3-2}-a, and that the light is linearly polarized in one fiber, whereas it is circularly polarized in the other one. 
The detailed field distribution in Fig.~\ref{coupledmode} shows the Electric field intensity of three components in Cartesian coordinate.

\afterpage{\clearpage}

\begin{figure}[htbp]
\centering\includegraphics[width=0.9\linewidth]{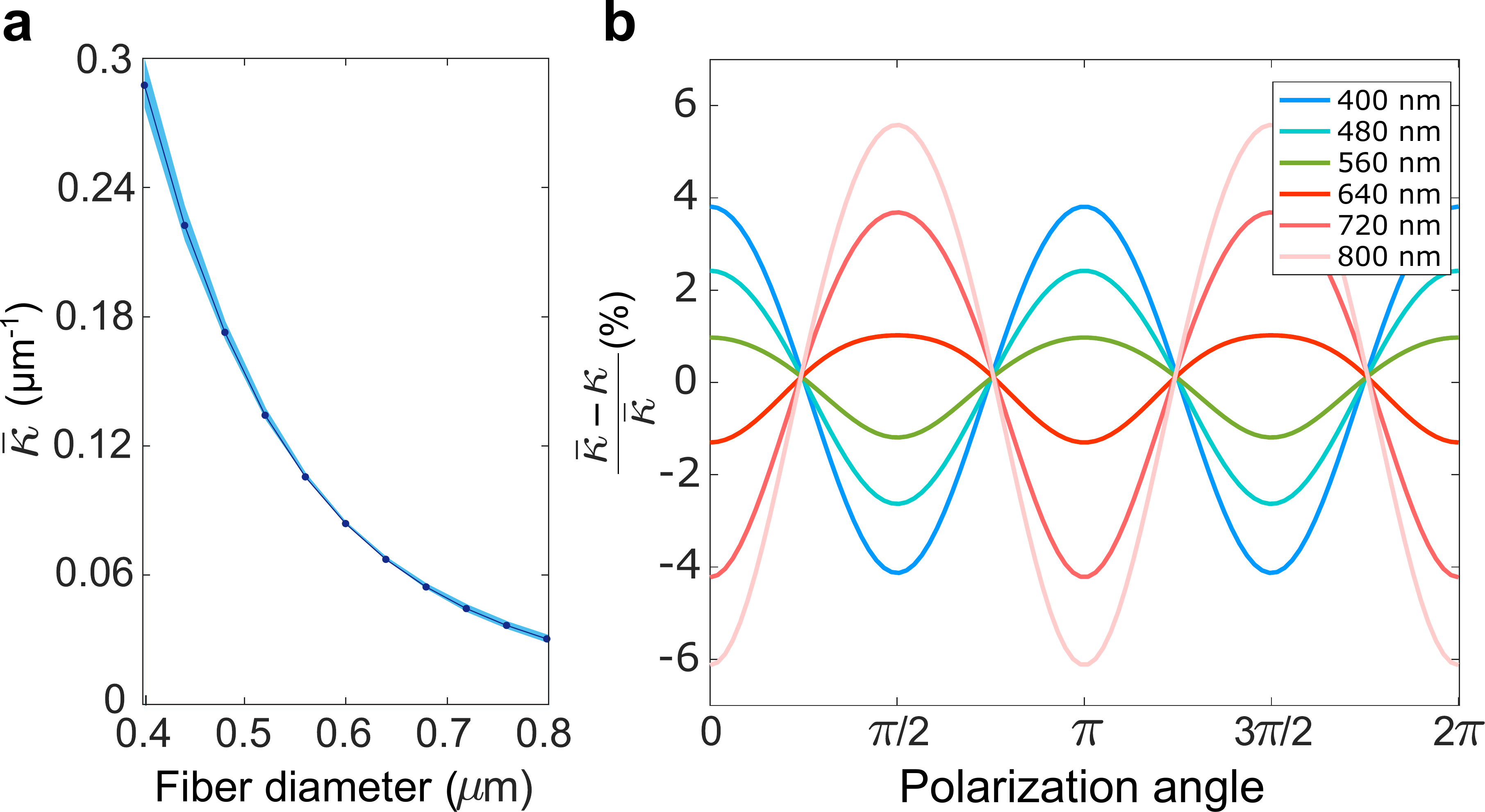}
\caption{a- Coupling coefficient $\kappa$ (at 800~nm) between nanofibers with varying diameter and averaged over the polarization degree of freedom. 
The colored area refers to the amplitude of the variation with the polarization. 
b- Relative change of the coupling coefficient as a function of the polarization angle with a fiber diameter varying from 500 nm to 900 nm.}\label{figure3-3}
\end{figure}

The field density in the transverse section in this configuration is presented in Fig.~\ref{figure3-2}-b.
We numerically calculated the coupling strength as a function of the fiber diameter averaged over the polarization degree of freedom $\bar\kappa = \langle \kappa \rangle_{\varphi}$, where $\varphi$ is the angle of the polarization vector with the $x$ axis.
Results are presented in Fig.~\ref{figure3-3}-a.
We see that $\bar\kappa$ decays exponentially with the fiber diameter. 
The larger the fibers are, the smaller their evanescent part of the field is. 
This decay is exponential so it is for the overlap of the fields. This strong dependency illustrates well the general interest to work with fiber at subwavelength scale rather than micrometric scale. 
Focusing now on the polarization dependency of the coupling strength we show two cases in Fig.~\ref{figure3-2}-b : 
(i) the polarization of the linearly polarized field in fiber 2 is along the $x$ axis (along the direction that connects the centers of the two fibers)  ($\varphi=0$) (upper panels of Fig.~\ref{figure3-2}-b) and 
(ii) the polarization of the linearly polarized field in fiber 2 is normal to the $x$ axis  ($\varphi=\pi/2$) (lower panels of Fig.~\ref{figure3-2}-b). 
For both cases the field density is presented for the linearly polarized field in fiber 2 (left panels), for the circularly polarized field in fiber 1 (central panels) and for their overlap $\mathbf{E}_1^*\cdot\mathbf{E}_2$ appearing in Eq. (\ref{k}), where  $\mathbf{E}_1^* \cdot \mathbf{E}_2=\mathbf{E}_{1x}^*\cdot \mathbf{E}_{2x}+\mathbf{E}_{1y}^*\cdot \mathbf{E}_{2y}+\mathbf{E}_{1z}^*\cdot \mathbf{E}_{2z}$.
The results shown are for a fiber diameter of 500 nm and wavelength of 800 nm.
 We see that even if the overlap intensity distribution is different from one case to the other,  their average magnitude and then the coupling coefficient are similar in the two cases as shown in Fig.\ref{figure3-2}-b.  
 
Actually, the coupling coefficient $\kappa$ is found to be only slightly dependent on the polarization. 
This variation depends on the fiber diameter, as visible in the b panel of Fig.~\ref{figure3-3} but leads to a marginal relative change.
 In the realization of the nanofiber twisted loop below, we use a fiber diameter of 500 nm for which the relative change is estimated to be less than $\pm 3\%$. 
This is illustrated in Fig.~\ref{figure3-3}-a by the colored region surrounding the mean coupling strength $\bar\kappa$, corresponding to the amplitude of the variation with respect to the polarization. 
Given that, we can reasonably neglect the effect due to polarization and approximate $\kappa\approx\bar\kappa$.

\begin{figure}
    \centering
    \includegraphics[width=0.5\linewidth]{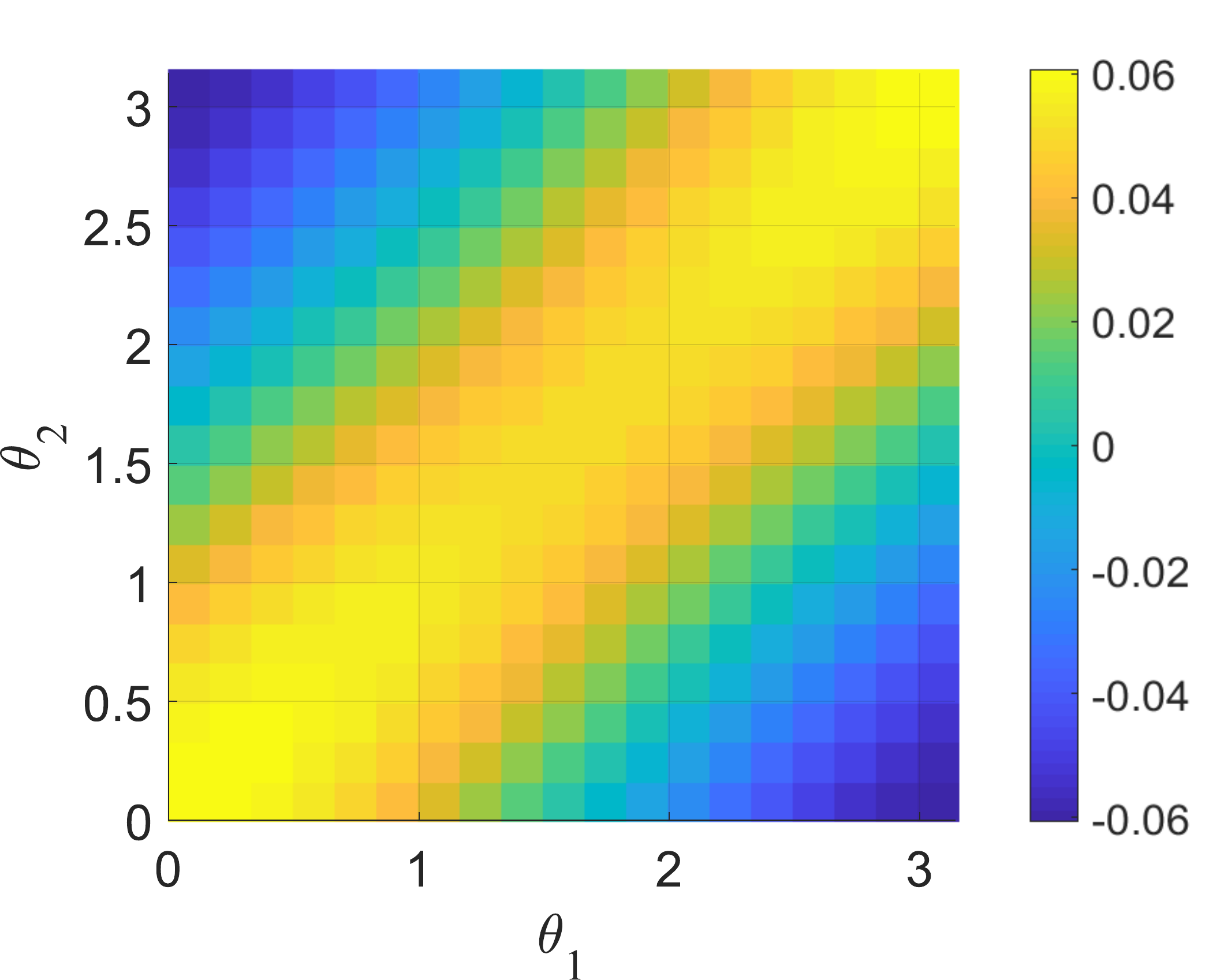}
    \caption{Directional coupling coefficient $\kappa/k$ as a map of the polarization. $\theta_1$ and $\theta_2$ are the angle of linearly polarized light in fiber 1 and fiber 2.}
    \label{fig:kappa}
\end{figure}

In Fig.\ref{fig:kappa}, we show the directional coupling coefficient $\kappa/k$ as a map of the polarization.
We assume the mode in both fibers are linearly polarized with polarization angle $\theta_1$ and $\theta_2$ respectively, where $\theta_1$ and $\theta_2$ $\in~[0,\pi]$.
While the polarization angle $\theta_1$ and $\theta_2$ changes from 0 to $\pi$, the directional coupling coefficient $\kappa/k$ varies from -0.06 to 0.06.

Therefore, the changing of polarization angle within the fiber should change the value of directional coupling coefficient $\kappa/k$.
In an optical fiber, the polarization of the propagating light changes when the birefringence induced by stress or fiber cross-section geometry changes.
This is something that we need to consider about when we further analyze the mode coupling of nanofiber structures.

\section{\label{sec:level2}Nanofiber interferometer}

We realized an optical nanofiber by pulling a commercial single-mode fiber to reach a 500~nm diameter over a length of 1~mm following \cite{hoffman2014ultrahigh}.
The transition between the commercial single mode fiber and the single mode nanofiber is adiabatic and its transmission is over 95\% \cite{joos2018polarization}.

To make a twisted loop structure presented in Fig.\ref{figure3-1}-a,  we first make a ring in the nanofiber region. 
Then, by fixing one side of the nanofiber, and rotating the other side, we can slowly increase the length of the entwined part. 
This is a well-known mechanical phenomenon studied in many contexts \cite{goriely1998nonlinear}.
Increased torsion will reduce the size of the loop and bend it locally.
At some point the bending exceeds the fiber tolerance and it breaks. 
In the experiment we carefully choose to remain below this threshold.

With this geometry, the system corresponds to a Sagnac interferometer \cite{culshaw2005optical}. 
As represented in Fig.\ref{figure3-1}-b, light propagating towards the loop finds two counter-propagating optical paths.
After the entwined region, part of the light is transferred into the clockwise path (red arrow) with a coefficient of $r$, whereas the remaining propagates along the anti-clockwise path with a coefficient of $t$. 
This two beams propagate separately accumulating a phase of $e^{i \beta L_r}$, where $\beta$ is the propagation constant, and $L_r$ is the length of the ring.
 Then, both paths interfere back into the entwined region, which acts as a beam-splitter, as represented in Fig.\ref{figure3-1}.b.  
The amplitude of output electromagnetic field can be written as:
\begin{equation}
 A_{\text{out}}=[r^2 +t^2]e^{i\beta L_r}A_0,
 \label{eq:Electromagnetic field Amplitude}
\end{equation}
with $t=\cos(\kappa z)$, and $r=-i\sin(\kappa z)$, as shown above, 
which leads to the following transmittance for the device:
\begin{equation}
 T_{\text{int}}=\left|\frac {A_{\text{out}}}{A_0}\right|^2=\left|r^2 +t^2\right|^2.
 \label{eq:transmission_sagnac}
\end{equation}
As mentioned above, we control the length of the coupling region by varying the torsion applied on the nanofiber, which ultimately tunes the reflection and transmission coefficients of our device.
Applied mechanical stress will be translated into an optical response, from reflective to transmissive.

\begin{figure}
    \centering
    \includegraphics[width=\linewidth]{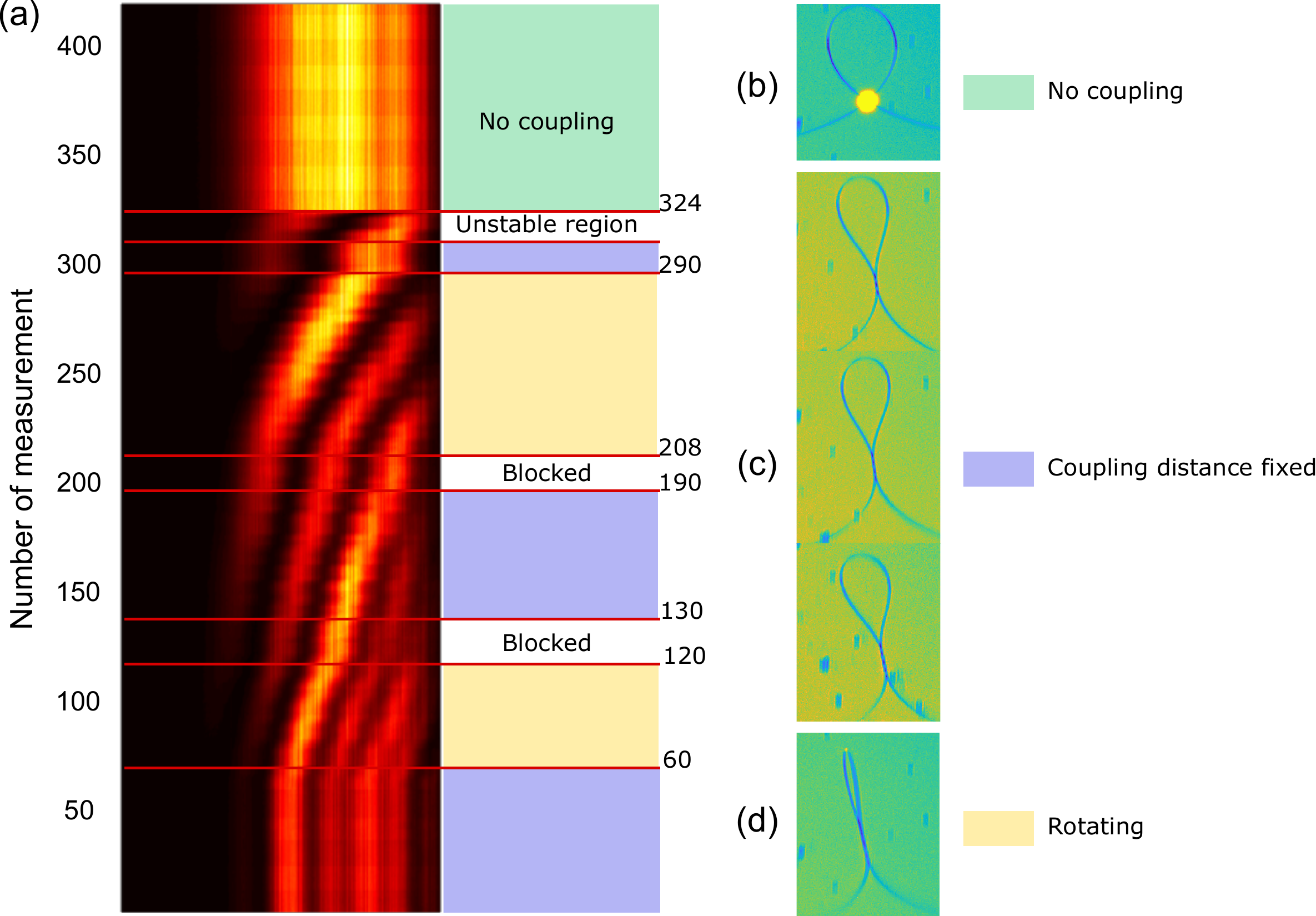}
    \caption{The transmission spectrum and the corresponding image of twisted loop. 
    a, the map of transmission spectrum while varying the torsion applied on the nanofiber. 
    b, image of nanofiber structure when there is no coupling (measurement No. 324 and plus).
    c, images of nanofiber twised loop when the coupling distance is measured to be fixed (measurement No. 1-59, 130-189 and 290-323).
    d, image of nanofiber twisted loop when it's rotating and the coupling distance is changing (measurement No. 60-119).}
    \label{fig:Spectrum-camera}
\end{figure}

\begin{figure}[htbp]
\centering\includegraphics[width=0.9\linewidth]{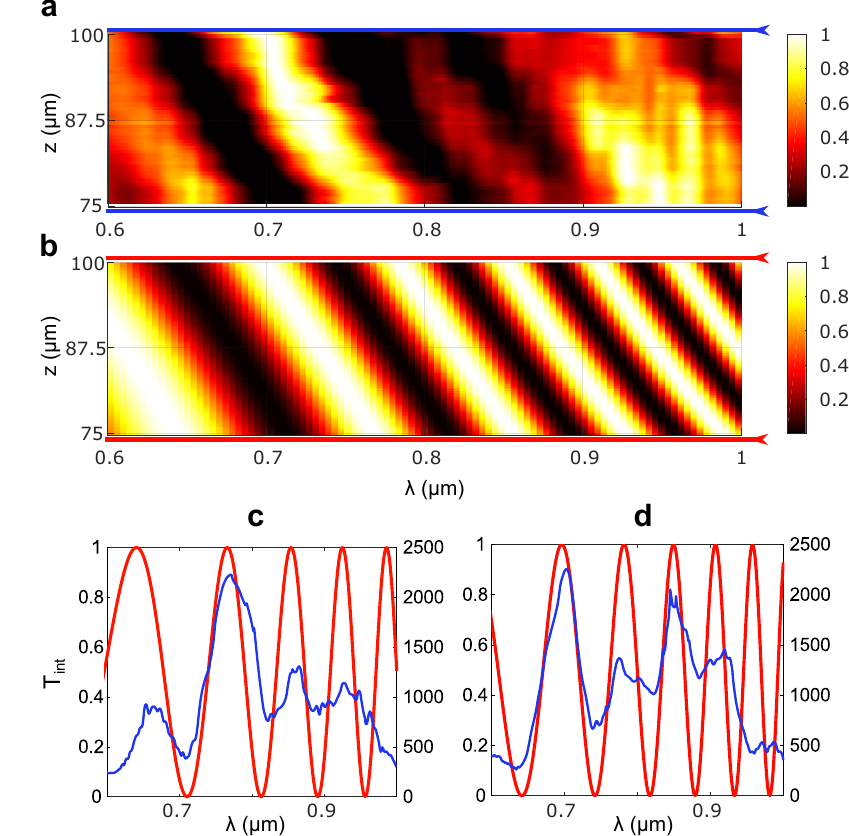}
\caption{Transmission spectrum through twisted loop structure. 
a and b- Experimental data and numerical simulation of the transmission spectrum of broad light source through the nanofiber interferometer as a function of the wavelength and coupling distance.
c and d- Experimental data (blue) and numerical simulation (red)  of the transmission spectrum for an entwined length of 75 $\mu$m (panel c) and 100 $\mu$m (panel d) respectively. 
The experiment data shown are smoothed using rloess methods with a span of 1\%.}\label{figure3-4}
\end{figure}

To probe the system transmittance we use a fiber-coupled linearly polarized super-continuous white laser (NKT Photonics SuperK COMPACT). 
The laser beam is coupled into the single mode fiber (SM800) with the twisted nanofiber region in the middle. 
The output signal is sent to a spectrometer.

To show the relationship between the transmission spectrum and the movement of the nanofiber twisted loop, we measured the transmission spectrum and its corresponding image of twisted loop in camera each time we rotate one side of the fiber 5 degrees.
We plot the transmission spectrum as a function of wavelength in Fig.\ref{fig:Spectrum-camera}-a and the ordinate is the number of measurements.
The length of entwined part is measured by analyzing the image taken with a calibrated camera, corresponding to the coupling length.
The camera is calibrated with a resolution test target (Thorlabs) at the focus plane.
By measuring the coupling length in each photo taken by camera, we find the coupling length changes periodically instead of continuously when we rotate one side of the twisted loop.
When the coupling length is fixed, the rotating of fiber end lower the tension in the system.
The coupling length starts to decrease when the loop begins to rotate, and the coupling length stops decreasing after the loop rotates 180 degrees and released one crossing in the entwined part.
In Fig.\ref{fig:Spectrum-camera}-c, we show the image of twisted loop with 3, 4 and 5 crossing.
At the alternation of these two states, the state of the fiber is unstable.
To study the influence of coupling length on the transmission spectrum, we focus on the part when the coupling length is changing or the part when loop is rotating.

Fig.~\ref{figure3-4}-a shows a map of the transmission spectrum through the nanofiber twisted loop structure as a function of the entwined region size.

To understand the spectrum obtained for a given entwined region length, also presented in Figs.~\ref{figure3-4}-c and d in blue, one has to note that when the fiber diameter is fixed, the extension of the evanescent part of the field increases with the wavelength. Accordingly the coupling strength $\kappa$ and so the effective reflection coefficient $r$ change too as shown by the red curves in Fig.~\ref{figure3-4}-c and d. In consequence, the spectrum for a fixed coupling length leads to the interference pattern visible in Fig.~\ref{figure3-4} and agrees with our description of the device as an interferometer.  
Moreover, increasing the coupling length leads to a shift of the interference to smaller wavelengths.
In Fig.~\ref{figure3-4}-b we represent the same map as in Fig.~\ref{figure3-4}-a, calculated from Eq. \ref{k}.
Despite the variability of many experimental parameters, our theoretical model shows good agreement with the experimental data.
This agreement can be verified quantitatively in panel c and d, where we show the measured (blue) and calculated (red) spectrum for a coupling length of respectively 75 $\mu$m and 100 $\mu$m with no adjustable parameters.

However, the contrast of the observed spectra is relatively low (0.2 to 0.6). 
Besides, the contrast seems to be different between nearby peaks, which is different from the numerical simulation. 
It is more obvious in Fig.\ref{figure3-4}-d, the experimentally measured transmission spectrum shows four peaks. The peaks at 700 nm and and 850 nm have higher contrast while the peaks at 790 nm and 920 nm have lower contrast.
To understand the limiting factor of this phenomenon, we did numerical simulations of transmission spectrum to study the effect of loss during the mode coupling. 
In our two fibers coupling model, the energy transfer between the two fibers were regarded as lossless.
In reality, the surface of tapered nanofiber is not purely smooth. 
There are a lot of small defects which scatter the light into free space as we can see in the camera after a nanofiber is fabricated.
For twisted nanofiber structure, the losses come from both the coupling between two fibers at the twisted part and the light propagating in a bending fiber.
Therefore, we add parameters representing coupling loss and propagation loss based on the equation.\ref{eq:Electromagnetic field Amplitude}.
We have:
\begin{equation}
 A_{\text{out}}=[((1-\rho_r) r)^2 +((1-\rho_t) t)^2](1-\rho)e^{i\beta L_r}A_0,
\end{equation}
where $\rho_r$ is the coupling loss for reflection coefficient $r$, $\rho_t$ is the coupling loss for transmission coefficient $t$, $\rho$ is the propagation loss for ring part.

\begin{figure}
    \centering
    \includegraphics[width=0.5\linewidth]{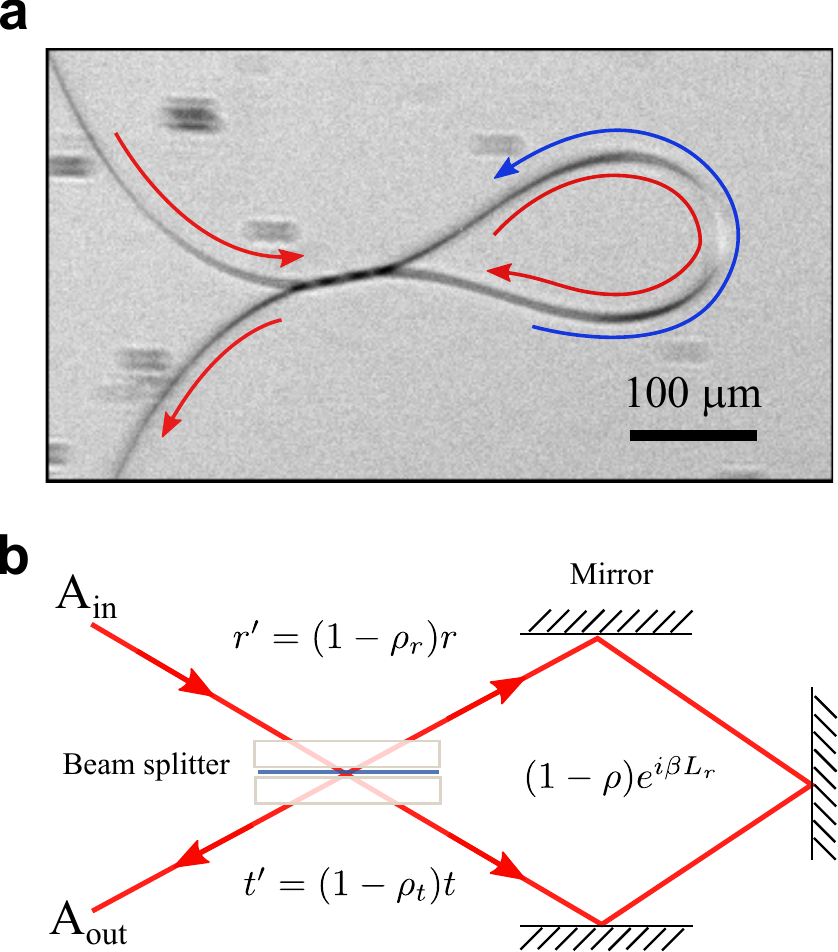}
    \caption{a, Optical microscope image of a nanofiber twisted loop. b, Schematic of introducing losses in the Sagnac interferometer equivalent optical setup. The transmission coefficient with loss is labeled as $t'$. The reflection coefficient with loss is labeled as $r'$.}
    \label{fig:Loss for Sagnac}
\end{figure}

\begin{figure}
    \centering
    \includegraphics[width=0.8\linewidth]{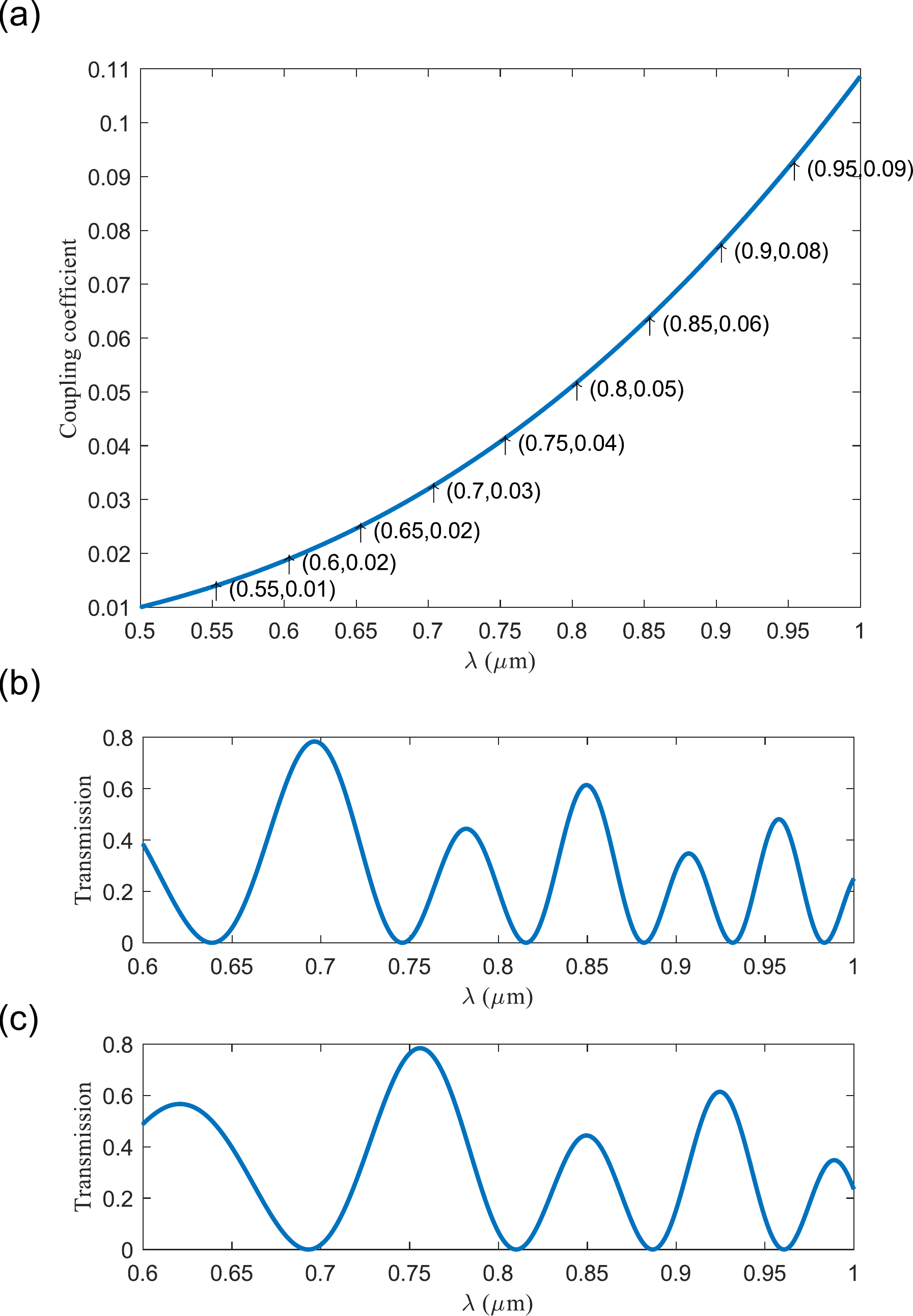}
    \caption{Transmission spectrum through twisted loop structure when coupling loss is taken into consideration. (a) Coupling coefficient as a function of wavelength. (b) Transmission spectrum when $z=100 \mu m$. (c) Transmission spectrum when $z=75 \mu m$.}
    \label{figure3-4-2}
\end{figure}

When we take the coupling loss into consideration, we found that the loss is higher for longer wavelength. In figure \ref{figure3-4-2}-a, we plot the directional coupling coefficient $\kappa$ as a function of wavelength.
The fiber diameter is fixed at 500 nm.
The longer wavelength gives higher coupling coefficient. 
For the same coupling length, we can imagine the longer wavelength bouncing back and forth more times than the shorter wavelength in the entwined region. 
Thus for the longer wavelength, there will be more loss. 

Besides, there is unbalanced loss between reflection coefficient $r$ and transmission coefficient $t$, which caused the contrast difference between nearby peaks.
In Fig.\ref{figure3-4-2}-b,c, we plot the transmission spectrum when coupling length $z=100~\mu m$ and $z=75~\mu m$ with 10 percent loss for reflection coefficient $r$ and 2 percent loss for transmission coefficient $t$.
We take the one with 100 $\mu m$ coupling length as an example, the peaks at 700 nm and and 850 nm have higher contrast while the peaks at 790 nm and 920 nm have lower contrast, which is similar as the measurement.
Therefore, we think the contrast difference between nearby peaks comes from the unbalanced loss between reflection coefficient $r$ and transmission coefficient $t$.
In reality, this can be due to the geometric asymmetry of nanofiber structure.

By tuning the coupling length, we can adjust the high contrast region to the wavelength we need, as we can see in the Fig.\ref{figure3-4}-a, the high contrast region shifts from 700 nm to 750 nm when we change the coupling length from 100 $\mu m$ to 75 $\mu m$.

\section{\label{sec:Nanofiberresonator}Nanofiber resonator}

Optical ring resonators can have many applications, such as optical add-drop filters, modulation, switching and dispersion compensation devices. 
To fabricate the knot structure as presented in Fig.~\ref{figure3-1}-c, we carefully made a large knot centered on the nanofiber region and by precise control of the spacing between the two displacement platforms used to pull the fiber, we can decrease the diameter of the knot to tens of micrometers ($\approx 11~\mu$m). 
As in the case of the twisted loop, the knotted loop induced an important bending of the nanofiber.
Accurate control of the size of the knot allows us to avoid breaking it. 

With this geometry, the system acts as a resonator.
As represented in Fig.~\ref{figure3-1}-d, the light injected in the device will either be directly reflected to the output with a coefficient $r$, or transmitted in the loop with a coefficient $t$.
In contrast with the twisted loop, there is only one optical path within the loop.
Moreover, the light circulating inside the knot will split again everytime it passes through the coupling region: some will go to the output, the rest will stay inside the knot. 
We represent in Fig.~\ref{figure3-1}-d the corresponding optical setup. 
It is remarkable that the change of the topology of the loop formed, twisted or knotted, completely changes the behavior of the device. 
Schematically, passing from one device to the other one is equivalent to rotating by 90$^\circ$ the beam splitter mimicking the fiber coupling region as presented in Fig.~\ref{figure3-1}-b and Fig.~\ref{figure3-1}-d. 

\begin{figure}[htbp]

\centering\includegraphics[width=0.6\linewidth]{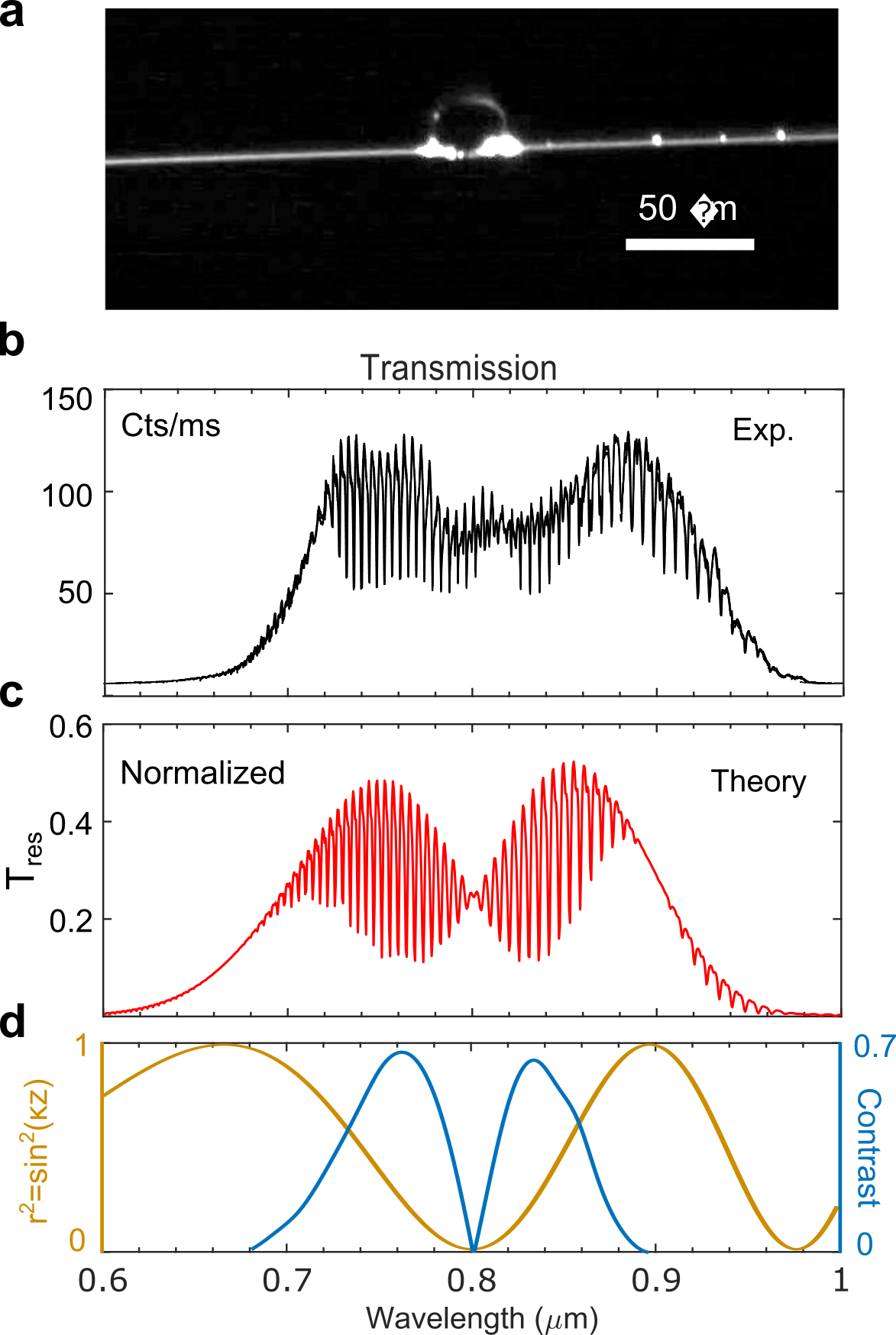}
\caption{Transmission spectrum through a knot structure.
a- Image of the nanofiber knot recorded by an optical microscope.
b- Experimental transmission spectrum of a broad light source through the nanofiber resonator as a function of the wavelength. We measure $\Delta\upsilon_{\text{FSR}}$=2.21 THz.
c- Calculated transmittance of the devices as given in eq. (\ref{tres}). 
d- Left axis is the reflectance of the contact zone and right axis is the visibility of resonance between wavelength 0.68 $\mu m$ and 0.9 $\mu m$. 
}
\label{figure3-5}
\end{figure}

Along the propagation in a loop, the field undergoes losses with a rate $\rho$ due to scattering. 
In Fig. \ref{figure3-5}-a we present an optical microscopy image of the knotted loop when light is propagating on the fiber. 
Bright spots on the fiber are due to the scattering of imperfections on the fiber surface.
Given the significant evanescent component of the field, any defects located close to the surface will strongly scatter the propagating light.
However, most of the losses come from the knotted region itself as visible in Fig. \ref{figure3-5}-a. 
They would drastically be reduced in a clean room environment.
Indeed, when we fabricate the knot manually, the knot was gradually tightened into small size. Thus, the entwined region sweeps several centimeters of fiber.
Impurities on the surface are blocked by the knot and inevitably accumulate there. 

After one lap in the loop we have $B'_0=(1-\rho)e^{i\beta L_k}t^2B_0$, where $L_k$ is the length of the ring. 
Assuming the reflection coefficient $r=- i \sin(\kappa z)$ and transmission coefficient $t=\cos(\kappa z)$, then we get the equation giving the amplitude of the electromagnetic field at the output:
\begin{equation}
\begin{aligned}
A_{\text{out}}=A_0[ r + (1-\rho)t^2 e^{i \beta L_k}  + (1-\rho)^2t^2 r e^{2i \beta L_k} \\+ (1-\rho)^3t^2 r^2 e^{3i \beta L_k} + \cdots ],
\end{aligned}
\end{equation}
leading to the following transmittance $ T_{\text{res}}$ of the device:
\begin{equation}
 T_{\text{res}}=\left|\frac {A_{\text{out}}}{ A_0}\right|^2=\left|r + \frac {(1-\rho)t^2e^{i \beta L_k} }{1- (1-\rho)re^{i \beta L_k}}\right|^2. \label{tres}
\end{equation}

In contrast to the nanofiber interferometer case, the size of the loop plays a major role here as it dictates the amount of phase accumulated after one lap in the device.
Optical resonances appear when the light traversing the loop accumulates a phase integer multiple of $2\pi$.  

To characterize the system we use the same setup as for the twisted loop.
Fig. \ref{figure3-5}-b is the transmission spectrum of the knot for a given diameter. Within the spectrum region ($\approx$ 700 nm to $\approx$ 950 nm), many fine peaks are observed, revealing the resonant wavelengths.

In Fig. \ref{figure3-5}-c, we present the calculated transmission spectrum, which agrees well with the experimental data, for its three main features.
Firstly, the wide spectral Gaussian envelop; this is measured beforehand by recording the spectrum of our laser transmitted through a fiber without the loop. The Gaussian envelop is observed due to the joint of two effect. In the short wavelength region, light enters the SM800 fiber as multimode, and higher order modes were cleaned at the the nanofiber region. For the long wavelength region, it's due to the sensitivity of the Si-detector.
Secondly, the fine peaks; they exhibits matched free spectral range (FSR) and amplitudes.
Finally, the larger scale contrast modulation ; its maximum around 750 nm and 850 nm are faithfully reproduced.
To explain this contrast modulation, we represent in Fig. \ref{figure3-5}-d the reflectance of the entwined region (i.e. without knot) calculated for the same condition. 
As observed previously, it depends strongly on the wavelength. 
For instance, at $\lambda\approx 800$ nm, the reflectance $|r|^2$ is zero; it corresponds to a scenario without beamsplitter.
As 100\% of the light leaves the ring after one lap, there cannot be interferences, and the resonance peaks contrast vanishes accordingly. 
Similarly when $r^2\sim 1$, around 900 nm and 670 nm, the entwined region acts as a simple mirror instead of a beam-splitter, and the resonance peaks fade as well, since no light gets inside the ring.

The length of the ring $L_k$ determines the interval between the peaks in the spectrum known as FSR and given by $\Delta\upsilon_{\text{FSR}}=c/(n_{\text{eff}}L_k)$, where $n_{\text{eff}}$ is the effective refractive index of air clad optical fiber. 
The analysis of the spectrum gives $\Delta\upsilon_{\text{FSR}}$=2.21 THz, which corresponds to a cavity length of about 108~$\mu$m. 
The finesse varies slightly along the spectrum reaching 8 from 820 nm to 860 nm. 
In this range, we measure a quality factor of 1300. 
It agrees  well with the calculation which predicted a $\Delta\upsilon_{\text{FSR}}$ of 2.136 THz, a finesse of 7.5 and a quality factor of 1300. 
These calculations have been done with an estimate of 35\% losses, extracted from experimental data.  
In the next section, we push forward the analysis of the spectrum and we identify an original birefringence effect.

\subsection*{Birefringence induced by ovalization under bending}
\begin{figure}[htbp]
\centering\includegraphics[width=0.7\linewidth]{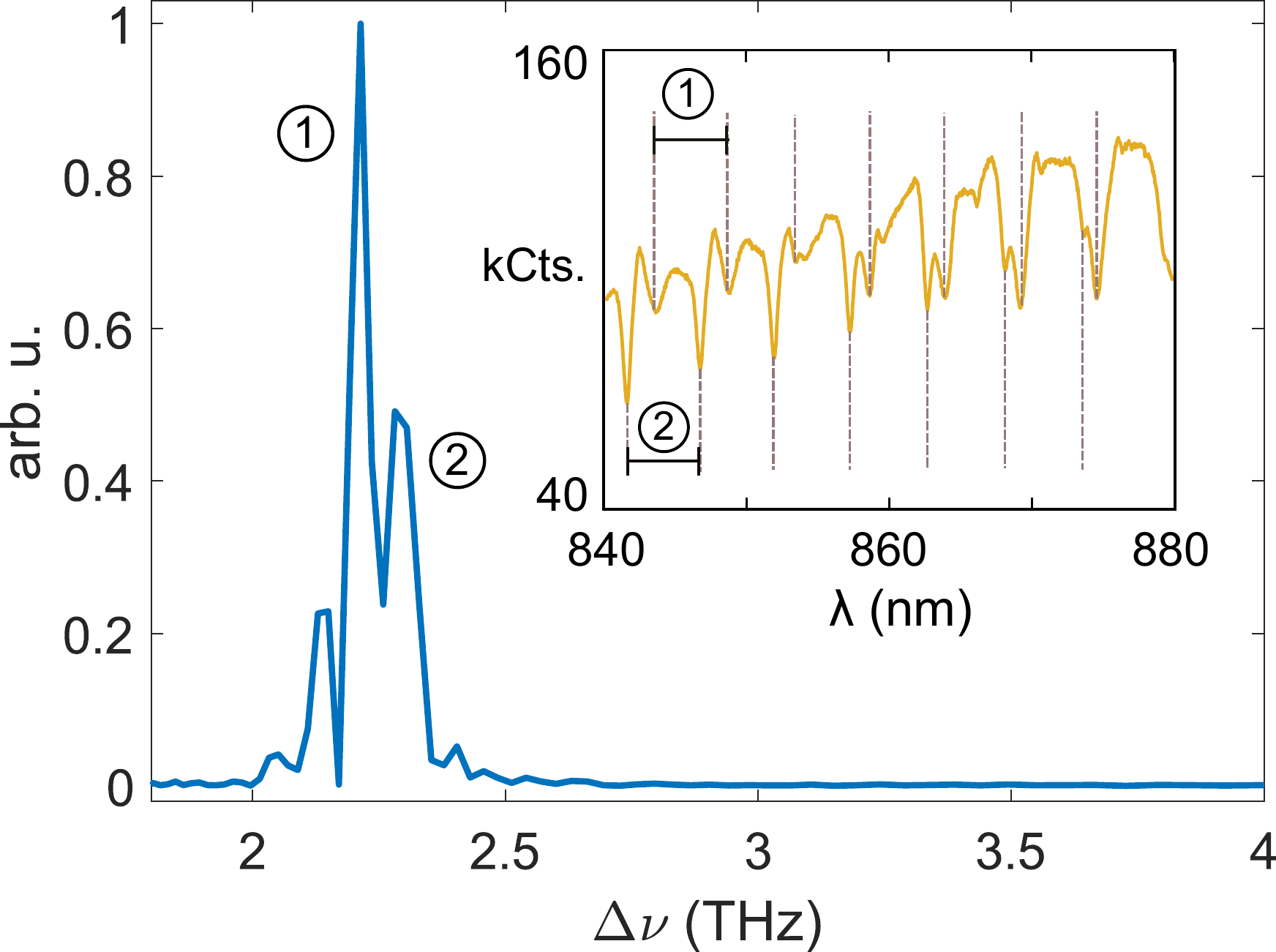}
\caption{Blue: Fourier transform of experimental transmission spectrum. 
Inset yellow: zoom of the Fig 6-b.
Two different sets of peaks can be seen.
The corresponding $\Delta\upsilon_{\text{FSR}}$ are 2.21 THz and 2.28 THz (peaks location in the blue curve).
This splitting corresponds to different polarization modes undergoing birefringence along the propagation in the structure.}\label{figure3-6}
\end{figure}

A keener look at the data reveals that two different resonance modes (see insert curve in Fig.~\ref{figure3-6}) contribute to the spectrum, otherwise we would observe regularly spaced peaks.
The FSR for the different modes is slightly offset.
When the peaks positions of the two modes are staggered, we can clearly see the two discrete peaks.
To identify the two values of FSR, we took the Fourier transform of the spectrum, which is plotted in Fig.\ref{figure3-6}. 
There, we clearly see two distinct resonant contributions labeled 1 and 2. 
This shift between the FSR values corresponds to a relative variation of the indices between the two propagating modes of $\Delta=(n_1-n_2)/n_1=3.1\%$, with $n_1$ and $n_2$ the effective refractive indices of both modes.

To determine the origin of this lifting of degeneracy, we will now focus on the mechanical properties of the knot.
As for the twisted loop we have here a nanofiber under strong bending constraints.
This bending implies stress and related mechanical effect that we assume at the origin of the mode splitting. 
We investigate two possible sources of birefringence: (i) stress-induced birefringence ($B_s$) and (ii) ovalization-induced birefringence ($B_o$) \cite{okamoto2006fundamentals}. 
Stress-induced birefringence is a well-known phenomenon in optical fiber \cite{ulrich1980bending} whereas ovalization-induced  birefringence is directly linked to the diameter of the fiber considered here.

Ovalization of an elastic rod is a well-known phenomenon in mechanical engineering \cite{wierzbicki1997simplified}, and can be commonly experienced when bending an elastic tube. 
The perfect circular section of the tube will change due to the bending to an oval section with its short axis along the direction of the bending. 
Such a mechanism is commonly neglected in optical fibers given the rigidity of standard fibers in which a bending force will lead to stress-induced birefringence and break the fiber before leading to significant ovalization of the section. 
However, in our case, with fiber of nanometric diameter, the mechanical properties are very different.
Moreover, the electromagnetic field is significantly less sensitive to stress, since it is largely localized outside the fiber. 
The stress only affects the part of the mode confined inside the silica. 
We evaluate the relative change of indices due to stress to be of $\Delta\approx1.7\%$. This is not enough to explain the different values of FSR we observed, and leads us to consider ovalization, or Brazier effect, as a significant factor in the degeneracy lifting.   

\begin{figure}[htbp]
\centering\includegraphics[width=0.8\linewidth]{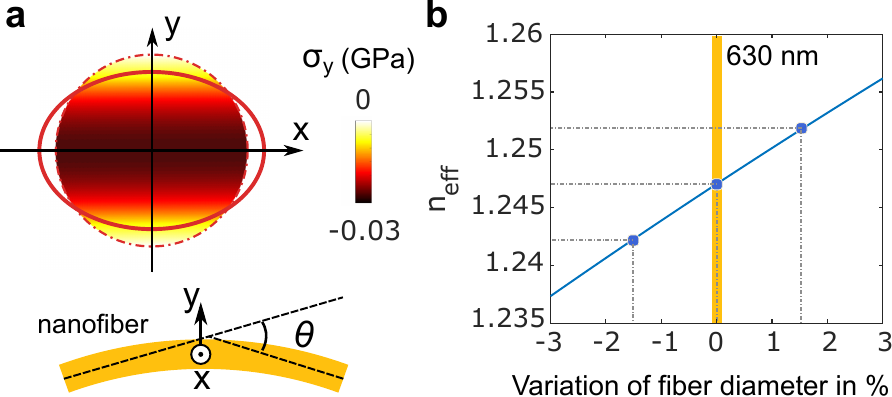}
\caption{Brazier effect on a bent optical fiber. 
a- Schematic of the fiber cross-section ($x-y$ plane) ovalization. 
$\sigma_y$ is the $y$ component of the bending induced stress, represented with the color scale within the fiber cross-section.
$\theta$ is the angle of rotation of the fiber ends.
b- Relative refractive index variation as a function of the fiber diameter change in percent compared with an initial diameter of 630 nm.}\label{figure3-8}
\end{figure}

We observe that ovalization or Brazier effect also participates in the birefringence. 
Brazier effect is a mechanical deformation that causes the ovalization of the cross-section of a bent tube.
 As shown in the Fig.~\ref{figure3-8}-a, $\sigma_y$ is defined as the $y$ component of the bending induced stress. The color in the fiber cross-section shows the value of $\sigma_y$ given by \cite{ulrich1980bending}:
 \begin{equation}
    \sigma_y = K^2 (E/2) (x^2-a^2)
\end{equation}
where $E$ is the Young's modulus of silica, $K=1/R$ is the curvature of the longitudinal axis, and $a$ is the fiber radius. 
 The value of $\sigma_y$ varies along the $y$ axis from 0 GPa at the fiber surface to -0.03 GPa at $y=0$ (corresponding to the compression stress), which gives an increased circumferential strain at fiber surface along the $x$ axis and a reduced circumferential strain at fiber surface along the $y$ axis \cite{wang2018cross}. 
 This effect caused the ovalization in the plane perpendicular to the bending axis. 
The relative displacement of two axes can be approximated as \cite{wierzbicki1997simplified}:  
\begin{equation}
    \overline{\delta}=0.553 aK
\end{equation}
where $a$ is the initial fiber radius before bending. 
Taking into account the measured loop radius ($R\approx11 \mu m$) and the fiber diameter  we found a reduction of the fiber diameter of  1.5\% parallel to the bending direction and an increase of 1.5\% in the normal direction. 
In contrast to standard fibers, this small change in the geometry will have strong impact given the transverse distribution of the field. 
A difference of 3\% of the radius, leads to a difference of effective refractive index of $(n_{+1.5\%}-n_{-1.5\%})/n=0.8\%$, as shown in Fig.\ref{figure3-8}-b.
 The joint effect of stress and ovalization give a birefringence of 2.5\%, which is in reasonable agreement with the observed value ($\Delta=3.1\%$). 
The small discrepancy between these two values is likely due to the fact that  the loop is not perfectly circular (as visible in Fig.\ref{figure3-5}-a ) and therefore leads to a non-homogeneous ovalization and stress effect along the ring.

\section*{Conclusion}
\addcontentsline{toc}{section}{Conclusion} 
In this chapter, we have reported the fabrication and characterization of two optical devices based on looped nanometrical optical fibers. 
The nature of the devices changes with the topology of the loop. 
\begin{itemize}
 \item The first device is a twisted loop. 
 
 The entwined part of the loop can be treated with the coupled mode theory and can be seen as a tunable beam splitter. 
We showed that a twisted loop creates a Sagnac interferometer, of which the dephasing is tuned by the torsion applied on the nanofiber. 

 \item The second device is a ring cavity, simply made out of a nanofiber knot. 

We analyzed its spectral response and found a finesse of 8 and a quality factor of 1300. 
Unlike in common resonators, the coupling efficency into the cavity strongly depends on the wavelength, which modulates the visibility of its resonances. 
It is reproduced accurately by our theoretical model.
\end{itemize}

Both setups, Sagnac interferometer and Fabry--Perot resonator are essential to photonics applications and optics in general. 
Their miniaturized versions presented here pave the way toward their integration in photonic circuits. 
 A refined analysis of the cavity spectrum revealed that birefringence of a bent nanofiber is also affected by ovalization of its profile, and not only by stress as a normal fiber would be.
With both devices, we showed how sensitive nanofibers are to mechanical constraints. 
\chapter{Radiation force detection with a nanofiber sensor}
\markboth{NANOFIBER SENSOR}{}
\label{chap:standing_wave}
Publication: \textbf{Ding, C.}, Joos, M., Bach, C., Bienaimé, T., Giacobino, E., Wu, E., Bramati, A. and Glorieux, Q.. Nanofiber based displacement sensor. APPLIED PHYSICS B, 126(6), (2020).\\

Interestingly, a nanofiber has remarkable mechanical properties, including a small mass, on the order of 0.1 $\mu$g/m and a high sensitivity to vibrations and weak forces.
In this chapter, we combine the optical and mechanical properties of a nanofiber to create a unique sensing device for displacement measurements and optomechanical applications.
This work is following the original idea of Arno Rauschenbeutel and build on the preliminary work did by Maxime Joos.
We present here an optimized system for measuring nanometric displacements of an optical nanofiber.

\section{\label{sec:level1}Introduction}
Position sensing with micro or nano-meter resolution is  widely used both in scientific research and for industrial applications \cite{muschielok2008nano,andrecka2009nano}. 
 In order to probe weak forces the miniaturization of the devices towards the nanoscale is one of the main approaches \cite{de2017universal,de2018eigenmode}. 
Nanowires \cite{feng2007very}, carbon nanotubes \cite{conley2008nonlinear}, and graphene \cite{eichler2011nonlinear} were successfully demonstrated as potential materials for position sensing.
Several hybrid systems were also used including nanomechanical oscillators coupled to a whispering gallery mode resonator \cite{anetsberger2009near}, superconducting microwave cavities cooled by radiation pressure damping \cite{teufel2008dynamical}, silicon nanowire mechanical oscillators for NMR force sensing \cite{nichol2012nanomechanical} and nanomechanical oscillators combined with single quantum objects \cite{arcizet2011single}.

Pico-meter sensitivity has been achieved using a quantum point contact as a scanned charge-imaging sensor (sensitivity of 3 pm/$\sqrt{\text{Hz}}$ \cite{cleland2002nanomechanical}) or using a nanorod placed within a Fabry-Pérot microcavity (sensitivity of 0.2 pm/$\sqrt{\text{Hz}}$ \cite{favero2009fluctuating}). 

Here, we propose to use an optical nanofiber as a nanometric displacement sensor which is all optical and can operate at room temperature and in free space.
It is designed by combining the optical and mechanical properties of a nanofiber.
By depositing a nanoparticle on top of a nanofiber and placing this system in an optical standing wave positioned transversely as a ruler, we report an absolute sensitivity up to 1.2~nm/$\sqrt{\text{Hz}}$.
Using the Mie scattering theory, we evaluate the force induced on the nanofiber by an external laser and demonstrate that a sensitivity of 1~pN  can be achieved. 
Force sensing with pico-newton precision can be used as a sensor for optomechanics \cite{larre2015optomechanical}, biophysics investigations \cite{freikamp2016piconewton} and molecule mechanics \cite{finer1994single}.

In this chapter, we first introduce the experimental setup and describe the calibration procedure that is used to measure the sensitivity.
We then present the experimental Allan deviation of our device, measured up to 10 minutes with a $\text{100 }\mu \text{s}$ integration time.
In the second part, we present a possible application of our sensor to detect radiation pressure force.
We propose a mechanical model to estimate the displacement of our sensor under an optically induced force, and we show that the full Mie scattering theory must be taken into account for accurate predictions.
Finally we model the two orthogonal polarization as a signature of the anisotropy of the system.

\section[Absolute displacement measurement of an optical nanofiber]{\label{sec:level2}Measuring the absolute displacement of an optical nanofiber}
\begin{figure}[]
\centering\includegraphics[width=1\linewidth]{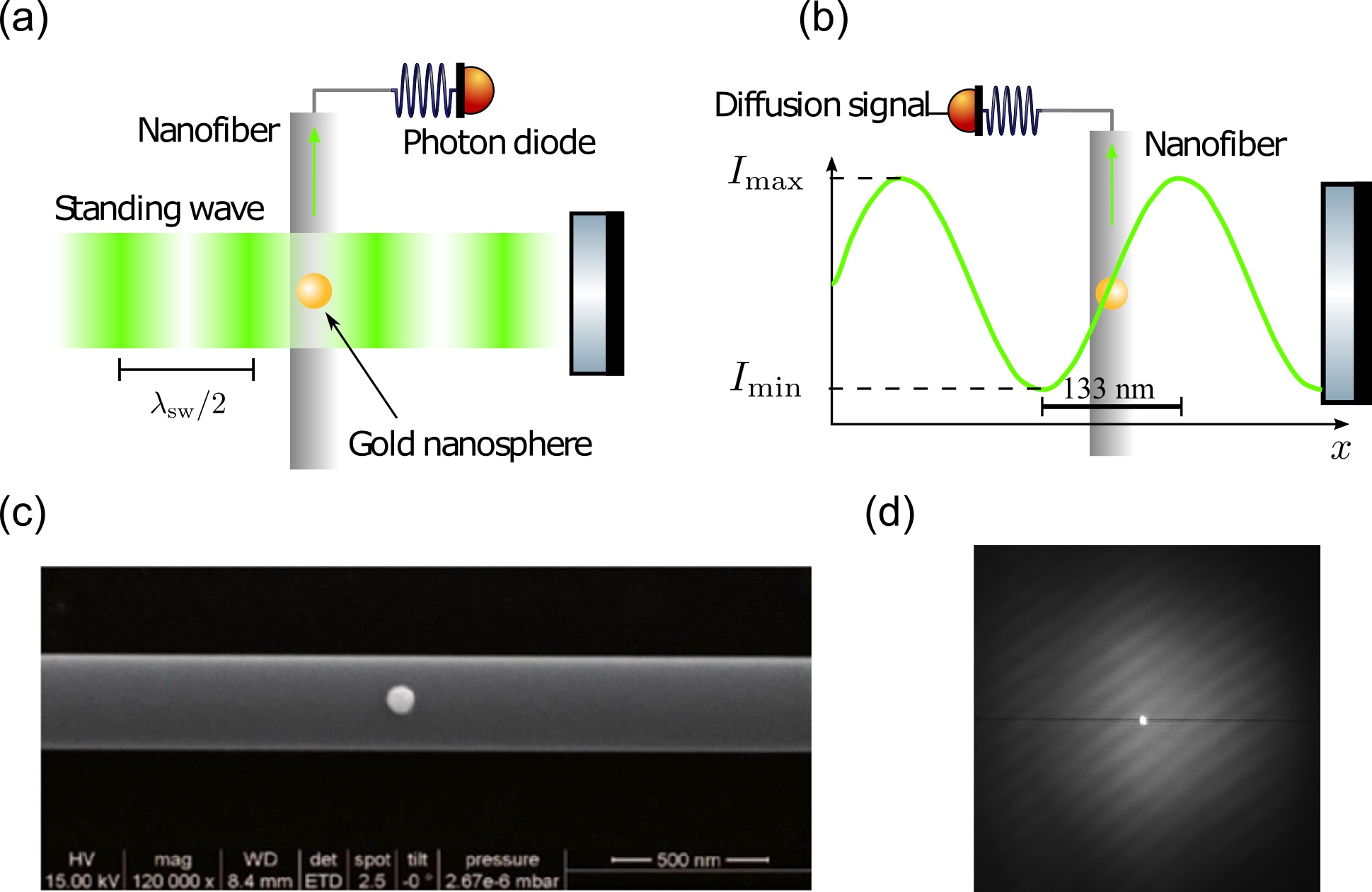}
\caption{a) Sketch of the experimental setup. A gold nano-sphere on the waist of an optical nanofiber is placed in a standing wave ($\lambda_\text{sw}$ is 532 nm). 
The diffusion signal is collected via the fiber.
b) Ruler resolution. Maximum and minimum  intensity are marked as $I_\text{max}$ and $I_\text{min}$, with 133 nm interval. 
c) Scanning electron microscope (SEM) image of a single gold nano-sphere deposited on nanofiber.
d) Optical microscope image of a gold-nano-sphere on the nanofiber within the standing wave.
(a,b is adapted from Maxime Joos \cite{joos2018dispositifs})}\label{figure1}
\end{figure}

In this chapter, we report on the use of an optical nanofiber as a nanometric displacement sensor.
We fabricated a tapered optical nanofiber with a diameter of 300~nm.
The diameter is chosen based on the guiding wavelength at 532 nm to achieve the best evanescent coupling at the fiber surface, which is introduced in section \ref{Dipole coupled to nanofiber}.
More than 95\% transmission is detected, following the method detailed in \cite{ward2014contributed} and used in \cite{joos2018polarization} as introduced in section \ref{Fabrication of the optical nanofiber}.
To observe scattering, we overlapped a transverse standing wave with the nanofiber, by retro-reflecting a standard laser diode at 532~nm on a mirror (see Fig.~\ref{figure1}a)).
The mirror is mounted on a piezo-transducer, so that we can scan it in the longitudinal axis to move the standing wave envelope around the nanofiber position. 
In order to guarantee a high contrast of the standing-wave, we verified that the coherence length of the laser is much longer than the distance between the mirror and the nanofiber.
In practice, this is not a stringent condition for standard narrow linewidth laser diode. 

In this configuration the light scattered inside the guided mode of the nanofiber is only due to the rugosity of the nanofiber and is extremely low.
To guarantee a high signal to noise ratio on the detection and therefore a high displacement sensitivity, we have deposited a gold nano-sphere on the nanofiber to enhance scattering (see Fig.~\ref{figure1}c).
The nano-sphere has a radius of 50~nm.
The deposition is done by filling a highly dilute solution of gold nano-spheres inside a micro-pipette and touching repeatedly the nanofiber with the meniscus at the extremity of the pipette.
We continuously inject light inside the nanofiber to monitor the scattered light from the deposition region. 
As seen in Fig.~\ref{figure1}-d, the scattered light increases dramatically after the deposition of the gold nano-sphere  \cite{yalla2012efficient}.

The nanofiber is mounted on a custom holder, which has a bending piezo transducer at one side, in order to control its tension.
Finally, we placed the system under vacuum to avoid dust particle deposition and perturbations from air-flows. 
The scattered light guided into the fiber is detected with an avalanche photodiode to have high detection efficiency (50$\%$) and fast response up to 100 MHz.


\begin{figure}[htbp]
\centering\includegraphics[width=1\linewidth]{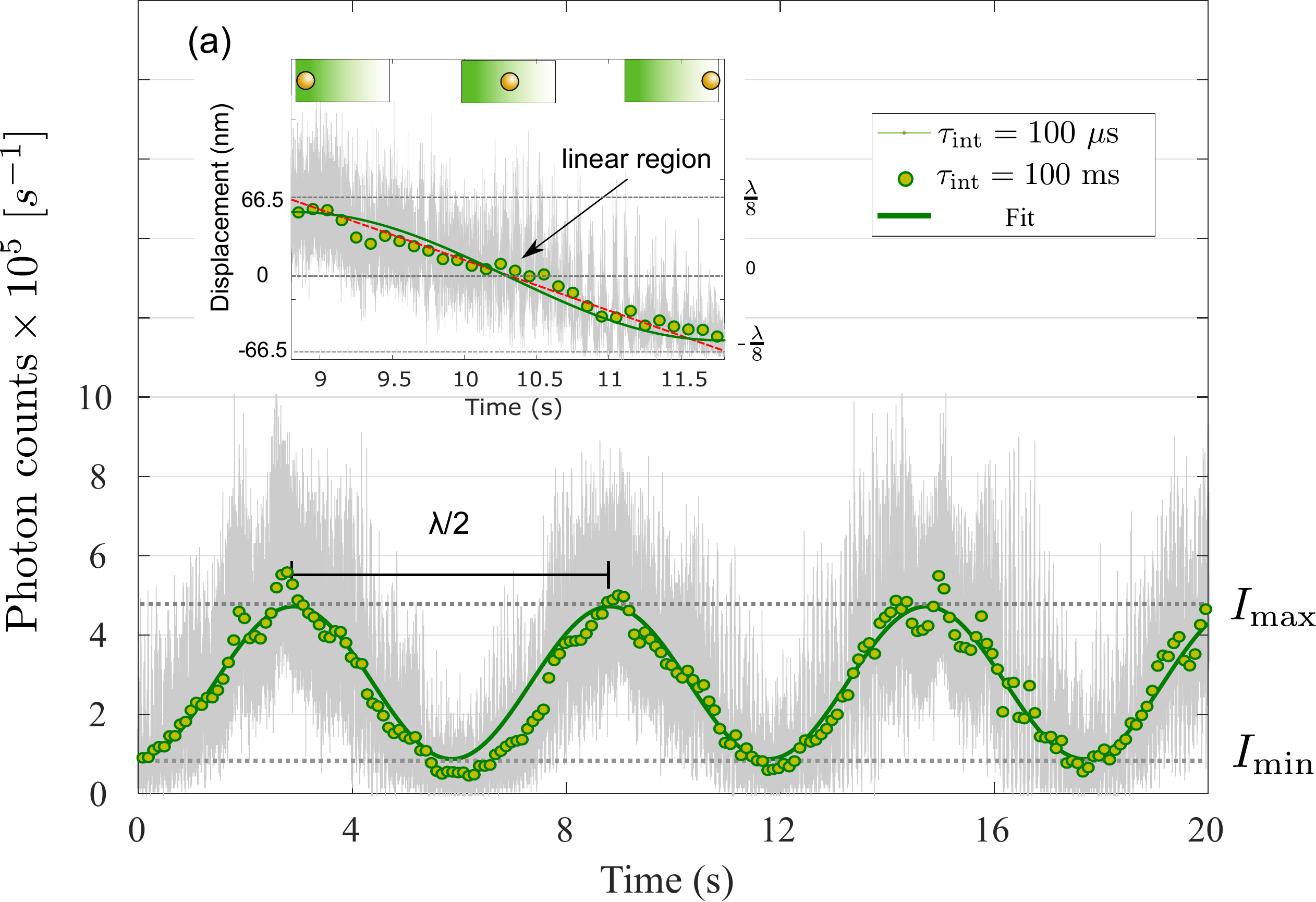}
\caption{Collected photon counts per second while scanning the standing wave.
Integration time is $\tau_\text{int}=100~\mu \text{s}$ for grey lines and $\tau_\text{int}=100$~ms for  green dots.
The solid green line is a sinusoidal fit of the data with $100$~ms integration time to determine the visibility.
$I_\text{max}=4.7$ and $I_\text{min}=0.87$.
Inset: the maximum resolution region is fitted with a linear trend. }\label{standingwave}
\end{figure}

To calibrate the displacement, we first scan the standing wave around the nano-sphere.
The signal oscillates with a period corresponding to $\lambda_\text{sw} / 2 = 266$ nm.
The visibility of the standing wave is defined as $v=(I_\text{max}-I_\text{min})/(I_\text{max}+I_\text{min})$,
where $I_\text{min}$ and $I_\text{max}$ are the minimum and maximum detected intensity.

The time window for photon counting adds a high frequency filter on the detected signal, which is defined as integration time.
A sinusoidal fit of the experimental data is presented in Fig.~\ref{standingwave} and gives a visibility of 0.71 calculated using the the measurement when the integration time is 100 ms.
In Fig.~\ref{standingwave}, we can see the vibrations in the environment through the sensing system.
These vibrations have high amplitude which widened the scan curve.
For this measurement a relatively long integration time (100 ms) has been used and therefore high frequency noise reduces the contrast below 1.
For small displacements ($d\ll \lambda/4$), the most sensitive region of the standing wave is the linear part (see Fig.~\ref{figure1}b).
In this region, the calibration of the displacement as a function of the scattered light is obtained by fitting the region with a linear model.


\section{\label{sec:level3}Sensitivity of the system}
The sensitivity of our system is linked to two main factors: at short integration time, the noise  essentially comes from the fluctuations of the scattered light and the mechanical oscillations of the nanofiber and, at long time scale, we observed a drift caused by a temperature shift in the environment which displaces the standing wave.

\begin{figure}[htbp]
\centering\includegraphics[width=0.8\linewidth]{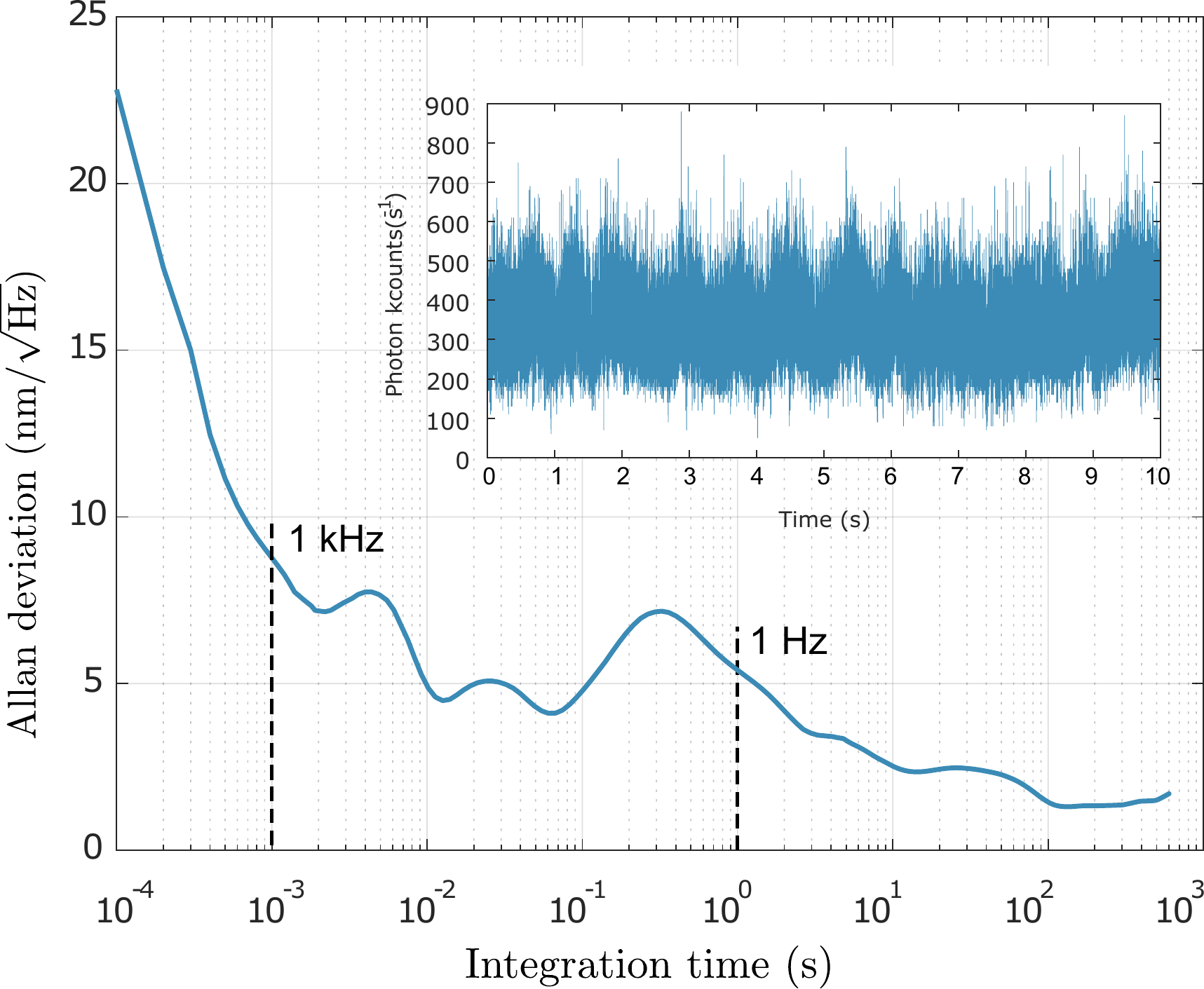}
\caption{Allan deviation.
Inset: scattering signal versus time.}\label{AllanDevnm}
\end{figure}

A quantitative description of the sensitivity is given in Fig.~\ref{AllanDevnm} by presenting the Allan deviation.
The Allan deviation is typically used to measure frequency stability in oscillators but it can also be implemented in the time domain.
Differently from standard deviation which gives only one value for a given integration time, Allan deviation gives access to the whole noise spectrum at different integration time in our system.
We follow the equation given by \cite{allan1981modified}.
$\Omega(\tau)$ is the measured time history of photon counts with a sample period of $\tau_\text{0}$. 
We divide the data sequence (photon arrival time on the photodiode) into clusters of time $\tau$.
The averaging time is set as $\tau=m\tau_\text{0}$, where the averaging factor $m$ is a group of integers from 1 to $n$. $n=100$ was used in calculation.
The Allan deviation $\sigma(\tau)$ is calculated using averages of the output rate samples over each time cluster. 
\begin{equation}
\sigma(\tau)=\sqrt{\frac{1}{2 \tau^{2}}\langle \left(x_{k+2 m}-2 x_{k+m}+x_{k}\right)^{2}\rangle},
\end{equation}
or in practice:
\begin{equation}
\sigma(\tau)=\sqrt{\frac{1}{2 \tau^{2}(N-2 m)} \sum_{k=1}^{N-2 m}\left(x_{k+2 m}-2 x_{k+m}+x_{k}\right)^{2}}
\end{equation}
where $x(t)=\int^{t} \Omega\left(t^{\prime}\right) d t^{\prime}$ is the integrated photon counts. $N$ is the sample points. 

According to the fit in Fig.~\ref{standingwave}, we got the correspondence between fiber displacement and photon count, which is about 2.9 kcounts/nm.
Therefore, the Allan deviation is given in unit nm/$\sqrt{\text{Hz}}$ in Fig.~\ref{AllanDevnm}.

On Fig.~\ref{AllanDevnm}, three main regions can be identified: i) at  integration time shorter than 1~ms (frequency higher than 1~kHz) where the sensitivity increases with the integration time (lower $\sigma$), ii) at integration time between 1~ms and 1~s (between  1~kHz and 1 Hz) where we observe a sinusoidal noise with characteristics bumps in the Allan deviation signature of the mechanical modes of the nanofiber, and iii) at integration time longer than 1~s (frequency lower than 1~Hz) where the sensitivity increases again.
Finally, at longer integration times than 3 minutes (not shown), the sensitivity decreases again (higher $\sigma$) due to the thermal drift in the standing wave phase.
We can extract two key figures of merit from these data. 
The resolution at 1~s integration time is 5.1~nm/$\sqrt{\text{Hz}}$, and the maximum absolute resolution is 1.2~nm/$\sqrt{\text{Hz}}$.
Comparing to the previous work, the displacement sensing system has higher visibility and higher resolution.

\section[Nanofiber displacement driven by force]{Displacement of the nanofiber driven by externally applied force}
\label{Displacement of the nanofiber driven by externally applied force}
In this section, we propose one possible application of our nanofiber displacement sensor for optomechanics experiments.
We assess the possibility to detect the radiation pressure force, by estimating theoretically the displacement of the nanofiber when it is driven by an externally applied force and we investigate in particular optically-induced forces based on the Mie theory.

We model the nanofiber as an elastic string of circular section (radius $a$) with the two ends fixed.
The location of the applied force on the nanofiber is an important parameter for the sensitivity to the force.
Here, we assume that the force is applied at the middle of the nanofiber waist, i.e. in the most sensitive region, however the model can be extended easily to other positions.
First, we take the initial tension on the nanofiber to be zero.
The expression of the displacement as function of the applied force $F$ is then  given by:
\begin{equation}
\delta=\left[\frac{F}{8 \pi a^{2} E_Y}\right]^{1 / 3} L,
\label{eq1}
\end{equation}
where $E_Y$ is the Young modulus of silica, and $L$ is the length of the nanofiber region \cite{bellan2005measurement, timoshenko1968elements}.

It is interesting to evaluate if a radiation pressure force applied by an external pushing laser could be observed using this system with the sensitivity measured earlier.
A coarse evaluation can be made using a simple model by considering the nanofiber with radius $a$ as a beam with squared section $(2a)^2$ and estimate the force by the differential Fresnel reflections on the sides of the silica beam.
In this geometrical approach, the pushing laser is described by its diameter $d$, its power $P$ and its intensity $I = P/(\pi d^2/4)$.
If $d \gg 2a$, the illuminated section of the nanofiber is $2a \times d$ and the force is 
\begin{equation}
    F_{\text{go}} = \frac{2 I}{c}\ 2R\ 2ad = 8\frac{ P}{c \pi d}\ 2R\ 2a,
\end{equation}
with $c$ the speed of light in vacuum and $R$ the normal incidence reflection coefficient.
With this model, we can estimate the force of a flat top beam of $P = 100\text{ mW}$ and diameter $d=10\text{ }\mu\text{m}$ on a \textit{nanofiber} of squared section $2a = 500\text{ nm}$ made in silica (optical index of $1,5$ which induces a Fresnel reflection coefficient $R = 0,04$):
\[
F_{go} \sim 3 \text{ pN}.
\]

However, if this simple model is instructive to obtain an order of magnitude of the applied force, it must be refined to take into account the interference and polarization effects.
In the following, we use the Mie theory of diffusion to give a more precise estimation of the optical force.
We assume a cylindrical nanofiber (radius $a$ and index of refraction $n_1$) illuminated on normal incidence by a plane wave (with Gaussian envelope of width $d$). 

\begin{figure}[htbp]
\centering\includegraphics[width=0.6\linewidth]{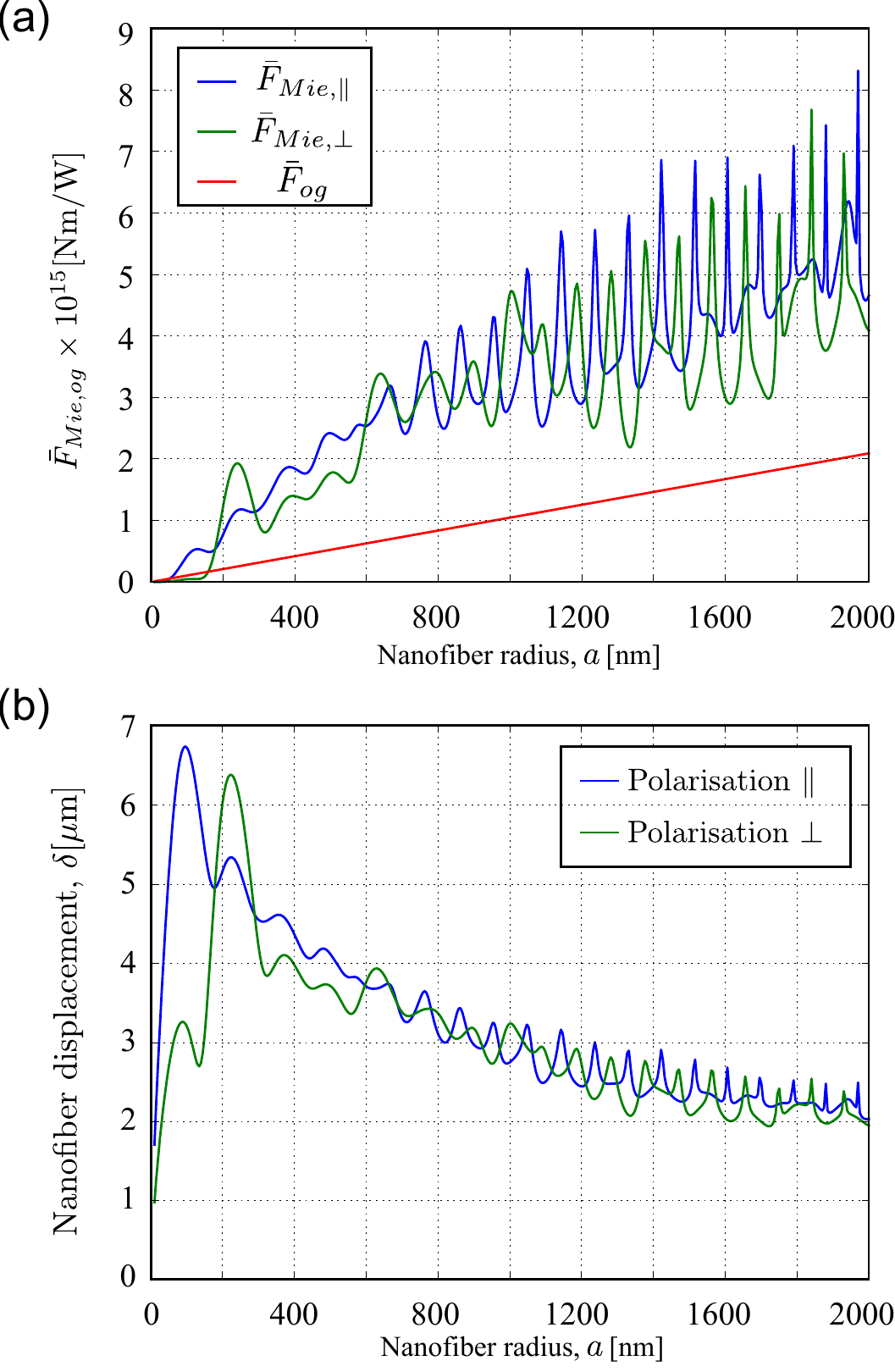}
\caption{Estimation of the radiation force on a silica nanofiber.
(a) Radiation force $\bar{F}_\text{rad}$ on a nanofiber per unit of intensity and per unit of length of illuminated fiber as a function of the nanofiber radius.
(b) Deflection of the nanofiber under the effect of the radiation force $\delta$ as a function of the nanofiber radius. 
Parameters: flat-top laser beam ($\lambda$ = 780 nm), with power $P$ = 100 mW and diameter $d$ = 10 $\mu$m, the Young's modulus of silica $E_\text{Young}$ = 70 GPa and the length of the nanofiber $L_\text{0}$ = 10 mm.
This figure is adapted from Maxime Joos \cite{joos2018dispositifs}.}\label{figure5}
\end{figure}
Two linear polarization can be used, either parallel or perpendicular to the nanofiber axis.
The force given by the Mie theory is  \cite{mitri2017radiation,loo2019imaging}:
\begin{equation}
    F_{\text{Mie}} = \frac{2I}{c} a d Y,
    \label{eq:Frad_Mie}
\end{equation}
where $Y$ is a dimensionless quantity which depends on the polarization.
For a linear  polarization parallel to the nanofiber $Y = Y_{||}(n_1, k, a)$ is given by
\begin{equation}
    Y_{||}(n_1, k, a) = -\frac{1}{ka}\text{Re}\left[ \sum_{l=-\infty}^{+\infty} b_l(b_{l+1}^* + b_{l-1}^* + 2)\right].
\end{equation}
The geometric coefficient $b_l = b_l(n_1, k, a)$ is given by 
\begin{equation}
b_l = \frac{n_1J'_l(n_1ka) J_l(ka) - J_l(n_1ka) J'_l(ka)}{n_1J'_l(n_1ka) H^{(1)}_l(ka) - J_l(n_1ka) H'^{(1)}_l(ka)},
\label{eq:Mie-coeff}
\end{equation}
with $H_l^{(1)}$ is the first order Hankel function and  $J_l$ is the first order Bessel function.
Similarly, for a linear polarization perpendicular to the nanofiber, we have
\begin{equation}
    Y_{\bot}(n_1, k, a) = -\frac{1}{ka}\text{Re}\left[ \sum_{l=-\infty}^{+\infty} a_l(a_{l+1}^* + a_{l-1}^* + 2)\right],
\end{equation}
with
\begin{equation}
a_l = \frac{J'_l(n_1ka) J_l(ka) - n_1J_l(n_1ka) J'_l(ka)}{J'_l(n_1ka) H^{(1)}_l(ka) - n_1J_l(n_1ka) H'^{(1)}_l(ka)}.
\label{eq:Mie-coeff2}
\end{equation}.

This model is compared numerically to the geometric model in Fig.~\ref{figure5}a). 
It can be observed that the geometric model, represented by the red line in the figure clearly under-estimate the force for both polarizations above a nanofiber diameter of 200~nm.
Compared to the geometrical optics model, the Mie scattering model also predicts oscillations in the radiation force as a function of the increasing radius of the nanofiber, which corresponds to an interference between the specularly reflected waves from the edge of the nanofiber and those reflected by its dielectric core.
In addition, the radiation force shows dependence on the incident polarization which is characteristic of the anisotropy of the system \cite{joos2019complete}.

In the following, we use the full model of Mie scattering to estimate the displacement.
We inject the expression of the force in the displacement of the nanofiber given by Eq. (\ref{eq1}) and we trace the displacement as function of the nanofiber radius in the absence of tension $T_0$.
In this configuration, we predict a displacement on the order of 5 $\mu$m for both polarizations at fiber diameter around 400~nm. While $5 \, \mu\text{m}$ far exceeds the linear dynamical range introduced in the first part of this paper, we can easily reduce the power of the beam generating the radiation pressure force such that the beam displacement is compatible with the requisite range of our experimental technique (displacement smaller than 266 nm).
The minimal force that can be detected with our device is then on the order of 1~pN.
Finally, to extend our model, we can include a non-zero initial tension on the fiber.
In presence of an initial tension $T_0$, a correction term can be added to Eq. (\ref{eq1}) to determine the force $F$ required to induce a displacement  $\delta$ of the nanofiber:
\begin{equation}\label{equation6}
F=8 \pi a^{2} Y_E\left[\frac{\delta}{ L}\right]^{3}+4 T_{0}\frac{\delta}{ L}.
\end{equation}
The first term is linked to the mechanical properties of silica as the second one is a consequence of the nanofiber tension.
As $\delta / L \ll 1$ in our experiment, tension might play an important role in the ability of our system for detecting weak force applied on the fiber. 
Experimentally, we have anticipated this situation by adding a bending piezoelectric transducer to the nanofiber holding mount.
This open the way to future investigations of the tension as well as the potential role of damping due to air around the nanofiber.



\section{Experimental procedure}
\begin{figure}
    \centering
    \includegraphics{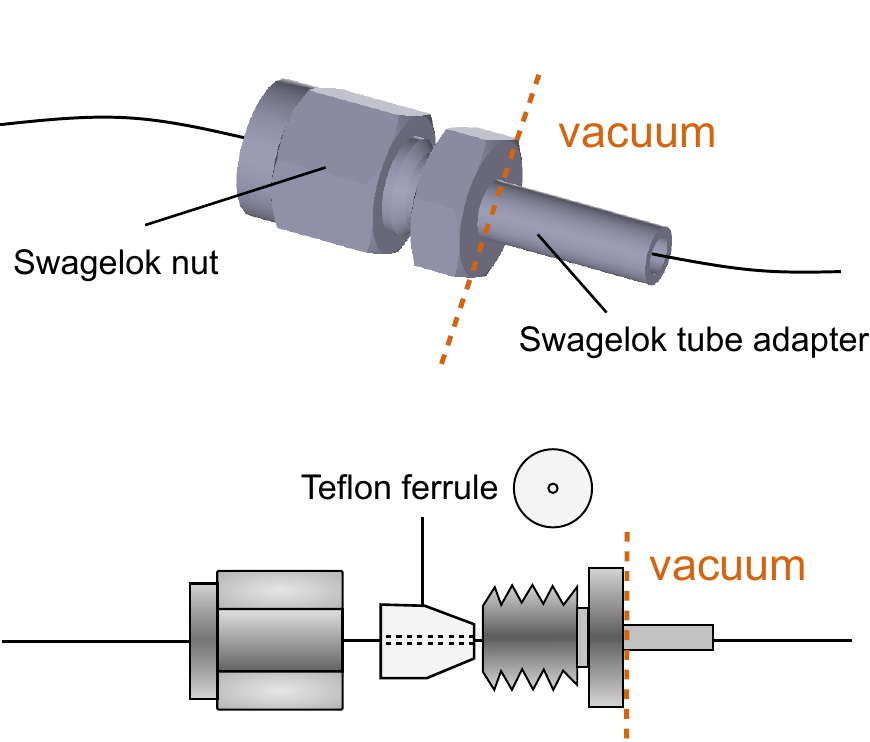}
    \caption{Assembly diagram for the Teflon fiber feedthrough on vacuum chamber.}
    \label{fig:swagelok}
\end{figure}

To install the nanofiber inside the vacuum chamber, we designed a mechanical piece working as a vacuum port that can hold the nanofiber and two feedthrough for optical fiber.
The optical fiber feedthrough is designed based on swagelok system, as shown in Fig.\ref{fig:swagelok}, following Baptiste Gouraud \cite{gouraud2016optical}.
The stainless steel fitting ferrule is replaced by one made with Teflon.
For a commercial optical fiber, the hole diameter on the Teflon ferrule was chosen to be 0.3 mm to achieve good transmission at the same time.
With this diameter, optical fiber without cladding can pass through the hole easily, and the part with cladding can be squeezed by Teflon producing a good seal.

The process of installing the nanofiber with a gold nanorod in the vacuum chamber is as following.

\begin{figure}
    \centering
    \includegraphics[width=\linewidth]{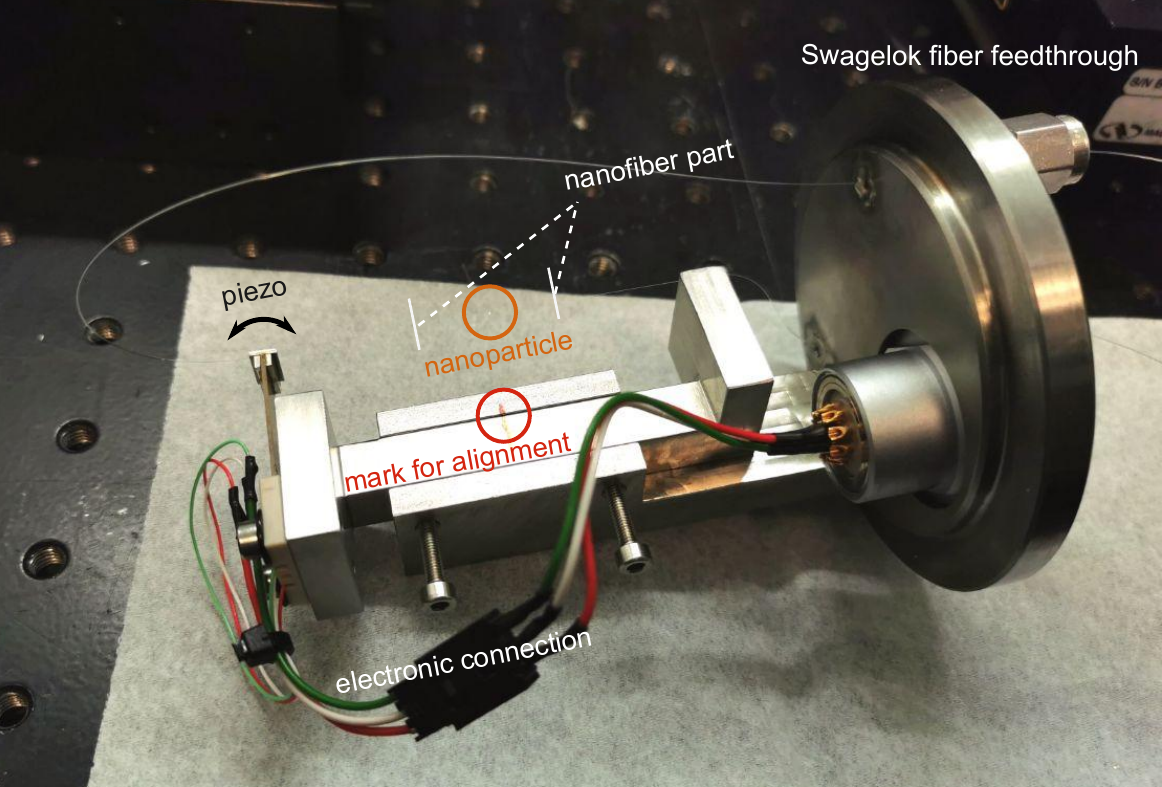}
    \caption{Tension-adjustable nanofiber holder for vacuum chamber.}
    \label{fig:lid}
\end{figure}

The first step is the nanofiber preparation. 
We fabricate the nanofiber and deposit a single gold nanosphere on the nanofiber, as introduced in section \ref{Fabrication of the optical nanofiber} and \ref{particle deposition}.
The whole process needs to be done in the air flow box.
The detailed process of the nanofiber fabrication and particle deposition are introduced in chapter 1.
We install the nanofiber on the fiber holder with particle aligned with the location mark, shown in Fig. \ref{fig:lid}, so that we can always have the particle around the same place in the vacuum chamber. 
One side of the fiber holder is equipped with a bending piezo with $\pm 450~\mu m$ traveling range to adjust the tension of the fiber.
We clean the two stages that will hold the fiber with isopropanol.
The two sides of the nanofiber are glued to the stage with UV curing optical adhesive.

The second step is to install the fiber holder on the vacuum port. 
The design of the nanofiber holder for vacuum chamber is shown in Fig. \ref{fig:lid}.
The fiber holder is installed on a column attached to the vacuum port aligned with the location mark on the column.
We pass the two fiber ends through the swagelok tube adapter.
Then, we need to clean both sides of the optical with isopropanol again before adding the Teflon ferrule and the swagelok nut.
We leave both sides of the fiber with appropriate length.
Also, we need to leave the fiber in a stable state with relatively large curvature to maintain high transmission.
The swagelok nut should not be fully tightened at this step.
Then we plug the electronic connection and fix the free wire to avoid touching the optical nanofiber.
We need to do all these steps with great care to avoid fiber breaking.
We transfer the mechanical piece with fiber holder into a container about the same size as the vacuum chamber so that the nanofiber can be well protected before we put it into vacuum chamber. 
And also we can check the fiber transmission (The transmission is about 80\% after we deposit a gold nanosphere) and adjust the electronic wire to avoid it touching the nanofiber.

The third step is to transfer the mechanical piece with fiber holder into the vacuum chamber.
We transfer the vacuum port into the vacuum chamber.
We slightly adjust the port to have no electronic wires on the way of standing wave.
Then, we fix the vacuum port on the chamber.
Between the vacuum chamber and the vacuum pump, there is a valve and a pressure detector.
To avoid the dust in the pipe blew onto the nanofiber, the pipe needs to be pumped with valve closed first.
Then we vacuum the vacuum chamber. 
We carefully tighten the Swagelok nut while detecting the nanofiber transmission and pressure in the vacuum chamber.
Pressing the Teflon ferrule too much will not help to lower the pressure, so we stop before the transmission is going to drop.

The last step is the alignment of the sensing system.
Thanks to the location marks, the nanosphere is already within the view of camera at this step.
The focus plane of the camera needs to be adjusted to set the focus on the gold nanosphere.
So we just need to adjust the laser at wavelength 532 nm to pass the gold nanosphere. 
By detecting the nanofiber guided signal, we can optimize the alignment.
Then we add the dichroic mirror at the other side of vacuum chamber to reflect back the laser.
To have the standing wave with best visibility, the reflected laser beam needs to be aligned with the incoming laser and the focus plane needs to be on the mirror.
The nanofiber guided signal is doubled after adding the dichroic mirror.

\begin{figure}
    \centering
    \includegraphics[width=\linewidth]{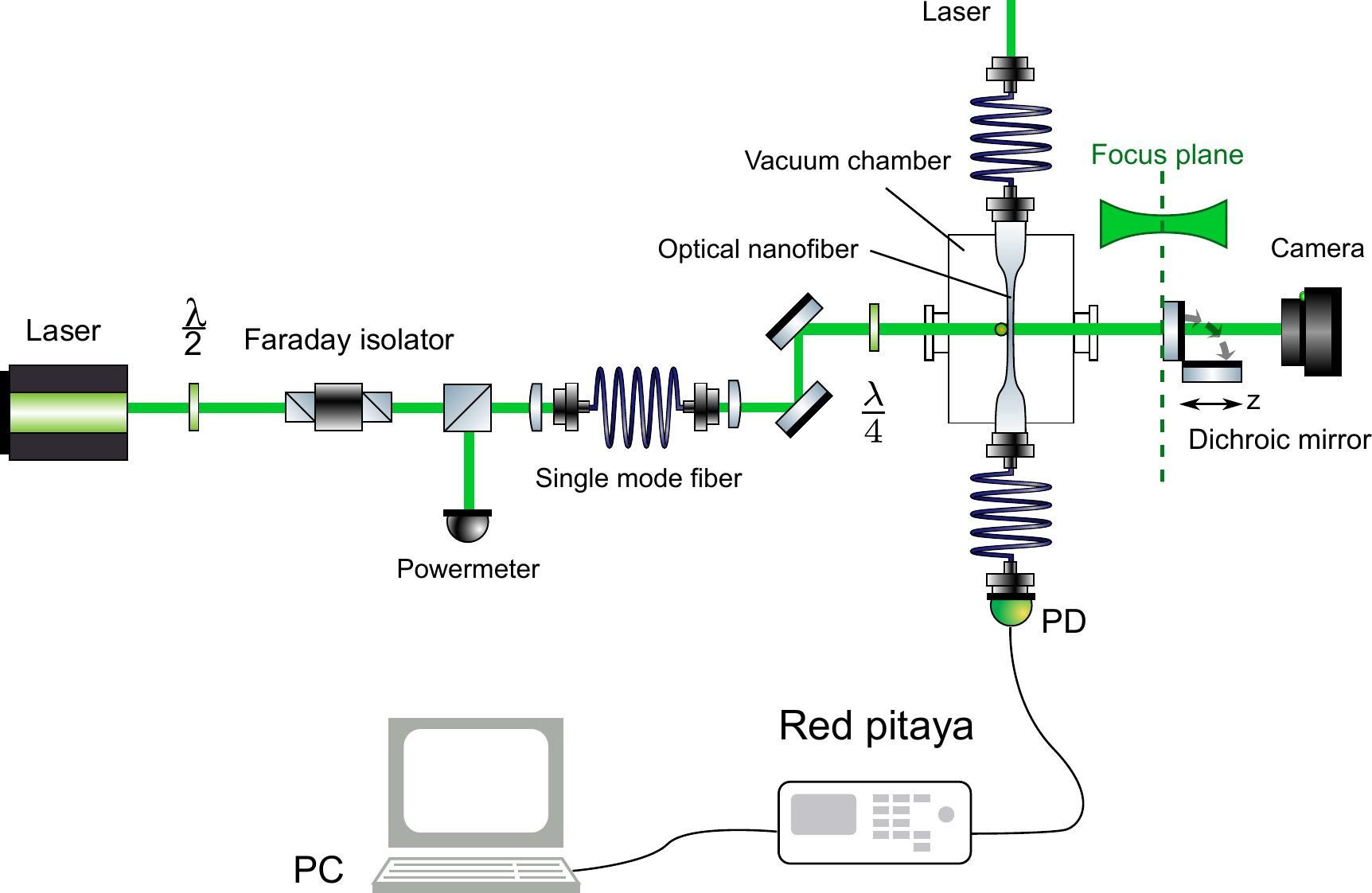}
    \caption{Experiment setup of displacement measurement.}
    \label{fig:standing wave experiemnt setup}
\end{figure}

The experiment setup of the nanofiber displacement measurement is shown in Fig.\ref{fig:standing wave experiemnt setup}.
The reflected light is isolated with an Faraday isolator.
The well isolated laser was sent to the gold nanosphere through two mirrors to gain the alignment freedom in $x-y$ plane.
We flip down the dichroic mirror to have a clear view of the scattered light from gold nanosphere and the location of laser beam in the camera.
We need to adjust these two mirrors to have the laser beam aligned with the gold nanosphere and along the $z$ axis of the camera.
The fine tuning of these two mirrors are guided by the scattered light from the gold nanosphere detected with the photodiode.
A quarter waveplate is installed before the vacuum chamber to adjust the excitation polarization. 
We maximize the signal on the photodiode to align the laser and the gold nanosphere in the transverse plane.

The piece of single mode fiber and the beam splitter connected to the optical powermeter after the Faraday isolator is used for checking the alignment of the standing wave.
The single mode fiber is used as a target for the reflected light to ensure the overlap of the light coupled out and coupled in.
With the dichroic mirror flipped up, the light reflected back can be optimized by adjusting the angle of dichroic mirror and the $z$ position of the out coupling lens of the single mode fiber.
The $z$ position of the out coupling lens of the single mode fiber decides the focus plane of the standing wave.
As we shown in the Fig.\ref{fig:standing wave experiemnt setup}, the focus plane of the standing wave need to be on the surface of dichroic mirror.
To find the best $z$ position for the out coupling lens, we check the scattered light from the gold nanosphere detected with the photodiode.
The best position is when the maximum power of the scattered light from the gold nanosphere with the standing wave is about double the one without diachroic mirror.
The maximum power of the scattered light from the gold nanosphere illuminated by the standing wave is measured by scanning the dichroic mirror in $z$ axis.
The dichroic mirror is mounted on a piezo.
The voltage of the piezo is connected to a voltage amplifier controlled by a programmed Arduino.
The visibility of the sensing system is highly related to the alignment of the standing wave.

The time resolution of the sensing system is further improved by replacing the photodiode by a single photon detector (which based on avalanche photodiode with 65\% photon detection efficiency at wavelength from 400 nm to 1060 nm and a dead time of 22 ns).
To know the damage threshold of the guiding light at 532 nm wavelength in our homemade optical nanofiber with 300 nm diameter within a the vacuum chamber, we did the following test.
We slowly increase the power of coupled light and when the power is 15 $\mu $W, the nanofiber fused. 
This is due to the gold nanosphere has been heated too much and melt the nanofiber.

To acquire scattering signal from the gold nanosphere, we need to respect the damage threshold of the nanofiber.
We use a photodetector with femtowatt sensitivity (Thorlabs PDF10A2) to read the weak scattered signal guided to one side of the optical nanofiber, which is a low-frequency device (20 Hz).
This limits the time resolution of our sensing system.
When we replace it by a single photon detector, the detection frequency has been increased and the sensitivity as well.
The data of photon number in time scale is recorded with the single photon detector and then sent to a programmed Red Pitaya which is connected to PC with local area network connection.
We can have 10 $\mu$s time resolution with real time data collection.
This allows us to obtain the mechanical movement of the optical nanofiber with frequency higher than kHz.

By scanning the dichroic mirror, we obtain the photon counts per second as a function of mirror displacement.
We move the dichroic mirror so that the gold nanosphere is in the linear region as shown in Fig.\ref{standingwave}.

\begin{figure}
    \centering
    \includegraphics[width=\linewidth]{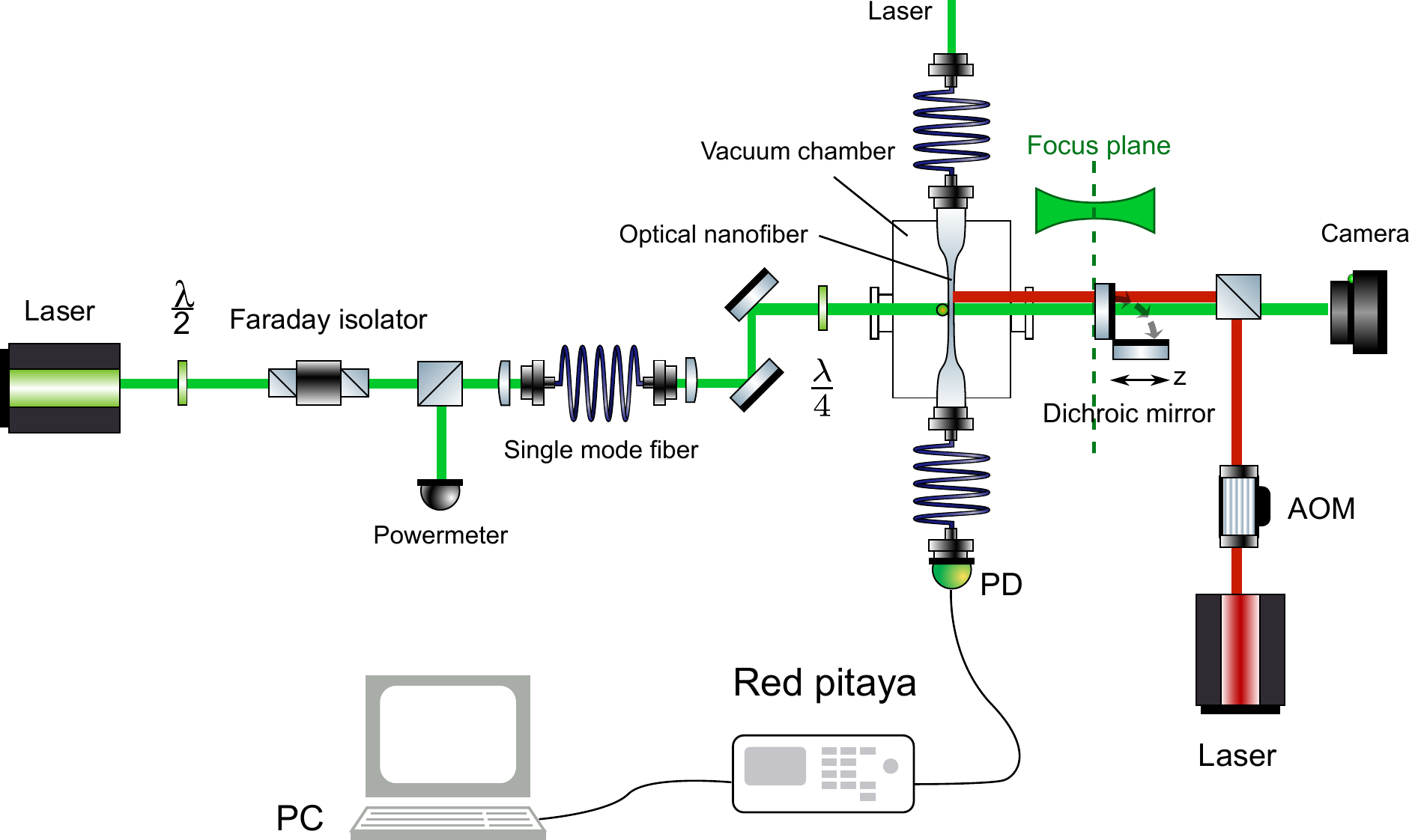}
    \caption{Radiation pressure detection experiment setup. The nanofiber is placed horizontally (along $x$ axis). A 780 nm laser line is added additionally on the position sensing system. An acousto-optic modulator (AOM) is used to modulate the intensity of the beam. The beam focuses on the nanofiber waist and is vertical to the nanofiber. To avoid heating the gold nanosphere, the focus position on the nanofiber is shifted tens of $\mu m$ in the $x$ axis along the longitudinal direction of the nanofiber.}
    \label{fig:radition pressure measurement}
\end{figure}

\section{Radiation pressure detection}
In section \ref{Displacement of the nanofiber driven by externally applied force}, we introduced the displacement of a silica nanofiber driven by the external force and estimated the situation when a radiation force is applied on the nanofiber waist.
As shown in Fig.\ref{figure5}, when the nanofiber has a radius about 150 nm, which is the parameter we use for the displacement sensing system, the displacement of the nanofiber under the effect of the radiation force from a flat-top laser beam at 780 nm with power of 100 mW and diameter about 10 $\mu m$ applied on the center of the nanofiber waist can reach few $\mu m$.
In this section, we introduce the experiment on radiation pressure detection.

The experiment setup for detecting radiation pressure is shown in Fig.\ref{fig:radition pressure measurement}.
The idea is to use an external laser beam to push the nanofiber at certain frequency.
The displacement of the nanofiber is recorded with the position sensing system.
We expect to detect an extra signal with applied frequency.
Based on the position sensing system, we add a 780 nm laser line.
An acousto-optic modulator (AOM) is used to modulate the intensity envelope of the beam as a function of time $t$ : 
$I (t) = A_\text{peak} \sin{(2 \pi \frac{t}{T_p})}$ .
The period $T_p$ and peak amplitude $A_\text{peak}$ are adjustable with function generator.
We focus the beam on the nanofiber waist and incident angle is vertical to the nanofiber. 
To avoid heating the gold nanosphere, the focus position on the nanofiber is shifted tens of $\mu m$ in the $x$ axis or the longitudinal direction of the nanofiber.

\section{Mechanical vibration detection}
With our displacement sensing system, a resolution of 1.2~nm/$\sqrt{\text{Hz}}$ is obtained, corresponding to the sensibility of an externally applied force at the nanofiber waist of 1 pN.
That's a good sign for applying the sensing system to the radiation pressure detection.
Here we study the displacement of the nanofiber in frequency domain. 
As we can see in the inset image in Fig. \ref{AllanDevnm}, the movement of the nanofiber is a mix of periodic and random vibration, and it can be continuous or intermittent.
For a periodic vibration, the motion repeating itself after certain periods of time $T$.

By adding a periodically modulated radiation force, the displacement measurement can be replaced by vibration measurement in the frequency domain.
And we can target at the resonance frequency of the fiber and extract the signal from background noise.

\subsection{Standing wave sensing system}
We use the same experiment system to measure the mechanical resonance of the nanofiber.
The experiment setup is shown in Fig.\ref{fig:standing wave experiemnt setup}.
The position of the dichroic mirror is set to have the gold nanosphere in the sensitive region.
As we introduced before, our sensing system can record the displacement of the nanofiber with frequency higher than kHz.

Although we have the information of the fiber displacement, there is other information hidden in the signal's frequency content, for example, the mechanical resonance of the nanofiber triggered by the vibration in the environment.
In order to extract frequency information from information of fiber displacement, the Fourier
Transform (FT) and Wavelet Transform (WT) are used for transforming mathematically our view of the
signal from time-based to frequency-based.
By FT, a raw time domain signal is broken down into constituent sinusoids of different frequencies.
Consequently the frequency-amplitude representation obtained by FT presents a frequency component for each frequency that exists in the signal.
The serious drawback of FT is that the information in time domain is lost in transforming to the information in frequency.
So FT is appropriate for the situation when the frequency does not change a lot over time.
However, for the spectrum characteristics of a time varying signal, for example, if we want to distinguish the resonance spectrum of the nanofiber when it's pushed by a radiation force, FT will not be a perfect choice.
The frequency generated at certain time will be easily covered by the background noise in larger time scale during the whole data acquisition.

\begin{figure}
    \centering
    \includegraphics[width=0.8\linewidth]{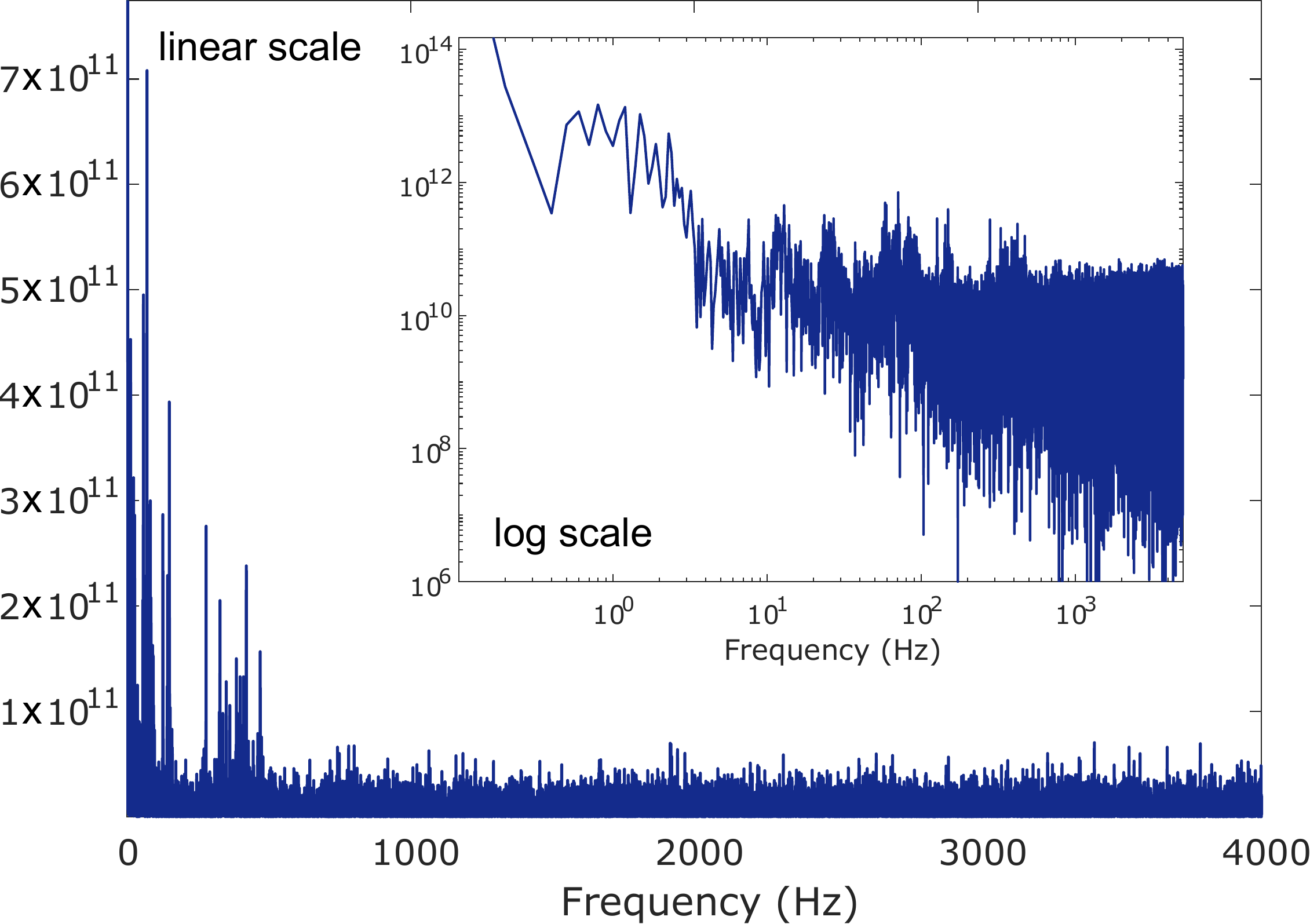}
    \caption{Fourier transform of the signal acquired within 10 seconds with integration time $\tau_\text{int}=$100 $\mu m$. Inset: the Fourier transform in log scale.}
    \label{fig:FT_background}
\end{figure}

In Fig.\ref{fig:FT_background}, we show the Fourier transform of the signal acquired within 10 seconds with integration time $\tau_\text{int}=$100 $\mu s$ when there is only the background noise.
The resonance frequency shown in this figure comes from the background mechanical vibration in the lab.

Therefore, it would be better to analyze the time localized signal with spectrum that also vary with time.
We introduce another mathematics approach that can transform signal from time-based to frequency-based while keep the information in time domain, which is the Wavelet Transform (WT).

A wavelet is a mathematical function that is used to divide a given function or continuous-time signal into different scale components \cite{burrus2013wavelets}.
Wavelet transform provides a time-frequency window that can be modulated. The width of the window changes with the frequency. 
When the frequency increases, the width of the time window becomes narrower to improve the resolution. The average value of the amplitude of the wavelet in the entire time range is 0, and it has limited duration and sudden frequency and amplitude.
We use the continuous wavelet analysis to deal with the 1-D signals we recorded.

\begin{figure}
    \centering
    \includegraphics[width=\linewidth]{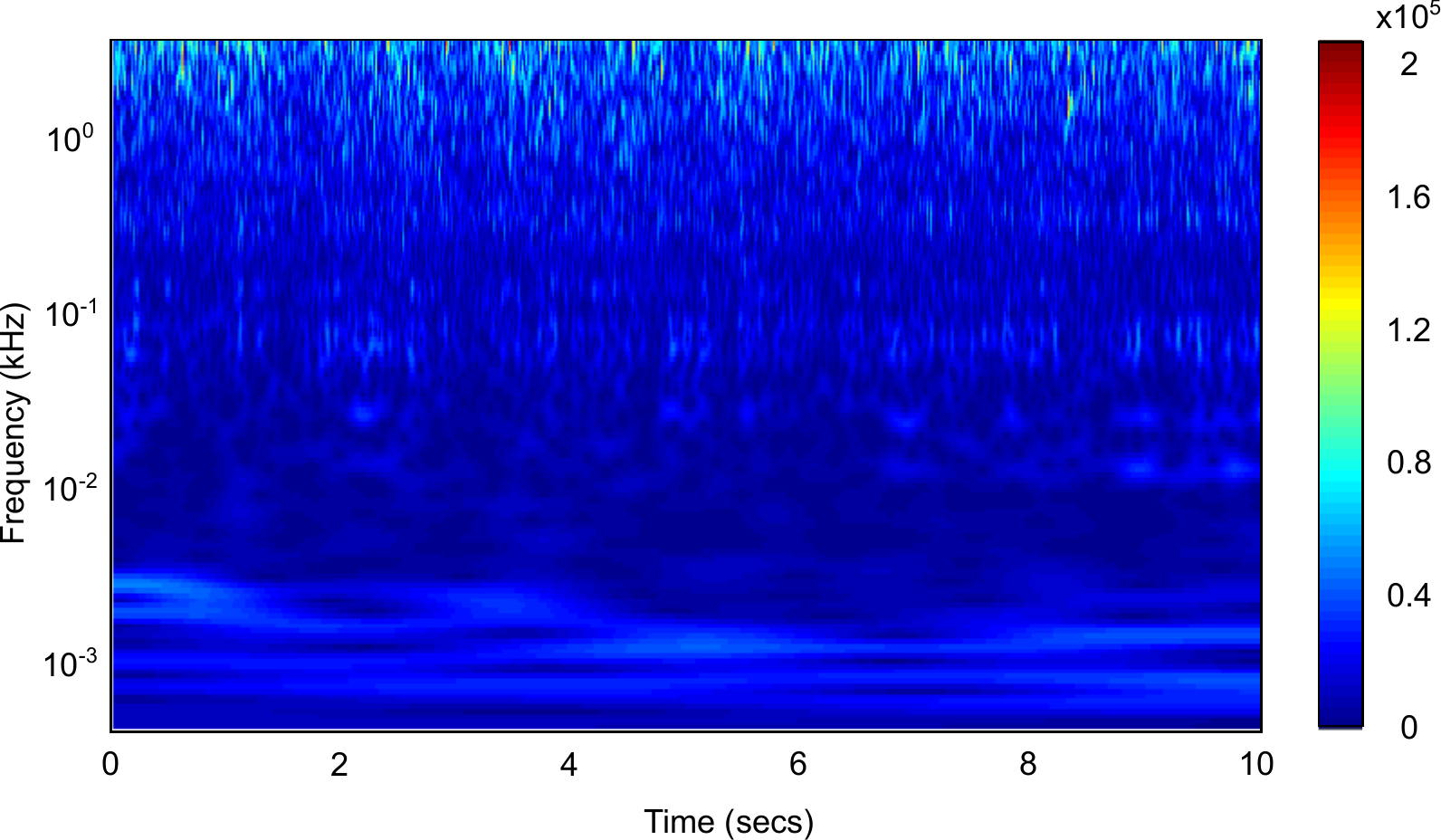}
    \caption{Wavelet transform of the signal acquired within 10 seconds with integration time $\tau_\text{int}=$100 $\mu s$. }
    \label{fig:WT_background}
\end{figure}

We show in Fig.\ref{fig:WT_background} the WT of the signal acquired within 10 seconds with integration time $\tau_\text{int}=$100 $\mu m$, which is using the same set of data as the Fourier transform in Fig.\ref{fig:FT_background}.
With wavelet transform approach, we manage to extract the spectrum of the signal over time.
The maximum distinguishable frequency we have is 4 kHz, which is limited by the time resolution of photon number detection system.
In principal it can be further improved by using a Time-Correlated Single Photon Counting (TCSPC) system with picosecond resolution.

\begin{figure}
    \centering
    \includegraphics[width=\linewidth]{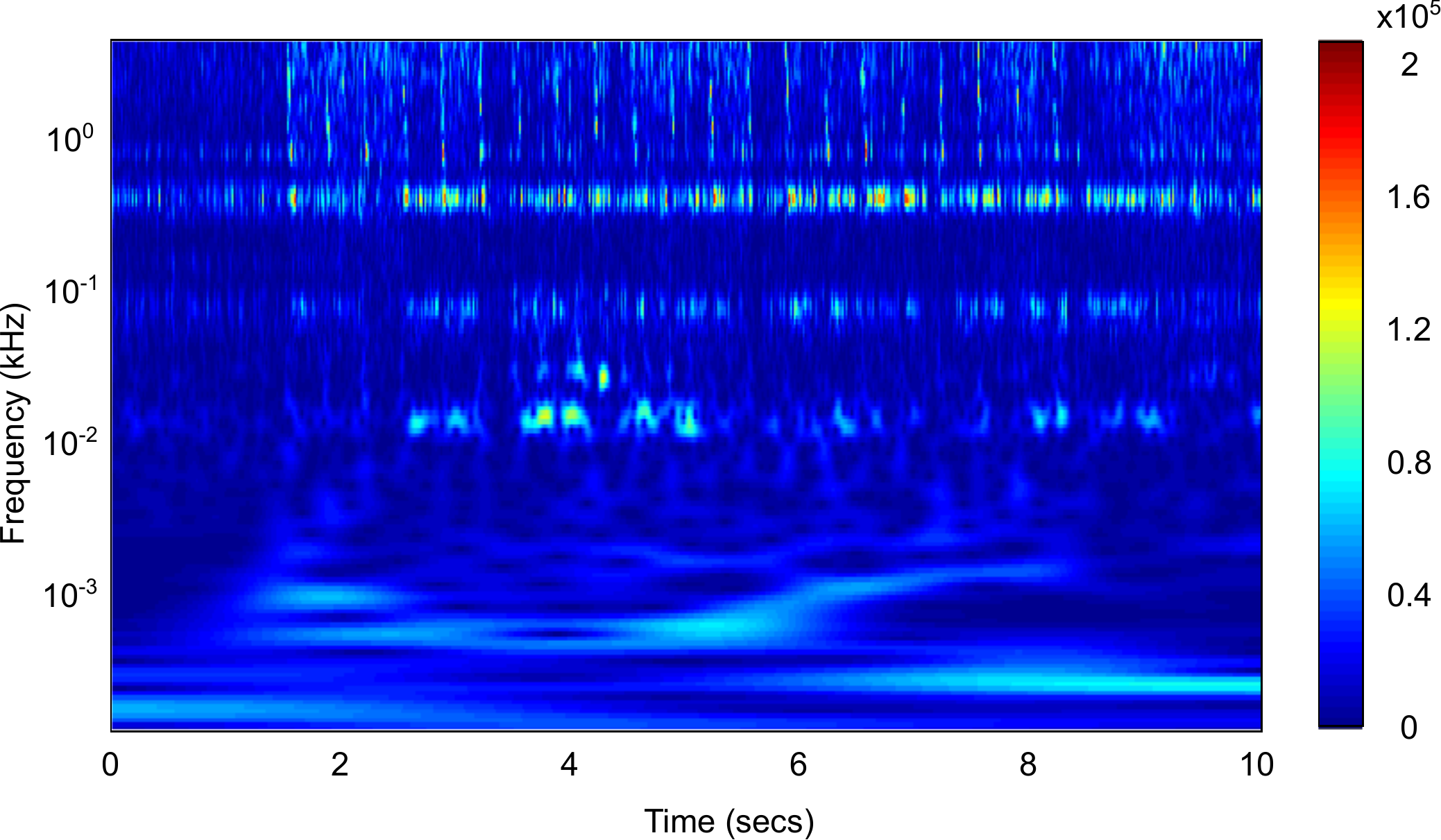}
    \caption{Wavelet transform of the signal acquired within 10 seconds with integration time $\tau_\text{int}=$100 $\mu s$. Each 1 second apart, we increase the tension applied on the nanofiber.}
    \label{fig:WT_Increasing tension}
\end{figure}

The resonance frequency of the string produce wave is decided by the length and tension of the string.
Similar effect as stringed instrument, the increase of tension on optical nanofiber increase its resonance frequency.
To find the resonance frequency of the nanofiber, we increase continuously the tension applied on the nanofiber at 1 second intervals to trace the frequency shift.
The tension was added to the optical nanofiber by the bending piezo on one side of the fiber holder, introduced in previous sections.
Based on Mersenne's laws, the relation between the resonance frequency of the string can be given by the tension $T$, the linear density of the string $\mu$ and the length of the string $L$:
\begin{equation}
f=\frac{1}{2 L} \sqrt{\frac{T}{\mu}}.
\end{equation}

The WT of the signal is acquired within 10 seconds with integration time $\tau_\text{int}=$100 $\mu s$, as shown in Fig.\ref{fig:WT_Increasing tension}.
The added tension causes the resonance to increase because it takes a greater wave speed to move the wire.
Therefore, we expect to see the shift of frequency towards higher frequency, and this can give us the information of fiber resonance frequency as well.
However, in Fig.\ref{fig:WT_Increasing tension}, the frequency excited by added tension seems to be stable.
The obtained frequency when we add tension is 400 Hz, 800 Hz, 1.2 kHz, 1.6 kHz, 2 kHz, 2.4 kHz, appears to be a harmonic spectrum.
Since the distinguishable frequency of our system is limited, and the resonance frequency of a nanofiber can be higher, the resonance frequency might be out of our detection range.

We didn't find the resonance frequency of the nanofiber. It might be due to the limit of the detection range or the sensitivity of this system is not enough for the radiation force we applied.
Based on the suggestion by Arno Rauschenbeutel, homodyne detection system is a better choice for detecting the frequency of the fiber deflection. 
It measures the phase changing due to the birefringence caused by the bending of optical nanofiber, which gives higher sensitivity than the displacement system.
Therefore, in the next section, we introduce our attempt to add a homodyne detection system on the original displacement system to have better resolution in vibration sensing.

\subsection{Homodyne detection system}
Homodyne detection system can extract information of the phase variation with high sensitivity.
When a nanofiber is triggered by arbitrary mechanical vibration, the nanofiber starts to swing in the direction perpendicular to the fiber longitudinal axis, or alternately twist clockwise or counterclockwise taking the fiber as axis.
When the mechanical vibration is the same frequency as the nanofiber resonance frequency, the amplitude of the movement becomes larger.
The sensing system based on standing wave is based on the displacement of the nanofiber.
An absolute sensitivity up to 1.2~nm/$\sqrt{\text{Hz}}$ offers quite good visibility to arbitrary movement in the direction perpendicular to the fiber longitudinal axis, but will not give as much sensitivity to the twisting of the nanofiber.
In both movement, extra birefringence is induced since bending the fiber or twisting the fiber will create asymmetric stress distribution on the fiber cross-section.
Therefore, Homodyne detection system can be sensitive to the vibration in both case.
Although it can not give the information of fiber displacement like the standing wave sensing system does, it can offer the information of fiber vibration more precisely in general.

\begin{figure}[htbp]\label{homodyne}
\centering\includegraphics[width=0.9\linewidth]{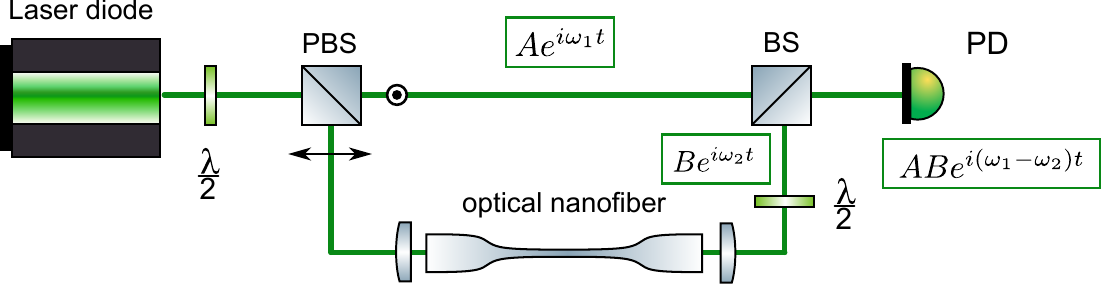}
\caption{Schematic of experiment system for measuring phase variation with homodyne detection. The interference of two beams was detected by a photodiode (Thorlabs DET36A/M 25 MHz). It gives the information of the phase shift $e^{i(\omega_1-\omega_2)t}$ within the nanofiber compared to the reference beam.}
\label{fig:homodyne}
\end{figure}

In Fig.\ref{fig:homodyne}, we show the sensing system with homodyne detection.
A linearly polarized beam was divided into horizontally-polarized and vertically-polarized beams with a polarized beam splitter.
The vertically-polarized light acts as the reference beam that propagating in the free space.
The reference beam carrying the phase information in free space is represented by $A e^{i \omega_1 t}$.
The horizontally-polarized light is coupled into the optical nanofiber and then coupled back to the free space followed by a half waveplate to rotate the polarization at the fiber output to match with the reference beam at the beam splitter.
The signal beam carrying the phase information of the nanofiber is represented by $B e^{i \omega_2 t}$.
The two beams interfere with each other after the beam splitter and carry the information of phase variation in the optical nanofiber generated by the induced birefringence due to the bending of the fiber, which can be described as $AB e^{i (\omega_1 - \omega_2) t}$.

With the homodyne detection system we will be able to monitor the movement of optical nanofiber, and the displacement sensing system will provide the information in absolute position.

\section*{Conclusion}
\addcontentsline{toc}{section}{Conclusion} 

First, we have presented a position sensor based on a gold nano-sphere deposited on a nanofiber and  placed within a standing wave.

This sensing system can measure the displacement of a nanofiber driven by externally applied force and vibrations.
We experimentally calibrated our sensing system and obtained a resolution of 1.2~nm/$\sqrt{\text{Hz}}$.
We proposed a mechanical model to estimate the response of our sensor to optically-induced pressure force and we found that the sensitivity corresponding to the sensibility of an externally applied force at the nanofiber waist is 1~pN.
This optical nanofiber sensor paves the way to realize integrated optomechanical research using this platform, such as the demonstration of superfluidity of light in a photon fluid \cite{larre2015optomechanical,fontaine2018observation}.

Second, based on the position sensor, we extract the spectrum over time, which can be used for mechanical resonance detection.

Third, we introduced a nanofiber homodyne detection system for measuring the mechanical resonance.

\chapter{Study of difference frequency generation with diamond vacancy}
\markboth{DIFFERENCE FREQUENCY GENERATION}{}
\label{chap:DFG}
Quantum emitters can be used to generate a qubit carrying information in a quantum network.
Light sources that is restrict to emit light in the form of unit energy or single photons are quantum emitters.
The most popular ones include atoms, ions, defects in diamond lattice and semiconductor quantum dots \cite{geng2016localised, manceau2014effect,vezzoli2015exciton,pierini2020highly}.
The wavelength of the emitted single photons distributes around visible-range according to the difference between energy levels.

The communication between nodes in a quantum network is normally realized by optical fiber. 
As a medium for telecommunication and computer networking, the advantage of optical fiber lies in its flexibility and low-loss communication.
For the typical fused silica glass fibers we use today, it is especially advantageous for long-distance communications with infrared light which has loss generally below 1 dB/km and a minimum loss at 1550 nm (0.2 dB/km).
It is a much lower attenuation compared to electrical cables.
This allows to span long distances with few repeaters. 

Here comes the problem that the communication between quantum nodes is suffering from the mismatch between the frequency of emitted photons and low-loss telecommunication-band region of silica. 

Quantum frequency conversion can solve this problem by alternating the visible photons to communication-friendly telecommunication band while preserving the other optical and quantum properties. 
We are interested in learning the possibility of achieving high conversion efficiency under room temperature with color centers in diamond considering their excellent stability for optical properties under room temperature.

\section{Color centers in diamond}
Color centers in diamond is one of the most promising candidates for efficient and stable single photon generation.
With a laser beam, one can repeatedly excite electrons at the color center. 
Each time, it emits a single photon in a specific quantum state when it decays back to its ground state.
There are differences between the diamond structure defects formed by different elements. 
The mechanisms behind are also slightly different.

\begin{itemize}
    \item Nitrogen Vacancy color center in diamond

The nitrogen-vacancy center is a point defect in the diamond lattice. 
The simplified atomic structure is shown in Fig. \ref{fig:NV}. 
It consists of a substitutional nitrogen atom with an adjacent carbon vacancy and has C$_{3v}$ symmetry, with the symmetry axis oriented along the crystallographic [111] axis, as shown in Fig.~\ref{fig:NV}-a).

\begin{figure}
    \centering
    \includegraphics[width=0.8\linewidth]{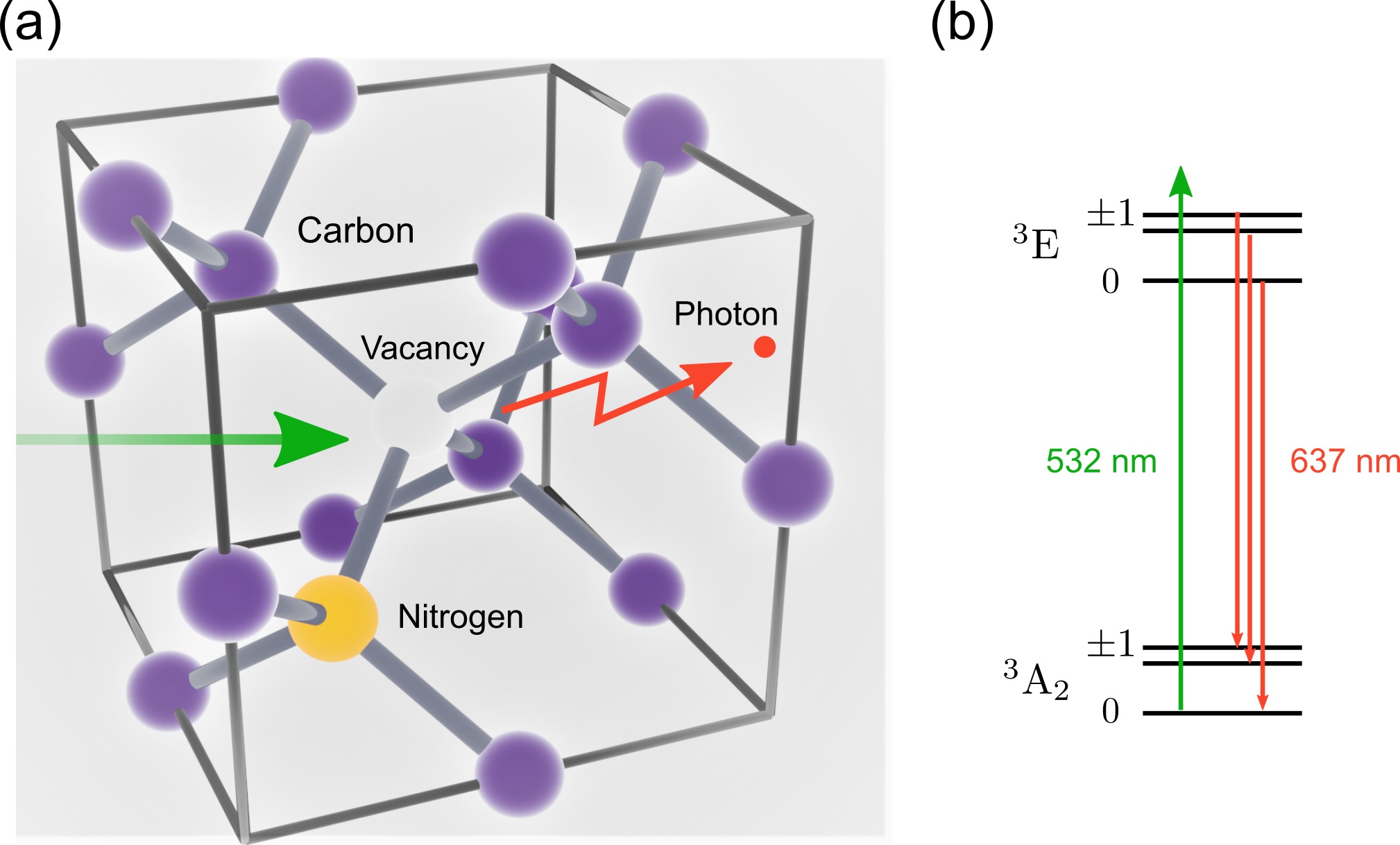}
    \caption{Simplified atomic structure of the NV$^{-}$ center}
    \label{fig:NV}
\end{figure}

The absorption spectrum of the NV color center is very wide, so there is no strict requirement in choosing the excitation laser.
The laser with wavelength around 500~nm can be used to excite the NV color center.
NV color center has two different charge states: neutral (NV$^0$) or negative (NV$^-$).
NV color center with neutral charge state has 5 electrons with spin $S=1/2$.
If NV$^0$ captures one electron in the lattice, it becomes NV$^{-}$, which has 6 electrons with spin $S=1$.
The spectrum of NV color center includes an intensity peak at 637 nm (zero phonon line for NV$^-$) and another peak at 575 nm (zero phonon line for NV$^0$). 
The phonon sideband of NV$^0$ expands to 750 nm, and the phonon sideband of NV$^-$ expands to 800 nm.
Therefore, the whole emission spectrum of NV$^{-}$ centers covers a wavelength range of about 100 nm. 
The zero phonon transition between the excited state and the ground state with energy level difference of $ E_1 - E_0$. 
The strength of the zero-phonon transition arises in the superposition of all of the lattice modes.
In the spectrum, it appears as a peak (called zero phonon line or ZPL) corresponding to the energy level difference between the excited state and the ground state.
The phonon sideband of NV color center in diamond is strongly suppressed at low temperatures, and the ZPL will be more obvious.
Besides, NV color center ensemble in nanodiamond has larger ZPL linewidth than the ZPL linewidth of NV color center in bulk diamond, which is the non-uniform broadening due to the presence of stress near the color center.

    \item Silicon Vacancy color center in diamond

Silicon Vacancy color center is another type of defect in the diamond lattice.
The lattice structure of a silicon vacancy color center is composed of one silicon atom with two adjacent carbon vacancy and has D$_{3d}$ symmetry, as shown in Fig.~\ref{fig:SiV}.

\begin{figure}
    \centering
    \includegraphics[width=0.6\linewidth]{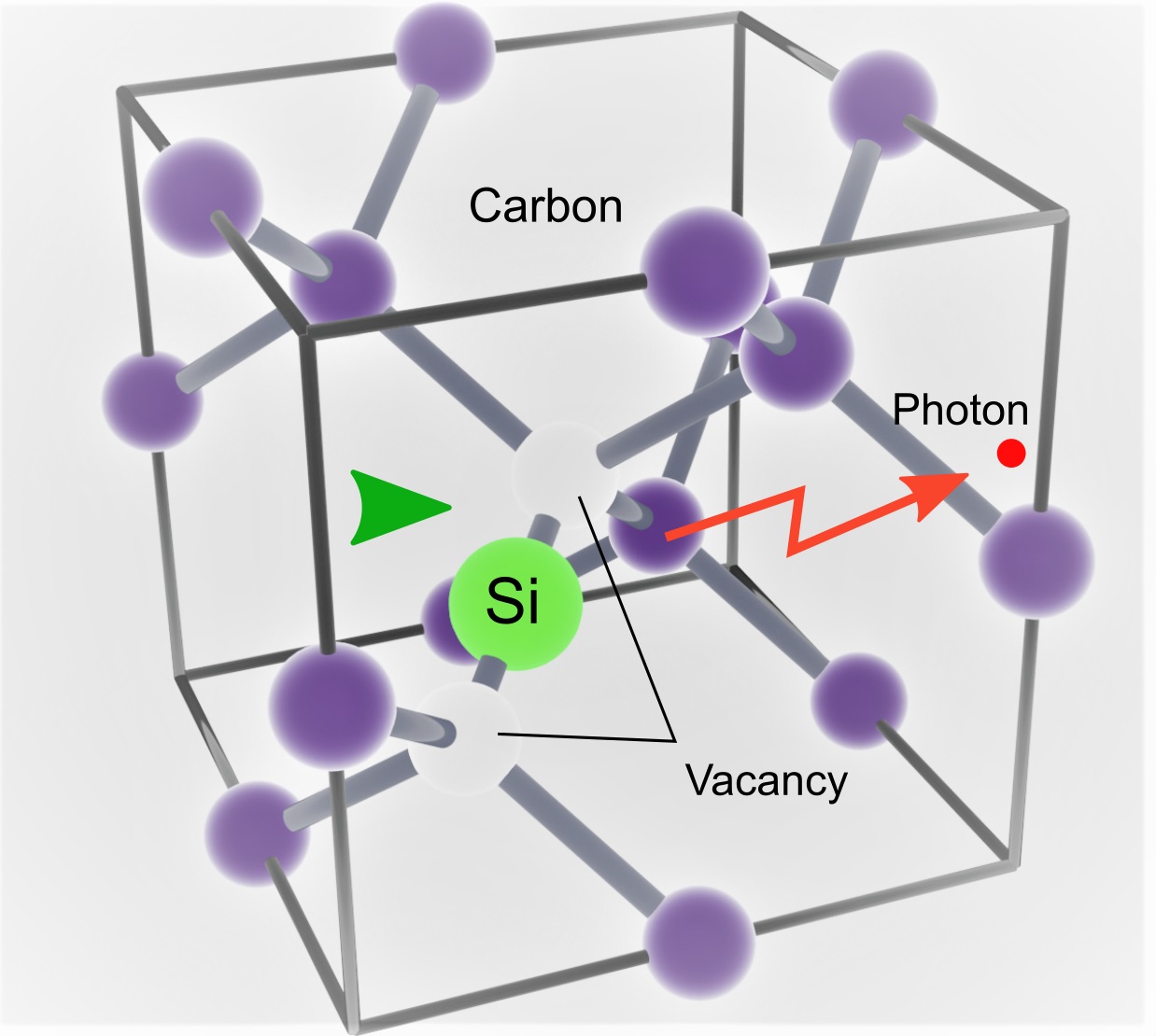}
    \caption{Simplified atomic structure of the SiV center}
    \label{fig:SiV}
\end{figure}

At room temperature, the SiV color center emits much more of its emission (approximately 70\%) into its zero phonon line at 738 nm than most other optical centers in diamond, such as the nitrogen-vacancy center (approximately 4\%).
Therefore, a significantly narrow spectral linewidth (2 nm) is achievable for SiV in bulk diamond under room temperature. 
\end{itemize}

So far, the frequency down-conversion working at a single-photon level from the visible to the telecommunication wavelengths has been actively studied by using nonlinear optical media \cite{dreau2018quantum}.
The difficulty of this method is to minimize the loss of photons during the conversion process while preserving the quantum properties.

The coupling of single color centers, like a nitrogen-vacancy (NV) \cite{kurtsiefer2000stable} or silicon-vacancy \cite{neu2011single} defect in diamond lattice, to an optical microcavity is considered as an essential building block for applications in quantum information \cite{prawer2008diamond, aharonovich2011diamond} or magnetometry \cite{acosta2010broadband}.
For example, the zero phonon line of such color centers in diamond can be resonantly enhanced by coupling to a diamond microcavity with an enhancement factor of 12 \cite{faraon2011resonant}.

However, the Purcell enhancement of spontaneous emission is highly related to the emission bandwidth.
The Purcell factor of an optical cavity can be approximated as:
\begin{equation}
    F \approx \left[4 g^{2} /(\kappa \gamma)\right] \kappa / \gamma^{\star}=4 g^{2} /\left(\gamma \gamma^{\star}\right),
\end{equation}
where $g$ is the emitter-cavity coupling rate, $\kappa$ the cavity loss rate, $\gamma$ the emitter's natural linewidth, and $\gamma^{\star}$ the homogeneously broadened linewidth.

Here, we take NV center as an example.
The NV center’s broad emission bandwidth spanning over 100 nm under room temperature is a drawback for coupling to narrow cavity modes. 
To get a sharp zero phonon line with about 20 GHz linewidth, the diamond needs to be cooled to below 10 K in a continuous-flow liquid-helium cryostat during the experiment \cite{faraon2011resonant}.


In this chapter, we numerically calculate the efficiency of the frequency down-conversion from color centers in diamond under room temperature without an optical microcavity in order to see the possibility to enhance the conversion efficiency with a broad emission bandwidth, especially for the NV color center and SiV color center that we are interested in.

The conversion efficiency is decided by the phase matching between the signal and pump wavelength, which has limited convertible linewidth.
This is bound to significantly lower the conversion efficiency during the frequency down-conversion process with a narrow bandwidth pump laser due to the corresponding narrow phase matching linewidth of signal.

Our simulation focuses on the possibility of conversion efficiency improvement and spectral compression based on wide-bandwidth pump source.

\section{Difference frequency generation}
The optical field incident upon a second-order nonlinear medium consists of two distinct frequency components, whose electric field strength is represented as:
\begin{equation}
    \tilde{E}(t) = E_{1}e^{-i \omega_{1} t} + E_{2}e^{-i \omega_{2} t} + c.c.
\end{equation}

Since 
\begin{equation}
    \tilde{P}(t) = \epsilon_0 [\chi^{(1)} \tilde{E}(t) + \chi^{(2)}\tilde{E}^{2}(t) +  \chi^{(3)}\tilde{E}^{3}(t) + \dotsc ] ,
\end{equation}
where $ \epsilon_0$ is the permittivity of free space,
the second-order contribution to the nonlinear polarization is:
\begin{equation}
\begin{aligned}
    \tilde{P}^{(2)}(t) =  &\epsilon_0\chi^{(2)}\tilde{E}^{2}(t) \\
    = &\epsilon_0 \chi^{(2)} [E_{1}^2e^{-2i \omega_{1} t} + E_{2}^2e^{-2i \omega_{2} t} + 2E_{1}E_{2}e^{-i (\omega_{1}+\omega_{2}) t} + 2E_{1}E_{2}^*e^{-i (\omega_{1}-\omega_{2}) t} \\
    + &c.c] + 2 \epsilon_0 \chi^{(2)} [E_{1}E_{1}^* + E_{2}E_{2}^*] .
\end{aligned}
\end{equation}

We divide the second-order contribution to the nonlinear polarization into several terms based on the converted frequency. 
$\tilde{P}^{(2)}(t) = \sum_{n} P(\omega_n) e^{-i \omega_{n} t}$, where $\omega_n$ is the frequency of the converted photon. 
Each term corresponds to a second-order nonlinear process, including Second~harmonic~generation~(SHG), Sum~frequency~generation~(SFG) and Difference~frequency~generation~(DFG), as shown in Table. \ref{tab:Nonlinear polarization terms}.

\begin{table}[]
    \centering
    \begin{tabular}{|c|c|}
    \hline \text { Nonlinear polarization term } & \multicolumn{1}{|c|} {\text { Corresponding nonlinear process }} \\
    \hline $P(2\omega_{1})=\epsilon_0 \chi^{(2)} E_{1}^2$ & SHG of $\omega_{1}$ \\
    \hline $P(2\omega_{2})=\epsilon_0 \chi^{(2)} E_{2}^2$ & SHG of $\omega_{2}$\\
    \hline $P(\omega_{1}+\omega_{2})=2\epsilon_0 \chi^{(2)} E_{1}E_{2}$ & SFG \\
    \hline $P(\omega_{1}-\omega_{2})=2\epsilon_0 \chi^{(2)} E_{1}E_{2}^*$ & DFG \\
    \hline $P(0)=2 \epsilon_0 \chi^{(2)} (E_{1}E_{1}^* + E_{2}E_{2}^*)$ & Optical~rectification \footnote{An electro-optic effect} \\
    \hline
    \end{tabular}
    \caption{The nonlinear polarization terms and their corresponding nonlinear process.}
    \label{tab:Nonlinear polarization terms}
\end{table}

Both the SHG and SFG will generate photon at higher frequency, only DFG can realize frequency conversion from visible range to infrared.
In the process of DFG, the atom first absorbs a photon of frequency $\omega_1$ and jumps to the highest virtual level.
This level decays by a two-photon emission process that is stimulated by the presence of the $\omega_2$ field.
Two-photon emission can occur even if the $\omega_2$ field is not applied.
The generated fields in such a case are very much weaker since they are created by spontaneous two-photon emission from a virtual level.
The DFG term can be described with effective nonlinear coefficient $d_\text{eff}$:
\begin{equation}
    P(\omega_{1}-\omega_{2}) = 4\epsilon_0 d_\text{eff} E_{1}E_{2}^*.
\end{equation}

\section[Conversion efficiency of DFG]{Conversion efficiency of difference frequency generation}
Spectral compression of single photons by using sum frequency generation has already be determined \cite{lavoie2013spectral}, while there is no good solution for spectral compression during difference frequency generation.
Difference frequency generation (DFG) as one of the three wave mixing process is the reverse of sum frequency generation (SFG).

In order to study the influence of emission bandwidth of different single photon emitters, it's essential to calculate numerically the conversion efficiency of the DFG process.
Let's consider the situation that two optical waves at frequency $\omega_{1}$ and $\omega_{2}$ interact in a lossless nonlinear optical medium to generate a wave with frequency $\omega_{3}=\omega_{1}-\omega_{2}$.

The coupled amplitude equation can be described with the population of signal, pump and idler photons \cite{boyd2019nonlinear}:

\begin{equation}\frac{d A_{3}}{d z}=\frac{2 i \omega_{3}^{2} d_{\mathrm{eff}}}{k_{3} c^{2}} A_{1} A_{2}^{*} e^{i \Delta k z},\end{equation}
\begin{equation}\frac{d A_{2}}{d z}=\frac{2 i \omega_{2}^{2} d_{\mathrm{eff}}}{k_{2} c^{2}} A_{1} A_{3}^{*} e^{i \Delta k z},\end{equation}
where the $A_1,~A_2,~A_3$ are the amplitude of photons at $\omega_{1}, \omega_{2}, \omega_{3}$ respectively, and the wavevector mismatch is:
\begin{equation}\Delta k=k_{1}-k_{3}-k_{2}.\end{equation}

In the case of perfect phase matching, $\Delta k= 0$.

The relation between the intensity of signal ($I_1$), pump ($I_2$) and idler ($I_3$) beam is described with the equation calculated referring to the SFG process in \cite{boyd2019nonlinear}:
\begin{equation}
\left|I_{3}\right|=\frac{2 \pi^{2} L^{2} \chi_\text{eff}^{2} n_{2} c \varepsilon_{0}}{n_{1} n_{3} \lambda_{3}^{2}}\left|I_{1} \| I_{2}\right| \sinc^{2}\left(\frac{\Delta k}{2} L\right).
\end{equation}

The conversion efficiency of the DFG $\eta$ is oscillating with the phase matching between pump and signal $\Delta k$ times the crystal length $L$.
In a periodically poled nonlinear crystal, a period of twice the coherent buildup length $L_{coh}$ is chosen to compensate for the influence of wave vector mismatch, defined as $\Lambda$.
The field amplitude grows monotonically with the propagation distance.
With an appropriate choice of the poling period ($\Lambda$), one can practically realize the quasi-phase-matching of desired non-linear interaction between the pump, signal and idler photons.
In experiment, this quasi-phase-matching is tuned with modified refractive index and thermal expansion under temperature regulation.

\section{Difference frequency generation with PPLN}
Periodically poled lithium niobate (PPLN), especially the one fabricated into waveguide, has a conversion efficiency 4 orders of magnitude more than that obtained employing bulk crystals \cite{tanzilli2001highly, tanzilli2002ppln}.

\begin{figure}
    \centering
    \includegraphics[width=0.8\linewidth]{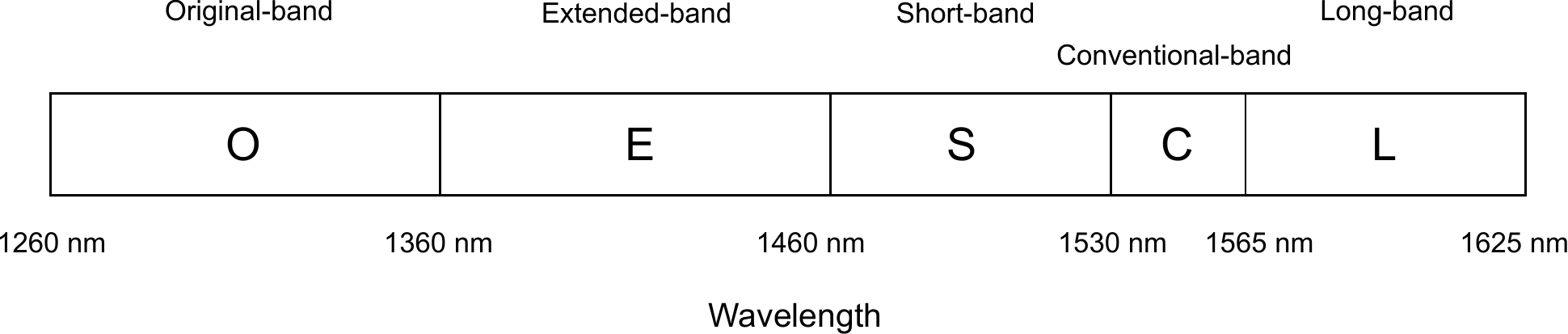}
    \caption{Definition of different bands in telecommunication wavelength.}
    \label{fig:telecom bands}
    \vspace*{\floatsep}
    \centering
    \includegraphics[width=0.8\linewidth]{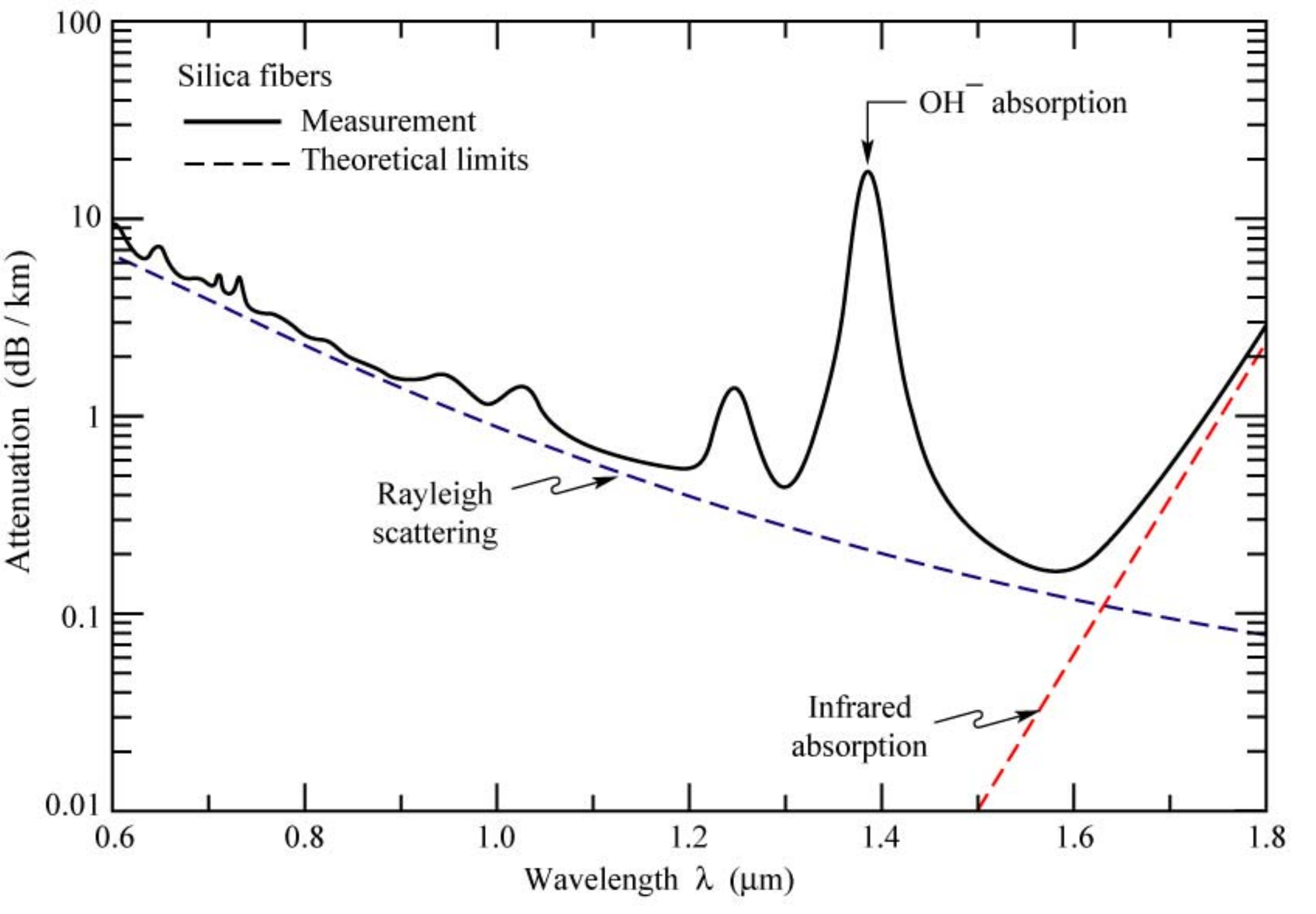}
    \caption{Measured attenuation in silica fibers (solid line and theoretical limits (dashed lines) given by Rayleigh scattering in the short-wavelength region and by molecular vibrations (infrared absorption) in the infrared region, adapted from Fig.22.2 of \cite{schubert2000light}.}
    \label{fig:Silica optical fiber attenuation}
\end{figure}

As shown in Fig. \ref{fig:telecom bands}, the telecommunication wavelength was divided into telecom O, E, S, C, L bands, corresponding to the original-band, extended-band, short-band, conventional-band and long-band.
The transmission loss of different bands in optical fiber vary between 0.2 dB/km to 0.8 dB/km.
In general, longer wavelengths show lower losses in transmission.
At telecom E band, the transmission loss is higher due to the OH$^-$ absorption. 
The minimum loss in this band is about 1 dB/km.

\subsection{DFG based on NV color center}
To convert single photon signal from NV color center which has ZPL at 637 nm to telecommunication frequency, the pump wavelength can be chosen at 1040 nm to get 1550 nm converted light at telecommunication C band (0.2 dB/km).
Under room temperature, the suitable period calculated for PPLN crystal is about 11 $\mu$m.

\begin{figure}
    \centering
    \includegraphics[width=0.8\linewidth]{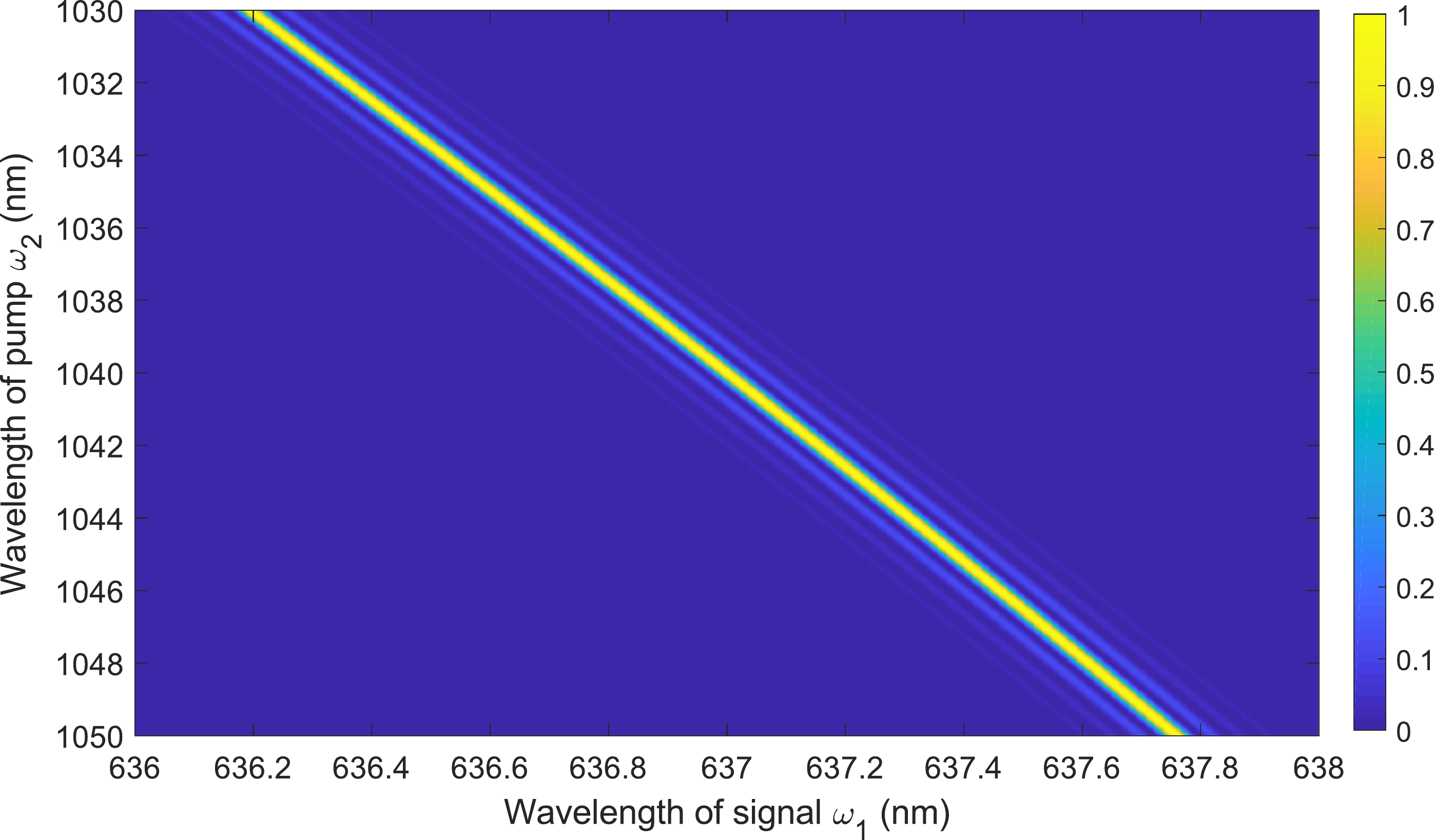}
    \caption{Conversion efficiency as a function of $\omega_{1}$ and $\omega_{2}$. $\omega_{1}$ and $\omega_{2}$ are the wavelength of signal and pump respectively with 50 mm long and 11 $\mu$m period PPLN.}
    \label{fig:637&1040}
\end{figure}

\begin{figure}
    \centering
    \includegraphics[width=0.8\linewidth]{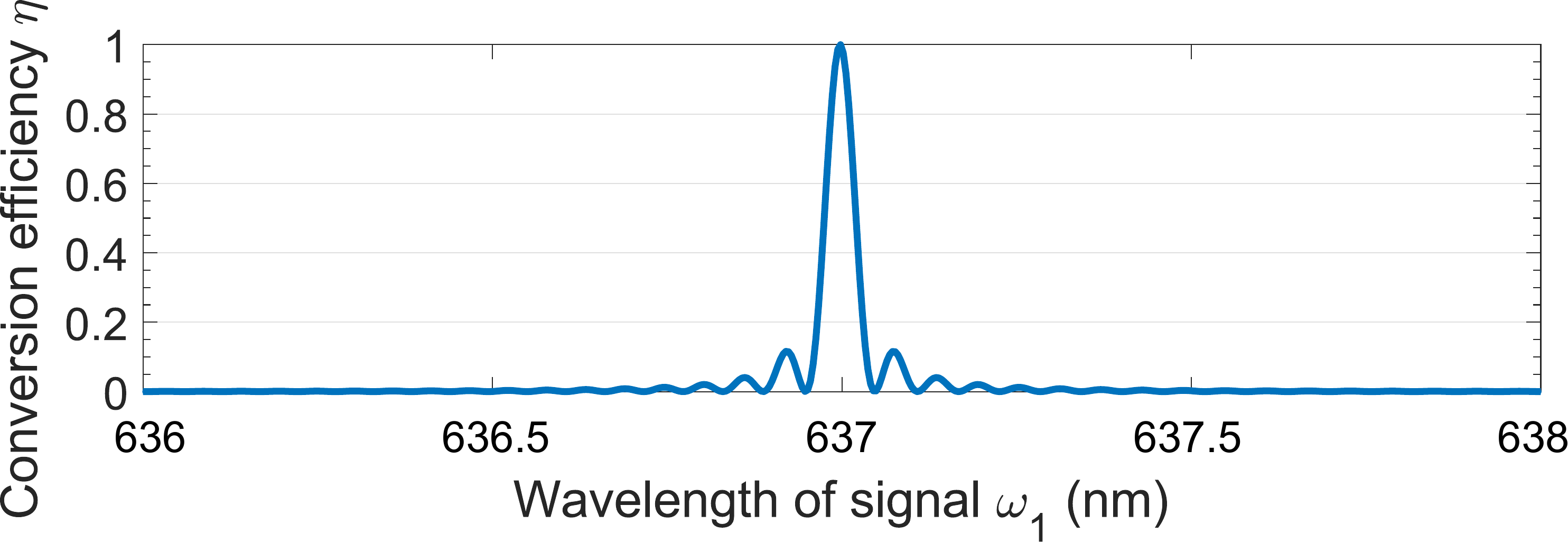}
    \caption{The oscillation of conversion efficiency $\eta$ with the signal frequency when pump wavelength is fixed at 1040 nm.}
    \label{fig:637&1040cut}
\end{figure}

In Fig. \ref{fig:637&1040}, we show the conversion efficiency as a function of signal wavelength and pump wavelength considering a PPLN with 50 mm length and 11 $\mu m$ period for poling.
When the signal wavelength shifts from 636 nm to 638 nm, the wavelength of the pump meeting the quasi phase matching condition varies from 1030 nm to 1050 nm.
For a pump with fixed wavelength, the convertible signal wavelength is quite narrow.
The Fig. \ref{fig:637&1040cut} shows the oscillation of conversion efficiency $\eta$ with the signal frequency when pump wavelength is fixed at 1040 nm.
Only the photons between 637$\pm0.1$ nm are able to be converted with high efficiency, while the other wavelengths will not be able to join this frequency conversion process.
Considering the emission spectrum of NV, a lot of energy will not be converted.

\subsection{DFG based on SiV color center}
We did the conversion efficiency calculation based on the SiV color center in bulk diamond, whose emission spectrum has relatively narrow linewidth, which is about 2 nm.

To convert single photon signal from SiV color center which has ZPL at 738 nm to telecommunication frequency, the pump wavelength can be chosen at 1403 nm so that we can get 1560 nm converted light at telecommunication C band.
Under room temperature, the suitable period calculated for PPLN crystal is 16.8 $\mu$m.

\begin{figure}
    \centering
    \includegraphics[width=\linewidth]{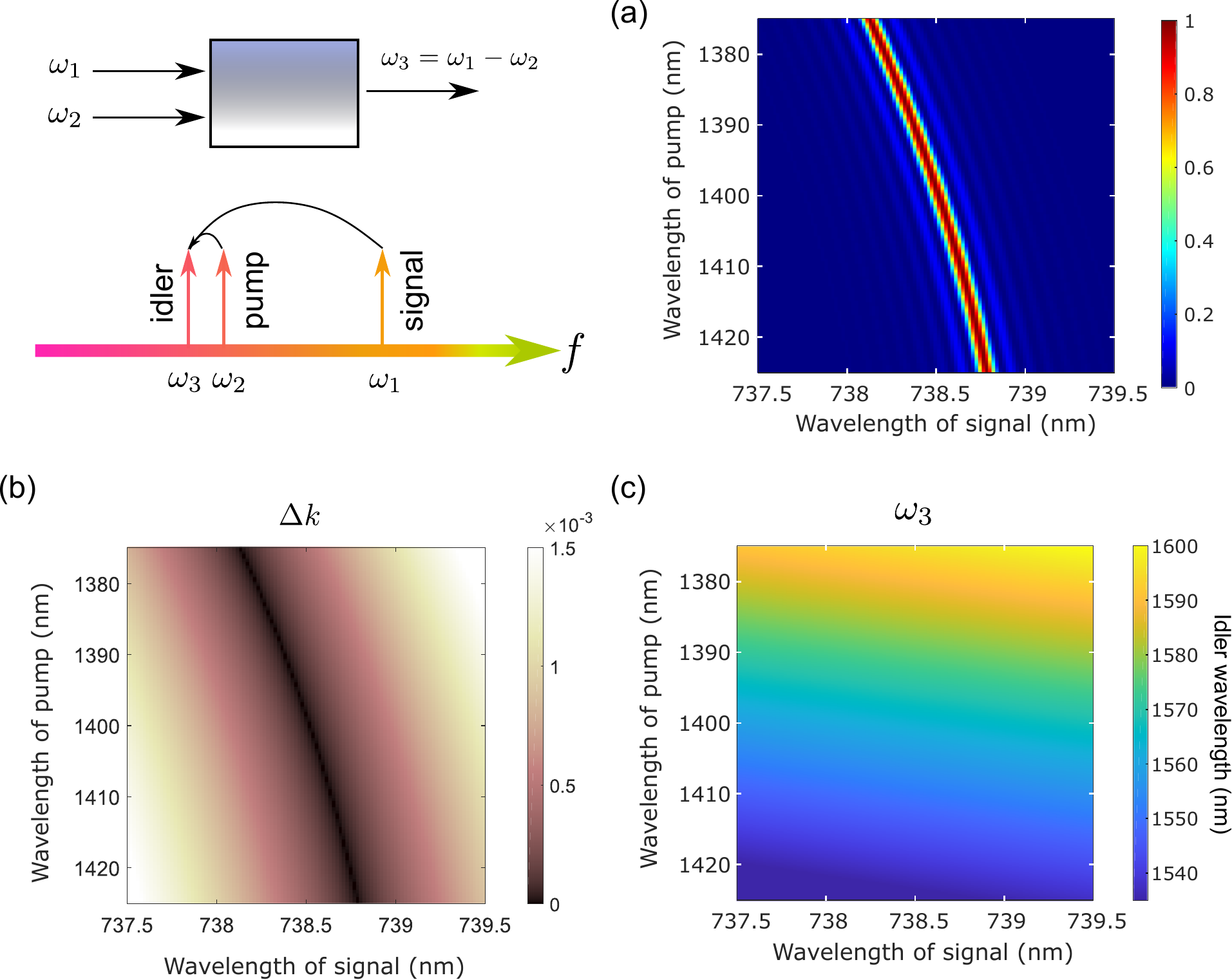}
    \caption{(a) Conversion efficiency as a function of the wavelength of signal ($\omega_{1}$) and pump ($\omega_{2}$) respectively with 50 mm long and 16.85 $\mu$m period PPLN. In practice, the PPLN period can be finely adjusted with temperature. (b) Phase matching described with $\Delta k$ as a function of $\omega_{1}$ and $\omega_{2}$. (c) Wavelength of idler light ($\omega_{3}$) as a function of $\omega_{1}$ and $\omega_{2}$.}
    \label{fig:738&1400}
\end{figure}

In Fig. \ref{fig:738&1400}-a, we show the conversion efficiency as a function of signal wavelength centered at 738.5 nm and pump wavelength centered at 1400 nm considering a PPLN with 50 mm length and 16.85 $\mu m$ period.
The phase matching is described with $\Delta k$ as a function of $\omega_{1}$ and $\omega_{2}$, shown in Fig. \ref{fig:738&1400}-b.
And the wavelength of converted light as a function of $\omega_{1}$ and $\omega_{2}$ is shown in Fig. \ref{fig:738&1400}-c.

\begin{figure}
    \centering
    \includegraphics[width=\linewidth]{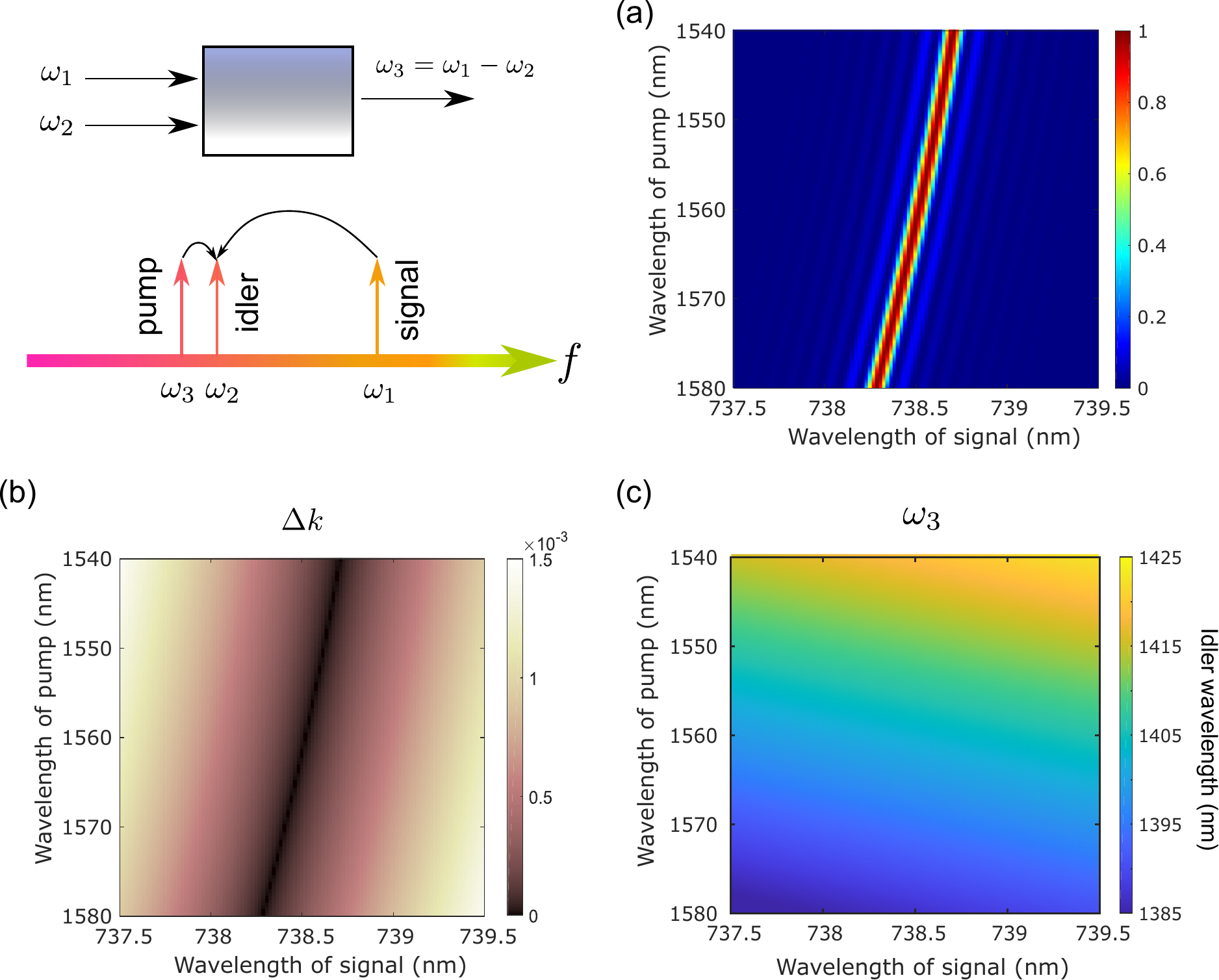}
    \caption{(a) Conversion efficiency as a function of the wavelength of signal ($\omega_{1}$) and pump ($\omega_{2}$) respectively with 50 mm long and 16.85 $\mu$m period PPLN. (b) Phase matching described with $\Delta k$ as a function of $\omega_{1}$ and $\omega_{2}$. (c) Wavelength of idler light ($\omega_{3}$) as a function of $\omega_{1}$ and $\omega_{2}$.}
    \label{fig:738&1560}
\end{figure}

If we use 1506 nm as pump wavelength, we can convert the 738.5 nm signal to telecom E band instead of telecom C band with slightly higher transmission loss of about 1 dB/km.
In Fig. \ref{fig:738&1560}-a, we show the conversion efficiency as a function of signal wavelength centered at 738.5 nm and pump wavelength centered at 1560 nm considering a PPLN with 50 mm length and 16.85 $\mu m$ period.
The phase matching is described with $\Delta k$ as a function of $\omega_{1}$ and $\omega_{2}$, shown in Fig. \ref{fig:738&1560}-b.
And the wavelength of converted light as a function of $\omega_{1}$ and $\omega_{2}$ is shown in Fig. \ref{fig:738&1560}-c.

\begin{figure}
    \centering
    \includegraphics[width=\linewidth]{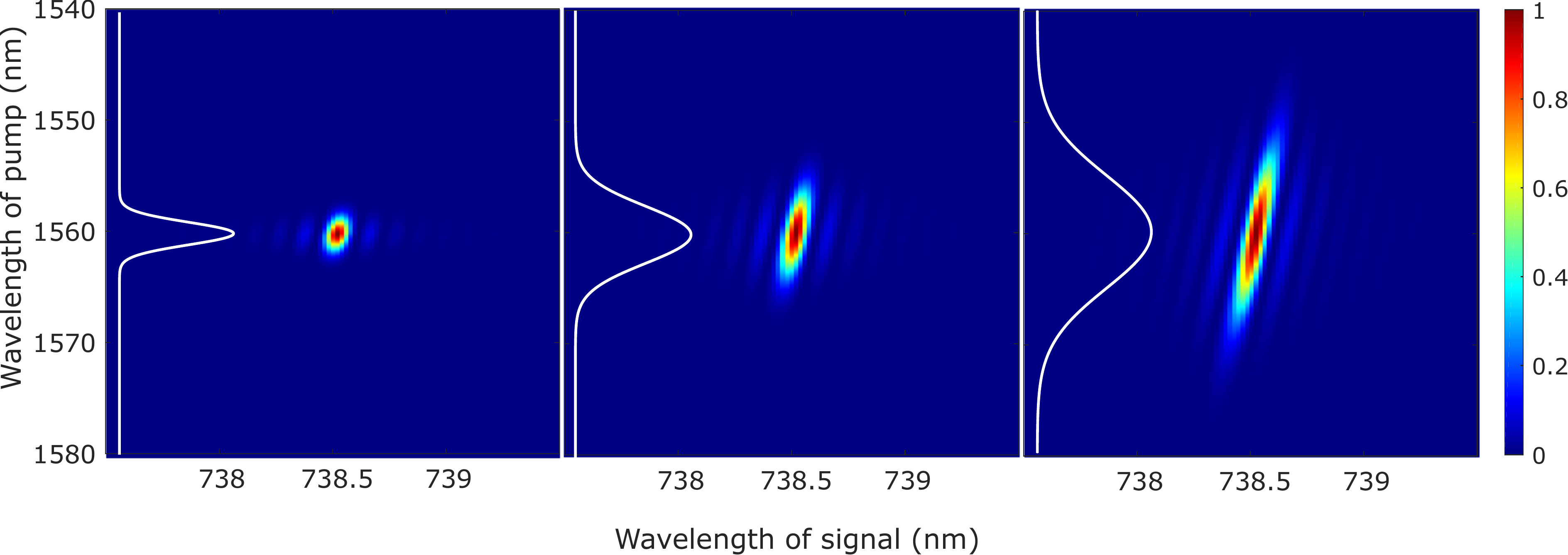}
    \caption{Conversion efficiency when converting signal from SiV color center to 1400 nm (telecom E band) using a pump with a width $\Delta v_{\mathrm{p}}=1,740 \pm 50 \mathrm{GHz}$ FWHM varying from 2.5 nm to 10 nm centered at 1560 nm. We use experimentally measured fluorescence spectrum of single SiV color center in nanodiamond at room temperature.}
    \label{fig:738&1560bandwidth}
\end{figure}

We analysis the conversion efficiency of fluorescence from single SiV color center converted to 1400 nm at telecom E band and its dependence on the pump linewidth.
We calculated the conversion efficiency using the experimentally measured fluorescence spectrum of a single SiV color center in nanodiamond at room temperature.
The spectrum linewidth of the SiV color center in nanodiamond we measured is about 5 nm centered at 738.5 nm instead of 2 nm centered at 738 nm.
The optical coherence property of defects in nanodiamonds is worse than that in bulk diamond, causing the spectral diffusion of ZPL (or spectral jumping of ZPL frequency) \cite{santori2010nanophotonics}.
When we vary the spectrum full width at half maximum (FWHM) of the pump from 2.5 nm to 5 nm and 10 nm centered at 1560 nm (308 GHz, 616 GHz, 1 THz in frequency), the map of conversion efficiency as a function of pump wavelength and signal wavelength is shown in Fig. \ref{fig:738&1560bandwidth}.
When the pump's linewidth becomes larger, the converted signal increased by factor 10.

\section{Difference frequency generation with PPKTP}

\begin{figure}
    \centering
    \includegraphics[width=\linewidth]{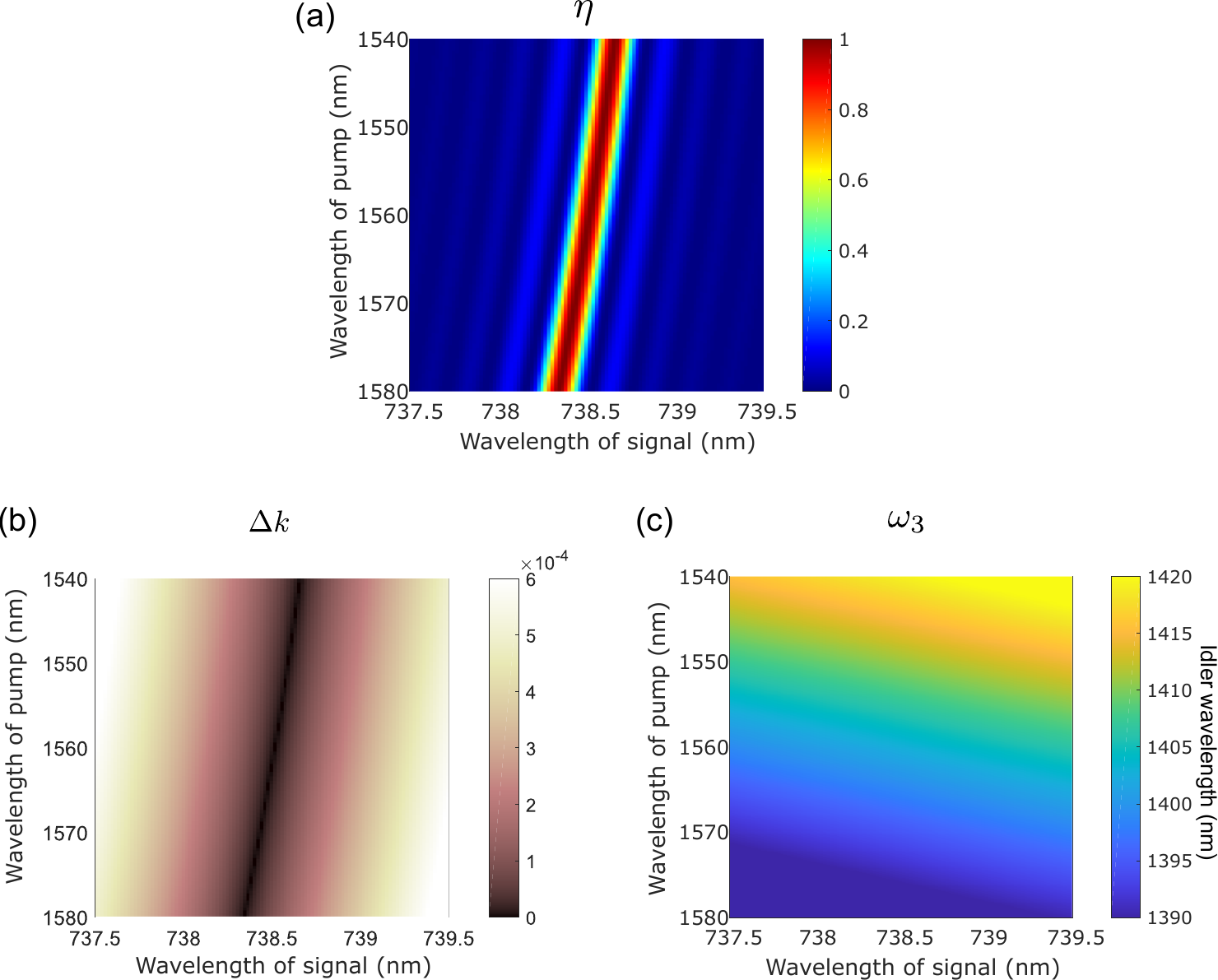}
    \caption{(a) Conversion efficiency as a function of the wavelength of signal ($\omega_{1}$) and pump ($\omega_{2}$) respectively with 50 mm long and 29.8 $\mu$m period PPKTP. (b) Phase matching described with $\Delta k$ as a function of $\omega_{1}$ and $\omega_{2}$. (c) Wavelength of idler light ($\omega_{3}$) as a function of $\omega_{1}$ and $\omega_{2}$.}
    \label{fig:738&1560KTP}
\end{figure}

We calculated the phase matching condition of defects in diamond while using two different nonlinear medias, the periodically poled lithium niobate (PPLN) and periodically poled potassium titanyl phosphate (PPKTP).
Since the light refraction in the two crystals is different (lithium niobate is uniaxial crystal and potassium titanyl phosphate is biaxial crystal), we want to understand how different nonlinear crystals affect the phase matching condition and whether we find a numerical solution to increase the conversion efficiency and bandwidth compression.

For calculating the conversion efficiency of different material, the difference lays in the material dispersion and thermal expansion.
The thermal expansion was found to be governed by parabolic dependence on temperature:
\begin{equation}
    L=L_0[1 + \alpha ( T - 25 ^{\circ} \text{C}) + \beta ( T- 25 ^{\circ} \text{C}].
\end{equation}
For material KTP: 
\begin{equation}
    \alpha = (6.7 \pm 0.7) \times 10^{-6} ,~
    \beta = (11 \pm 2) \times 10^{-9}.
\end{equation}

In Fig. \ref{fig:738&1560KTP}-a, we show the conversion efficiency as a function of signal wavelength centered at 738.5 nm and pump wavelength centered at 1560 nm considering a PPKTP with 50 mm length and 29.8 $\mu m$ period.
The phase matching is described with $\Delta k$ as a function of $\omega_{1}$ and $\omega_{2}$, shown in Fig. \ref{fig:738&1560KTP}-b.
And the wavelength of converted light as a function of $\omega_{1}$ and $\omega_{2}$ is shown in Fig. \ref{fig:738&1400}-c.

\section{Conversion efficiency calibration of PPLN waveguide}
In this section, we introduce the experiment result of frequency down conversion with a PPLN waveguide.
We realized frequency down conversion of light at 637 nm to telecommunication wavelength of optical fiber for simulating the nonclassical emission of nitrogen vacancy color center ZPL with pump wavelength at 1064 nm.

By mixing high-power pump at $\lambda$=1064 nm, we realized the frequency down conversion of the low-power signal source at $\lambda$=637 nm to $\lambda 2=\left(\lambda_{1}^{-1}-\lambda_{p}^{-1}\right)^{-1}=1587 \quad \mathrm{nm}$. 

For a PPLN waveguide with a length of $L$=4.6 cm, the $\hat{a}_{1}(L)$ and $\hat{a}_{2}(L)$ is are given by:

\begin{equation}\begin{array}{l}
\hat{a}_{1}(L)=\cos (\gamma L) \hat{a}_{1}(0)-\sin (\gamma L) \hat{a}_{2}(0) \\
\hat{a}_{2}(L)=\sin (\gamma L) \hat{a}_{1}(0)+\cos (\gamma L) \hat{a}_{2}(0)
\end{array}\end{equation}
where the coupling constant $\gamma=\sqrt{\eta_{n o r} P_{p}}$. 

The expected value of the number of photons at the output $\left\langle\hat{n}_{2}(L)\right\rangle$ can be obtained according to the following equation:

\begin{equation}\frac{\left\langle\hat{n}_{2}(L)\right\rangle}{\left\langle\hat{n}_{1}(0)\right\rangle}=\frac{\left\langle\hat{a}_{2}^{+}(L) \hat{a}_{2}(L)\right\rangle}{\left\langle\hat{a}_{1}^{+}(L) \hat{a}_{1}(L)\right\}}=\sin ^{2}(\gamma L)=\sin ^{2}\left(\sqrt{\eta_{\text {nor }} P_{p}} L\right)\end{equation} \label{eq: pump power dependence}

When $\sqrt{\eta_{\operatorname{nor}} P_{p}} L=\pi/2$, the parametric gain is close to 0, and all photons at $\omega_1$ are converted to $\omega_2$.

\subsection{Experiment approach}
In the experiment, we need to couple the signal and pump light into the waveguide with low-loss.
To test the mode shape of the PPLN waveguide, we first couple light to the waveguide through an optical fiber placed face to face in front of the waveguide. 
The output of the waveguide is monitored with a powermeter.
The coupling efficiency we obtained is about 70\%.
When we couple the space light to the waveguide from the lens, and the efficiency obtained is much lower. 
The main reason is that the mode at the output end after single-mode fiber filtering is closer to the standard Gaussian mode than the mode of the spatial light.

High coupling efficiency from free space to the waveguide requires the matching between the mode of the waveguide and the mode of the signal and pump in the free space.
Therefore, we added two telescopes to adjust the beam sizes of signal and pump in free space.
Cylindrical lens are used to optimize the beam shape to match with the mode of the PPLN waveguide.
In the end, a coupling efficiency of about 50\% is achieved for both signal and pump.

The PPLN waveguide we used is type 0, which means the polarization of the signal photons and pump photons need to be the same, either vertical or horizontal to the waveguide.
In the experiment, we used two half-waveplates to control the polarization of signal and pump.

The output signal is filtered first with a Pellin Broca Prism, then with longpass filters.
The signal at 1560 nm is detected with a indium gallium arsenide (InGaAs) photodiode.

The temperature of the PPLN waveguide is controlled with a customized oven with fine temperature control from room temperature up to 200 $^{\circ}$C.
To achieve the best quasi-phase matching, we measured the temperature dependence of the conversion efficiency and find the maximum conversion efficiency when the temperature of the PPLN waveguide is 154 $^{\circ}$C.

The thermal expansion of the crystal changes the phase-matching condition as well as the coupling of the pump and signal beam.
The coupling efficiency is sensitive to the temperature shift.
Therefore, optimizing the coupling is necessary during the temperature dependence measurement.

\begin{figure}
    \centering
    \includegraphics[width=0.8\linewidth]{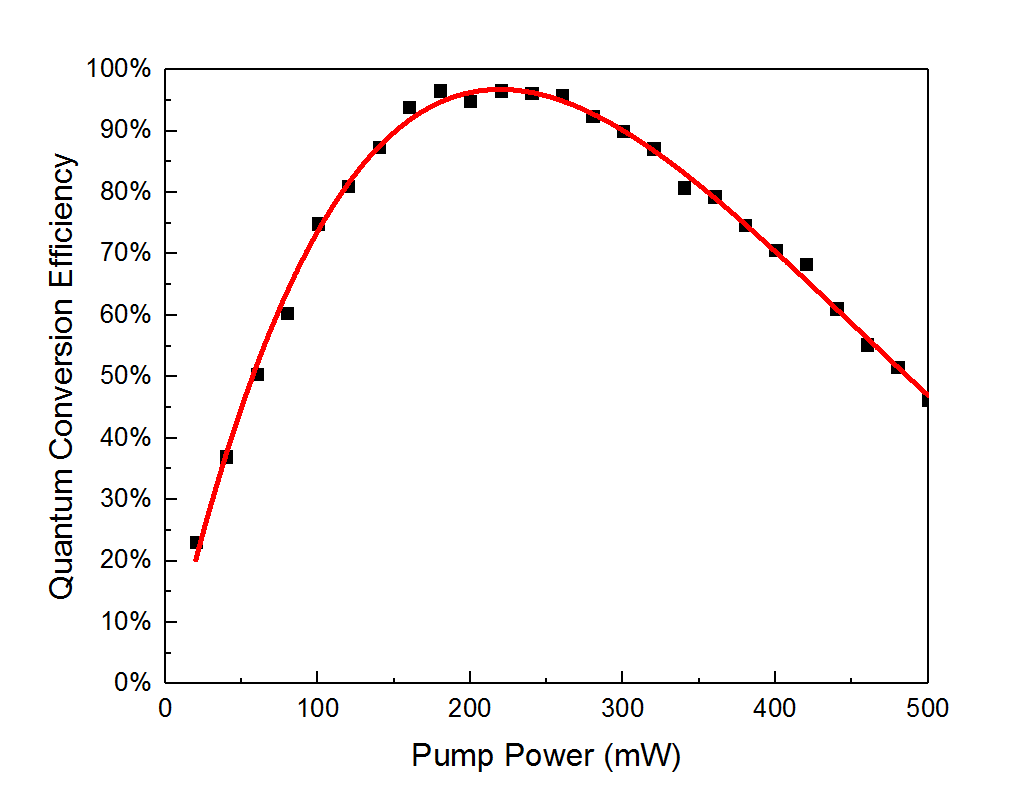}
    \caption{The pump power dependence of quantum conversion efficiency, where the solid line is the fitting of the data according to equation.\ref{eq: pump power dependence}.
    When $P_{p}=220$ mW, we obtained the highest quantum conversion efficiency, which is 96.7\%.}
    \label{fig:pump power dependence of quantum conversion efficiency}
\end{figure}

\subsection{Result and discussion}

As it is obtained that the PPLN waveguide achieves the best quasi-phase matching when temperature is 154 $^{\circ}$C.
The conversion efficiency is measured when the temperature of the PPLN waveguide is fixed at 154 $^{\circ}$C.

We fixed the power of the signal at 50 $\mu$W and varied the power of the pump from 25 mW to 500 mW.
The power of the converted photons varies with the power of the pump light, and the curve of conversion efficiency is shown with quantum conversion efficiency which takes the energy difference between the photons into consideration:

In Fig.\ref{fig:pump power dependence of quantum conversion efficiency}, we show the quantum conversion efficiency as a function of the pump power, fitted with the equation.\ref{eq: pump power dependence}.
The quantum conversion efficiency reaches its maximum when the power of pump is $P_{p}=220$ mW.
The acquired conversion efficiency is expressed with quantum frequency conversion efficiency which is 96.7\%. 
The normalized conversion efficiency of the waveguide $\eta_{\text {nor }}=53.00 \% / \mathrm{W} / \mathrm{cm}^{2}$ can be obtained by the fitting of the curve.

\section*{Conclusion}
\addcontentsline{toc}{section}{Conclusion} 
In this chapter, we proposed numerical simulation for quantum frequency conversion from visible range to telecommunication wavelength.
We calculated the phase matching condition for nitrogen vacancy color center and silicon vacancy color center with PPLN or PPKTP waveguide. 
The calculation shows the opposite effect on bandwidth compression or expanding when using pump of wavelength longer or shorter than idler signal. 
Besides, by using a wide-bandwidth pump, we can increase the convertible bandwidth to 2 nm centered at 738.5 nm for silicon vacancy color center and get a conversion efficiency of more than 50\% with bandwidth compression.

Besides, we experimentally realized the frequency down conversion of light at 637 nm to telecommunication wavelength of optical fiber corresponding to the zero phonon line of the emission from nitrogen vacancy color center.
The quantum frequency conversion efficiency we obtained is 96.7\%.
\chapter{Four wave mixing in suspended core fiber for single photon frequency conversion}
\markboth{FOUR WAVE MIXING}{}
\label{chap:FWM}
Parametric down-conversion is commonly chosen to generate efficient frequency down conversion by using nonlinear media lacking inversion symmetry.
Some materials with symmetry can also achieve relatively high frequency conversion efficiency, which mainly bases on their third-order nonlinear susceptibility, for example, a fused-silica optical fiber.
In this chapter, we introduce a novel system which uses a noble-gas-filled three holes suspended-core fiber (SCF) for the frequency down conversion of single photons with four-wave mixing.

In this chapter, we introduce a possible solution for experimentally realizing single photon frequency conversion from visible range to telecommunication range with suspended core fiber using four wave mixing.

The suspended core fiber has a cross section with a sub-micro core surrounded by three holes.
The interesting point of using the suspended core fiber as the non-linear waveguide is that the surrounding holes can be filled with gas to induce non-linearity.
The type and the pressure of gas modify the nonlinear refractive index of the fiber, therefore providing on demand phase matching.
In this chapter, we will first introduce the nonlinear process in a solid-core fused-silica fiber.
Then we will discuss the four wave mixing with fused-silica fiber and explain the influence of gas pressure filling the suspended-core fiber on the phase matching condition of the four wave mixing process.

\section{Nonlinear process in a solid-core fused-silica fiber}
Parametric amplification in a glass fiber absorbs two pump photons and, through the $\chi^{(3)} E ^3$ term in the polarization expansion, creates one photon higher in frequency than the pump and one photon lower in frequency than the pump.
We refer to the low frequency wave as the Stokes wave and the higher frequency wave as the anti-Stokes wave.
The third-order susceptibility $\chi^{(3)}$ is a complex number with the real part leading to the parametric gain and the imaginary part giving Raman Stokes gain and Raman anti-Stokes absorption.
The second-order susceptibility $\chi^{(2)}$ is zero in glass because of inversion symmetry.

\begin{figure}
    \centering
    \includegraphics[width=0.8\linewidth]{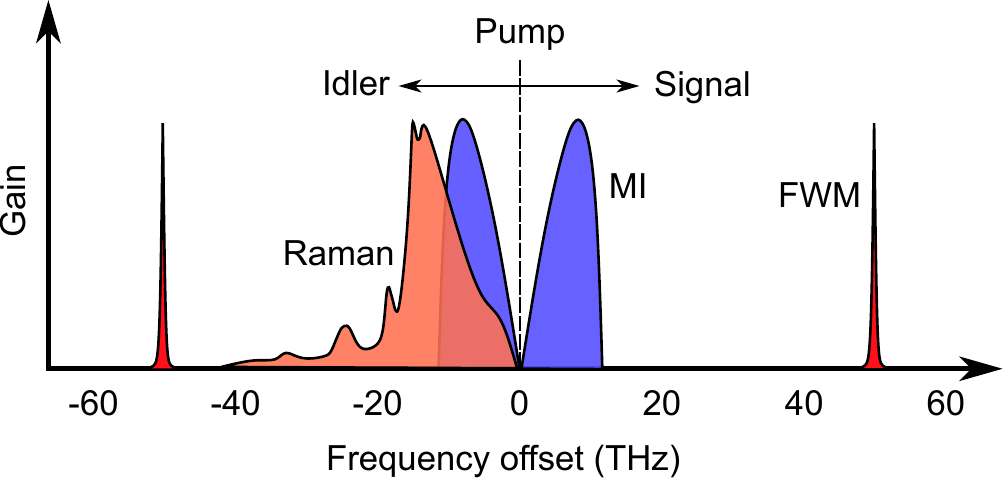}
    \caption{Normalized gain spectrum for Raman scattering, modulational instability (MI) and four-wave mixing (FWM) in a solid-core fused-silica fiber adapted from \cite{hollenbeck2002multiple}. The sideband at higher frequency is named as "signal" and the one at lower frequency is named as "idler".}
    \label{fig:raman_scattering}
\end{figure}

The nonlinearity in solid-core-fused-silica fiber is provided by the third-order nonlinear susceptibility $\chi^{(3)}$ of silica, which is always present in centro-symmetric materials. 
In $\chi^{(3)}$-based sources, twin beams can be generated through the interaction of four waves \cite{rarity2005photonic}. 
The two possible nonlinear generation processes are four-wave mixing (FWM) and modulational instability (MI).
FWM leads to narrow sidebands with a large separation in frequency from the pump, whereas MI generates broad sidebands in the vicinity of the pump as shown in Fig.\ref{fig:raman_scattering} \cite{rarity2005photonic}.
However, additional uncorrelated photons are generated due to Raman scattering giving increased frequency downshifted (Stokes) noise \cite{hollenbeck2002multiple}.
This is the major drawback of $\chi^{(3)}$-based correlated photon sources and is a serious problem for applications in quantum optics.
Of course, the reverse process can happen as well and generates a frequency up-shifted (anti-Stokes) photon, however, it has a much lower probability to occur.
In the case of fused silica, the gain for Raman scattering extends over a broad bandwidth of about 45 THz with a strong peak near -13 THz when pump and Stokes wave are co-polarized, as shown in Fig.~\ref{fig:raman_scattering}.

Frequency conversion from visible to telecom wavelength with a near infrared (NIR) pump has a frequency offset of more than 100 THz.
The twin beam generated in FWM regime can have frequency offset large enough to avoid the influence of the Raman noise.

\section{Dispersion in fused-silica fiber}

The dispersion in optical fiber comes from four different reasons.

In an optical fiber allowing multimode propagation, since the effective reflective index of different modes are different due to the proportion of mode propagating in the core or cladding, the group velocity itself may be different from mode to mode, causing the multimode dispersion.

Besides, in birefringence fibers (also in bending single mode fibers), there is polarization related dispersion caused by the difference between the orthogonally polarized HE$^x_{11}$ and HE$^y_{11}$ modes as discussed in chapter \ref{chap:Nanofiber_interferometer_and_resonator}.

Material refractive index varies with wavelength and therefore causes the group velocity to vary.
It is classified as material dispersion. 
The wavelength dependence of refractive index can be expressed by Sellmeier’s equation \cite{arosa2020refractive}: 

\begin{equation}n^{2}(\lambda)=1+\frac{B_{1} \lambda^{2}}{\lambda^{2}-C_{1}}+\frac{B_{2} \lambda^{2}}{\lambda^{2}-C_{2}}+\frac{B_{3} \lambda^{2}}{\lambda^{2}-C_{3}},\end{equation}
where $n$ is the refractive index, $\lambda$ is the guiding wavelength, and $B_1, B_2, B_3, C_1, C_2, C_3$ are experimentally determined Sellmeier coefficients of material.
The coefficient for fused silica is shown in Table \ref{tab:Sellmeier coefficients for fused silica}.

\begin{table}[]
    \centering
    \begin{tabular}{|c|c|}
    \hline \text { Coefficient } & \multicolumn{1}{|c|} {\text { Value }} \\
    \hline $\mathrm{B}_{1}$ & 0.696166300 \\
    \hline $\mathrm{B}_{2}$ & 0.407942600 \\
    \hline $\mathrm{B}_{3}$ & 0.897479400 \\
    \hline $\mathrm{C}_{1}$ & $4.67914826 \times 10^{-3} \mathrm{\mu m}^{2}$ \\
    \hline $\mathrm{C}_{2}$ & $1.35120631 \times 10^{-2} \mathrm{\mu m}^{2}$ \\
    \hline $\mathrm{C}_{3}$ & $0.979340025 \times 10^{2} \mathrm{\mu m}^{2}$ \\
    \hline
    \end{tabular}
    \caption{The Sellmeier coefficients for fused silica.}
    \label{tab:Sellmeier coefficients for fused silica}
\end{table}


Waveguide dispersion is the result of wavelength-dependence of the propagation constant of the optical waveguide. 
It is important in single-mode waveguides. 
The larger the wavelength, the more the fundamental mode will spread from the core into the cladding.
This causes the fundamental mode to propagate faster.



For an ideal single mode waveguide, material dispersion and waveguide dispersion are the dominant factors.
The sum of material dispersion and waveguide dispersion is called chromatic dispersion, given as $\sigma=\sigma_{m}+\sigma_{w}$.
The material and waveguide dispersion gives $\Delta k_{m}$ and $\Delta k_{w}$ respectively.

In optical fiber, the material and waveguide dispersion can effectively cancel each other out to produce a zero-dispersion wavelength (ZDW).
The ZDW of a waveguide can be controlled by changing the core diameter and relative refractive-index difference between core and cladding.

\section[FWM in fused-silica fiber]{FWM in fused-silica fiber: theoretical considerations}

The Kerr nonlinearity of a system can be quantified by the nonlinear index, which relates the refractive index change to the optical intensity.

\begin{equation}\tilde{n}(\omega, I)=n(\omega)+n_{2} I=n+\bar{n}_{2}|E|^{2},\end{equation}
where $n$ is the linear refractive index and $\bar{n}_{2}$ is the (second order) nonlinear-index coefficient, which describes the instantaneous nonlinearity and is related to $\chi^{(3)}$ for a linearly polarized field via $\bar{n}_{2}=3 /(8 n) \operatorname{Re}(\chi^{(3)})$.

The third order non-linearity process in a fused-silica fiber allows the spontaneous conversion of two photons at pump wavelength to a pair of photons, which are defined as signal and idler.
The energy is conserved with 2$\omega_p=\omega_s+\omega_i$.
The phase matching condition for this FWM process is given by the peak power of the pump beam $P_p$ and the effective nonlinearity of the fiber mode $\gamma$ and the wave vectors of pump, signal and idler:
\begin{equation}
    2k(\omega_p)-2\gamma P_p=k(\omega_s)+k(\omega_i).
\end{equation}

\begin{equation}\gamma_{j}=\frac{n_{2} \omega_{j}}{c A_{\mathrm{eff}}},\end{equation}
$A_{\mathrm{eff}}=\left\langle|F|^{2}\right\rangle^{2} /\left\langle|F|^{4}\right\rangle $ is the effective mode area, $n_{2}$ is the nonlinear refractive index, $c$ is the speed of light.

The FWM process can be realized in silica fiber with seed photons or without.
In the case of sending photons at any one of the sideband as seed, the FWM process is a classic non-linear phenomenon.
The unseeded FWM can be regarded as seeded by the vacuum fluctuation, and be described by quantum optics.
The unseeded FWM generates a pair of photons at the two sidebands.

\begin{figure}
    \centering
    \includegraphics[width=\linewidth]{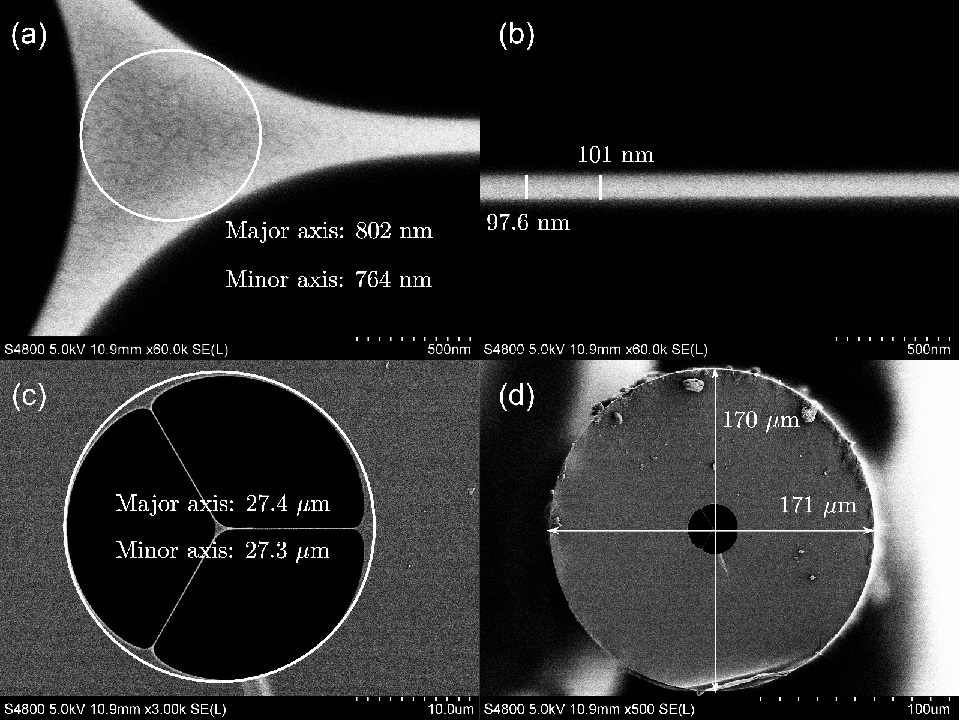}
    \caption{The cross-section of suspended-core fiber under scanning electronic microscope(SEM). a, fiber core with triangle shape; b, the thickness of the wall separating the holes is about 100 nm; c, three holes structure; d, the whole fiber cross-section with fiber diameter about 170 $\mu$m. The SEM image of the suspended core fiber is offered by Jonas Hammer from Max Planck Institute for the Science of Light.}
    \label{fig:PCF_crosssection}
\end{figure}

The cross-section microscope image of the fiber we used is shown in Fig.~\ref{fig:PCF_crosssection}.abcd.
It is composed of three holes around the fiber center at each 120 degrees.
The three hole walls are connected at the center of the fiber and form a sub-micro waveguide with triangle cross-section.

\begin{figure}
    \centering
    \includegraphics[width=0.9\linewidth]{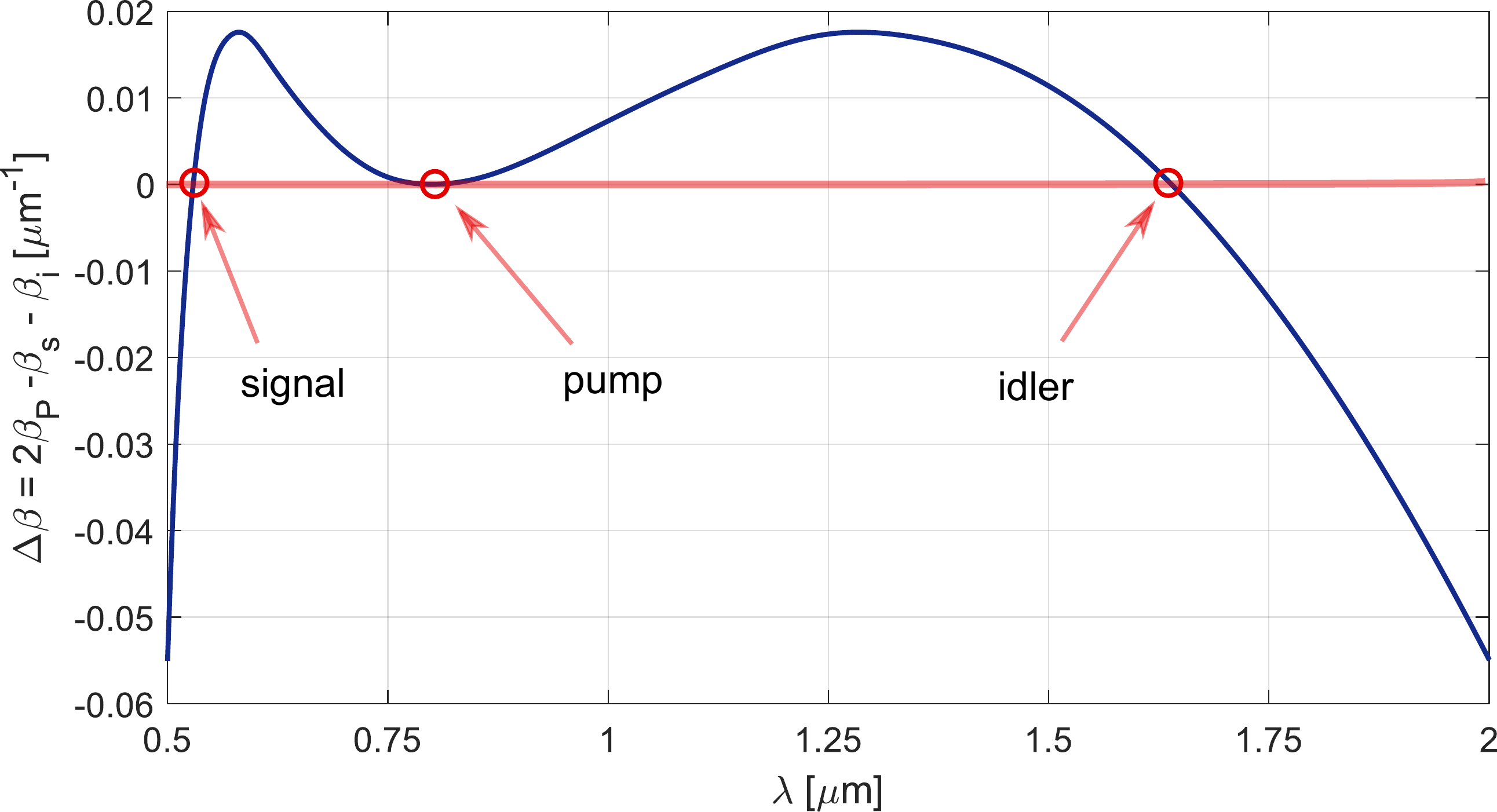}
    \caption{Phase matching of FWM as a function of wavelength in an idealized (symmetric) air-cladding optical fiber with a core diameter of 790nm.}
    \label{fig:790phasematch}
    \vspace*{\floatsep}
    \centering
    \includegraphics[width=0.7\linewidth]{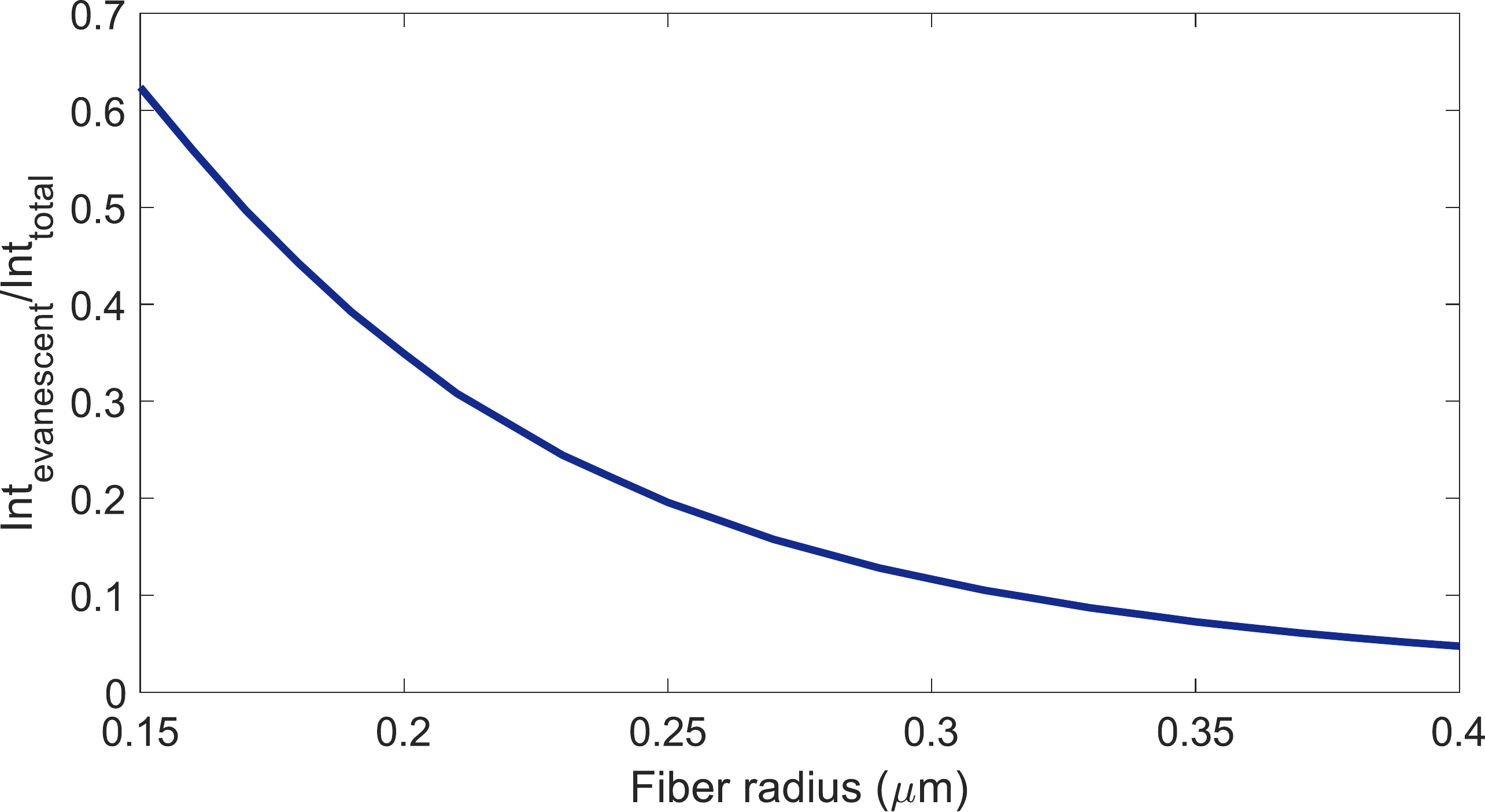}
    \caption{Ratio of the evanescent field among the total intensity of guided light as a function of fiber radius $a$. The guiding wavelength is set as 637 nm.}
    \label{fig:eva_a_637}
    \vspace*{\floatsep}
    \centering
    \includegraphics[width=0.6\linewidth]{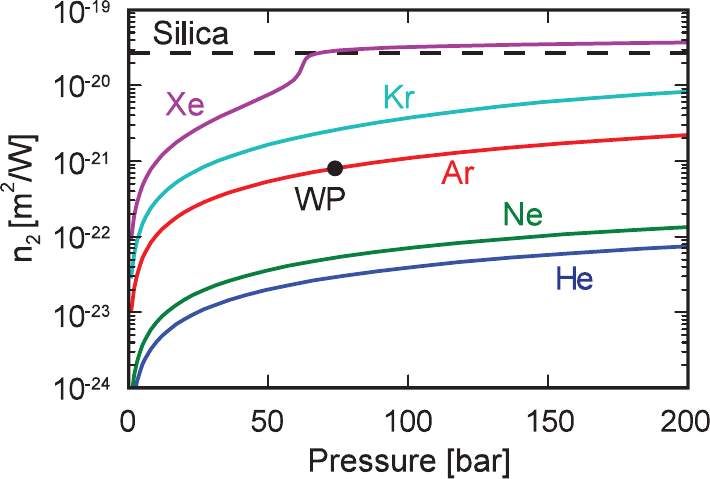}
    \caption{Pressure variation of the nonlinear refractive index at 800 nm for various noble gases at
room temperature (293 K). (Reprinted from Fig.2.11 in Tunable Twin Beams Generated in Hollow-Core Photonic Crystal Fibers by Martin Arnold Finger, Max Planck Institution for the science of light(2017))}
    \label{fig:gas pressure}
\end{figure}

Fig.~\ref{fig:790phasematch} is the calculated by fixing the pump wavelength at 0.8 $\mu$m and shifting the signal wavelength from 0.5 $\mu$m  to 2 $\mu$m.
And the phase mismatch is therefore between the pump wavelength (0.8 $\mu$m), signal wavelength ($x$ axis) and corresponding idler wavelength.
The points when the phase mismatch is 0 satisfy the phase matching condition.
We have two groups of signal and idler that matches the condition.
Here we choose the one when signal is in visible range and idler is in infrared range.
The point at the middle is when signal, pump and idler are all 0.8 $\mu$m.
In Fig.~\ref{fig:790phasematch}, we show the calculated phase matching given by the difference between propagation constant while varying wavelength: $\Delta \beta=2\beta_{p}-\beta_{s}-\beta_{i}$. 
It is calculated in an idealized (symmetric) air-cladding optical fiber with a core diameter of 790 nm, roughly corresponds to the fiber we used in Fig.~\ref{fig:PCF_crosssection}.
It's calculated based on the effective refractive index $n_\text{eff}$ of the waveguide as a function of $\lambda$.
The wavelengths of the intersection points when the curve intersects with line $\Delta \beta = 0$ are: 800 nm, 530 nm and 1.62 $\mu$m, corresponding to two pump photons, signal and idler.
Here we use the blueshifted sideband at 530 nm as the signal and the redshifted sideband at 1.62 $\mu m$ as idler, marked in the Fig.~\ref{fig:790phasematch}.
As we show in Fig.\ref{fig:PCF_crosssection}(a), the core of the suspended core fiber we used has approximate size of about 790 nm. 
Here we assume the phase matching wavelength for this suspended core fiber is close to our calculation of a symmetric air-cladding optical fiber with a core diameter of 790 nm.
With this phase matching wavelength, we can convert photons from visible range (530 nm) to infrared range (1.62 $\mu$m) with pump wavelength at 800 nm.

We also studied the possibility using an optical nanofiber as the nonlinear medium.
The good point is that it's possible to have the single photon emitter deposited directly on the nanofiber and the system will be very integrated with both single photon generation and frequency conversion on a piece of optical nanofiber.
The premise for achieving this is that the fiber diameter needed for strong evanescent coupling at the emitter's wavelength has to be close to the fiber diameter needed for meeting the phase matching condition in FWM.
According to the calculation for the phase matching when pumping at 800 nm and generated idler at telecom wavelength, it requires a fiber diameter of about 800 nm.
The wavelength of the signal photons are better to be infrared to achieve high coupling efficiency.
For a emitter emitting light around 600 nm, the best coupling appears when the nanofiber diameter is below 300 nm.
Therefore, it's not suitable for the case when the frequency conversion goes from visible to telecom wavelength.
In Fig. \ref{fig:eva_a_637}, we show the coupling efficiency of 637 nm light as a function of fiber diameter.

The advantage of using the central part of the suspended core fiber instead of an optical nanofiber is that: 
first, we can realize the fine tuning of phase matching, the feasibility has been recently proven by \cite{hammer2020broadly}.
The holes within the suspended core fiber can be filled with noble gas.
The changing of refractive index shifts the group velocity dispersion of the gas-filled suspended core fiber.
Therefore, changing the gas pressure changes the refractive index of the gas around the core.
Second, high conversion efficiency due to unlimited optical path lengths.
Third, the risk of scattering loss during propagation due to pollution is lower.
It will be easier to handle outside the clean room.


In Fig.\ref{fig:gas pressure}, we show the nonlinear refractive index for several noble gases as a function of gas pressure.
The nonlinear refractive index of silica is also plotted as a comparison.
The gas pressure varies from 0 to 200 bar.
As shown in the figure, the nonlinear refractive index of the gas increases significantly following the rising of pressure.
The nonlinear refractive index discribes the instantaneous nonlinearity of the filling gas.
The joint effect of linear refractive index and nonlinear refractive index changes causes the shift of Kerr nonlinearity of the gas-filled suspended core fiber.

\section{Experiment setup}
\begin{figure}
    \centering
    \includegraphics[width=\linewidth]{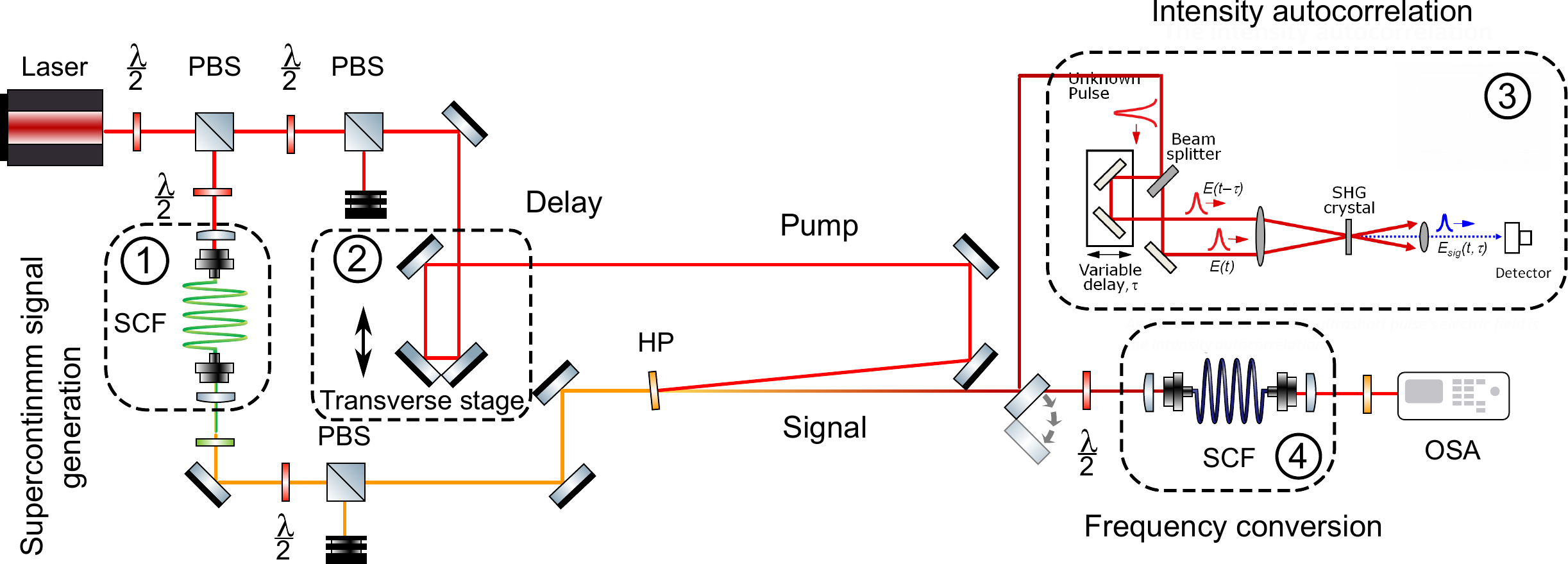}
    \caption{The experiment setup for FWM with SCF. The polarization beam splitter is written as PBS. The OSA corresponds to the optical spectrum analyzer. The laser beam was divided into the pump line and signal line. The two lines are later join together at a dichronic filter and sent to a SCF for frequency conversion. The super-continuum signal generation, delay, intensity autocorrelation and frequency conversion systems are marked with frames with dashed line.}
    \label{fig:setup}
\end{figure}

The experiment setup for FWM is shown in Fig.~\ref{fig:setup}.
The general idea of this setup is that the pump and the signal are coupled into the suspended core fiber to generate idler by FWM.
This system contains some important parts, which are marked in Fig.~\ref{fig:setup}, corresponding to (1)the super-continuum signal generation, (2)delay, (3)intensity autocorrelation and (4)frequency conversion.
We will introduce them in detail in the following paragraphs.

In our setup, both the pump and signal are pulsed.
The pump and signal are generated originally with the same pulselaser at 800 nm with 700 fs pulsewidth and 80 MHz repetition rate and has an average power above 1 W.
The laser beam was divided into two paths with adjustable intensity in each path with three sets of half-waveplates and polarized beam splitters, which are pump line and signal line respectively.

We generate a pulse signal with a wavelength from 400 nm to infrared (above 1000 nm).
This is realized by using the super-continuum generation, marked as number 1 in Fig.~\ref{fig:setup}.
Here we use the SCF as a medium to generate super-continuous signal.
When the ultrashort pulses at the zero dispersion wavelength (ZDW) reaches a certain peak power, the non-linear refractive index starts to play an important role which leads to nonlinear process including self-phase modulation, modulational instability, soliton generation and fission and cross phase modulation and expand the linewidth \cite{dudley2010supercontinuum,dudley2006supercontinuum}.
Here, we use the SCF with ZDW of 800 nm, matching the laser wavelength.
Using a super-continuum signal can be helpful to search for the phase matching wavelength of the SCF and adjust the phase matching point with the gas pressure.
Later, the signal line will be replaced by a single photon generation system based on CdSe-CdS dot-in-rods emitting photons around 580 nm wavelength.
The optical property of CdSe-CdS dot-in-rods has studied by previous colleagues in our group \cite{manceau2014effect, vezzoli2015exciton,manceau2018cdse}.
In order to have super-continuum signal, the optical power of the laser beam coupled into the SCF needs to reach a certain threshold.
The mode size of the SCF we used is 2 $\mu m$.
The coupling efficiency from free space to the SCF is above 50\%.
With optical power around 1 W, we manage to get super-continuum signal at SCF output.

The pump is generated from the same pulse.
Therefore, it is centered at 800 nm with 700 fs pulsewidth and 80 MHz repetition rate.
The average pump power needed to get enough peak power for frequency conversion is 200 $\mu$W.
On the contrary, the average pump power I got is 10 mW, which is much higher than the power needed for FWM with high conversion efficiency.

The signal line and the pump line join together after a dichroic filter.
We add two mirrors in both signal and pump line for the alignment of the two beams.

We add a delay in pump line, marked (2) in Fig.~\ref{fig:setup}, with a retro-reflector mounted on a transverse stage.
This is to synchronize the pump and signal pulses at the FWM system by changing the length of the pump line.
A delay adjustment with 0.1 mm resolution is required to get good overlap between two pulses of different wavelengths.

To realize this, firstly, we send both the pump and the signal to a photodiode placed at the location of the SCF for frequency conversion.
The output signal of the photodiode is sent to an oscilloscope.
We can read the pump pulse and signal pulse with the oscilloscope and adjust the delay in pump line to roughly synchronize these two pulses with 6 cm resolution, which is limited by the time resolution of oscilloscope.
In order to further increase the resolution, an intensity auto-correlation measurement is necessary.
In our experiment, an auto-correlator was used to measure the auto-correlation between the two paths, marked (3) in Fig.~\ref{fig:setup}.
It was linked to the pump and signal lines through a flip mirror.
The intensity autocorrelation is based on the second harmonic generation of the input signal.
When the input beams are synchronized, there will be only one second harmonic generation output.
However, when we have two input signals corresponding to the pump pulse and signal pulse that are not synchronized, the output of the intensity autocorrelation will include the second harmonic generation of the pump pulse and signal pulse respectively.
Besides, when the delay between these two pulses is small, it will generate an extra peak in between which is the second harmonic generation between the two input pulses (or sum frequency generation if the two input pulses have different wavelengths).
The nonlinear crystal in the auto-correlator we used was suitable only for infrared wavelengths.
Therefore, we added a filter in the signal line and synchronized the pulses at 850 nm in the supercontinuum signal line with the pump pulses using the autocorrelator.
We need to be aware of the fact that the pulses at visible range will be shifted in time scale due to the dispersion in SCF.
The delay length needs to be further optimized based on the output spectrum after the frequency conversion with the SCF.
The output spectra is measured with an optical spectrum analyzer (OSA) at the fiber output.

The frequency conversion part marked (4) in Fig.~\ref{fig:setup} is composed of a three-holes SCF and two coupling lens.
The three holes within the SCF can be filled with noble gas.
Since the refractive index of the SCF is dependent on the pressure of the filled noble gas.
In this way, we can control the phase matching of the FWM within the SCF by tuning the gas pressure.
In practice, the SCF needs to be fixed within a high pressure chamber using gas system (Swagelok) equipped with a pressure controllable gas input.
The SCF will be fixed with two v-groove clamps inside the chamber.
Since the gas will be both inside and outside the SCF, the pressure will be balanced.
To couple the light into the SCF, the two sides of the SCF need to be equipped with lens to have a good match between beam waist and the mode size of SCF.
In our system, the coupling efficiency is about 50 \%.

The bottom path is used to generate supercontinuous signal with a photonic crystal fiber (PCF).
By filtering this supercontinuous signal, we can get a pulsed signal at expected wavelength. 
FWM process happens in the SCF where the pump and signal need to be synchronized.


A high peak power is the advantage of using a femtosecond pulse laser in our experiment.
However, there are also drawbacks using femtosecond pulse laser.
The narrow pulse width brings trouble in the synchronicity between pump and seed.
Since the wavelength difference between pump and seed is large, the walk off between different wavelength within the fiber will be significant, which limits the interaction length and the conversion efficiency.
the increasing of length of optical path in suspended-core fiber will not further increase the conversion efficiency.
A continuous wave laser with high power can be a solution.


This cooperate experiment with the group of Nicolas Joly in Max Planck Institution for the science of light.
Since the suspended core fiber has a different cross section from the ideal cylindrical fiber.
The calculation we did on the phase matching will not completely match with the fiber we used in experiment.
With the fabricated suspended core fiber, we didn't observe the expected evidence of generated idler photons in spectrum.
Therefore, we didn't put data of the spectrum here.
There will be more work to be done in the future on the technique.
The good news is that, the group of Nicolas Joly recently published their work on four wave mixing with the suspended core fiber, but for different phase matching wavelength \cite{hammer2020broadly}.
The gas pressure was tuned up to 25 bar to have on demand phase matching.
It is a promising solution for on demand single photon frequency conversion.

\section*{Conclusion}
\addcontentsline{toc}{section}{Conclusion} 
In this chapter, we introduced a possible solution for single photon frequency conversion by four-wave-mixing in a suspended-core fiber.
The single photons are emitted from CdSe-CdS quantum dots.
Based on the core size of the suspended core fiber we used, through seeded four wave mixing with input single photons at 530 nm, photons at 800 nm will be converted into photons at 1.62 $\mu m$.
The phase matching is adjustable by controlling the pressure of the injected noble gas in the holes of the suspended-core fiber.
This technique is proposed to be implemented for on demand frequency conversion.

\chapter*{Conclusion}
\markboth{CONCLUSION}{}
\addcontentsline{toc}{chapter}{Conclusion} 
In this thesis, we discussed theory, experiment and applications of optical waveguides and single nano emitters in non-linear and quantum optics and explained the physics behind.

In part one, we introduced several applications of the optical nanofibers based on different properties, including the coupling between the dipole on the nanofiber surface, the coupling between two parallel nanofibers and the mechanical property of optical nanofiber, corresponding to chapter 2-4 respectively.

Our first experiment on the optical nanofiber indicates that a linear dipole located in an evanescent field can emit elliptical polarizations.
A Berek compensator was used to eliminate the influences of the birefringence of the optical fiber between the location of the coupled dipole and the fiber output.
Therefore, the fiber output can preserve the guided polarization from the coupled dipole.  
We deposited a group of gold nanorod and analyzed the guided polarization.
The experimental result has shown a deterministic relationship between the geometry of the dipole (nanorod) and the polarization emitted that the orientation of the dipole (nanorod) defines the ellipticity of the polarization of the guided light and the azimuthal position controls the main axis of polarization.
This opens the way to alternative methods for polarization control of light sources at the nanoscale.

The second experiment on the optical nanofiber shows the fabrication and characterization of two optical devices based on looped nanometrical optical fibers. 
The nature of the devices changes with the topology of the loop. 
The first device is a twisted loop. 
The entwined part of the loop can be treated with the coupled mode theory and is shown to be a tunable beam splitter. 
We showed that a twisted loop creates a Sagnac interferometer, of which the dephasing is tuned by the torsion applied on the nanofiber. 
The second device is a ring cavity, simply made out of a nanofiber knot. 
We analyzed its spectral response and found a finesse of 8 and a quality factor of 1300. 
Unlike in common resonators, the coupling efficency into the cavity strongly depends on the wavelength, which modulates the visibility of its resonances. 
It is reproduced accurately by our theoretical model.

Both devices are essential blocks for photonics applications. 
Their miniaturized versions presented here pave the way toward their integration in photonic circuits. 
A refined analysis of the cavity spectrum revealed that birefringence of a bent nanofiber is also affected by ovalization of its profile, and not only by stress as a normal fiber would be.
With both devices, we showed how sensitive nanofibers are to mechanical constraints. 

The third application of optical nanofiber is a displacement sensor based on a gold nano-sphere deposited on a nanofiber and  placed within a standing wave. 
During this thesis, this device was implemented and calibrated to have a resolution down to 1.2~nm/$\sqrt{\text{Hz}}$. 
We also proposed a mechanical model to estimate the response of our sensor to optically-induced pressure force and we found that the sensitivity corresponding to the sensibility of an externally applied force at the nanofiber waist is 1~pN.
We also introduced a nanofiber homodyne detection system for measuring mechanical vibration detection.
This optical nanofiber sensor is an exciting tool to realize integrated optomechanical research using this platform, such as the demonstration of superfluidity of light in a photon fluid \cite{larre2015optomechanical,fontaine2018observation}.\\

In the second part of the manuscript about single emitters, we presented the results we obtained for improving the photoluminescence quantum yield of lanthanide-doped upconversion nanoparticles, and lower the attenuation during transportation of single photons through frequency conversion.

In Chapter 5, we realized the enhancement of luminescence from single lanthanide-doped upconversion nanoparticles with surface plasmon resonance from gold nanorods.
With the help of atomic force microscope manipulation, gold nanorods are assembled with Yb$^{3+}$/Er$^{3+}$/Mn$^{2+}$ co-doped NaYF$_4$ nanocrystals into hybrid dimers.
We sheltered the NaYF$_4$ nanocrystals with 7 nm shell silicon to avoid quenching.
We studied the enhancement dependence on the shape of gold nanorods, and got the conclusion that the gold nanorods with larger scattering cross section offer stronger near field enhancement.
The excitation polarization dependence measurement shows the upconversion luminescence enhancement is the strongest when the excitation polarization is along the longitudinal axis of gold nanorod.
With gold nanorods of 46.7 nm diameter, we obtained a plasmonic enhancement up to 110 times.
Besides, the upconversion luminescence enhancement arises from the joint effect of excitation enhancement (raising of plasma radiation rate) at low excitation power and quantum efficiency enhancement at high excitation power.


In Chapter 6 and 7, we proposed numerical simulation and possible experimental solutions for quantum frequency conversion. 
The emitters we studied include defects in nanodiamond which are promising single photon sources at room temperature and quantum dots (II-VI) CdSe-CdS dot-in-rods.

In Chapter 6 we gave the numerical study of difference frequency generation with defects in diamond, specifically Nitrogen vacancy and Silicon vacancy color centers in nanodiamond.
We calculated the phase matching condition of these two types of defects in diamond while using different nonlinear media, the periodically poled lithium niobate (PPLN) and periodically poled potassium titanyl phosphate (PPKTP).
The calculation shows the opposite effect on bandwidth compression or expanding when using pump of wavelength longer or shorter than idler signal. 
Besides, by using a wide-bandwidth pump, we can increase the convertible bandwidth to 2 nm centered at 738 nm for silicon vacancy color center and obtain a conversion efficiency of more than 50\% with bandwidth compression.

In Chapter 7, we introduced a solution for single photon frequency conversion with four wave mixing in suspended core fiber.
The single photons are emitted from CdSe-CdS quantum dots.
Based on the core size of the suspended core fiber we used, through seeded four wave mixing with input single photons at 530 nm, photons at 800 nm will be converted into photons at 1.62 $\mu m$.
The phase matching is adjustable by controlling the pressure of the injected noble gas in the holes of the suspended-core fiber.
This technique is proposed to be implemented for on demand frequency conversion.


\bibliographystyle{unsrt}
\bibliography{refs}

\begin{thebibliography}{100}

\bibitem{kao1966dielectric}
K~Charles Kao and George~A Hockham.
\newblock Dielectric-fibre surface waveguides for optical frequencies.
\newblock In {\em Proceedings of the Institution of Electrical Engineers},
  volume 113, pages 1151--1158. IET, 1966.

\bibitem{tong2003subwavelength}
Limin Tong, Rafael~R Gattass, Jonathan~B Ashcom, Sailing He, Jingyi Lou,
  Mengyan Shen, Iva Maxwell, and Eric Mazur.
\newblock Subwavelength-diameter silica wires for low-loss optical wave
  guiding.
\newblock {\em Nature}, 426(6968):816--819, 2003.

\bibitem{warken2007ultradunne}
Florian Warken et~al.
\newblock {\em Ultrad{\"u}nne Glasfasern als Werkzeug zur Kopplung von Licht
  und Materie}.
\newblock PhD thesis, Universit{\"a}ts-und Landesbibliothek Bonn, 2007.

\bibitem{hoffman2014ultrahigh}
JE~Hoffman, S~Ravets, JA~Grover, P~Solano, PR~Kordell, JD~Wong-Campos,
  LA~Orozco, and SL~Rolston.
\newblock Ultrahigh transmission optical nanofibers.
\newblock {\em AIP advances}, 4(6):067124, 2014.

\bibitem{praeger2012fabrication}
M~Praeger, E~Saleh, A~Vaughan, WJ~Stewart, and WH~Loh.
\newblock Fabrication of nanoscale glass fibers by electrospinning.
\newblock {\em Applied Physics Letters}, 100(6):063114, 2012.

\bibitem{quintero2009laser}
F{\'e}lix Quintero, Juan Pou, Rafael Comesa{\~n}a, Fernando Lusqui{\~n}os,
  Antonio Riveiro, Adrian~B Mann, Robert~G Hill, Zoe~Y Wu, and Julian~R Jones.
\newblock Laser spinning of bioactive glass nanofibers.
\newblock {\em Advanced Functional Materials}, 19(19):3084--3090, 2009.

\bibitem{quintero2020continuous}
F~Quintero, J~Penide, A~Riveiro, J~Del~Val, R~Comesa{\~n}a, F~Lusqui{\~n}os,
  and J~Pou.
\newblock Continuous fiberizing by laser melting (cofiblas): Production of
  highly flexible glass nanofibers with effectively unlimited length.
\newblock {\em Science advances}, 6(6):eaax7210, 2020.

\bibitem{nayak2007optical}
KP~Nayak, PN~Melentiev, M~Morinaga, Fam Le~Kien, VI~Balykin, and K~Hakuta.
\newblock Optical nanofiber as an efficient tool for manipulating and probing
  atomic fluorescence.
\newblock {\em Optics Express}, 15(9):5431--5438, 2007.

\bibitem{vetsch2010optical}
E~Vetsch, D~Reitz, G~Sagu{\'e}, R~Schmidt, ST~Dawkins, and A~Rauschenbeutel.
\newblock Optical interface created by laser-cooled atoms trapped in the
  evanescent field surrounding an optical nanofiber.
\newblock {\em Physical Review Letters}, 104(20):203603, 2010.

\bibitem{nayak2008single}
Kali~P Nayak and K~Hakuta.
\newblock Single atoms on an optical nanofibre.
\newblock {\em New Journal of Physics}, 10(5):053003, 2008.

\bibitem{corzo2019waveguide}
Neil~V Corzo, J{\'e}r{\'e}my Raskop, Aveek Chandra, Alexandra~S Sheremet,
  Baptiste Gouraud, and Julien Laurat.
\newblock Waveguide-coupled single collective excitation of atomic arrays.
\newblock {\em Nature}, 566(7744):359--362, 2019.

\bibitem{nieddu2016optical}
Thomas Nieddu, Vandna Gokhroo, and S{\'\i}le~Nic Chormaic.
\newblock Optical nanofibres and neutral atoms.
\newblock {\em Journal of Optics}, 18(5):053001, 2016.

\bibitem{schroder2012nanodiamond}
Tim Schr{\"o}der, Masazumi Fujiwara, Tetsuya Noda, Hong-Quan Zhao, Oliver
  Benson, and Shigeki Takeuchi.
\newblock A nanodiamond-tapered fiber system with high single-mode coupling
  efficiency.
\newblock {\em Optics express}, 20(10):10490--10497, 2012.

\bibitem{liebermeister2014tapered}
Lars Liebermeister, Fabian Petersen, Asmus~v M{\"u}nchow, Daniel Burchardt,
  Juliane Hermelbracht, Toshiyuki Tashima, Andreas~W Schell, Oliver Benson,
  Thomas Meinhardt, Anke Krueger, et~al.
\newblock Tapered fiber coupling of single photons emitted by a
  deterministically positioned single nitrogen vacancy center.
\newblock {\em Applied Physics Letters}, 104(3):031101, 2014.

\bibitem{yalla2012efficient}
Ramachandrarao Yalla, Fam Le~Kien, M~Morinaga, and K~Hakuta.
\newblock Efficient channeling of fluorescence photons from single quantum dots
  into guided modes of optical nanofiber.
\newblock {\em Physical review letters}, 109(6):063602, 2012.

\bibitem{fujiwara2011highly}
Masazumi Fujiwara, Kiyota Toubaru, Tetsuya Noda, Hong-Quan Zhao, and Shigeki
  Takeuchi.
\newblock Highly efficient coupling of photons from nanoemitters into
  single-mode optical fibers.
\newblock {\em Nano letters}, 11(10):4362--4365, 2011.

\bibitem{pollinger2009ultrahigh}
Michael P{\"o}llinger, Danny O’Shea, Florian Warken, and Arno Rauschenbeutel.
\newblock Ultrahigh-q tunable whispering-gallery-mode microresonator.
\newblock {\em Physical review letters}, 103(5):053901, 2009.

\bibitem{russell2011sub}
Laura Russell, Kieran Deasy, Mark~J Daly, Michael~J Morrissey, and
  S{\'\i}le~Nic Chormaic.
\newblock Sub-doppler temperature measurements of laser-cooled atoms using
  optical nanofibres.
\newblock {\em Measurement Science and Technology}, 23(1):015201, 2011.

\bibitem{zhang2011micro}
Lei Zhang, Jingyi Lou, and Limin Tong.
\newblock Micro/nanofiber optical sensors.
\newblock {\em Photonic Sensors}, 1(1):31--42, 2011.

\bibitem{han2014side}
Yuqi Han, Zhe Chen, Dong Cao, Jianhui Yu, Haozhi Li, Xiaoli He, Jun Zhang,
  Yunhan Luo, Huihui Lu, Jieyuan Tang, et~al.
\newblock Side-polished fiber as a sensor for the determination of nematic
  liquid crystal orientation.
\newblock {\em Sensors and Actuators B: Chemical}, 196:663--669, 2014.

\bibitem{joos2018polarization}
Maxime Joos, Chengjie Ding, Vivien Loo, Guillaume Blanquer, Elisabeth
  Giacobino, Alberto Bramati, Valentina Krachmalnicoff, and Quentin Glorieux.
\newblock Polarization control of linear dipole radiation using an optical
  nanofiber.
\newblock {\em Physical Review Applied}, 9(6):064035, 2018.

\bibitem{ding2019fabrication}
Chengjie Ding, Vivien Loo, Simon Pigeon, Romain Gautier, Maxime Joos, E~Wu,
  Elisabeth Giacobino, Alberto Bramati, and Quentin Glorieux.
\newblock Fabrication and characterization of optical nanofiber interferometer
  and resonator for the visible range.
\newblock {\em New Journal of Physics}, 21(7):073060, 2019.

\bibitem{ding2020nanofiber}
Chengjie Ding, Maxime Joos, Constanze Bach, Tom Bienaim{\'e}, Elisabeth
  Giacobino, E~Wu, Alberto Bramati, and Quentin Glorieux.
\newblock Nanofiber based displacement sensor.
\newblock {\em Applied Physics B}, 126(6), 2020.

\bibitem{xue2017tuning}
Yingxian Xue, Chengjie Ding, Youying Rong, Qiang Ma, Chengda Pan, E~Wu, Botao
  Wu, and Heping Zeng.
\newblock Tuning plasmonic enhancement of single nanocrystal upconversion
  luminescence by varying gold nanorod diameter.
\newblock {\em Small}, 13(36):1701155, 2017.

\bibitem{morrissey2013spectroscopy}
Michael~J Morrissey, Kieran Deasy, Mary Frawley, Ravi Kumar, Eugen Prel, Laura
  Russell, Viet~Giang Truong, and S{\'\i}le Nic~Chormaic.
\newblock Spectroscopy, manipulation and trapping of neutral atoms, molecules,
  and other particles using optical nanofibers: a review.
\newblock {\em Sensors}, 13(8):10449--10481, 2013.

\bibitem{hoffman2015rayleigh}
Jonathan~E Hoffman, Fredrik~K Fatemi, Guy Beadie, Steven~L Rolston, and Luis~A
  Orozco.
\newblock Rayleigh scattering in an optical nanofiber as a probe of
  higher-order mode propagation.
\newblock {\em Optica}, 2(5):416--423, 2015.

\bibitem{solano2017optical}
Pablo Solano, Jeffrey~A Grover, Jonathan~E Hoffman, Sylvain Ravets, Fredrik~K
  Fatemi, Luis~A Orozco, and Steven~L Rolston.
\newblock Optical nanofibers: a new platform for quantum optics.
\newblock In {\em Advances In Atomic, Molecular, and Optical Physics},
  volume~66, pages 439--505. Elsevier, 2017.

\bibitem{love1986quantifying}
JD~Love and WM~Henry.
\newblock Quantifying loss minimisation in single-mode fibre tapers.
\newblock {\em Electronics Letters}, 22(17):912--914, 1986.

\bibitem{snyder2012optical}
Allan~W Snyder and John Love.
\newblock {\em Optical waveguide theory}.
\newblock Springer Science \& Business Media, 2012.

\bibitem{petersen2014chiral}
Jan Petersen, J{\"u}rgen Volz, and Arno Rauschenbeutel.
\newblock Chiral nanophotonic waveguide interface based on spin-orbit
  interaction of light.
\newblock {\em Science}, 346(6205):67--71, 2014.

\bibitem{neugebauer2019emission}
Martin Neugebauer, Peter Banzer, and Sergey Nechayev.
\newblock Emission of circularly polarized light by a linear dipole.
\newblock {\em Science advances}, 5(6):eaav7588, 2019.

\bibitem{saleh2019fundamentals}
Bahaa~EA Saleh and Malvin~Carl Teich.
\newblock {\em Fundamentals of photonics}.
\newblock john Wiley \& sons, 2019.

\bibitem{born1999principles}
Max Born and Emil Wolf.
\newblock Principles of optics, 7th (expanded) edition.
\newblock {\em United Kingdom: Press Syndicate of the University of Cambridge},
  461, 1999.

\bibitem{olivard1999measurement}
P~Olivard, PY~Gerligand, B~Le~Jeune, J~Cariou, and J~Lotrian.
\newblock Measurement of optical fibre parameters using an optical polarimeter
  and stokes-mueller formalism.
\newblock {\em Journal of Physics D: Applied Physics}, 32(14):1618, 1999.

\bibitem{vetsch2012nanofiber}
Eugen Vetsch, Samuel~T Dawkins, Rudolf Mitsch, Daniel Reitz, Philipp
  Schneeweiss, and Arno Rauschenbeutel.
\newblock Nanofiber-based optical trapping of cold neutral atoms.
\newblock {\em IEEE Journal of Selected Topics in Quantum Electronics},
  18(6):1763--1770, 2012.

\bibitem{le2004field}
Fam Le~Kien, JQ~Liang, K~Hakuta, and VI~Balykin.
\newblock Field intensity distributions and polarization orientations in a
  vacuum-clad subwavelength-diameter optical fiber.
\newblock {\em Optics Communications}, 242(4-6):445--455, 2004.

\bibitem{kimble2008quantum}
H~Jeff Kimble.
\newblock The quantum internet.
\newblock {\em Nature}, 453(7198):1023, 2008.

\bibitem{albrecht2013coupling}
Roland Albrecht, Alexander Bommer, Christian Deutsch, Jakob Reichel, and
  Christoph Becher.
\newblock Coupling of a single nitrogen-vacancy center in diamond to a
  fiber-based microcavity.
\newblock {\em Physical Review Letters}, 110(24):243602, 2013.

\bibitem{le2004atom}
Fam Le~Kien, VI~Balykin, and K~Hakuta.
\newblock Atom trap and waveguide using a two-color evanescent light field
  around a subwavelength-diameter optical fiber.
\newblock {\em Physical Review A}, 70(6):063403, 2004.

\bibitem{ding2010ultralow}
Lu~Ding, Cherif Belacel, Sara Ducci, Giuseppe Leo, and Ivan Favero.
\newblock Ultralow loss single-mode silica tapers manufactured by a
  microheater.
\newblock {\em Applied Optics}, 49(13):2441--2445, 2010.

\bibitem{goban2012demonstration}
A~Goban, KS~Choi, DJ~Alton, D~Ding, C~Lacro{\^u}te, M~Pototschnig, T~Thiele,
  NP~Stern, and HJ~Kimble.
\newblock Demonstration of a state-insensitive, compensated nanofiber trap.
\newblock {\em Physical Review Letters}, 109(3):033603, 2012.

\bibitem{nayak2014optical}
KP~Nayak, Pengfei Zhang, and K~Hakuta.
\newblock Optical nanofiber-based photonic crystal cavity.
\newblock {\em Optics letters}, 39(2):232--235, 2014.

\bibitem{takashima2016detailed}
Hideaki Takashima, Masazumi Fujiwara, Andreas~W Schell, and Shigeki Takeuchi.
\newblock Detailed numerical analysis of photon emission from a single light
  emitter coupled with a nanofiber bragg cavity.
\newblock {\em Optics Express}, 24(13):15050--15058, 2016.

\bibitem{schell2015highly}
Andreas~W Schell, Hideaki Takashima, Shunya Kamioka, Yasuko Oe, Masazumi
  Fujiwara, Oliver Benson, and Shigeki Takeuchi.
\newblock Highly efficient coupling of nanolight emitters to a ultra-wide
  tunable nanofibre cavity.
\newblock {\em Scientific Reports}, 5:9619, 2015.

\bibitem{cai2000observation}
Ming Cai, Oskar Painter, and Kerry~J Vahala.
\newblock Observation of critical coupling in a fiber taper to a
  silica-microsphere whispering-gallery mode system.
\newblock {\em Physical Review Letters}, 85(1):74, 2000.

\bibitem{aoki2006observation}
Takao Aoki, Barak Dayan, Elizabeth Wilcut, Warwick~P Bowen, A~Scott Parkins,
  TJ~Kippenberg, KJ~Vahala, and HJ~Kimble.
\newblock Observation of strong coupling between one atom and a monolithic
  microresonator.
\newblock {\em Nature}, 443(7112):671, 2006.

\bibitem{ward2014contributed}
JM~Ward, A~Maimaiti, Vu~H Le, and S~Nic Chormaic.
\newblock Contributed review: Optical micro-and nanofiber pulling rig.
\newblock {\em Review of Scientific Instruments}, 85(11):111501, 2014.

\bibitem{okamoto2006fundamentals}
Katsunari Okamoto.
\newblock {\em Fundamentals of optical waveguides}.
\newblock Academic press, 2006.

\bibitem{kien2020coupling}
Fam~Le Kien, Lewis Ruks, Sile~Nic Chormaic, and Thomas Busch.
\newblock Coupling between guided modes of two parallel nanofibers.
\newblock {\em arXiv preprint arXiv:2007.11311}, 2020.

\bibitem{goriely1998nonlinear}
Alain Goriely and Michael Tabor.
\newblock Nonlinear dynamics of filaments: Iv spontaneous looping of twisted
  elastic rods.
\newblock In {\em Localization And Solitary Waves In Solid Mechanics}, pages
  235--254. World Scientific, 1999.

\bibitem{culshaw2005optical}
Brian Culshaw.
\newblock The optical fibre sagnac interferometer: an overview of its
  principles and applications.
\newblock {\em Measurement Science and Technology}, 17(1):R1, 2005.

\bibitem{ulrich1980bending}
Rashleigh Ulrich, SC~Rashleigh, and W~Eickhoff.
\newblock Bending-induced birefringence in single-mode fibers.
\newblock {\em Optics letters}, 5(6):273--275, 1980.

\bibitem{wierzbicki1997simplified}
Tomasz Wierzbicki and Monique~V Sinmao.
\newblock A simplified model of brazier effect in plastic bending of
  cylindrical tubes.
\newblock {\em International Journal of Pressure Vessels and Piping},
  71(1):19--28, 1997.

\bibitem{wang2018cross}
Chunge Wang, Zhiyuan Zhang, Ruixue Zhai, Gaochao Yu, and Jun Zhao.
\newblock Cross-sectional distortion of lsaw pipes in over-bend straightening
  process.
\newblock {\em Thin-Walled Structures}, 129:85--93, 2018.

\bibitem{muschielok2008nano}
Adam Muschielok, Joanna Andrecka, Anass Jawhari, Florian Br{\"u}ckner, Patrick
  Cramer, and Jens Michaelis.
\newblock A nano-positioning system for macromolecular structural analysis.
\newblock {\em Nature Methods}, 5(11):965, 2008.

\bibitem{andrecka2009nano}
Joanna Andrecka, Barbara Treutlein, Maria Angeles~Izquierdo Arcusa, Adam
  Muschielok, Robert Lewis, Alan~CM Cheung, Patrick Cramer, and Jens Michaelis.
\newblock Nano positioning system reveals the course of upstream and
  nontemplate dna within the rna polymerase ii elongation complex.
\newblock {\em Nucleic acids research}, 37(17):5803--5809, 2009.

\bibitem{de2017universal}
Laure~Mercier de~L{\'e}pinay, Benjamin Pigeau, Benjamin Besga, Pascal Vincent,
  Philippe Poncharal, and Olivier Arcizet.
\newblock A universal and ultrasensitive vectorial nanomechanical sensor for
  imaging 2d force fields.
\newblock {\em Nature Nanotechnology}, 12(2):156, 2017.

\bibitem{de2018eigenmode}
Laure~Mercier de~L{\'e}pinay, Benjamin Pigeau, Benjamin Besga, and Olivier
  Arcizet.
\newblock Eigenmode orthogonality breaking and anomalous dynamics in multimode
  nano-optomechanical systems under non-reciprocal coupling.
\newblock {\em Nature Communications}, 9(1):1401, 2018.

\bibitem{feng2007very}
XL~Feng, Rongrui He, Peidong Yang, and ML~Roukes.
\newblock Very high frequency silicon nanowire electromechanical resonators.
\newblock {\em Nano Letters}, 7(7):1953--1959, 2007.

\bibitem{conley2008nonlinear}
William~G Conley, Arvind Raman, Charles~M Krousgrill, and Saeed Mohammadi.
\newblock Nonlinear and nonplanar dynamics of suspended nanotube and nanowire
  resonators.
\newblock {\em Nano letters}, 8(6):1590--1595, 2008.

\bibitem{eichler2011nonlinear}
Alexander Eichler, Joel Moser, Julien Chaste, Mariusz Zdrojek, I~Wilson-Rae,
  and Adrian Bachtold.
\newblock Nonlinear damping in mechanical resonators made from carbon nanotubes
  and graphene.
\newblock {\em Nature Nanotechnology}, 6(6):339, 2011.

\bibitem{anetsberger2009near}
Georg Anetsberger, Olivier Arcizet, Quirin~P Unterreithmeier, R{\'e}mi
  Rivi{\`e}re, Albert Schliesser, Eva~Maria Weig, J{\"o}rg~P Kotthaus, and
  Tobias~J Kippenberg.
\newblock Near-field cavity optomechanics with nanomechanical oscillators.
\newblock {\em Nature Physics}, 5(12):909--914, 2009.

\bibitem{teufel2008dynamical}
JD~Teufel, JW~Harlow, CA~Regal, and KW~Lehnert.
\newblock Dynamical backaction of microwave fields on a nanomechanical
  oscillator.
\newblock {\em Physical Review Letters}, 101(19):197203, 2008.

\bibitem{nichol2012nanomechanical}
John~M Nichol, Eric~R Hemesath, Lincoln~J Lauhon, and Raffi Budakian.
\newblock Nanomechanical detection of nuclear magnetic resonance using a
  silicon nanowire oscillator.
\newblock {\em Physical Review B}, 85(5):054414, 2012.

\bibitem{arcizet2011single}
Olivier Arcizet, Vincent Jacques, Alessandro Siria, Philippe Poncharal, Pascal
  Vincent, and Signe Seidelin.
\newblock A single nitrogen-vacancy defect coupled to a nanomechanical
  oscillator.
\newblock {\em Nature Physics}, 7(11):879--883, 2011.

\bibitem{cleland2002nanomechanical}
AN~Cleland, JS~Aldridge, DC~Driscoll, and AC~Gossard.
\newblock Nanomechanical displacement sensing using a quantum point contact.
\newblock {\em Applied Physics Letters}, 81(9):1699--1701, 2002.

\bibitem{favero2009fluctuating}
Ivan Favero, Sebastian Stapfner, David Hunger, Philipp Paulitschke, Jakob
  Reichel, Heribert Lorenz, Eva~M Weig, and Khaled Karrai.
\newblock Fluctuating nanomechanical system in a high finesse optical
  microcavity.
\newblock {\em Optics Express}, 17(15):12813--12820, 2009.

\bibitem{larre2015optomechanical}
Pierre-{\'E}lie Larr{\'e} and Iacopo Carusotto.
\newblock Optomechanical signature of a frictionless flow of superfluid light.
\newblock {\em Physical Review A}, 91(5):053809, 2015.

\bibitem{freikamp2016piconewton}
Andrea Freikamp, Anna-Lena Cost, and Carsten Grashoff.
\newblock The piconewton force awakens: quantifying mechanics in cells.
\newblock {\em Trends in cell biology}, 26(11):838--847, 2016.

\bibitem{finer1994single}
Jeffrey~T Finer, Robert~M Simmons, and James~A Spudich.
\newblock Single myosin molecule mechanics: piconewton forces and nanometre
  steps.
\newblock {\em Nature}, 368(6467):113--119, 1994.

\bibitem{joos2018dispositifs}
Maxime Joos.
\newblock {\em Dispositifs hybrides: nanoparticules coupl{\'e}es {\`a} une
  nanofibre optique}.
\newblock PhD thesis, Sorbonne universit{\'e}, 2018.

\bibitem{allan1981modified}
David~W Allan and James~A Barnes.
\newblock A modified allan variance with increased oscillator characterization
  ability.
\newblock In {\em Proceedings of the 35th Annual Frequency Control Symposium},
  volume~5, pages 470--475, 1981.

\bibitem{bellan2005measurement}
Leon~M Bellan, Jun Kameoka, and Harold~G Craighead.
\newblock Measurement of the young’s moduli of individual polyethylene oxide
  and glass nanofibres.
\newblock {\em Nanotechnology}, 16(8):1095, 2005.

\bibitem{timoshenko1968elements}
Stephen Timoshenko and Donovan~Harold Young.
\newblock {\em Elements of strength of materials}.
\newblock Van Nostrand New York, 1968.

\bibitem{mitri2017radiation}
FG~Mitri.
\newblock Radiation force and torque of light-sheets.
\newblock {\em Journal of Optics}, 19(6):065403, 2017.

\bibitem{loo2019imaging}
Vivien Loo, Guillaume Blanquer, Maxime Joos, Quentin Glorieux, Yannick
  De~Wilde, and Valentina Krachmalnicoff.
\newblock Imaging light scattered by a subwavelength nanofiber, from near field
  to far field.
\newblock {\em Optics Express}, 27(2):350--357, 2019.

\bibitem{joos2019complete}
Maxime Joos, Alberto Bramati, and Quentin Glorieux.
\newblock Complete polarization control for a nanofiber waveguide using the
  scattering properties.
\newblock {\em Optics Express}, 27(13):18818--18830, 2019.

\bibitem{gouraud2016optical}
Baptiste Gouraud.
\newblock {\em Optical Nanofibers Interfacing Cold Atoms-A Tool for Quantum
  Optics}.
\newblock PhD thesis, Sorbonne universit{\'e}, 2016.

\bibitem{burrus2013wavelets}
C~Sidney Burrus.
\newblock Wavelets and wavelet transforms.
\newblock 2013.

\bibitem{fontaine2018observation}
Q~Fontaine, T~Bienaim{\'e}, S~Pigeon, E~Giacobino, A~Bramati, and Q~Glorieux.
\newblock Observation of the bogoliubov dispersion in a fluid of light.
\newblock {\em Physical Review Letters}, 121(18):183604, 2018.

\bibitem{geng2016localised}
Wei Geng, Mathieu Manceau, Nancy Rahbany, Vincent Sallet, Massimo De~Vittorio,
  Luigi Carbone, Quentin Glorieux, Alberto Bramati, and Christophe Couteau.
\newblock Localised excitation of a single photon source by a nanowaveguide.
\newblock {\em Scientific Reports}, 6:19721, 2016.

\bibitem{manceau2014effect}
M~Manceau, S~Vezzoli, Q~Glorieux, F~Pisanello, E~Giacobino, Luigi Carbone,
  M~De~Vittorio, and A~Bramati.
\newblock Effect of charging on cdse/cds dot-in-rods single-photon emission.
\newblock {\em Physical Review B}, 90(3):035311, 2014.

\bibitem{vezzoli2015exciton}
Stefano Vezzoli, Mathieu Manceau, Godefroy Lem{\'e}nager, Quentin Glorieux,
  Elisabeth Giacobino, Luigi Carbone, Massimo De~Vittorio, and Alberto Bramati.
\newblock Exciton fine structure of cdse/cds nanocrystals determined by
  polarization microscopy at room temperature.
\newblock {\em ACS nano}, 9(8):7992--8003, 2015.

\bibitem{pierini2020highly}
Stefano Pierini, Marianna d'Amato, Mayank Goyal, Quentin Glorieux, Elisabeth
  Giacobino, Emmanuel Lhuillier, Christophe Couteau, and Alberto Bramati.
\newblock Highly photo-stable perovskite nanocubes: towards integrated single
  photon sources based on tapered nanofibers.
\newblock {\em arXiv preprint arXiv:2005.09359}, 2020.

\bibitem{dreau2018quantum}
Ana{\"\i}s Dr{\'e}au, Anna Tchebotareva, Aboubakr El~Mahdaoui, Cristian Bonato,
  and Ronald Hanson.
\newblock Quantum frequency conversion of single photons from a
  nitrogen-vacancy center in diamond to telecommunication wavelengths.
\newblock {\em Physical Review Applied}, 9(6):064031, 2018.

\bibitem{kurtsiefer2000stable}
Christian Kurtsiefer, Sonja Mayer, Patrick Zarda, and Harald Weinfurter.
\newblock Stable solid-state source of single photons.
\newblock {\em Physical Review Letters}, 85(2):290, 2000.

\bibitem{neu2011single}
Elke Neu, David Steinmetz, Janine Riedrich-M{\"o}ller, Stefan Gsell, Martin
  Fischer, Matthias Schreck, and Christoph Becher.
\newblock Single photon emission from silicon-vacancy colour centres in
  chemical vapour deposition nano-diamonds on iridium.
\newblock {\em New Journal of Physics}, 13(2):025012, 2011.

\bibitem{prawer2008diamond}
Steven Prawer and Andrew~D Greentree.
\newblock Diamond for quantum computing.
\newblock {\em Science}, 320(5883):1601--1602, 2008.

\bibitem{aharonovich2011diamond}
Igor Aharonovich, Andrew~D Greentree, and Steven Prawer.
\newblock Diamond photonics.
\newblock {\em Nature Photonics}, 5(7):397, 2011.

\bibitem{acosta2010broadband}
VM~Acosta, E~Bauch, A~Jarmola, LJ~Zipp, MP~Ledbetter, and D~Budker.
\newblock Broadband magnetometry by infrared-absorption detection of
  nitrogen-vacancy ensembles in diamond.
\newblock {\em Applied Physics Letters}, 97(17):174104, 2010.

\bibitem{faraon2011resonant}
Andrei Faraon, Paul~E Barclay, Charles Santori, Kai-Mei~C Fu, and Raymond~G
  Beausoleil.
\newblock Resonant enhancement of the zero-phonon emission from a colour centre
  in a diamond cavity.
\newblock {\em Nature Photonics}, 5(5):301, 2011.

\bibitem{lavoie2013spectral}
Jonathan Lavoie, John~M Donohue, Logan~G Wright, Alessandro Fedrizzi, and
  Kevin~J Resch.
\newblock Spectral compression of single photons.
\newblock {\em Nature Photonics}, 7(5):363, 2013.

\bibitem{boyd2019nonlinear}
Robert~W Boyd.
\newblock {\em Nonlinear optics}.
\newblock Academic press, 2019.

\bibitem{tanzilli2001highly}
S{\'e}bastien Tanzilli, Hugues De~Riedmatten, Wolfgang Tittel, Hugo Zbinden,
  Pascal Baldi, Marc De~Micheli, Daniel~Barry Ostrowsky, and Nicolas Gisin.
\newblock Highly efficient photon-pair source using periodically poled lithium
  niobate waveguide.
\newblock {\em Electronics Letters}, 37(1):26--28, 2001.

\bibitem{tanzilli2002ppln}
S~Tanzilli, Wolfgang Tittel, Hugues De~Riedmatten, Hugo Zbinden, Paolo Baldi,
  M~DeMicheli, Da~B Ostrowsky, and Nicolas Gisin.
\newblock Ppln waveguide for quantum communication.
\newblock {\em The European Physical Journal D-Atomic, Molecular, Optical and
  Plasma Physics}, 18(2):155--160, 2002.

\bibitem{schubert2000light}
E~Fred Schubert, Thomas Gessmann, and Jong~Kyu Kim.
\newblock Light emitting diodes.
\newblock {\em Kirk-Othmer Encyclopedia of Chemical Technology}, 2000.

\bibitem{santori2010nanophotonics}
C~Santori, PE~Barclay, KM~C Fu, RG~Beausoleil, S~Spillane, and M~Fisch.
\newblock Nanophotonics for quantum optics using nitrogen-vacancy centers in
  diamond.
\newblock {\em Nanotechnology}, 21(27):274008, 2010.

\bibitem{rarity2005photonic}
JG~Rarity, J~Fulconis, J~Duligall, WJ~Wadsworth, and P~St~J Russell.
\newblock Photonic crystal fiber source of correlated photon pairs.
\newblock {\em Optics Express}, 13(2):534--544, 2005.

\bibitem{hollenbeck2002multiple}
Dawn Hollenbeck and Cyrus~D Cantrell.
\newblock Multiple-vibrational-mode model for fiber-optic raman gain spectrum
  and response function.
\newblock {\em JOSA B}, 19(12):2886--2892, 2002.

\bibitem{arosa2020refractive}
Yago Arosa and Ra{\'u}l de~la Fuente.
\newblock Refractive index spectroscopy and material dispersion in fused silica
  glass.
\newblock {\em Optics Letters}, 45(15):4268--4271, 2020.

\bibitem{hammer2020broadly}
Jonas Hammer, Maria~V Chekhova, Daniel~R H{\"a}upl, Riccardo Pennetta, and
  Nicolas~Y Joly.
\newblock Broadly tunable photon-pair generation in a suspended-core fiber.
\newblock {\em Physical Review Research}, 2(1):012079, 2020.

\bibitem{dudley2010supercontinuum}
John~M Dudley and James~Roy Taylor.
\newblock {\em Supercontinuum generation in optical fibers}.
\newblock Cambridge University Press, 2010.

\bibitem{dudley2006supercontinuum}
John~M Dudley, Go{\"e}ry Genty, and St{\'e}phane Coen.
\newblock Supercontinuum generation in photonic crystal fiber.
\newblock {\em Reviews of modern physics}, 78(4):1135, 2006.

\bibitem{manceau2018cdse}
Mathieu Manceau, Stefano Vezzoli, Quentin Glorieux, Elisabeth Giacobino, Luigi
  Carbone, Massimo De~Vittorio, J-P Hermier, and Alberto Bramati.
\newblock Cdse/cds dot-in-rods nanocrystals fast blinking dynamics.
\newblock {\em ChemPhysChem}, 19(23):3288--3295, 2018.

\end{thebibliography}
\end{document}